\documentclass[12pt,letterpaper]{article}
\usepackage[cp1251]{inputenc}
\usepackage[english,russian]{babel}
\usepackage[T2A,T1]{fontenc}
\usepackage{amssymb,amsmath,mathbbol}
\usepackage{color}
\usepackage{graphicx}
\usepackage[cmtip,arrow]{xy}
\usepackage{pb-diagram,pb-xy}
\usepackage{cmap} 
\usepackage[linktocpage=true]{hyperref}
\makeatletter
\newcounter{rusection}
\@addtoreset{section}{rusection} 
\newcounter{rufootnote}
\newcounter{rupage}
\@addtoreset{page}{rupage} 
\@addtoreset{footnote}{rufootnote} 
\newcounter{ruequation}
\@addtoreset{equation}{ruequation} 
\newcounter{rufigure}
\@addtoreset{figure}{rufigure} 
\newcounter{rutable}
\@addtoreset{table}{rutable} 
\makeatother
\def\adimH{\mathsf{K}} 
\def\alert{\emph} 
\def\aNumbers{\N} 
\def\baseformN{\mathrm{R}} 
\def\rankG{\baseformN}
\def\bmat{\begin{pmatrix}}
\def\dimirr{\mathrm{d}} 
\def\efacdetA{E} 
\def\emat{\end{pmatrix}}
\def\esp{\Pi} 
\def\coordname{\text{\textbf{\textit{site}}}}
\def\FBX{\mathsf{E}} 
\def\FBT{\mathsf{W}} 
\def\C{\mathbb{C}} 
\def\classN{\mathsf{m}} 
\def\e{\mathrm{e}}
\def\F{\mathbb{F}} 
\def\GAP{\rmfamily{GAP~}}
\def\Magma{\rmfamily{Magma~}}
\def\Hspace{\mathcal{H}} 
\def\id{\mathbf{1}} 
\def\idmat{\mathrm{I}} 
\def\ig{\gamma} 
\def\iG{\Gamma} 
\def\iGN{{\cabs{\iG}}} 
\def\iGX{\iG^{\X}} 
\def\irrepsN{N_{\mathrm{Irr}}} 

\def\ls{\sigma} 
\def\lS{\Sigma} 
\def\lSN{{\cabs{\lS}}} 
\def\lSX{\lS^{\X}} 
\def\N{\mathbb{N}} 
\def\matones{\mathrm{J}} 
\def\natmod{\mathsf{H}} 
\def\NF{\mathcal{F}} 
\def\Oppbare{{^\mathrm{c}}}
\def\Partransportt{\chi} 
\def\Partransportx{\mu} 
\def\period{\mathcal{P}} 
\def\Prob{\mathrm{\mathbf{P}}} 
\def\Q{\mathbb{Q}} 
\def\R{\mathbb{R}} 
\def\regrep{\mathrm{P}} 
\def\Repbare{\rho}
\def\repq{\mathrm{U}} 
\def\repirr{D} 

\def\runisymb{\mathsf{r}} 
\def\sg{\mathsf{f}} 
\def\sG{\mathsf{F}} 
\def\sGloc{\sG_{\mathrm{loc}}} 
\def\sGN{{\cabs{\sG}}} 
\def\Time{\mathcal{T}} 
\def\tin{0} 
\def\tfin{T} 
\def\TprimG{\mathsf{T'}} 
\def\transmatr{\mathrm{T}} 
\def\transmatrprim{\mathrm{T'}} 
\def\wbrl{\left(}
\def\wbrr{\right)}
\def\wg{\mathsf{g}} 
\def\wG{\mathsf{G}} 
\def\wGN{\mathsf{M}} 
\def\wmult{} 
\def\ws{\omega} 
\def\wS{\Omega} 
\def\wSord{\wS} 
\def\wSN{\mathsf{N}} 
\def\x{\mathsf{x}} 
\def\X{\mathrm{X}} 
\def\XN{\mathsf{n}} 
\def\XT{\mathrm{M}} 
\def\Z{\mathbb{Z}} 
\def\ZR{\mathrm{ZR}} 
\newcommand{\AltG}[1]{\mathsf{A}_{#1}} 
\newcommand{\Aut}[1]{\mathrm{Aut}\vect{#1}} 
\newcommand{\baseform}[1]{\mathcal{A}_{#1}} 
\newcommand{\baseformcoegen}[1]{\mathit{a}_{#1}} 
\newcommand{\barket}[1]{\left|#1\right\rangle} 
\newcommand{\braket}[1]{\left\langle#1\right\rangle} 
\newcommand{\bornform}[1]{\mathcal{B}_{#1}} 
\newcommand{\bornformequi}[1]{\mathcal{\widetilde{B}}_{#1}} 
\newcommand{\cabs}[1]{\left|#1\right|} 
\newcommand{\cconj}[1]{\overline{#1}} 
\newcommand{\class}[1]{K_{#1}} 
\newcommand{\comm}[2]{\left[{#1},{#2}\right]} 
\newcommand{\coord}[2]{\coordname\!\vect{{#1},{#2}}} 
\newcommand{\CyclG}[1]{\Z_{#1}} 
\newcommand{\DihG}[1]{\mathsf{D}_{#1}} 
\newcommand{\esymm}[2]{\esp_{#1}\vect{#2}}
\newcommand{\fra}[2]{\frac{\textstyle{#1}}{\textstyle{#2}}}
\newcommand{\inner}[2]{\left\langle#1\mid#2\right\rangle} 
\newcommand{\innerstandard}[2]{\left(#1\mid#2\right)} 
\newcommand{\invar}[3]{\mathrm{#1}\vect{#2,#3}} 
\newcommand{\invarL}[2]{\mathrm{L}_{#1}\vect{#2}} 
\newcommand{\invarQ}[3]{\mathrm{Q}_{#1}\vect{#2,#3}} 
\newcommand{\IrrRep}[1]{\mathbf{#1}} 
\newcommand{\Math}[1]{$#1$} 
\newcommand{\Mathh}[1]{$$#1$$} 
\newcommand{\Mone}[1]{\bmat#1\emat} 
\newcommand{\Mtwo}[4]{\bmat#1&#2\\#3&#4\emat} 
\newcommand{\Mthree}[9]{\bmat#1&#2&#3\\
 #4&#5&#6\\
 #7&#8&#9\emat} 
\newcommand{\Nei}[1]{\mathrm{N}\vect{#1}} 
\newcommand{\onetonset}[1]{\vec{#1}}
\newcommand{\Opp}[1]{{#1}\Oppbare}
\newcommand{\ordset}[1]{\left[#1\right]} 
\newcommand{\Perm}[1]{\mathrm{Sym}\left(#1\right)} 
\newcommand{\PermRep}[1]{\mathbf{\underline{#1}}} 
\newcommand{\ProbBorn}[2]{\Prob\!\vect{#1,#2}} 
\newcommand{\QuatG}[1]{\mathsf{Q}_{#1}} 
\newcommand{\Rep}[1]{\Repbare\!\left(#1\right)} 
\newcommand{\runi}[1]{\runisymb_{#1}} 
\newcommand{\SL}[1]{\mathsf{SL}\!\vect{#1}} 
\newcommand{\SU}[1]{\mathsf{SU}\vect{#1}} 
\newcommand{\set}[1]{\left\{#1\right\}} 
\newcommand{\SymG}[1]{\mathsf{S}_{#1}} 
\newcommand{\transpose}[1]{{#1}^{\mathrm{T}}}
\newcommand{\UG}[1]{\mathsf{U}\!\vect{#1}} 
\newcommand{\vect}[1]{\left(#1\right)} 
\newcommand{\Vtwo}[2]{\bmat#1\\#2\emat} 
\newcommand{\Vthree}[3]{\bmat#1\\#2\\#3\emat} 
\newcommand{\welem}[2]{\wbrl{#1},~{#2}\wbrr} 
\newcommand{\ZRcomm}[2]{\left[\baseform{#1},\baseform{#2}\right]} 
\begin{document} 
\date{}
\selectlanguage{english}
\title{
Classical and Quantum Discrete Dynamical Systems}
\author{Vladimir V. Kornyak\\Joint Institute for Nuclear Research, Dubna, Russia}
\maketitle
\def\abstractname{Abstract}
\begin{abstract}
\noindent
We study deterministic and quantum dynamics from a constructive ``finite'' point of view, 
since the introduction of the continuum or other actual infinities in physics creates 
serious conceptual and technical difficulties, without any need for these con\-cepts 
for physics as an empirical science.
Particular attention is paid to the symmetry properties of discrete systems.
For a consistent description of the symmetries of dynamical systems at different 
time instants and the symmetries of various parts of such systems, 
we introduce discrete analogs of gauge connections.
These gauge structures are particularly important to describe the quantum behavior.
The sym\-me\-ries govern the fundamental properties of the behavior of dynamical systems.
In particular, we can show that the moving soliton-like structures are inevitable in 
a deterministic (classical) dynamical system, whose symmetry group breaks the set 
of states into a finite number of orbits of the group.
We argue that the quantum behavior is a natural consequence of symmetries of
dynamical systems.
This behavior is a result of the fundamental inability to trace the identity of 
indistinguishable objects during their evolution.
Information is only available on invariant statements and values related to such objects.
Using mathematical arguments of a general nature one can show that 
any quantum dynamics can be reduced to a sequence of permutations. 
The quantum interferences occur in the invariant subspaces of permutation 
representations of the symmetry groups of dynamical systems.
The observables can be expressed in terms of permutation invariants.
We also show that in order to describe quantum phenomena it is sufficient
to use cyclotomic fields instead of nonconstructive field of complex numbers. 
The cyclotomic fields are simplest number systems suitable for quantum mechanics.
As such they are minimal extensions of natural numbers.
The finite groups of symmetries play the central role in this review.
In physics there is an additional reason for such groups to be of interest.
Numerous experiments and observations in particle physics point to the importance 
of finite groups of relatively low orders in a number of fundamental processes.
The origin of these groups has no explanation within presently recognized theories, 
such as the Standard Model.
\end{abstract}
\tableofcontents
\section{Introduction}
There are many arguments, both pragmatic and conceptual, in favor of the fact that the discrete or even finite mathematics is more suitable for describing physical reality than continuous mathematics.
This is particularly pronounced at the fundamental level. 
In particular, the study of physical processes, 
whose space-time scales are close to the Planck values, almost
inevitably leads to the need to introduce discrete structures.
As an example, one can mention the ``holographic principle'' by G. ’t Hooft \cite{tHooft} 
that emerged while studying black hole thermodynamics.
According to this principle, all physical information contained in
some region of space is completely defined by  discrete data at 
the two-dimensional boundary of this region. 
The information density of these data is, at most, one bit per the Planck area (the ``Bekenstein bound'').
From a more speculative point of view, the entire Universe is a finite informational structure on a closed 
two-dimensional lattice --- the ``cosmological horizon'' --- while observed three dimensions arise only as 
a result of the effective description at macroscopic scales and low energies.
\par
The study of discrete systems is also important from the practical point of view: many physical objects, 
for example, nanostructures, are in fact discrete rather than continuous entities, even though the space-time
scales appropriate for their description are much larger than the fundamental scales.
\par
From a conceptual point of view, the question of ``whether the real world is discrete or continuous'' or
even ``finite or infinite'' refers exclusively to metaphysics, since neither empirical observations nor logical
arguments are able to justify a definite choice --- this is a matter of faith, taste, or habit. 
Generally, in practice the idea of the continuity arises from  macroscopic experience as an approximate 
(``averaged'', ``thermodynamic'') description of large sets of discrete elements.
A typical example of a physical system that generates an empirical idea of continuity is a liquid.
Fluid dynamics is described by partial differential equations such as the Navier-Stokes equations, 
while at a more fundamental level, the liquid is a collection of molecules, and at an even more fundamental level 
it is a combination of quantum particles that obey the discrete laws.%
\footnote{
Description of a physical system is called \emph{phenomenological}, 
if it is known or assumed that it can be, at least in principle, derived
from a more fundamental description by using approximations and simplifying assumptions.
In the case of hydrodynamics, a chain of such simplifications, starting from the quantum-mechanical description, is easy to trace:
WKB approximation 
\Math{\rightarrow} Liouville equation for the distribution function 
\Math{\rightarrow} expansion of the latter over the orders of moments 
\Math{\rightarrow} ``closure hypotheses'' 
(phenomenological relations, expressing the ``higher'' moments through the ``lower'' ones and, thus, 
allowing to ``cut'' an infinite chain of equations for the moments) 
\Math{\rightarrow} hydrodynamic equations.
}
The mathematical concept of continuum is a logical refinement of empirical idea of continuity.
\par
Of course, discrete and continuous mathematical theories differ significantly 
and their utility in physics depends on the specific historical circumstances.
The continuity allows to create models of physical systems based on differential
equations,  for the study and solution of which, a large variety of methods have been developed.
An empirical equivalent of the concept of derivative is the assumption that small changes in 
physical quantities are approximately proportional to small changes in the coordinates 
in which these values are specified.
It is clear that such an hypothesis considerably simplifies physical models.
In fact, since the time of Newton and until the advent of modern computers 
analysis and differential geometry were the only means of mathematical study of physical systems.
A heuristic power of the notion of continuity was highly appreciated by Poincar\'{e}
(although he did not recognize the fundamental validity of this concept).
In his book ``The value of science''
(\cite{PoincareVal}, pp 80--81 of the English translation)
Poincar\'{e} writes:
``The sole natural object of mathematical thought is the whole 
number. It is the external world which has imposed the continuum upon us, 
which we doubtless have invented, but which it has forced us to invent. 
Without it there would be no infinitesimal analysis; all mathematical science 
would reduce itself to arithmetic or to the theory of substitutions.
\par 
On the contrary, we have devoted to the study of the continuum 
almost all our time and all our strength.~ \textellipsis
\par
Doubtless it will be said that outside of the whole number there is no rigor, 
and consequently no mathematical truth; that the whole number hides everywhere, 
and that we must strive to render  transparent the screens which cloak it, 
even if to do so we must resign  ourselves to interminable repetitions. 
Let us not be such purists and let us be grateful to the continuum, 
which, if all springs from the whole number, was alone capable of making so much proceed therefrom.''
\par
Nowadays, analogue technologies almost everywhere are replaced by digital ones.
Due to computers the real possibilities of discrete mathematics in applications
increased substantially.
Accordingly, a ``discrete'' style of thinking is becoming more popular.
An important advan\-tage of the discrete description is its conceptual ``economy'' in the Occam sense:
there is no need for ``extra entities'' based on ideas of actual infinity, such
as the ``Dedekind cuts'', ``Cauchy sequences'', etc.
Moreover, discrete mathematics is richer in content
than continuous: the continuity ``smooths
out'' fine details of structures.
A simplification of the description, achieved by introducing derivatives
into mathematical structures, may be quite expensive.
To illustrate this statement, one can compare the lists
of simple Lie and simple finite groups.
Due to differentiability and, hence, possibility to introduce Lie algebras, 
the classification of simple Lie groups is a relatively easy task, 
which was solved in the early 20th century.
The result of the classification is four infinite series and five exceptional groups.
Classification of finite simple groups is an extremely difficult problem, 
the solution of which is believed to have been completed by 2004.
The list of finite simple groups consists of 18 infinite sequences 
(16 of which contain direct analogs of all simple Lie groups) and 26 sporadic groups.
\par
The present review begins with a general discussion of discrete dynamical systems.
The most fundamental concepts are discrete time and a set of states evolving in time.
Space is viewed as an additional structure allowing to organize the set of states 
as a set of functions on the points of space with values in some set called the set of local states.
We discuss the symmetries of space and local states and how these symmetries can be combined 
into a group of symmetries of the system as a whole.%
\footnote{If the problem is approached from the opposite side, the space can be constructed 
from a given full set of states as a set of cosets of some subgroup of the full symmetry group, 
and the group of spatial symmetries, as part of the full group effectively acting on this set of cosets. 
However, in this review we will not develop this theme.}
\par
An important feature of systems with symmetries is the presence of degrees of freedom 
in their description associated with the arbitrariness in the choice of coordinate systems. 
If this choice is ``global'', i.e. established once and for all the evolution of the system, 
or (if there is a nontrivial spatial structure) for different parts of the system, it can not affect the objectively observable data.
The possibility of a ``local'' choice of coordinate systems leads to physically observable gauge effects.
Natural mathematical framework, appropriate for describing these effects, is the theory of connections in fiber bundles.
Therefore, in the present review we use a suitably modified for the discrete case notions of fiber bundles and connections in these bundles.
For illustration, we show how, starting from the constructions we have introduced,
one can build the continuum versions of the Abelian (electrodynamics) and non-Abelian (Yang–Mills) gauge theories.
\par
Using a simple model, we show how the specific space-time structures, such as ``limiting velocity'' and ``light cones'', can arise in discrete dynamics.
\par
Next, we discuss features of the behavior of  classical deterministic systems.
We show that, in the presence of symmetries in such systems, moving whith maintaining the shape (soliton-like) structures appear quite naturally.
We also discuss the idea of G. ’t Hooft on how the reversibility of fundamental processes
observed in nature may arise, and show that in the case of discrete deterministic systems, 
(a priori rare) reversibility inevitably results from the evolution as an effectively observable phenomenon.
\par
The most fundamental --- in fact decisive --- role symmetries play in the phenomena of quantum behavior. 
The distinctive feature of quantum behavior is its universality.
This behavior is demonstrated by systems of completely different physical nature and scale: 
from subatomic elementary particles to large molecules.%
\footnote{In particular, in \cite{fullerinterferEn} the observations of  quantum interference 
between the fullerene molecules \Math{C_{60}} are described, and recently
analogous results for a much larger molecule --- the phthalocyanine derivative with 
the gross formula \Math{C_{48}H_{26}F_{24}N_8O_8} --- has been reported \cite{phthalointerfer}.}
Universality of a phenomenon (i.e. independence from a physical substrate) suggests that 
the theory describing it has to be based on some a priori mathematical principles.
We show that in the case of quantum mechanics, such a leading mathematical principle is symmetry.
Our arguments are based on the fact that the quantum behavior is demonstrated only by the systems 
containing indistinguishable particles: any deviation from the exact identity of particles destroys quantum interferences.
Indistinguishability of the elements of a system means that they lie on the same orbit of the symmetry group of the system.
For systems with symmetries, only independent of ``relabeling''  ``homogeneous'' elements, 
i.e. \emph{invariant}, relations and statements are objective.
For example, no objective meaning can be attributed to electric potentials \Math{\varphi} and \Math{\psi} 
or space points denoted by the vectors \Math{\mathbf{a}} and \Math{\mathbf{a}}.
Only some of their combinations, for instance, those denoted by \Math{\psi-\varphi} or \Math{\mathbf{b}-\mathbf{a}} 
(or in a more general group-theoretic notation \Math{\varphi^{-1}\psi} and \Math{\mathbf{a^{-1}b}}), have objective sense.
\par
Unitary operators in Hilbert spaces underlie the quantum formalism.
These operators belong to the general unitary group, or more rigorously, to a unitary representation of 
the group of automorphisms (i.e. transformations that preserve the Hermitian inner product) of the Hilbert space.
To make the quantum concepts constructive, we can replace the general unitary group by unitary 
representations of \emph{finite} groups without any risk to distort the \emph{physical} content 
of the problem because, as noted above, the metaphysical distinction between 
``\emph{finite}'' and ``\emph{infinite}'' can not have any empirically observable consequences.
Moreover, there are reliable experimental evidences that some fundamental processes 
are underlain by finite groups of relatively low orders.%
\footnote{Appendix \hyperref[Enappflavor]{E} provides a brief overview of how finite symmetries appear in flavor physics phenomenology.}
\par
Using the fact that \emph{any linear representation} of a finite group is a subrepresentation of a permutation representation, we show that any quantum mechanical problem can be reduced to \emph{permutations}, and quantum observables can be expressed in terms of \emph{permutation invariants}.
A more detailed analysis of the ``permutation'' approach shows that the complex numbers in the quantum mechanical formalism must be replaced by the cyclotomics.
If this modification to the formalism is accepted, then
\begin{itemize}
	\item
quantum amplitudes acquire a simple and natural meaning:
they are projections onto invariant subspaces of the vectors of multiplicities
(``occupation numbers'') of elements of the set on which the group acts by permutations;
	\item 
the Born probabilities are rational numbers, in full accordance with the natural for finite sets ``frequency interpretation'' of probability;
	\item
quantum effects arise from the fundamental impossibility to trace identity of indistinguishable objects in the process of evolution: the only available information is about invariant combinations of such objects.
\end{itemize}
\par
Some of the topics (structural analysis of discrete relations, cellular automata, local quantum models on graphs, etc.), 
that deviate a few from the main subject of the review, as well as some technical information, are presented in Appendices.
\par
To ensure the integrity of presentation and unity of notation, 
let us describe from the very beginning the most fundamental concepts used in this review.
\paragraph{Natural numbers.}
The base number system to start with is the \emph{semiring of natural numbers}
\begin{equation*}
	\N=\set{0,1,2,\ldots}.
\end{equation*}
We emphasize that we include zero in the set of natural numbers.
Natural numbers are most natural in discrete mathematics, where they act as the counters of elements of discrete systems and combinations of those elements.
The other constructive numbers can be derived from the set \Math{\N} by purely mathematical means.
For example, we consider the \emph{ring of integer numbers} 
\Math{\Z = \set{\ldots,-1,0,1,\ldots}} as an \emph{algebraic extension}
of the semiring \Math{\N} generated by \emph{primitive square root of unity};%
\footnote{
This definition, as will be seen later, is motivated by quantum mechanics.
Another common way of introducing integers is interpreting them as equivalence classes 
of pairs of natural numbers: the equivalence of pairs \Math{\vect{n,m}\mathbf{\sim}\vect{n',m'}} 
is defined via  \Math{n+m'=n'+m}, where \Math{n,m,n',m'\in\N}. 
}
the field of rational numbers \Math{\Q}, in turn, is interpreted as the field of fractions of  \Math{\Z}, etc.
\paragraph{Fiber bundles.} 
For systems with symmetries, especially when researching their quantum behavior, 
the notion of a fiber bundle plays an important role.
By definition,%
\footnote{
In the literature, the theory of fiber bundles is usually presented
as an integral part of differential geometry \cite{Kobayashi,Sulanke}.
However, the elements of this theory, that are essential for us, do not require the
notion of differentiability and are easily transferred to the discrete and finite cases.
} 
the \emph{fiber bundle} is a structure of the form
\begin{equation*}
	\vect{E,X,S,G,\pi},
\end{equation*}
where
\begin{description}
	\item
	\Math{E} --- a set called a \emph{(total) space} of the fiber bundle; 
	\item 
\Math{X} --- a set called a \emph{base} of the bundle;	 
	\item 
\Math{S} --- a set called a \emph{typical (model) fiber} of the bundle;	 
	\item 
\Math{G} --- a group of transformations of a typical fiber \Math{S}
called a \emph{structure group} of the bundle;	 
	\item
\Math{\pi} --- a map \Math{E\rightarrow{}X} called a \emph{projection} of the bundle.
\end{description}
\par
The projection \Math{\pi:E\rightarrow{}X} is characterized by the following property.
If \Math{x} is a point of the base, i.e., \Math{x\in{}X},
then, the set \Math{S_x=\pi^{-1}\vect{x}} called a \emph{fiber over}  \Math{x} is isomorphic to the typical fiber \Math{S}.
Isomorphism here means that there is an element \Math{g_x} of the structure group  \Math{G}, such that \Math{S_xg_x^{-1}=S}.
\par
The role of the structure group in physics is that it allows one to ``compare'' the data relating to  different points in time and space by using gauge connections.%
\footnote{
We will consider the notion of connection, which substantially depends on the structure of the base, after more specific discussion of discrete versions of the base.
}
\par
A map \Math{\sigma: X\rightarrow{}E} such that \Math{\pi\circ{}\sigma=\mathrm{id}_X} is called a \emph{section of the bundle} \Math{\vect{E,X,S,G,\pi}}.
The set of all sections can be symbolically represented as \Math{\prod\limits_{x\in{}X}S_x}.
Thus, for finite \Math{X} and \Math{S} the total number of all possible sections is \Math{\cabs{S}^{\cabs{X}}}.
\par
The bundle is called \emph{principal}, if a typical fiber is the
structure group itself.
Symbolically, the principal bundle can be written as
\begin{equation*}
	\vect{P,X,G,\rho},
\end{equation*}
where \Math{P} is a total space, and \Math{\rho} is a projection.
It is assumed that the structure group \Math{G} acts on the fiber \Math{G}
by the right (or left) translations.
\section{Classical and Quantum Dynamics}
\subsection{Basic Concepts and Structures}
The most fundamental concept of dynamics is \emph{time}.
In the discrete version the time \Math{\Time} is a sequence of integers labeling a sequence of events, for example,
\begin{equation}
	\Time=\Z \text{~~or~~} \Time=\ordset{\tin,1,\ldots,\tfin},
	\label{Entime} 
\end{equation}
where \Math{\tfin} is some natural number.
\par
We assume that the \emph{states} of a dynamical system form a finite set of some elements 
\begin{equation}
	\wS=\set{\ws_1,\ldots,\ws_{\wSN}}.
	\label{Enws} 
\end{equation}
We also suppose the presence of a nontrivial \emph{symmetry group} acting on the set of states \Math{\wS}
\begin{equation}
	\wG=\set{\wg_1,\ldots,\wg_{\wGN}}\leq\Perm{\wS}.
	\label{EnwG}
\end{equation}
Without loss of generality, we can assume that the group \Math{\wG} acts on \Math{\wS} \emph{faithfully} 
(otherwise, \Math{\wG} can be replaced by the quotient group by the kernel of the action) and 
\emph{transitively} (otherwise, the orbits of the action can be considered individually%
\footnote{
In some cases, as for example, when describing the discussed below ``solitons'' in discrete systems, we have to consider sets of orbits, i.e., nontransitive actions. 
}).
\par
\emph{Evolution} or \emph{trajectory} of the dynamical system is a sequence of states, indexed by time
\begin{equation}
	h=\cdots\rightarrow{}w_{t-1}\rightarrow{}w_{t}\rightarrow{}w_{t+1}\rightarrow\cdots,
	 ~\text{where}~w_i\in\wS.
	\label{Enevolution}
\end{equation}
\par
Here we are faced with the problem of identifying the ``indistinguishable'' objects 
(more formally, the elements of \eqref{Enws} lying on the same orbit of group \eqref{EnwG}) at different times.%
\footnote{
In the Appendix ``Art Combinatoria'' to the book \cite{Weyl} H. Weyl 
(author of the idea of gauge invariance in physics) discusses in detail the problems of individualization 
of indistinguishable objects and their identification in the process of evolution in time.
}
The fact is that for registration and identification of individually indistinguishable objects some individually identifiable marks (letters, numbers) are used.
It is clear that there is no absolute way to associate such distinguishable labels with indistinguishable elements.
To fix such elements is possible only relative to some complementary system \cite{Shafarevich}, which manifests itself in various contexts as a \emph{``reference frame''} or \emph{``observer''} or \emph{``measuring device''}.
\par  
Let us consider more carefully the notions of a reference frame and the change of coordinates in the finite case.
For a start we fix some (arbitrary) \emph{order} on the initially unordered set \Math{\wS}, i.e. we assume that \Math{\wS} is a vector (in this context, the word ``vector'' means simply an ordered set, also called ``tuple'') of the form 
\begin{equation*}
	\wSord=\vect{\ws_1,\ldots,\ws_{\wSN}}.
\end{equation*}
We will call \Math{\wSord} the \emph{canonical set of states}.
The action of the symmetry group \Math{\wG} on the set \Math{\wSord} has the form%
\footnote{In this review we will mainly write actions of groups on the right.
This convention is more natural and intuitive.
It is widely used in mathematics and, in particular, in basic systems of computer algebra oriented to the problems of group theory, \GAP and \Magma.
}
\Mathh{\wSord{}g=\vect{\ws_1g,\ldots,\ws_kg,\ldots,\ws_{\wSN}g}
=\vect{\ws_{i_1},\ldots,\ws_{i_k},\ldots,\ws_{i_\wSN}}.}
It is convenient to label the elements \Math{\ws_k\in\wSord} by the natural numbers \Math{k}.
Let us introduce the function \Math{\coord{v}{V}}, that for a given element \Math{v},
which is included into an ordered set \Math{V} ones and only once, outputs its position in the set; for example, \Math{\coord{\ws_k}{\wSord} = k} and \Math{\coord{\ws_{i_k}}{\wSord{}g} = k}.
\par 
Position of an element in an ordered set will be called a ``coordinate''.
Applying the function \Math{\coordname} to the elements of the set \Math{\wSord}, we obtain the vector  
 \begin{equation}
	R=\vect{1,2,\ldots,\wSN},
	\label{Enmodrefframe}
\end{equation}
which we call the ``canonical reference frame''.
That is, we mark in a certain fixed way the states from the set \Math{\wS} by natural numbers. 
This will allow us to control the individuality of states in the evolution of a dynamical system.
The action of the element of the group \Math{g\in\wG} on the states in terms of coordinates can be written as a permutation
\Mathh{\Vtwo{1~\cdots{}k\cdots~\wSN}{1g\cdots{}kg\cdots\wSN{}g}.}
This ``contravariant'' action transforms one state to another: \Math{\ws_kg\rightarrow\ws_j}.
Corresponding ``covariant'' transformation of the reference frame has the form
\begin{equation}
	S=Rg^{-1}=\vect{1g^{-1},2g^{-1},\ldots,\wSN{}g^{-1}}.
	\label{Enframechange}
\end{equation}
This means that if some element \Math{x} of the set of states in the canonical reference frame \Math{R} has a coordinate \Math{x_{R}=\coord{x}{R}}, then its coordinate in the frame \Math{S} will be \Math{x_{S}=\coord{x}{S}}.
Operationally ``getting coordinates'' means that we are able to recognize \Math{x} among the elements of \Math{S} and count the number of steps passed to that element. 
This procedure is similar to the ``coordinatization'' \cite{Shafarevich} of points of a vector space.
For example, the coordinate of the point of real line (one-dimensional space \Math{\R}) is the ``distance'', which is necessary to pass from the origin to that point.
The origin and the ``scale of step length'' (metric) depend on the choice of coordinate system.
\par
It is clear that according to \eqref{Enframechange} it is possible to construct all other frames starting from the canonical reference frame \eqref{Enmodrefframe} and that the transitions between arbitrary reference frames are described by 
\Math{S'=Sh^{-1}},  \Math{h\in\wG}. 
\par
Let us return to evolution \eqref{Enevolution}.
We do not know \emph{a priori} which one of the objects \Math{a\in\wS} or \Math{b\in\wS}
at the instant \Math{t-1} is ``predecessor'' of some object \Math{c} at the instant \Math{t}, 
if \Math{a} and \Math{b} lie in the same group orbit, i.e., \Math{b=ag} for some \Math{g\in\wG}.
Such identification is realized by introducing an additional structure called the \emph{gauge connection} or \emph{parallel transport}.
\par
Consider the bundle whose base is time \eqref{Entime}, typical fiber is canonical set of states \eqref{Enws}, and structure group is group \eqref{EnwG} 
\begin{equation}
	\vect{\FBT,\Time,\wSord,\wG,\tau}.
	\label{Enbundleovert}
\end{equation}
Here \Math{\FBT} and \Math{\tau} are total space and projection, respectively.
Connection defines isomorphism between fibers of this bundle over different moments of time:
 \Mathh{\wSord_{t_2}=\wSord_{t_1}\Partransportt\vect{t_1,t_2},~\text{where}~
  \Partransportt\vect{t_1,t_2}\in\wG.}
An important principle in physics is the \emph{gauge invariance} --- the ability to set the 
reference frames in the fibers over different points of the base independently of each other.
In the present case this principle leads to the following transformation rule for the connections
\begin{equation}
		\Partransportt\vect{t_1,t_2}
		\rightarrow{}g\vect{t_1}^{-1}\Partransportt\vect{t_1,t_2}g\vect{t_2},
		~~g\vect{t_1},g\vect{t_2}\in\wG.\label{Enconnruleovert}
\end{equation}
Here \Math{g\vect{t_1}} and \Math{g\vect{t_2}} are \emph{arbitrary} elements of the group 
\Math{\wG}, i.e., the values of an arbitrary section of the principal \Math{\wG}-bundle over \Math{\Time}.
Obviously, the connection between arbitrary points in time can be written as the product of connections corresponding to elementary steps in time:
\Mathh{\Partransportt\vect{t_1,t_2} = \Partransportt\vect{t_1,t_1+1}\cdots\Partransportt\vect{t,t+1}
\cdots\Partransportt\vect{t_2-1,t_2}.}
Thus, in the case at hand the connection can be determined by the function on pairs of successive moments of time, i.e. the function on the edges of a linear graph whose vertices are the points of time.
\par
Trajectories \eqref{Enevolution} of a dynamical system can be treated as the sections of bundle \eqref{Enbundleovert}.
Belonging of the trajectory \Math{h} to the set of sections can be written symbolically as
\begin{equation}
	h\in\prod\limits_{t\in\Time}\wSord_t.
	\label{Enevolutiongeneral}
\end{equation}
The classical approach is reduced to the study of individual trajectories under the assum\-ption of fixed for all times rules of identification of the states \Math{\ws_i}.
In this case, one can replace the bundle \eqref{Enbundleovert} by the Cartesian product\Math{\Time\times\wS} and interpret the trajectory as a function of time with values in the set of states: 
\Math{h\in\wS^\Time.}
Accounting local gauge invariance acquires physical significance, when the sets of different individual trajectories connecting the points of bundle are considered.
This is manifested particularly clearly in the Feynman path integral formulation of quantum mechanics.
\subsection{Dynamical Systems with Space}
In physics, the set of states \Math{\wS} usually has a special structure of the set of functions
\begin{equation}
\wS=\lSX,
\label{Enstateswithspace}
\end{equation}
where the sets of arguments
\begin{equation}
	\X=\set{\x_1,\ldots, \x_\XN}
	\label{Enspace}
\end{equation}
and values
\begin{equation}
	\lS=\set{\ls_1,\ldots,\ls_\lSN}
	\label{Enlocalstate}
\end{equation}
are called \emph{space} and \emph{local states}, respectively.
\par 
An important feature of dynamic systems with space is the possibility of introducing nontrivial \emph{gauge connections} between points of the space.
The gauge structures lead to observable physical effects:
the curvatures of nontrivial connections describe the forces in physical theories.
Another topic, in which the spatial structure is important, is the relation between
spin and statistics.
\par
We assume that both space \eqref{Enspace} and local states \eqref{Enlocalstate} have nontrivial symmetry groups:
\begin{align}
	\sG&=\set{\sg_1,\ldots,\sg_\sGN}\leq\Perm{\X}~~~\,\text{--- \emph{space symmetry},}
	\label{Enspacegroup}\\
	\iG&=\set{\ig_1,\ldots, \ig_\iGN}\leq\Perm{\lS}~~\text{--- \emph{internal symmetry}.}\nonumber
\end{align}
\par
To interpret the internal symmetry group \Math{\iG} as a local gauge group, 
let us introduce a bundle with the total space \Math{\FBX}, base \Math{\X},
typical fiber \Math{\lS}, and  projection \Math{\pi}:
\begin{equation}
	\vect{\FBX,\X,\lS,\iG,\pi}.
	\label{Enbundleoverx}
\end{equation}
Now, the set of states, instead of \eqref{Enstateswithspace}, takes the form of the set of sections of this bundle:
\begin{equation}
\wS=\prod\limits_{x\in\X}\lS_x.
\label{Enstateswithspacegauged}
\end{equation}
\par
Non-triviality of the space symmetry group \eqref{Enspacegroup}  implies the existence of some structure in the set \eqref{Enspace}.
In physics, the topological concepts are essential. They allow to classify space points according to their ``proximity'' to each other.  
\par
In the approach, known as \emph{general topology}, the concept of ``proximity'' of points is formalized by a family of open subsets satisfying certain axioms. 
These axioms are abstrac\-tions of the usual properties of continuum and not very suitable (though formally applicable) for discrete spaces.
\par
The most adequate in the discrete case topological structure is an \emph{abstract simplicial complex} \cite{Spanier}, which is defined by some collection \Math{K} of subsets of \Math{\X}.
In this context, the elements of \Math{\X} are called \emph{points} or \emph{vertices}.
The elements of \Math{K},  called \emph{simplices},%
\footnote{The intuitive idea of ``proximity'' of points corresponds to that the point 
from one simplex are ``closer to each other'' than those from different simplices.}
have the property that any their subsets are also simplices, i.e. belong to the set \Math{K}.
Also, any one-element subset of \Math{\X} must be a simplex.
Subsimplices of simplices are often called \emph{faces}.
Obviously, the structure of the complex \Math{K} is uniquely determined by the \emph{maximum by inclusion} simplices.
\emph{Dimension of the simplex} \Math{\delta\in{}K} is defined by \Math{\dim\delta=\cabs{\delta}-1}.
This definition is motivated by the fact that \Math{k+1} points, immersed in a general position into the Euclidean space \Math{\R^{n}}, form the vertices of a simplest convex \Math{k}-dimensional polyhedron, if \Math{n\geq{}k}.
\emph{Dimension of the complex} \Math{K} is defined as the maximum dimension of its simplices: \Math{\dim{}K=\max\limits_{\delta\in{}K}\dim\delta}.
One-dimensional simplicial complex is called a \emph{graph}, and its maximum simplices
(two-element subsets) are called \emph{edges}.
From the point of view of the abstract combinatorial topology the possibility of embedding into Euclidean spaces has no special significance.
It is important only how the simplices, that compose a complex, join (intersect) each other.
Note, however, that in the topological dimension theory there is the \emph{N\"{o}beling-Pontryagin theorem} from which it follows that any \Math{k}-dimensional simplicial complex can be embedded into the space \Math{\R^{2k+1}}.
In particular, any graph can be embedded into the three-dimensional space \Math{\R^{3}}.
\par
If we are not interested in the (irrelevant in the discrete context) issues of geometrical realization of complexes in Euclidean spaces, then any problems with data on arbitrary complexes can be reformulated on the graphs by using the notion of \emph{skeleton of a simplicial complex}. 
By definition, \Math{q}\emph{-dimensional skeleton} (\Math{q}\emph{-skeleton}) of a simplicial complex \Math{K} is a simplicial complex consisting of all \Math{p}-dimensional simplices of \Math{K}, where \Math{p\leq{}q}.
Any simplex of the complex \Math{K} corresponds to some complete subgraph of the 1-skeleton of \Math{K}.
Recall that a graph is called \emph{complete} if all pairs of its vertices are edges.
Thus, instead of an arbitrary complex \Math{K} we can consider its 1-skeleton, i.e., a graph.
\par
Graphs --- we will call them also \emph{lattices} --- are sufficient for all problems considered in this review.
In particular, they are sufficient for introducing gauge and quantum structures.
In order not to overload the text, we will use the same symbol \Math{\X} to denote the space, assuming that in addition to the vertices \Math{\set{\x_1,\ldots, \x_\XN}}
the edges (i.e., some set of two-element subsets of the form \Math{e_{ij}=\set{\x_i,\x_j}}) are also given.
The space symmetry group in this case is the group of automorphisms of the graph,
i.e., \Math{\sG=\mathrm{Aut}\vect{\X}}.
In principle, the automorphism group of a graph with \Math{\XN} vertices can have up to \Math{\XN!} elements.
However, there is a very efficient algorithm, developed and implemented by B. McKay
\cite{BMcKay}.
This algorithm creates a concise set of elements that generate the group \Math{\sG}.
The number of generators in real problems is usually 2 or 3, and can not exceed n - 1 by the construction of the algorithm.
\par
\begin{figure}[!h]
\centering
\begin{minipage}[t]{0.3\textwidth}
\begin{center}
{\small{}
Tetrahedrane \Math{C_4H_4}\\
\Math{\sG=\SymG{4}}\\
\Math{\sGloc=\SymG{3}\cong\DihG{6}}\\[10pt]
}
\includegraphics[width=0.5\textwidth]{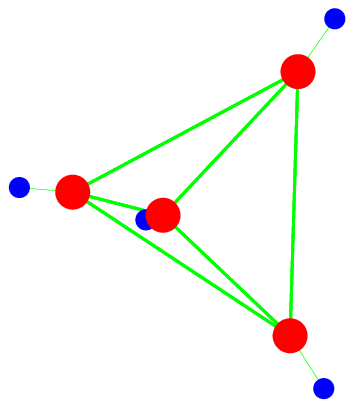}
\end{center}
\end{minipage}
~
\begin{minipage}[t]{0.3\textwidth}
\begin{center}
{\small{}
Cubane \Math{C_8H_8}\\
\Math{\sG=\SymG{4}\times\CyclG{2}}\\
\Math{\sGloc=\DihG{6}}\\[10pt]
}
\includegraphics[width=0.7\textwidth]{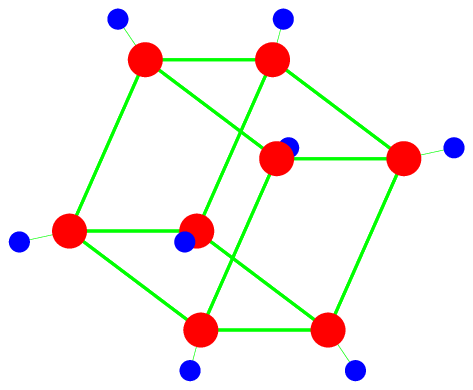}
\end{center}
\end{minipage}
~
\begin{minipage}[t]{0.3\textwidth}
\begin{center}
{\small{}
Dodecahedrane \Math{C_{20}H_{20}}\\
\Math{\sG=\AltG{5}\times\CyclG{2}}\\
\Math{\sGloc=\DihG{6}}\\[5pt]
}
\includegraphics[width=0.8\textwidth]{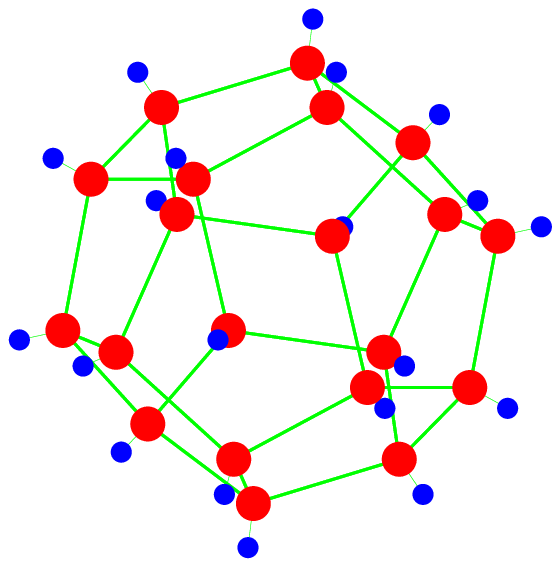}
\end{center}
\end{minipage}
\par
\vspace*{10pt}
\hspace*{10pt}
\begin{minipage}[t]{0.35\textwidth}
\begin{center}
{\small{}
Fullerene \Math{C_{60}}\\
\Math{\sG=\AltG{5}\times\CyclG{2}}\\
\Math{\sGloc=\CyclG{2}}\\[15pt]
}
\includegraphics[width=0.9\textwidth]{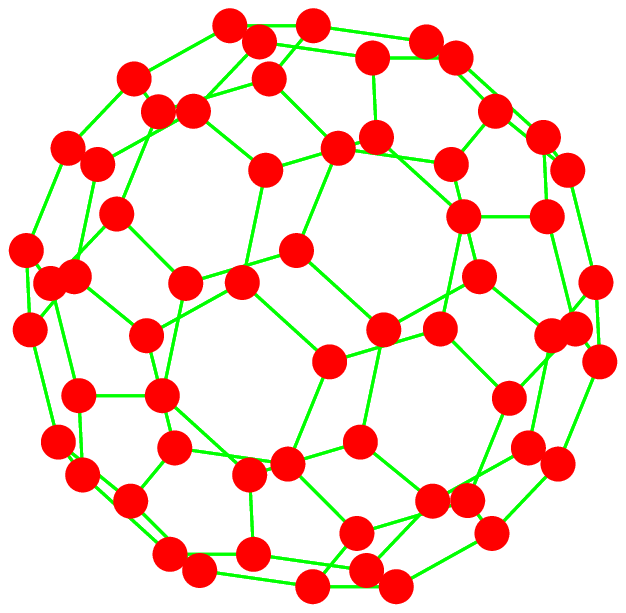}
\end{center}
\end{minipage}
\hspace{10pt}
\begin{minipage}[t]{0.5\textwidth}
\begin{center}
{\small{}
Toroidal graphene \Math{n\times{}m}\\
\Math{\sG=\DihG{n}\times\DihG{2m}}
\hspace{5pt}
\Math{\sG=\vect{\CyclG{}\times\CyclG{}}\rtimes\DihG{6}}\\
\Math{\sGloc=\CyclG{2}}
\hspace{35pt}
\Math{\sGloc=\DihG{6}}\hspace{10pt}~\\
\hspace{70pt}
\Math{n,m\rightarrow\infty}\\[3pt]
}
\includegraphics[width=0.55\textwidth]{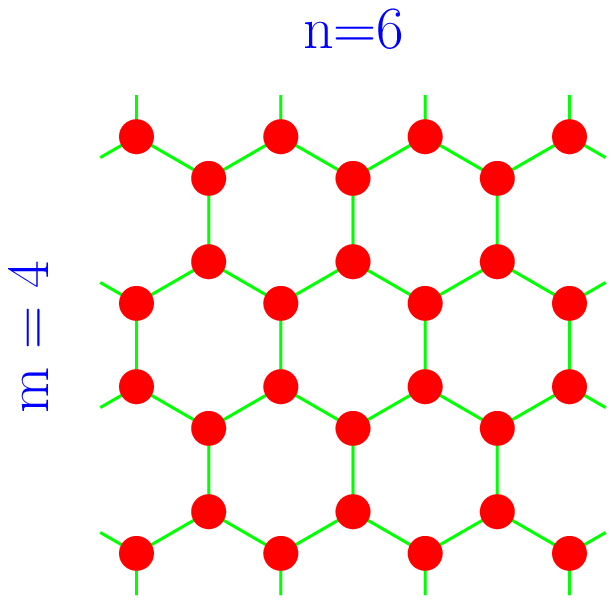}
\end{center}
\end{minipage}
\caption{Examples of (hydro)carbon nanostructures. \Math{\sG} is symmetry group of the graph of structure.
\Math{\sGloc} is symmetry group of neighborhoods of vertices.
Symbols \Math{\CyclG{k},\DihG{2k},\AltG{k}} denote the cyclic, dihedral and alternating groups, respectively.}
	\label{EnNanoCarbons}
\end{figure}

Let us give examples of lattices of sufficiently high symmetry, corresponding to real physical objects \cite{Kornyak08a}.
Figure \ref{EnNanoCarbons} shows the so-called \emph{Platonic hydrocarbons}:
\emph{tetrahedrane}, \emph{cubane}, and \emph{dodecahedrane}, as well as
\emph{fullerene} and \emph{graphene} --- the carbon structures, promising in nanotechnology.
\par
Note that cubane and dodecahedrane were artificially synthesized.
They are sufficiently stable though not found in nature. 
A tetrahedrane is unstable, but its stable analogs can be synthesized by attaching to the
tetrahedral carbon skeleton the tert-butyl \Math{\vect{CH_3}_3C} or
trimethylsilyl \Math{\vect{CH_3}_3Si} radicals instead of hydrogen atoms. 
The figure also shows complete symmetry groups of the graphs of corresponding objects \Math{\sG}, and their subgroups \Math{\sGloc}, that fix individual vertices.
Because of low complexity, the lattices presented in Fig. \ref{EnNanoCarbons} are
interesting from a methodological point of view for constructing simple models of the discrete dynamical systems.
Here, a respective graph is treated as a discrete space \Math{\X}; the symmetry group \Math{\sG} of the graph, as the full group of symmetries of space; and subgroup
\Math{\sGloc}, as a group of \emph{local symmetries of space}.
The group \Math{\sGloc} describes the symmetries of the neighborhood of a vertex of the graph.
It is an analog of infinitesimal symmetries in physical models based on differential equations.
If the laws of dynamics of discrete models are determined by local relations, i.e.,
relations defined on the neighborhoods of graphs (as, for example, in cellular automata or described in Appendix \hyperref[Enquantumonregulargraphs]{C} quantum models on regular graphs),
then, the natural requirement should be symmetry of these relations with respect to the group \Math{\sGloc}.
\par
Note that the graphs of the structures in Fig. \ref{EnNanoCarbons} are \emph{three-regular} (\emph{trivalent}).
The regularity of a graph is in some aspects similar to the isotropy and homogeneity
of the continuous space.
Reasoning speculatively, one would assume that the space at the Planck scale can be modeled by some \Math{k}-regular graph. 
However, macroscopically observable homo\-geneity and isotropy of the physical space may result from averaging over large ensembles of graphs of various structures.
Nevertheless, the assumption of  regularity considerably simplifies discrete physical models.
\par
Taking advantage of the fact that the space \Math{\X} has the structure of a graph, we can introduce connection by defining a \Math{\iG}-valued function on the edges \Math{\Partransportx\vect{\set{\x_i,\x_j}}\equiv\Partransportx\vect{\x_i,\x_j}}.
The connection has an obvious property 
\Math{\Partransportx\vect{\x_i,\x_j}=\Partransportx\vect{\x_j,\x_i}^{-1}}. 
\par
A connection \Math{\tilde{\Partransportx}\vect{\x_i,\x_j}} is called \emph{trivial}, if it can be expressed in terms of a function on the points (vertices
of a graph):
\begin{equation}
	\tilde{\Partransportx}\vect{\x_i,\x_j}=\alpha\vect{\x_i}\alpha\vect{\x_j}^{-1},~
	\alpha\vect{\x_i},\alpha\vect{\x_j}\in\iG.\label{Enconntriv}
\end{equation}
\par
The local gauge invariance implies the \emph{transformation rule for connections} \eqref{Enconnruleovert}
\begin{equation}
		\Partransportx\vect{\x_i,\x_j}\rightarrow
		\gamma\vect{\x_i}^{-1}\Partransportx\vect{\x_i,\x_j}\gamma\vect{\x_j},~~
		\gamma\vect{\x_i},\gamma\vect{\x_j}\in\iG,\label{Enconnruleoverx}
\end{equation}
which is similar to \eqref{Enconnruleovert}.
Here \Math{\gamma\vect{\x_i}} and \Math{\gamma\vect{\x_j}} are arbitrary elements of the 
group of internal symmetries
(we can consider that they fix local coordinates in the fibers \Math{\lS_{\x_i}} and \Math{\lS_{\x_j}}).
\par
Assuming \emph{connectivity} of the graph \Math{\X} (i.e., the presence of a path between two arbitrary vertices \Math{x_{i_1}} and \Math{x_{i_k}}) we can define the \emph{parallel transport}
\Mathh{\Partransportx\vect{x_{i_1},x_{i_2},\ldots,x_{i_k}}=
\Partransportx\vect{x_{i_1},x_{i_2}}\Partransportx\vect{x_{i_2},x_{i_3}}
\cdots \Partransportx\vect{x_{i_{k-1}},x_{i_k}},}
where each pair \Math{\set{\x_{i_{m}},\x_{i_{m+1}}}} is an edge.
The parallel transport depends on the path connecting \Math{x_{i_1}} and \Math{x_{i_k}}.  
\par
The parallel transport around a cycle of a graph, i.e., along the closed path
\Mathh{\Partransportx\vect{x_{i_1},x_{i_2},\ldots,x_{i_k},x_{i_1}}
=\Partransportx\vect{x_{i_1},x_{i_2}}\Partransportx\vect{x_{i_2},x_{i_3}}
\cdots \Partransportx\vect{x_{i_k},x_{i_1}}}
is called \emph{holonomy}.
In continuous space, an infinitesimal analog of holonomy is the \emph{curvature of
connection}, which describes the force fields in physical theories.
It is clear that any holonomy of the trivial
connection \eqref{Enconntriv} is trivial, i.e., is equal to identity of the
group:
\Math{\tilde{\Partransportx}\vect{x_{i_1},x_{i_2},\ldots,x_{i_k},x_{i_1}} = \id}.
This means, in particular, that trivial connections do not lead to any
physically observable effects.
\par
Note that in our considerations the time is ``more fundamental'' than the space.
That is, the space is not included in a single structure on an equal footing with the time, as is customary in relativistically invariant theories.
The introduction of local times in discrete models can be rather problematic and artificial due to the lack of continuous Lorentz symmetry.
However, later we will show that such concepts as the \emph{``limiting velocity''} and \emph{``light cone''} appear in the discrete dynamics quite naturally, and the Lorentz
symmetry may arise on a macroscopic scale as a continuum approximations by averaging over large clusters of discrete elements.
Besides, the global, i.e., without local reparametrizations,
time \Math{\Time}  (the \emph{``age of the universe as a whole''}) ensures a unified description of the dynamics independent of whether the states of a system are represented in  general form \eqref{Enws} or have special structure \eqref{Enstateswithspacegauged}, implying the existence of space.
\par
Description of the evolution of dynamical systems with a space should take into account the evolution of the space itself.
Here, due to the presence of spatial symmetry \Math{\sG}, the natural problem of identifying points of space at different moments of time arises as well.
To handle this problem, we present the space-time \Math{\XT} as a bundle \Math{\sigma\!: \XT\rightarrow\Time} with base \Math{\Time}, typical fiber \Math{\X}, and structure group \Math{\sG}: 
\begin{equation}
	\vect{\XT,\Time,\X,\sG,\sigma}.
	\label{Enxovert}
\end{equation}
It is clear that if the space \Math{\X} has the structure of a graph, and the time \Math{\Time} is discrete, then the space-time \Math{\XT} we have
just constructed is also a graph and we can use the constructions of connection and parallel transport similar to those given above.
\par
To include evolution of systems with space in the general scheme, which is appropriate for bundle \eqref{Enbundleovert}, the structure group \Math{\wG} of this bundle should be presented as a combination of spatial and internal symmetries: the space group \Math{\sG} permutes the points of bundle \eqref{Enxovert}, while the group of internal symmetries transforms the fibers in \eqref{Enbundleoverx}.
A more detailed analysis leads to the construction called the \emph{wreath product}  
\cite{Cameron,Dixon,Hall,Rotman}\label{Enwreathpage} 
\begin{equation}
	\wG=\iG\wr_\X\sG\cong\iGX\rtimes\sG\enspace.
	\label{Enwreath1}
\end{equation}
The last isomorphism means that the wreath product has the structure of a \emph{semidirect} product, i.e., the group \Math{\iGX} of \Math{\iG}-valued functions on the space \Math{\X} is a normal subgroup of the group \Math{\wG}.
\par
Formula \eqref{Enwreath1} shows that the group \Math{\wG} is the special
case of a split extension of the group \Math{\sG} by the group \Math{\iGX}.
Let us give explicitly  operations for the group \Math{\wG}, having  structure \eqref{Enwreath1}, in terms of operations for its constituents \Math{\sG} and \Math{\iG}.
The action of \Math{\wG} on the set of states \Math{\wS}, having the form \eqref{Enstateswithspacegauged}, is given by 
\begin{equation}
	\sigma\vect{x}\vect{\alpha\vect{x},a}=\sigma\vect{xa^{-1}}\alpha\vect{xa^{-1}},
	\label{Enwreathaction}
\end{equation}
where the pair \Math{\vect{\alpha\vect{x},a}} denotes an element of the group \Math{\wG,~x\in\X,~\sigma\vect{x}\in\wS,~a\in\sG}.
The product of elements \Math{\vect{\alpha\vect{x},a}\in\wG} and \Math{\vect{\beta\vect{x},b}\in\wG} has the form
\begin{equation}
	\vect{\alpha\vect{x},a}\vect{\beta\vect{x},b}=\vect{\alpha\vect{x}\beta\vect{xa},ab}.
\end{equation} 
For completeness, we give also the inverse element formula:
\begin{equation}
	\vect{\alpha\vect{x},a}^{-1}=\vect{\alpha\vect{xa^{-1}}^{-1},a^{-1}}.
	\label{Enwreathinverse}
\end{equation}
\par
Note that the group extensions form equivalence classes.
In particular, formulas (\ref{Enwreathaction}--\ref{Enwreathinverse}) can
be written in different equivalent ways.
A general discussion of the strucure of these equivalences, as well as the ways of unification of internal and spatial symmetries, is given in Appendix \hyperref[Ensymmetryunification]{A}.
\par
It is instructive to see how the gauge structures are implemented in continuous physical theories.
We will consider two examples, in which the constructions introduced above are used without assuming that they are discrete.
In both examples the set \Math{\XT} is the four-dimensional Minkowski space with points \Math{x=\vect{x^\nu}}, and the sets of local states \Math{\lS} are complex linear
spaces of various dimensions.
The sections \Math{\psi(x)} of these bundles are called fields of particles.  
\subsubsection{Electrodynamics. Abelian prototype of all gauge theories.} 
In this case the group of internal symmetries (the symmetries of Lagrangians and physical
observables) is Abelian unitary group \Math{\iG=\UG{1}}.
The sections of the principal \Math{\UG{1}}-bundle over the space-time may be represented as \Math{e^{-i\alpha(x)}.}
\par
Consider a parallel transport for two closely situated space-time points: \Mathh{\Partransportx\vect{x, x+\Delta{}x} = e^{-i\rho\vect{x, x+\Delta x}}.}
Fundamental transformation rule \eqref{Enconnruleoverx} for connection takes the form
\Mathh{\Partransportx'\vect{x, x+\Delta{}x} 
= e^{i\alpha\vect{x}}\Partransportx\vect{x, x+\Delta{}x}e^{-i\alpha\vect{x+\Delta{}x}}.}
Further, in accordance with the conjecture on which empirical applications of the differen\-tial calculus are based, we replace these expressions by the linear combinations of differences of coordinates. This leads to the approximations:  
\begin{align*}
	\Partransportx(x,x+\Delta{}x)=e^{-i\rho(x,x+\Delta{}x)}&\approx\id-i{A}(x)\Delta{}x,
	\\
	\Partransportx'(x,x+\Delta{}x)=e^{-i\rho(x,x+\Delta{}x)}&\approx\id-i{A'}(x)\Delta{}x,
	\\
	e^{-i\alpha(x+\Delta{}x)}&\approx{}e^{-i\alpha(x)}\vect{\id-i\nabla\alpha(x)\Delta{}x}.
\end{align*}
The coefficients \Math{{A}(x)} and \Math{{A'}(x)} introduced in these
approximations are called the \emph{connec\-tion 1-forms}.
Using commutativity of the group \Math{\UG{1}} we obtain the transformation rule for the Abelian connection:
\begin{equation*}
{A'}(x)={A}(x)+\nabla\alpha(x).
\end{equation*}
In the components, this formula has the form
\begin{equation*}
{A'_\nu}(x)={A_\nu}(x)
+\frac{\textstyle{\partial\alpha(x)}}{\textstyle{\partial x^\nu}}.	
\end{equation*}
The 1-form \Math{A(x)} with values in the Lie algebra of the group \Math{\UG{1}} is identified with the vector potential of the electromagnetic field, and its differential \Math{F=\vect{F_{\nu\eta}}=\mathrm{d}A} corresponds to the electric and magnetic force fields.
Recall that in the tensor notation  the differential \Math{F} of the form \Math{A} in local coordinates is written as
\Mathh{F_{\nu\eta} = \frac{\textstyle{\partial{}A_\eta}}{\textstyle{\partial x^\nu}}-
\frac{\textstyle{\partial{}A_\nu}}{\textstyle{\partial x^\eta}}.}
\par
To ensure gauge invariance one should replace partial by \emph{covariant} derivatives in the differential equations describing the fields \Math{\psi(x)}:
\Mathh{\partial_\nu\rightarrow{}D_\nu=\partial_\nu-iA_\nu(x).}
\par
In order to complete the construction, the equations that describe the evolution of the gauge field \Math{{A}(x)} itself need to be added. 
Exact forms of such equations do not follow directly from the gauge principle.
It is only clear that they must be gauge invariant.
In the case at hand, the simplest choice of gauge invariant equations leads to the \emph{Maxwell equations}:
\begin{align}
	\mathrm{d}F&=0\hspace*{20pt}\text{\emph{first pair}}\nonumber,\\	
	\mathrm{d}\star{}F&=0\hspace*{20pt}\text{\emph{second pair}}\label{Enm2nd}.	
\end{align}
Here \Math{\star} is the \emph{Hodge star operator}.
Note that Eq. \eqref{Enm2nd} corresponds to the Maxwell equations in vacuum. In the presence of current \Math{J} the \emph{second pair} of Maxwell equations assumes the form
\Mathh{\star{}\mathrm{d}\star{}F=J.}
Note also that the \emph{first pair} of equations is essentially an \textit{a priori} statement, that is, an identity which holds for any differential of exterior form. \Math{F} 
is a differential by definition.
\subsubsection{Non-Abelian gauge theories in continuous space-time.}
If the gauge group \Math{\iG} is a non-Abelian Lie group, then the above construction 
need to be slightly modified.
Again, replacement of the \Math{\iG}-valued parallel transport between two
closely situated space-time points \Math{x} and \Math{x+\Delta{}x} by the
linear combination of coordinate differences between
these points leads to the 1-form \Math{A=\vect{A_{\nu}}} with values in
the Lie algebra of the group \Math{\iG}:
\Mathh{\Partransportx\vect{x,x+\Delta{}x}\approx\id+A_\nu(x)\Delta{}x^\nu.}
We have used here the fact, that \Math{\Partransportx\vect{x,x}=\id}. 
The standard infinitesimal manipulations with transformation rule \eqref{Enconnruleoverx} for connection
\eqref{Enconnruleoverx}
\Mathh{\gamma\vect{x}^{-1}\Partransportx\vect{x,x+\Delta{}x}\gamma\vect{x+\Delta{}x}
~\longrightarrow~
\gamma\vect{x}^{-1}\vect{\id+A_\nu\vect{x}\Delta{}x^\nu}
\vect{\gamma\vect{x}+\frac{\textstyle{\partial{}\gamma\vect{x}}}
{\textstyle{\partial{}x^\nu}}\Delta{}x^\nu}}
yield the following transformation rule for the connection 1-form 
\begin{equation*}
{A'_\nu}\vect{x}=\gamma\vect{x}^{-1}{A_\nu}\vect{x}\gamma\vect{x}+\gamma\vect{x}^{-1}
\frac{\textstyle{\partial{}\gamma\vect{x}}}{\textstyle{\partial{}x^\nu}}.	
\end{equation*}
The connection curvature 2-form
\begin{equation}
	F=\mathrm{d}A+\left[A\wedge{}A\right]\label{En2form}
\end{equation}
is interpreted as \emph{physical force fields}.
The 1-form of the \emph{trivial} connection is
\Mathh{\widetilde{A}_\nu\vect{x}=\gamma_0\vect{x}^{-1}
\frac{\textstyle{\partial{}\gamma_0\vect{x}}}{\textstyle{\partial{}x^\nu}}.}
Curvature form \eqref{En2form} of such a connection is zero.
The trivial connection does not generate any physical forces. It is called \emph{flat} connection in the differential geometry.
\par
There are various versions of dynamical equations for the gauge fields \cite{Oeckl}.
The most important example is the \emph{Yang-Mills theory} based on the Lagrangian
\Mathh{L_{YM}=\mathrm{Tr}\left[F\wedge\star{}F\right].}
The Yang-Mills equations of motion have the form
\begin{align}
	\mathrm{d}F+\left[A\wedge{}F\right]&=0,\label{EnBianci}\\
	\mathrm{d}\star{}F+\left[A\wedge\star{}F\right]&=0.\nonumber
\end{align}
Here again Eq. \eqref{EnBianci} is an \textit{a priori} statement called the \emph{Bianchi identity}.
Note that the Maxwell equations are a special case of the Yang–Mills equations.
\par
It is instructive to look how the Yang-Mills Lagrangian can be represented on a discrete lattice.
For simplicity, let us consider a hypercube lattice in the Minkowski space-time \Math{\XT}.
Calculation  \cite{Oeckl} using the infinitesimal approximations shows that
\Math{L_{YM}\sim\sum_f\sigma\vect{\gamma_f}},
where the summation runs over all two-dimensional faces of the elementary hypercube.
The function \Math{\sigma} has the form 
\Mathh{
\sigma\vect{\gamma} = 2\dim\rho-\chi\vect{\rho\vect{\gamma}}
-\chi\vect{\rho^\dagger\vect{\gamma}}
\equiv2\chi\vect{\id}-\chi\vect{\rho\vect{\gamma}}
-\chi\vect{\rho^\dagger\vect{\gamma}},}
where \Math{\rho} is the fundamental representation of the gauge group
(\Math{\rho^\dagger} means the dual repre\-sentation),
\Math{\chi} is the character (the trace of the representation matrix),
\Math{\gamma_f} is a holonomy of the group \Math{\iG} around the face \Math{f}.
\par
The Yang–Mills theory uses the Hodge operator that transforms \Math{k}-forms to \Math{\vect{n-k}}-forms in the presence of metrics in the \Math{n}-dimensional space \Math{\XT}.
In topological applications 
In topological applications an important role is played by the so-called \emph{BF theory}.
This theory avoids the use of a metric by introducing an additional dynamical field \Math{B}, which is an \Math{\vect{n-2}}-form with values in the Lie algebra of the gauge group.
The gauge invariant Lagrangian of the BF theory has the form
\Mathh{L_{BF}=\mathrm{Tr}\left[B\wedge{}F\right].}
\subsubsection{Emergence of space in discrete dynamics.}
In some modern physical theories it is assumed that the space (and even the time) are derived concepts, i.e., they are absent in the fundamental formulation of the theory, but appear as certain approximate notions at the
macroscopic level.
An interpretation of the space in this spirit arises quite naturally.
As for the time,%
\footnote{The motivation to treat the time at the same level of fundamentality as the space comes from the special theory of relativity.}
the attempts to derive it from more fundamental concepts require extra efforts and use of some \textit{ad hoc} arguments. 
The problem of deriving space and time from more fundamental entities are discussed, in particular, in string theory \cite{Seiberg}, theory of causal sets \cite{Markopoulou} and theory of spin networks \cite{Baez}.
\par
Using a simple construction \cite{Kornyak11d} we will show how the space-time structures may arise in discrete dynamics. 
Suppose that there is a sequence of events of different types labeled by time moments and the ability to distinguish between those types.
More specifically, we consider the set of symbols \Math{\Sigma=\set{\sigma_{1},\sigma_{2},\ldots,\sigma_{N}}},
that denote elements of experience of some ``observer'', who is able to ``percept'' these elements and 
``memorize'' their sequences (streams) 
\begin{equation}
	h=s_0s_1\cdots{}s_t, \text{~where~} s_i\in\Sigma, t\in\N.
 \label{Enpath}	
\end{equation}
The concepts of ``causality'' and ``light cone'' arise quite naturally.
Inability to exceed the ``speed of light'' means impossibility to obtain more than \Math{t} symbols (``elements of perception'') in \Math{t} observations.
\par
\begin{figure}[!h]
\begin{center}
	\includegraphics[width=0.45\textwidth]{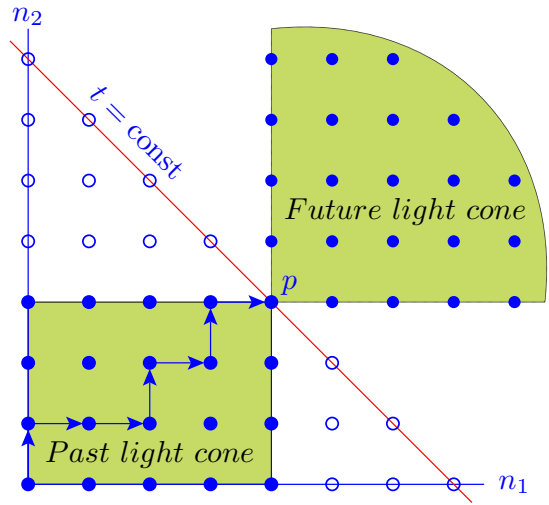}
\caption{Space-time point \Math{p} as a class of equivalence of paths with the same numbers 
\Math{n_1, n_2, \ldots, n_{N}}\label{Enlightcones}.}
\end{center}
\end{figure}

In the simplest approach, the counters of events of different types can be interpreted as spatial dimensions, and
the space-time point \Math{p} can be defined as an equivalence class of sequences with the same multiplicity of
each symbol, i.e., \Math{p} is a ``commutative monomial'' of the total degree \Math{t} described by the vector
of \Math{N} natural numbers:
\begin{equation}
	p=\vect{n_1,\ldots,n_{N+1}}^\mathrm{T},~n_1+\cdots+n_{N}=t,
	\label{Enpmonom}
\end{equation}
where \Math{n_i} is multiplicity of symbol \Math{\sigma_{i}} in a sequence \Math{h}.
In terms of commutative monomials the ``past light cone'' is the set of divisors of the
monomial \Math{p}, while the “future light cone” is the set of its
multiples (see Fig.  \ref{Enlightcones}). 
\par
Of course, the above scheme is only illustrative.
To adapt it to the description of more realistic models, it is necessary to introduce additional structures and, perhaps, another definition of the concept of space-time point, taking into account, in particular, the symmetry between the symbols of the set \Math{\Sigma}.
For example, it is necessary that in the limit of large \Math{n_i} (or --- depending on the definition of the concept of a point --- some other similar characteristics of sequence \eqref{Enpath}) the Lorentz (or Poincar\'{e}) symmetries arise as a continuum approximation.
\par
To illustrate these considerations, we construct a simple physical model.
Consider a set of two symbols \Math{\Sigma=\set{\sigma_{1},\sigma_{2}}} with the natural assumption that it is symmetric with respect to the permutation \Math{\sigma_{1}\rightleftarrows\sigma_{2}}.
This permutation generates the group \Math{\SymG{2}}.
Suppose now that according to \eqref{Enpmonom} a space-time point is characterized by the vector \Math{p=\Vtwo{n_1}{n_2}}.
The represention of \Math{\SymG{2}} in the module of such vectors is described by two matrices \Math{e=\Mtwo{1}{0}{0}{1}} and \Math{r=\Mtwo{0}{1}{1}{0}}.
To take into account the symmetry of symbols, the point \Math{p} must be expressed in terms of invariant submodules of this representation.
To do this, the representation should be decomposed into irreducible components.
The transition to the basis of invariant submodules can be done, for example, by the matrix 
\Math{T=\Mtwo{1}{1}{1}{\!-1}}.
The identity matrix \Math{e}, obviously, remains unchanged, and the reflection matrix takes the form  \Math{r'=\Mtwo{1}{0}{0}{-1}}.
The space-time point \Math{p} in the new basis becomes  \Math{p'=\Vtwo{n_1+n_2}{n_1-n_2}}, or, if we introduce the notations \Math{t=n_1+n_2} and \Math{x=n_1-n_2}, \Math{p'=\Vtwo{t}{x}}.
We will call the variable \Math{x} ``space'' (``spatial dimension'').
We see that the presence of the symmetry reduced the number of expected spatial dimensions and, in addition, the transition to invariant submodules forced us to introduce negative numbers.%
\footnote{It follows from \Math{r^2=\id}, that the eigenvalues of the reflection matrix are square roots of unity,
one of which, \Math{\epsilon=\e^{2\pi{}i/2}}, is \emph{primitive}. Thus, we obtain the ring \Math{\Z} as an 
\emph{algebraic extension} of the semiring \Math{\N} generated by the algebraic element \Math{\epsilon}.} 
\par 
Now add some physics by assuming that a sequence
of symbols \Math{h=s_0s_1\cdots{}s_t,~s_i\in\set{\sigma_{1},\sigma_{2}}} corresponds to the \emph{Bernoulli trials} scheme, i.e., in sufficiently long subsequences the symbols \Math{\sigma_{1}} and 
\Math{\sigma_{2}} appear with the fixed frequencies (probabilities) \Math{p_{1}} and \Math{p_{2}}, where 
\Math{p_{1}+p_{2}=1}.
Then, the probability of each individual sequence is described by the binomial distribution
\begin{equation}
P\vect{n_1,n_2}=\frac{\vect{n_1+n_2}!}{n_1!n_2!}p_1^{n_1}p_2^{n_2}.\label{Enbindistr}
\end{equation}
Introduce the quantity \Math{v=p_1-p_2} and call it ``velocity''.
The velocity is obviously limited in the magnitude: \Math{-1\leq{}v\leq1}.
Let us compute using Stirling's formula the asymptotic of \eqref{Enbindistr} for large \Math{n_1} and \Math{n_2}
in the vicinity of the stationary value of this expression.
We restrict ourselves to the second order Taylor expansion of the logarithm of \eqref{Enbindistr} in the vicinity of the stationary value. Replacing \Math{n_1} and \Math{n_2} by the variables \Math{t} and \Math{x}, and the probabilities \Math{p_{1}} and \Math{p_{2}}  by the velocity \Math{v}, we obtain
\begin{equation}
	P\vect{x,t}\approx\widetilde{P}\vect{x,t}
	=\frac{1}{\sqrt{1-v^2}}\sqrt{\frac{2}{\pi{}t}}
	\exp\set{-\frac{1}{2t}\vect{\frac{x-v t}{\sqrt{1-v^2}}}^2}\enspace.
	\label{EnPxt}
\end{equation}
Note that this expression contains ``relativistic'' fragment \Math{\frac{\textstyle{x-v{}t}}{\textstyle{\sqrt{1-v^2}}}}.
After the change of variables \Math{t=T+t'} and \Math{x=vT+x'}, Eq. \eqref{EnPxt} can be rewritten as
\begin{equation}
\widetilde{P}\vect{x,t}
=\frac{1}{\sqrt{1-v^2}}\sqrt{\frac{2}{\pi{}T}}
		\exp\set{-\frac{1}{2T}\vect{{\frac{x'-vt'}{\sqrt{1-v^2}}}}^2}+O\vect{\frac{t'}{T}}\enspace.
	\label{EnPxtT}
\end{equation}
If one assumes that \Math{t'\ll{}T} (\Math{1/T} is a ``Hubble constant'', whereas \Math{t'} is a ``typical time of observation'', then ``relativistic invariance'' is reproduced in \eqref{EnPxtT} with high accuracy.
The right hand side of equation \eqref{EnPxt} is the fundamental solution of the \emph{heat equation}%
\footnote{This equation is also called the \emph{diffusion equation} or the \emph{Fokker-Planck equation}, depending on the interpretation of the function \Math{\widetilde{P}\vect{x, t}}.}
\begin{equation*}
	\frac{\partial{}\widetilde{P}\vect{x,t}}{\partial{}t}
	+v\frac{\partial{}\widetilde{P}\vect{x,t}}{\partial{}x} 
	=\frac{\vect{1-v^2}}{2}
	\frac{\partial^2 \widetilde{P}\vect{x,t}}{\partial{}x^2}.
\end{equation*}
In the ``limit of the speed of light'' \Math{\cabs{v}=1} this equation turns into the \emph{wave equation}
\begin{equation*}
	\frac{\partial{}\widetilde{P}\vect{x,t}}{\partial{}t}\pm\frac{\partial{}
	\widetilde{P}\vect{x,t}}{\partial{}x}=0.
\end{equation*}
\par
The simplicity of this model makes it possible to compare exact combinatorial expres\-sions with their continuum approximation.
The comparison shows that the continuum approximations may introduce severe artifacts.
Consider the typical for mechanics problem of finding extremal trajectories connecting two fixed space-time points \Math{\vect{0,0}} and \Math{\vect{X,T}}.
Our version of the \emph{``least action principle''} will be search the maximum probability trajec\-tories.
According to the rule for calculating the \emph{conditional probability}, the probability of the trajectory that
links the points \Math{\vect{0,0}} and \Math{\vect{X,T}} and passes through some intermediate point \Math{\vect{x,t}} is given by 
\begin{align}
  P_{\vect{0,0}\rightarrow\vect{x,t}\rightarrow\vect{X,T}}
  &=\frac{P\vect{x,t}P\vect{X-x,T-t}}{P\vect{X,T}}\nonumber\\
	&=\frac{t!\vect{T-t}!\vect{\frac{T-X}{2}}!\vect{\frac{T+X}{2}}!}
	{\vect{\frac{t-x}{2}}!\vect{\frac{t+x}{2}}!
	\vect{\frac{T-t}{2}-\frac{X-x}{2}}!\vect{\frac{T-t}{2}+\frac{X-x}{2}}!T!}.
	\label{Enexactp}
\end{align}
The similar conditional probability for continuum approximation \eqref{EnPxt} takes the form
\begin{equation}
		\widetilde{P}_{\vect{0,0}\rightarrow\vect{x,t}\rightarrow\vect{X,T}}
		=\frac{T}{\sqrt{\frac{\pi}{2}\vect{1-v^2}tT\vect{T-t}}}
		\exp\set{-\frac{\vect{Xt-xT}^2}{2\vect{1-v^2}tT\vect{T-t}}}.
		\label{Enapproxp}
\end{equation}
There are significant differences between exact expression \eqref{Enexactp} and its continuous appro\-ximation \eqref{Enapproxp}:
\begin{itemize}
	\item 
The expression in \eqref{Enapproxp} contains \emph{artificial dependence} on the velocity \Math{v} (or equivalently on the probabilities \Math{p_1} and \Math{p_2}) unlike the exact probability given in \eqref{Enexactp}.
	\item
It is easy to check that in the case of exact expression \eqref{Enexactp} the same maximum probability is achieved on many different trajectories, while the functional in \eqref{Enapproxp} has only one extremal, namely, the straight line \Math{x = \frac{X}{T}t} as a \emph{deterministic trajectory}.
\par
Here we have an example that deterministic behavior may emerge as a result of approximation based on the law of large numbers.
\end{itemize}
\subsection{Features of Deterministic Dynamics} 
The results of this subsection are based on \cite{Kornyak07a,Kornyak07c,Kornyak08}.
\par
The absence of any regularity in the behavior of a dynamical system means that its trajectory may be represented with
indefinite probability by any section \eqref{Enevolutiongeneral} of bundle \eqref{Enbundleovert}.
Consider the evolution with a fixed number \Math{k} of steps in time: \Math{h_{k,t} = \vect{\ws_{t-k},\ldots,\ws_t}.}
These trajectories are elements of the set \Math{H_{k,t}=\prod\limits_{s=t-k}^t\wSord_{s}}.
The standard way to introduce regularity into the dynamics is to attach some non-negative numbers \Math{p\vect{h_{k,t}}} to the points of the set \Math{H_{k,t}}. These numbers will be called \emph{weights}.%
\footnote{By normalizing the \emph{weight function} \Math{p\vect{\cdot}} over all the elements of
the set \Math{H_{k,t}} one can obtain a \emph{probability distribution}.}
The weights can take arbitrary real values, however, in accordance with the ideology of this review, we prefer the natural numbers. 
If only 0 and 1 are taken to be the values for the weight function, then we obtain a \emph{characteristic
function} of some subset of the set \Math{H_{k,t}}.
In the classical case, the identification of the states of system with the elements of \Math{\wS} does not depend on time.
So the set of trajectories can be written as \Math{H_{k,t}=\wS^{k+1}\equiv{}H_{k}}.
Note that we  ``have lost'' dependence on the ``current time'' \Math{t}, so it is natural to assume that a characteristic function also depends only on the length of the trajectory.
Any subset of the Cartesian product of \Math{n} sets (not necessarily identical) is called \Math{n}-ary relation.%
\footnote{More detailed discussion of the discrete relations and their
applications to the structural analysis of dynamical systems, in
particular, cellular automata, see in Appendix \hyperref[Endiscreterelations]{B}.}
\Math{(k+1)}-ary discrete relation on the set \Math{H_{k}} will be called \emph{evolution relation of \Math{k}th order}.
Via characteristic function such a relation can be written as an equation:
\begin{equation}
	R\vect{\ws_{t-k},\ldots,\ws_t}=0.
	\label{Enevrelk}
\end{equation}
We will call a \Math{k}-th order evolution relation \emph{deterministic}, if Eq. \eqref{Enevrelk} can be solved%
\footnote{Relations of this type are called \emph{functional} (Appendix \hyperref[Endiscreterelations]{B}).}
for the variable \Math{\ws_t}.
Deterministic evolution equation of \Math{k}th order can be written as
\begin{equation}
	\ws_t=F\vect{\ws_{t-1},\ldots,\ws_{t-k}}.
	\label{Endetevrelk}
\end{equation}
Deterministic dynamical systems of the first order are most common in applications.
For any trajectory of such a system, its state at any given moment of time is
a \emph{function} of the state at the previous moment:
\begin{equation}
	w_{t}=F\vect{w_{t-1}}, ~w_{t},w_{t-1}\in\wS.
	\label{Endeterministic}
\end{equation}
Further in this subsection we consider only the first order systems, although, with some technical complications, the results, similar to those that will be discussed below, can also be obtained for more general deterministic dynamics \eqref{Endetevrelk}.
\par
Let us begin with some general remarks on the relationship between symmetry and dynamics in the deterministic case.
The symmetry group \Math{\wG} splits the set \Math{\wS} into disjoint orbits of finite sizes.
The functionality of relation \eqref{Endeterministic} immediately implies that
\begin{itemize}
	\item 
\emph{dynamical trajectories} pass the \emph{group orbits}	in the order in which the orbit sizes do not increase;
	\item \label{Enoncycle}
	any periodic trajectory passes through the orbits of the same size.
\end{itemize}
\subsubsection{Soliton-like structures in deterministic dynamics.} 
Let us turn to dynamical systems with a space.
One of the characteristic features of dynamics of such systems is the formation of structures moving in space with retaining their shapes.
We will show that such behavior is a natural consequence of the symmetries of space. 
\par
\begin{figure}[!h]
\centering
\includegraphics[width=0.65\textwidth]{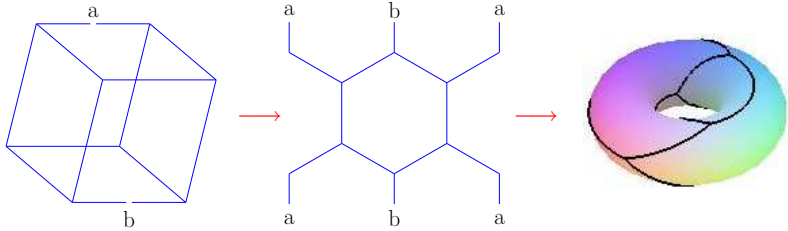} 
\caption{The graph of a cube forms a tetragonal lattice on a sphere (6 tetragons) and a hexagonal 
lattice on a torus (4 hexagon).}
	\label{EnCube-on-tor}
\end{figure}

Let us start with a simple and graspable example of a dynamical system.
As the space \Math{\X} we take a cube (for brevity, a ``cube'' here stands for the graph of its edges).
Note that a cube can be regarded as the simplest ``finite model of graphene'' obtained by closing a hexagonal sublattice into a torus, as shown in Fig. \ref{EnCube-on-tor}.
\par
The symmetry group of a cube is \Math{\sG=\SymG{4}\times\CyclG{2}}. The order of \Math{\sG} is 48. 
If \Math{\lS=\set{0,1}} is a set of states of the vertices of cube, then the total number of states of the model is \Math{\cabs{\lS^{\X}}=2^8=256}. 
For the sake of simplicity, we assume that the group of internal symmetries is trivial \Math{\iG=\set{\id}}; then, according to formula \eqref{Enwreath1} the total symmetry group of the model in fact coincides with the space symmetry group:
\Math{\wG=\iGX\rtimes\sG=\set{\id}^{\X}\rtimes\sG\cong\sG.}
The group \Math{\wG} splits the set of states \Math{\lS^{\X}} into 22 orbits, whose sizes and numbers are listed in the table
\begin{center}
\begin{tabular}{c|ccccccc}
size of orbits&1&2&4&6&8&12&24
\\\hline
number of orbits&2&1&2&2&5& 4& 6
\end{tabular}\enspace.
\end{center}
\par
Consider the deterministic dynamical system on a cube, namely, the symmetric \cite{Kornyak06b,Kornyak07b} binary trivalent cellular automaton (see Appendix \hyperref[Endiscreterelations]{B}) with the rule 86.
The number 86 is a decimal representation of the bit string 
(an ascending order of binary digits is assumed) from the last
column of the Table \ref{Enevol86}, which specifies the evolution
rule of the automaton.
\begin{table}[h] 
\begin{center}
\begin{tabular}[b]{cccc|l}
\Math{x_{1,i}}&\Math{x_{2,i}}&\Math{x_{3,i}}&\Math{x_i}&\Math{x'_i}
\\
\hline
0&0&0&0&\Math{0}\\
0&0&0&1&\Math{1}\\
1&0&0&0&\Math{1}\\
1&0&0&1&\Math{0}\\
1&1&0&0&\Math{1}\\
1&1&0&1&\Math{0}\\
1&1&1&0&\Math{1}\\
1&1&1&1&\Math{0}\\
\end{tabular}
\end{center}
\caption{Evolution rule 86. \Math{x_i} and \Math{x'_i} are previous and subsequent states of \Math{i}th vertex of the cube, respectively; \Math{x_{1,i},x_{2,i},x_{3,i}} are previous states of the vertices adjacent to the \Math{i}th vertex. The rule is symmetric with respect to all permutations of the vertices \Math{x_{1,i},x_{2,i},x_{3,i}}, so the first three columns contain the symmetrized combinations of values.}
\label{Enevol86}
\end{table}
\par
This rule can also be written in terms of ``Birth''/``Survival'' lists --- the style usual for automata such as Conway's Life --- as B123/S0, or as the polynomial over the finite field \Math{\F_2}
\Mathh{x'_i = x_i+x_{1,i}+x_{2,i}+x_{3,i}+x_{1,i}x_{2,i}+x_{1,i}x_{3,i}+x_{2,i}x_{3,i}+x_{1,i}x_{2,i}x_{3,i}.}
\par
\begin{figure}[!h]
\centering
\includegraphics[width=340pt]{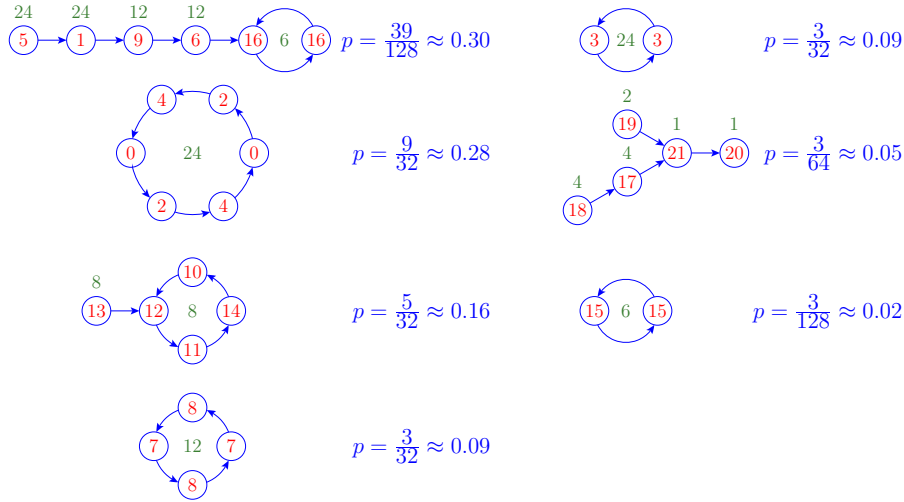}%
\caption{Equivalence classes of trajectories for symmetric binary celullar automaton on a cube with evolution rule  86.}
\label{EnPhasePortrait}
\end{figure}

The phase portrait of the automaton is shown in Fig. \ref{EnPhasePortrait}.
In this figure, the group orbits are represented by small circles containing the (red) orbit labels.%
\footnote{In fact, these labels are ordinal numbers which our computer program for calculating evolution of automata  assigns to the orbits as they appear.}
The (green) numbers above orbits and within cycles are orbit sizes.
Recall, that any cycle passes through orbits of equal size (see remark on page \pageref{Enoncycle}).
A (blue) rational \Math{p} denotes the weight of the corresponding element of the phase portrait.
In fact, \Math{p} is the probability that the trajectory that begins with randomly selected state will either be
an isolated cycle or be captured by a respective attractor, i.e., \Math{\textstyle{}p=
\frac{\textstyle{}\text{\emph{size of basin}}}{\textstyle{}\text{\emph{total number of states}}}},
where \emph{size of basin} is the sum of sizes of orbits included in a given structure.
\par
Generalizing this example, we see that if the group of symmetries of a \emph{deterministic} dynamical system splits the set of states into a \emph{finite} number of orbits, then \emph{any} trajectory will inevitably go in a finite number of time steps into a cycle over some sequence of orbits.
This just means the formation of \emph{soliton-like structures}.
Namely, consider the evolution
\begin{equation}
	\sigma_{t_0}(x)\rightarrow\sigma_{t_1}(x)=A_{t_1t_0}\vect{\sigma_{t_0}(x)}.
	\label{Enevol}
\end{equation}
If the states at different times \Math{t_0} and \Math{t_1} belong to \emph{the same orbit}, i.e., 
\Mathh{\sigma_{t_0}(x), \sigma_{t_1}(x)\in{}O_i\subseteq\lSX;}
then evolution \eqref{Enevol} can be replaced by a \emph{group action}
\begin{equation}
	\sigma_{t_1}(x)=\sigma_{t_0}(x)g,~~g\in{}\wG.
	\label{Enmove}
\end{equation}
Since an action of the symmetry group of a space is equivalent to a ``\emph{motion}'' in the space, 
then \eqref{Enmove} means that the initial state (``\emph{shape}'') \Math{\sigma_{t_0}(x)} is reproduced after some movement in the space.
\par
Let us give a few familiar examples of cycles over group orbits indicating respective symmetries (two of which are continuous):
\begin{itemize}
	\item \emph{traveling waves} \Math{\sigma\vect{x-vt}} in mathematical physics --- the Galilei group;
	\item ``\emph{generalized coherent states}'' in quantum physics --- unitary representations of com\-pact Lie
groups;
	\item ``\emph{spaceships}'' in cellular automata --- symmetries
of lattices (graphs). 
\end{itemize}
\par
As a more detailed example, let us consider the ``\emph{glider}'' --- one of the ``spaceships'' of Conway’s
automaton Life.
The space \Math{\X} of this automaton is a square lattice.
We assume that this lattice is implemented as a discrete torus of size \Math{N\times{}N}.
If \Math{N\neq4}, then the space symmetry group has the structure of a semidirect product of the two-dimensional translations \Math{\mathrm{T}^2=\Z_N\times\Z_N} and the dihedral group \Math{\DihG{8}\cong\Z_4\rtimes\Z_2}: 
\begin{equation*}
	\sG\cong\mathrm{T}^2\rtimes\DihG{8},\mbox{~~if~~} N = 3,5,6,\ldots,\infty.
\end{equation*}
In the case \Math{N=4} the translation subgroup \Math{\mathrm{T}^2=\Z_4\times\Z_4}
is no longer a normal subgroup, and the group \Math{\sG} will
acquire a somewhat more complicated structure \cite{Kornyak09c}:
\begin{equation*}
	\sG = \overbrace{
	\vect{\vect{\vect{\vect{\Z_2 \times \DihG{8}} 
	\rtimes \Z_2} \rtimes{\color{red}\Z_3}} \rtimes \Z_2}
	}^{
	\mbox{normal closure of~}\textstyle\mathrm{T}^2}
	\rtimes \Z_2.
\end{equation*}
The presence of additional factor \Math{\color{red}\CyclG{3}} in this expression is explained by the symmetry of the Dynkin diagram with four vertices
\Mathh{D_4=~~\text{\raisebox{-0.025\textwidth}{\includegraphics[width=0.065\textwidth]{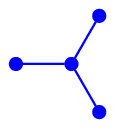}}},}
corresponding to the case \Math{N=4}.
\par
\begin{figure}[!h]
\centering
\includegraphics[width=0.7\textwidth]{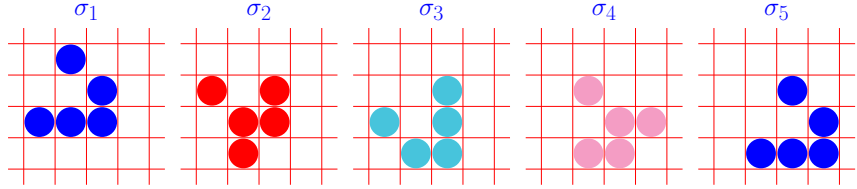}
\caption{Example of soliton-like structure.
The ``glider'' of the automaton Life is a cycle over \emph{two} orbits
of the group \Math{\sG=\mathrm{T}^2\rtimes\DihG{8}}: the  configurations \Math{\sigma_3} and  
\Math{\sigma_4} are obtained from \Math{\sigma_1} and  
\Math{\sigma_2}, respectively, by the same combination of one-step \emph{shift} down, 
 90\textdegree \emph{rotation}  and \emph{reflection}.}
	\label{EnGlider-2}
\end{figure}

The set of local states of the Life automaton cells has the form 
\Mathh{\lS=\set{0,1}\equiv\set{\text{``\emph{dead}''},\text{``\emph{alive}''}}.}
Since the local rule for this automaton is not symmetric with respect to the state per\-mutation 
\Math{0\rightleftarrows1}, the group of internal symmetries \Math{\iG} is trivial.
Thus, \Math{\wG = \set{\id}^{\X}\rtimes\sG\cong\sG.} 
According to \eqref{Enwreathaction}, the action of \Math{\wG} on functions \Math{\sigma\vect{x}\in\lS^{\X}} 
has the form \Math{\sigma\vect{x}g=\sigma\vect{xf^{-1}}}, where \Math{g = \vect{\id,f},~f\in\sG}.
Fig. \ref{EnGlider-2} reproduces (under the assumption \Math{N>4}) four steps of evolution of the glider.
\subsubsection{On reversibility in discrete deterministic systems.} 
A typical discrete deterministic system is \emph{irreversible}.
The phase portrait of such a system looks like that shown in Fig. \ref{EnPhasePortrait}.
We see that the figure contains several isolated and limit cycles.%
\footnote{Fixed points can be treated as cycles of unit length.}
In the continuous case, we can, in principle, increasing the accuracy of the description, restore the prehistory of any trajectory tending to a limit cycle, since a point of continuum may contain potentially infinite amount of information.
In the discrete case, the information about the states in which the system was before transition to a limit cycle is irretrievably lost.
This means that after a certain number of steps in time limit cycles become physically indistinguishable from isolated cycles and the dynamic system behaves just as a reversible system.
Perhaps such considerations can be used to explain the observed reversibility of the fundamental laws of nature.
\par
G. 't Hooft uses similar ideas 
to resolve the conflict between the irreversibility of gravity (the loss of information on the horizons of black holes) and reversibility (unitarity) of standard quantum mechanics (see, e.g., \cite{tHooft99,tHooft06}). 
't Hooft's approach is based on the assumptions:
\begin{itemize}
	\item
physical systems at the microscopic (Planck) scale are described by \emph{discrete degrees of freedom};
	\item
the states of these degrees of freedom form a \emph{primordial} basis of a Hilbert space (where non-unitary evolution is possible);
	\item
the primordial states form \emph{equivalence classes}: two states are equivalent if they transit into the same state after some lapse of time;
	\item
by construction, these equivalence classes form a basis of a Hilbert space with unitary evolution, which is described by the time-reversible Schr\"{o}dinger equation.
\end{itemize}	
In our terminology, this corresponds to the transition to limit cycles.
In a finite number of time steps a system completely ``forgets'' its ``precycle'' history.
\par
Even if the above considerations have anything to do with physical reality, the irreversi\-bility of this kind can hardly be observed experimentally.
A system should probably spend outside a cycle the time of the Planck order, i.e., approximately \Math{10^{-44}}  sec.
The shortest time interval fixed at present experimentally is equal to approximately \Math{10^{-18}} sec or \Math{10^{26}} Planck units.
\section{Constructive Description of Quantum Behavior}
This section is based on \cite{Kornyak11a,Kornyak11b,Kornyak12a,Kornyak12b}.
\par
\begin{figure}[!h]
\centering
\includegraphics[width=0.85\textwidth]{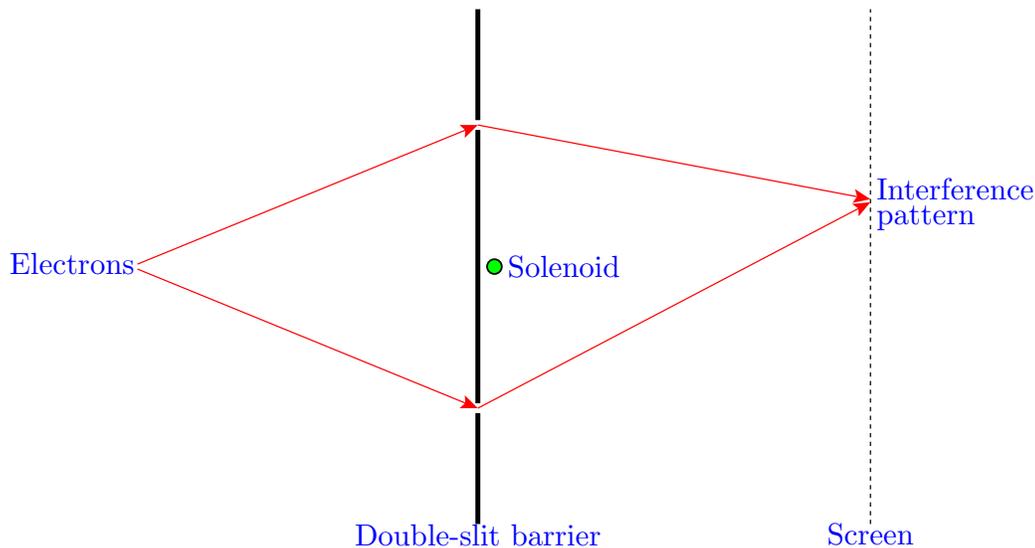}
\caption{Aharonov--Bohm effect.
Interference pattern is shifted when the solenoid is switched on despite the absence of electromagnetic forces outside the solenoid.}
	\label{EnAharonov-Bohm}
\end{figure}

The most famous empirical demonstration of quantum behavior is the \emph{double-slit ex\-periment}.
If it is carried out with charged particles in the installation, equipped with a solenoid, then one can observe the
\emph{Aharonov--Bohm effect} (Fig. \ref{EnAharonov-Bohm}), which demonstrates the role of gauge connections in quantum mechanics.
Charged particles are moving in the region, containing a perfectly shielded thin solenoid.
The interference pattern changes when the solenoid is turned on, despite the absence of electromagnetic forces acting on the particles.
The effect occurs because the working solenoid creates around itself the \Math{\UG{1}}-connection that changes the phases of the wave function of the particles.
\par
The most convenient way to quantize dynamical systems with spatial structure is the Feynman path integral method
\cite{Feynman}.
In fact, this method arose from the study of the two-slit experiment.%
\footnote{It is known that R. Feynman liked to say that ``all of quantum mechanics can be gleaned from carefully thinking through the implications of this single experiment''.}
According to Feynman's approach amplitude of the quantum transition from one state to another is calculated by summing the amplitudes along all possible classical paths connecting these states.
Amplitude along an individual path is evaluated as a product of the amplitudes of transitions between the nearest successive states on the path.
Namely, the amplitude along a path is represented as the exponential of the
action along that path
\begin{equation}
A_{\mathrm{U}(1)}~=~A_0\exp\vect{iS}=A_0\exp\vect{i\int\limits_0^T{}Ldt}.
\label{Enamplclass}
\end{equation}
The function \Math{L} depending on the \emph{first time derivatives} of states is called the \emph{Lagrangian}.
In the discrete time the exponential of integral turns into the product:
\Math{\exp\vect{i\int{}Ldt}\rightarrow
	\e^{{iL_{0,1}}}\ldots\e^{{iL_{t-1,t}}}
	\ldots\e^{{iL_{T-1,T}}}}
and expression for the amplitude takes the form
\begin{equation*}
	A_{\mathrm{U}(1)}~=~A_0
	\e^{{iL_{0,1}}}\ldots\e^{{iL_{t-1,t}}}
	\ldots\e^{{iL_{T-1,T}}}.
\end{equation*}
The factors \Math{\Partransportx_{t-1,t}=\e^{{iL_{t-1,t}}}} of this product are elements of the \emph{connection} with values in \emph{one-dimensional} unitary representation \Math{\UG{1}} of a circle,
i.e. the commutative Lie group \Math{\iG=S^1\equiv\R/\Z}, which is a group of internal symmetries for this case.
In line with our view of quantum behavior, 
this implies the existence of some set of \emph{local states} \Math{\lS} 
on which the Lie group \Math{S^1} acts, 
while the representation of this group \Math{\UG{1}} acts on the linear space (in the case at hand it is \Math{\C})
of \emph{amplitudes} attached to the states from \Math{\lS}.
Later, when discussing quantum behaviour from a general point of view, we will clarify the relationship between states and their amplitudes, i.e. numerical weights attributed to the states. 
Here, as usual, the states and the amplitudes on which the group \Math{S^1} and, respectively, the representation \Math{\UG{1}} act are unobservable; only invariants of these actions are observable.
Continuous version of that the connection elements are defined on pairs of states corresponding to the
ends of the elementary time intervals, i.e., on edges of an abstract graph%
\footnote{Recall that the definition of a connection as a function on \emph{pairs} of points (rather than on points) is required to ensure the nontriviality of connection.}
is the dependence of a Lagrangian on first time derivatives. 
\par
A natural generalization is to assume  that the group \Math{\iG} is not necessarily a circle and that its unitary representation \Math{\Rep{\iG}} is not necessarily one-dimensional.
In this case, the amplitude is a multi-component vector, which is suitable for describing particles with more complicated internal degrees of freedom.
The value of such a multicomponent amplitude on a path takes the form%
\footnote{In the noncommutative case it is important to observe the order of operators. The ordering of operators in \eqref{Enamplgen} corresponds to the tradition to
write matrices to the left of vectors.} 
\begin{equation}
	A_{\Rep{\iG}}=\Rep{\alpha_{T,T-1}}\ldots\Rep{\alpha_{t,t-1}}
	\ldots\Rep{\alpha_{1,0}}A_0,\hspace*{10pt}\alpha_{t,t-1}\in\iG.\label{Enamplgen}
\end{equation}
According to our ideology, we will assume that  \Math{\iG} is a finite group.
In this case there is no need to care about unitarity, since the linear representations of finite groups are automatically unitary.
It is clear that the standard quantization \eqref{Enamplclass} can be approximated using one-dimensional representations of cyclic groups of large enough periods.
\par
The Feynman rules, formulated in the abstract form, coincide in fact with the rules of matrix multiplication.
According to Feynman's rules, in order to obtain the amplitude of the transition from one state to another, one should connect these states by all possible paths, multiply the amplitudes along these paths and sum up all the products.
The coincidence of these instructions with the rules of matrix multiplication is obvious from the illustration on which the two-step evolution of a quantum system with two states (``single-qubit register'') is presented in parallel in the Feynman and matrix forms: 
\begin{center}
\begin{tabular}[t]{ccc}
\includegraphics[height=0.24\textwidth]{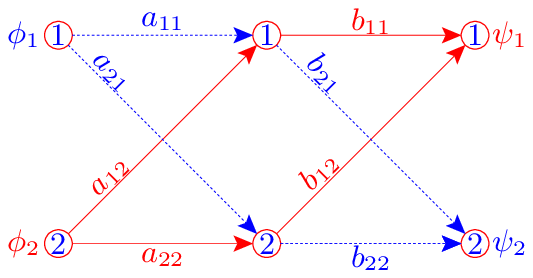}
&\raisebox{0.1\textwidth}{\Math{\sim}}
&
\includegraphics[height=0.24\textwidth]{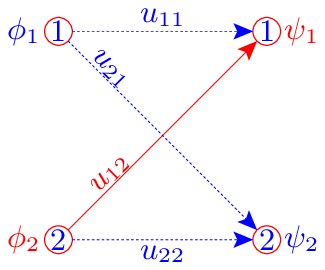}
\\
\Math{\Updownarrow}&&\Math{\Updownarrow}
\\
\Math{
BA=
\bmat
{\color{blue}b_{11}a_{11}+b_{12}a_{21}}
&
{\color{red}b_{11}a_{12}+b_{12}a_{22}}
\\[5pt]
{\color{blue}b_{21}a_{11}+b_{22}a_{21}}
&
{\color{blue}b_{21}a_{12}+b_{22}a_{22}}
\emat
}~~~~~~~~~
& \Math{\sim}~~ &
\Math{
U=
\bmat
{\color{blue}u_{11}}&{\color{red}u_{12}}
\\[5pt]
{\color{blue}u_{21}}&{\color{blue}u_{22}}
\emat
}~~~~~~~~
\end{tabular}
\end{center}
Following the Feynman rules, the transition from, say, the state \Math{\phi_2} to the state \Math{\psi_1} is given by the sum along two paths: \Math{b_{11}a_{12}+b_{12}a_{22}}.
But just the same expression is the element \Math{u_{12}} of the product of matrices \Math{U=BA}.
The general case of an arbitrary number of states and an arbitrary number of time steps is easily derived from this elementary example by mathematical induction on these numbers.
\par
In the case of non-commutative connection, formula \eqref{Enamplgen} corresponds to a non-Abelian gauge theory.
The above argument about the correspondence between the Feynman quanti\-zation and matrix multiplication is applicable here as well.
We only need to treat the evolution matrices \Math{A, B} and \Math{U} as block matrices with noncommuting elements, which are matrices from the representation \Math{\Rep{\iG}}.  
\par
The Feynman rules --- ``\emph{multiply successive events}'' and ``\emph{sum up alternative histories}''
--- are suitable to represent multiplication of matrices with arbitrary elements.
In the usual formulation of Feynman's quantization, elements of matrices of elementary quantum transitions have a very special form.
Namely, all the elements are complex numbers lying on the unit circle: \Math{\e^{iLdt}}.
These numbers can be treated as continuous analogs of roots of unity.
The fact that they cannot vanish means that it is impossible to realize systems that do not allow transitions between some states, for example, between states at widely separated points of space.  
Moreover, an important case of the trivial group of internal symmetries \Math{\iG=\set{\id}} is discarded.
In this case, quantum evolutions are described by permutation matrices.
An \Math{n\times{}n} permutation matrix contains \Math{n} units and \Math{n^2-n} zeros.
Zero, of course, can not be represented as the exponential of a number.
\par
The study of quantum behavior of discrete dynamical systems with spatial structure is a
combinatorially hard task. 
Constructing relevant models essentially depends on a number of assumptions that must be imposed on the structure of space and properties of quantum transitions to make the models tractable.
One of the approaches to constructing quantum models on discrete spaces is described in 
Appendix \hyperref[Enquantumonregulargraphs]{C}. 
Quantum transitions in these models are defined locally,  namely, in
the neighborhoods of regular graphs. This makes the models, in a certain sense, similar to cellular automata.
\par
Further, we will consider the problem of the quantum behavior of discrete systems in general, i.e., ignoring  possible presence of the space in the structure of a complete set of states.
Note that in many physical problems, for example, in quantum computing, identifying any spatial structure in the set of states is not essential.
\subsection{Quantization of Discrete Systems}
Thus, we will stick to the traditional formulations of quantum mechanics, in which the evolution of a quantum dynamical system from the initial to the final state is described by the \emph{matrix of evolution} \Math{U}:
\Math{\barket{\psi_{\tin}}\rightarrow\barket{\psi_{\tfin}}=U\barket{\psi_{\tin}}}. 
The evolution matrix can be represented as the product of the matrices corresponding to elementary time steps:
\Math{U=U_{\tfin\leftarrow\tfin-1}\cdots{}U_{t\leftarrow{}t-1}\cdots{}U_{1\leftarrow0}.}
\subsubsection{Standard and ``finite'' versions of quantum mechanics}
For a more formal justification of the legitimacy of replacement of infinite groups by finite ones in
physical problems, we can use the notion of the \emph{residual finiteness}.
A (infinite) group \Math{G} is called \emph{residually finite} \cite{Magnus}, if for every \Math{g\in{}G}, \Math{g\neq\id}, there exists a homomorphism \Math{\phi} from \Math{G} onto a finite group \Math{H}, such that \Math{\phi\vect{g}\neq\id}.
This means that any relation between the elements of \Math{G} can be modeled by a relation between
the elements of a finite group.
Here there is an analogy with the widely used technique in physics, when for solving some problem 
an infinite space is replaced, for example, by a torus whose size is sufficient to accomodate data relating to the problem. 
In fact, all constructive infinite groups used in physics are residually finite.
A typical example of an infinite residually finite group is the \emph{braid group}, which is popular in theoretical physics.
According to the standard construction \cite{PraSos}, the braid group on \Math{n} strands \Math{B_n} is generated by the elements \Math{b_1,\ldots,b_{n-1}}, satisfying the relations:
\begin{align}
b_{i}b_{i+1}b_{i}&=b_{i+1}b_{i}b_{i+1},~~i=1,\ldots,n-1~~
\text{~~~~~ \textit{Yang-Baxter equation},}\label{EnYB}\\
b_{i}b_{j}&=b_{j}b_{i},~~\cabs{i-j}\geq2~~
\hspace{55pt}\text{~~~~~~ \textit{far commutativity}.}\label{EnDC}	
\end{align}
Note that the braid group is the closest infinite ``relative'' of the permutation group \Math{\SymG{n}}, which is obtained from \Math{B_n} by adding the relations \Math{b_{i}^2=\id} to \eqref{EnYB} and \eqref{EnDC}.
\par
We now turn to the unitary groups that play a central role in quantum physics.
Here, since we start from the continuum, we need two steps to go from
the ``infinite'' to the ``finite'':
\begin{enumerate}
	\item
The theory of quantum computing \cite{Nielsen} says that there exist universal finite sets of unitary operators called \emph{quantum gates} that can generate dense subsets in the set of all unitary operators acting in the corresponding Hilbert space.
The quantum gates act in Hilbert spaces of low fixed dimensions	(single-, double-qubit gates, etc.),
but they can be used as the building blocks for constructing unitary matrices acting in Hilbert spaces of arbitrary dimension.
These matrices can be considered as the generators of a finitely generated group, which approximates to arbitrary precision (is densely embedded in) the general unitary group.
	\item 
According to the theorem of A.I. Mal'cev \cite{Malcev}, every finitely generated group of matrices over (any) field is \emph{residually finite}.
\end{enumerate} 
Thus,  unitary groups can be replaced by finite groups without any harm to physics as an
empirical science.
\par
In order to reproduce quantum mechanics in the constructive ``finite'' formulation, we will adhere to the Occam principle, i.e., we will introduce new elements of the description only if they are really needed.
Below is a comparison of the main elements of standard quantum mechanics with their constructive counterparts.
\begin{enumerate}
	\item
\begin{enumerate}
	\item 
Standard quantum mechanics deals with	\emph{unitary operators} \Math{U} acting in a \emph{Hilbert space}
\Math{\Hspace} over the field of complex numbers \Math{\C}.
The elements of this space \Math{\barket{\psi}\in\Hspace} are called ``\emph{states}'', ``\emph{state vectors}'', ``\emph{wave functions}'', ``\emph{amplitudes}'', etc. The operators \Math{U} belong to the general unitary group \Math{\Aut{\Hspace}} acting in \Math{\Hspace}. 
	\item 
In ``finite'' quantum mechanics,%
\footnote{The term ``finite quantum mechanics'' is  sometimes used in the literature in a significantly different way. So is called an approach (see, e.g., \cite{AthaFQM,FlorFQM,FlorNiFQM}) that emerged from the H. Weyl proposal \cite{WeylFQM} to replace the phase space \Math{\vect{p,q}} in quantum mechanics by a finite discrete torus.
We use here this term for lack of another sufficiently concise and meaningful.}
Hilbert space over the field \Math{\C} is replaced by \Math{\adimH}-dimensional Hilbert space \Math{\Hspace_\adimH} over an \emph{Abelian number field} \Math{\NF} --- i.e., an extention of rationals \Math{\Q} with Abelian Galois group  \cite{Shafarevich}.	
Operators \Math{U} belong now to a unitary representation \Math{\repq} of a \emph{finite group} \Math{\wG=\set{\wg_1,\ldots,\wg_{\wGN}}} in the space \Math{\Hspace_\adimH}.
The field \Math{\NF} is determined by the structure of the group \Math{\wG} and its representation \Math{\repq}.
In fact, \Math{\NF} is the \emph{minimal extension} of natural numbers, which is sufficient to describe the quantum behavior.
\end{enumerate}
	\item
In both versions of quantum mechanics, quantum \emph{particles} are associated with \emph{uni\-tary representations} of certain symmetry groups.
The representations, according to their dimensions, are called ``\emph{singlets}'',
 ``\emph{doublets}'', ``\emph{triplets}'', etc. Multidimensional repre\-sentations describe \emph{spin}.
	\item
Quantum \emph{evolution} is described by a unitary transformation of the \emph{initial} state vector \Math{\barket{\psi_{in}}} into the \emph{final} \Math{\barket{\psi_{out}}=U\barket{\psi_{in}}}.	
\begin{enumerate}
	\item 
In standard quantum mechanics, the elementary step of evolution	in continuous time is described by the \emph{Schr\"{o}dinger equation}
\Mathh{\displaystyle{}i\frac{\mathrm{d}}{\mathrm{d}t}\barket{\psi}=H\barket{\psi},}
where \Math{H} is an \emph{Hermitian} operator called \emph{energy operator} or \emph{Hamiltonian}.
	\item
In finite quantum mechanics, there are only a finite number of possible evolu\-tions:	
\Mathh{U_j\in\set{\repq\vect{\wg_1},
\ldots,\repq\vect{\wg_j},\ldots,\repq\vect{\wg_\wGN}}.}
It is clear that here there is no need for any analog of the Schr\"{o}dinger equation.
However, formally one can always introduce the Hamiltonian using the formula
\Math{H_j=i\ln{}U_j\equiv\sum\limits_{k=0}^{p-1}\lambda_k{}U_j^k},
where \Math{p} is \emph{period} of the operator \Math{U_j} (i.e., minimal \Math{p>0}, such that \Math{U_j^p=\idmat}), \Math{\lambda_k} are easily calculable coefficients.%
\footnote{These coefficients contain a \emph{non-algebraic} element, namely, the number \Math{\pi}, which is an \emph{infinite} sum of the elements from \Math{\NF}. In other words, \Math{\lambda_k} are elements of a \emph{transcendental extension} of the field \Math{\NF} --- logarithmic function is essentially a construction of continuous mathematics that deals with the actual infinities.
}
\end{enumerate}
	\item
In both versions of quantum mechanics, quantum mechanical \emph{experiment} (\emph{observa\-tion}, 
``\emph{measurement}''%
\footnote{This is not a quite relevant term since actually quantum-mechanical ``measurements'' are always reduced to counting numbers of certain events.}
) is reduced to comparison of the state \Math{\barket{\psi}} of the \emph{system} with the state \Math{\barket{\phi}} of the \emph{instrument}.
	\item
In both versions of quantum mechanics --- in accordance with the \emph{Born rule} --- the probability to detect a particle, described by the state \Math{\barket{\psi}}, by apparatus, configured to the state \Math{\barket{\phi}}, is equal to
\begin{equation}
\ProbBorn{\phi}{\psi} = \frac{\textstyle{\cabs{\inner{\phi}{\psi}}^2}}
{\textstyle{\inner{\phi}{\phi}\inner{\psi}{\psi}}}.
\label{EnBornEn}
\end{equation}
However, the case of finite quantum mechanics requires some \emph{conceptual refinement}.
In the ``finite'' background, the only meaningful interpretation of probability is the \emph{frequency interpretation}: probability is the ratio of \emph{favorable} combinations of elements of a system to the total number of considered combinations.
Since we are dealing with combinations of elements of finite sets, we expect that, if everything is arranged properly, formula \eqref{EnBornEn} should provide \emph{rational numbers}.
We will use this in further considerations as one of the guiding principles. 
	\item
Quantum \emph{observables} are described by \emph{Hermitian operators} in Hilbert spaces.	
In finite quantum mechanics, these operators can be expressed as elements of the \emph{group algebra} representation
\begin{equation*}
	A=\sum\limits_{k=1}^{\wGN}\alpha_k{}\repq\!\vect{\wg_k}.
\end{equation*}
Of course, the requirement of hermiticity may impose certain restrictions on the coefficients \Math{\alpha_k}. 
\end{enumerate}
Note that all other elements of the finite quantum theory can be obtained from the standard one simply by rewriting respective expressions.
For example, as is well known, the \emph{Heisenberg uncertainty principle} follows from the \emph{Cauchy} (–Bunyakovsky–Schwarz) \emph{inequality}
\begin{equation*}
	\inner{A\psi}{A\psi}\inner{B\psi}{B\psi}\geq\cabs{\inner{A\psi}{B\psi}}^2,
\end{equation*}
which holds for any Hilbert space over any field.
It is clear that the Cauchy inequality is equivalent to the standard property of any probability 
 \Math{\ProbBorn{A\psi}{B\psi}\leq1}.
\subsubsection{Permutations, representations and numbers}
It is easy to describe all the sets on which a finite group \Math{\wG=\set{\wg_1,\!\ldots\!,\wg_{\wGN}}} acts transitively.
Any such set \Math{\wS=\set{\ws_1,\!\ldots\!,\ws_\wSN}} is in one-to-one correspondence with the set of (\emph{right} \Math{H\backslash\wG} or \emph{left} \Math{\wG/H}) cosets of a subgroup \Math{H\leq\wG} \cite{Hall}.
The set \Math{\wS} is called a \emph{homogeneous space} of the group \Math{\wG} (or, briefly, \Math{\wG}-space).   
The action of \Math{\wG} on \Math{\wS} is \emph{faithful}, if the subgroup \Math{H} does not contain
a normal subgroup of the group \Math{\wG}.
We can write the action in the form of permutations
\begin{equation}
	\pi(g)=\dbinom{\ws_i}{\ws_ig}\sim\dbinom{Ha}{Hag},
	\hspace*{20pt}g,a\in{}\wG,~~~i=1,\ldots,\wSN,
	\label{Enperm}
\end{equation}
or, equivalently, in terms of \Math{\vect{0,1}}-matrices
\begin{equation}
\pi(g)\rightarrow\regrep(g)=
\Mone{\regrep(g)_{ij}},\text{~~where~~} \regrep(g)_{ij}=\delta_{\ws_ig,\ws_j};
~~ i,j=1,\ldots,\wSN.
\label{EnpermrepEn}
\end{equation}
Here \Math{\delta_{\alpha,\beta}} denotes the Kronecker delta on the set \Math{\wS}.
Mapping \eqref{EnpermrepEn} is called a \emph{permu\-tation representation}.
\par
The maximal transitive set \Math{\wS} is equivalent to the set of all elements of the group \Math{\wG}, i.e., to the set of cosets of the trivial subgroup \Math{H=\set{\id}}.
Appropriate action and matrix representation are called \emph{regular}.
One of the central theorems of the representation theory  (see Appendix \hyperref[Enirreps]{D}) 
states that \emph{any irreducible representation of a finite group is contained in the regular representation}. 
\par
Representation \eqref{EnpermrepEn} makes sense over any number system containing 0 and 1.
The most natural number system is the \emph{semiring of natural numbers} \Math{\N=\set{0,1,2,\ldots}.}
Having this semiring we can endow the elements of \Math{\wS} by \emph{counters} interpreting them as ``\emph{multiplicities of occurrence}'' (or ``\emph{occupation numbers}'') of elements \Math{\ws_i} in the state of a system that contains these elements.
This state can be written as a vector with natural component
\begin{equation}
	\barket{n} = \Vthree{n_1}{\vdots}{n_{\wSN}}.
\label{Ennatvect}	
\end{equation}
Thus, we come to the representation of the group \Math{\wG} in the \Math{\wSN}-dimensional \emph{module} \Math{\natmod_\wSN} over the semiring \Math{\N}.
Action \eqref{EnpermrepEn} on the vector \eqref{Ennatvect} is simply a permutation of its components.
For further development, we need to turn the module \Math{\natmod_\wSN} into \Math{\wSN}-dimensional Hilbert space by extending the semiring \Math{\N} to a field.
\par
The field of complex numbers \Math{\C} is considered to be the main field in representation theory (and hence, in quantum mechanics).
The reason for this choice is the algebraic closedness of the field \Math{\C}, which means, in particular, that no problem can arise in solving characteristic equations and, hence, in the entire linear algebra. 
However, the field \Math{\C} is excessively large --- most of its elements are non-constructive and, thus, useless in application to the empirical reality.
So we must examine the situation more carefully.
\par
First of all, we do not need to solve \emph{arbitrary} characteristic equations: 
any linear representation is a subrepresentation of some permutation representation, all eigenvalues of which are \emph{roots of unity}.
This is seen from easily calculable \emph{characteristic polynomial} of permutation matrix \eqref{EnpermrepEn}
\begin{equation*}
\chi_{\regrep(g)}\vect{\lambda}=\det\vect{\regrep(g)-\lambda\idmat}
=\vect{\lambda-1}^{k_1}\vect{\lambda^2-1}^{k_2}\cdots\vect{\lambda^n-1}^{k_n},
\end{equation*}
where \Math{k_i} is the number of cycles of length \Math{i} in permutation \eqref{Enperm}.
To ensure unitarity of representations, \emph{square roots} of their dimensions are used as normalizing factors.
In fact, the irrationalities of both types, i.e., roots of unity and square roots of integers have the same nature --- they are all \emph{cyclotomic integers}, i.e., integer linear combinations of the roots of unity. 
Using standard identities for roots of unity, the negative coefficients in the cyclotomic integers can always be replaced by positive.
Thus, the primary elements of the number system, that we are going to construct, are \emph{natural numbers} and linear combination with natural coefficients of \emph{roots of unity} of some degree  \Math{\period}, which depends on the structure of the group \Math{\wG}.
Everything else is constructed via completely formal mathematical procedures.
The degree \Math{\period} in the mathematical literature is usually called the \emph{conductor}.
Generally, the conductor is a multiple of some divisor of the group \emph{exponent}, which is defined as the least common multiple of the orders of elements of \Math{\wG}.
\par
From any set of cyclotomic integers one can construct an \emph{Abelian number field} \Math{\NF}, which contains all the elements of this set.
In particular, the minimal Abelian number field containing a given set of irrationalities can be computed by using the computer algebra system \GAP \cite{gapEn}.
The command \texttt{\textbf{Field(\textit{gens}\!)}} of this system returns the minimum field containing all the elements from the list of irrationalities \texttt{\textbf{\textit{gens}}}. 
The Kronecker–Weber theorem states that any Abelian (in particular, quadratic) extension of rational numbers is a subfield of a cyclotomic field.
\par
Let us dwell on the construction of cyclotomic fields. 
A \emph{\Math{\period}th root of unity} is any solution of the \emph{cyclotomic equation} \Math{r^\period=1}. 
Any root of unity is a power of a \emph{primitive} root of unity, i.e., root of unity whose period is equal exactly to  
\Math{\period}.
All primitive \Math{\period}th roots of unity (and only they) are solutions of the equation \Math{\Phi_\period\vect{r}=0}, 
where  \Math{\Phi_\period\vect{r}} is so-called \Math{\period}th \emph{cyclotomic polynomial}. \Math{\Phi_\period\vect{r}}  is an irreducible over \Math{\Z} divisor of the binomial \Math{r^\period-1}. 
Linear combinations of the roots of unity with integer coefficients form a ring of cyclotomic integers \Math{\aNumbers_\period}.
We introduced this notation since, in fact, for the coefficients of cyclotomic integers it is sufficient to take only \emph{natural} numbers. 
In the case \Math{\period\geq2},  negative integers can be introduced using the identity
\Math{\vect{-1}=\sum\limits_{k=1}^{p-1}\runi{}^{\frac{\period}{p}k}}, where \Math{p}  is an arbitrary divisor of the period \Math{\period}. 
The introduction of negative numbers is not conceptually necessary.
However, it gives technical advantages, allowing the use of advanced methods of polynomial algebra.
The \Math{\period}th \emph{cyclotomic field} \Math{\Q_\period} is introduced as the \emph{field of fractions} of the ring \Math{\aNumbers_\period}. 
The field \Math{\Q_\period} can be represented as
\begin{equation}
	\Q_\period=\Q\ordset{r}/\braket{\Phi_\period\vect{r}}.
	\label{Enpowerbas}
\end{equation}
From this representation and properties of cyclotomic polynomials it follows that the field \Math{\Q_\period} is a vector space (moreover, an algebra) of the dimension \Math{\varphi\vect{\period}} over the rationals \Math{\Q}.
Here \Math{\varphi\vect{\period}} is the \emph{Euler function} (also known as the \emph{totient}), defined as the number of positive integers not exceeding \Math{\period}  and coprime with \Math{\period}  
(by definition, 1 is  coprime with any integer).
A natural basis of the vector space \Math{\Q_\period} is the \emph{power basis}.
In this basis, the elements of \Math{\Q_\period} are represented as rational linear combinations of the powers of a primitive root of unity restricted by the degree of the cyclotomic polynomial.
This basis is \emph{integral} in the sense that any cyclotomic integer can be represented in it as an integral vector.
Note that in mathematics other integral bases are often used for representing cyclotomic numbers.
\par
An Abelian number field \Math{\NF} is a subfield of the field \Math{\Q_\period}, fixed by additional symmetries called the \emph{Galois automorphisms}.
The Galois automorphism \Math{*k}  is the mapping defined by the transformation \Math{\runi{\period}\rightarrow\runi{\period}^k}, where \Math{\runi{\period}} is a \Math{\period}th primitive root of unity,
\Math{k}  is a coprime with \Math{\period} integer within the range \Math{1\leq{}k<\period}.
\par
\begin{figure}
	\centering
\begin{tabular}{cc}
\hspace*{-10pt}\Math{\period=12}&\hspace*{-20pt}\Math{\period=7}\\[-10pt]
\hspace*{-10pt}\includegraphics[width=0.507\textwidth]{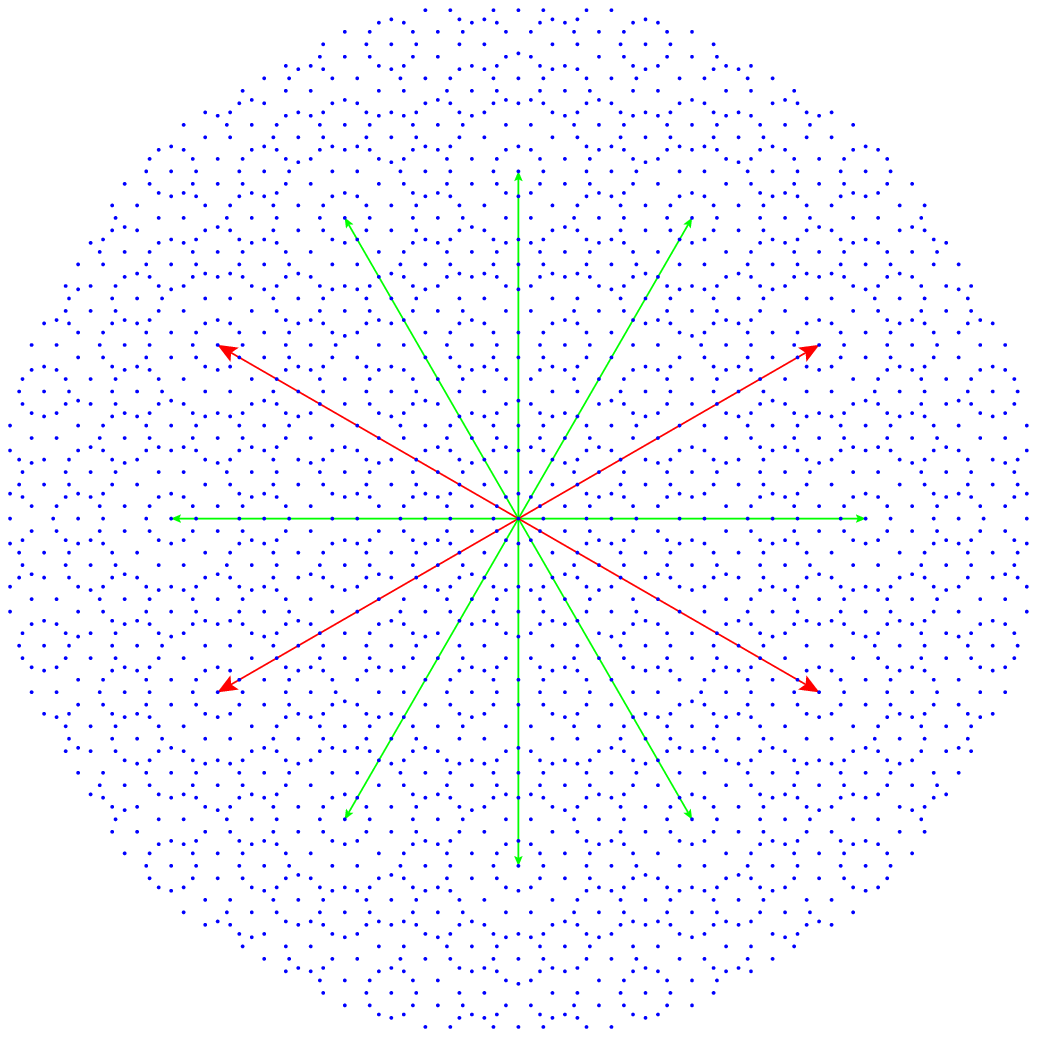}
&
\hspace*{-20pt}\includegraphics[width=0.507\textwidth]{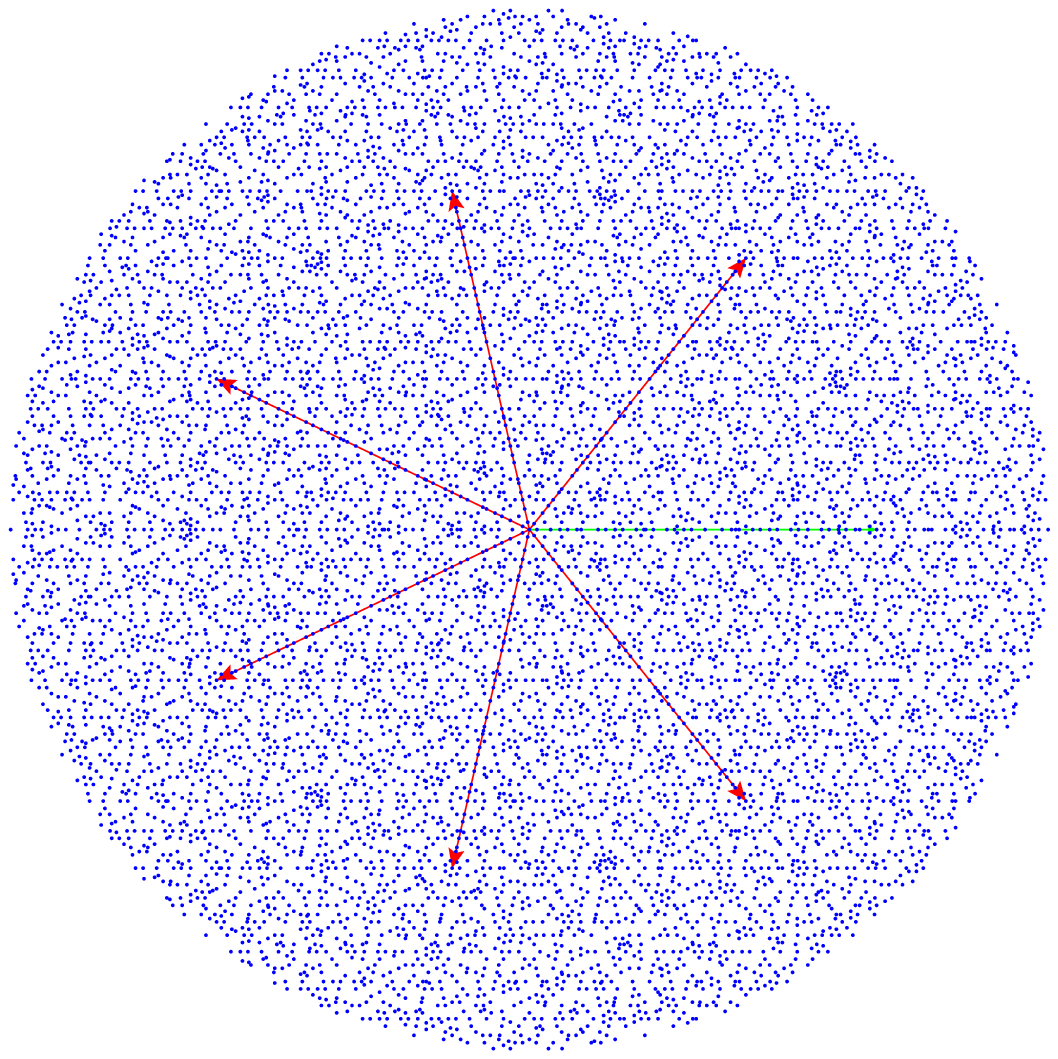}\\[-10pt]
\end{tabular}
\caption{Embedding \Math{\aNumbers_\period} in \Math{\C}.
Primitive roots are indicated by red arrows.} 
\label{EninC}		
\end{figure}

 
Cyclotomic numbers can be embedded in the complex field \Math{\C}, but this is not necessary.
Purely algebraic properties of cyclotomic numbers are sufficient for all manipulations in the Hilbert space \Math{\Hspace_\wSN}  and its subspaces.
For example, the complex conjugate of a cyclotomic number is defined by the transformation \Math{\cconj{\runi{\period}^m}=\runi{\period}^{\period-m}}.
However, for illustration, we present in Fig. \ref{EninC} examples of embedding of cyclotomic integers into the complex plane.
Clearly, the cyclotomic fractions, i.e., elements of the field \Math{\Q_\period}, are filling the complex plane everywhere densely if \Math{\period\geq3}.
By the way, this gives a trivial explanation of the emergence of complex numbers in quantum mechanics. 
\par
All irrationalities are intermediate elements of quantum description, disappearing in the final expressions for quantum observables.
This statement is a constructive refinement of the relationship between complex and real numbers in standard quantum mechanics, where intermediate values may be complex, but the observables must be real.
\subsubsection{Representation of quantum evolution by permutations}
From the above it follows that any \Math{\adimH}-dimensional representation \Math{\repq} can be extended to \Math{\wSN}-dimensional representation \Math{\widetilde{\repq}} corresponding to the \emph{permutation action} of the group \Math{\wG} on \Math{\wSN}-element set \Math{\wS=\set{\ws_1,\ldots,\ws_{\wSN}}}.
This means that \Math{\transmatr^{-1}\regrep\transmatr=\widetilde{\repq}}, where \Math{\regrep}  is  permutation representation \eqref{EnpermrepEn}, and \Math{\transmatr} is a transformation matrix.
It is clear that \Math{\wSN\geq\adimH}. 
If  \Math{\wSN=\adimH}, then \Math{\repq} is itself a permutation representation.
In the proper case, i.e., when \Math{\wSN>\adimH}, the permutation representation has the structure
\Mathh{
\transmatr^{-1}\regrep\transmatr
=\Vtwo{
\left.
\begin{aligned}
\!\IrrRep{1}&\\[-2pt]
&\hspace{8pt}\mathrm{V}
\end{aligned}
\right\}\Hspace_{\wSN-\adimH}
}{
\left.
\hspace{29pt}
{\repq}
\right\}\Hspace_{\adimH}
},\hspace{10pt}
\Hspace_{\wSN} = \Hspace_{\wSN-\adimH}\oplus\Hspace_{\adimH},
}	
where \Math{\IrrRep{1}} is the trivial one-dimensional representation, mandatory for any permutation representation;
the component \Math{\mathrm{V}} may be missing.
The data in the spaces  \Math{\Hspace_{\adimH}} and \Math{\Hspace_{\wSN-\adimH}}
are \emph{independent} since both spaces are invariant subspaces of \Math{\Hspace_{\wSN}}. So we can consider the data in \Math{\Hspace_{\wSN-\adimH}} as ``hidden parameters'' with respect to the data in \Math{\Hspace_{\adimH}}.
Thus, \emph{any quantum problem} in \Math{\adimH}-dimensional Hilbert space can be reformulated in terms of permutations of \Math{\wSN} objects.
\par
A trivial approach would be to set arbitrary (e.g., zero) data in the complementary subspace \Math{\Hspace_{\wSN-\adimH}}. 
This approach is not interesting since it is not falsifiable by means of the standard quantum mechanics. 
In fact, it leads to the standard quantum mechanics \emph{modulo} the empirically unobservable distinction between the ``finite'' and the ``infinite''. 
The only difference is technical: we can replace the linear algebra in the 
\Math{\adimH}-dimensional space \Math{\Hspace_{\adimH}} by permutations of \Math{\wSN} things.%
\footnote{One could assume that replacement of the linear algebra operations by much more simple manipulations with permutations may give some computational advantages. However, the difference between  \Math{\adimH} and \Math{\wSN} may be quite large.
For example, the Fischer-Griess Monster, the largest sporadic finite simple group, has minimal faithful linear representation of the dimension \Math{\approx2\cdot10^5}, whereas its minimal faithful permutation representation has the degree of about \Math{10^{20}}.}
\par
A more promising approach requires some changes in the concept of quantum amp\-litudes. We assume \cite{Kornyak12b} that
they are projections onto invariant subspaces of vectors of multiplicities
(``occupation numbers'') of elements of the set \Math{\wS} on which the group \Math{\wG} acts by permutations.
Thus, we assume that quantum states of a system and apparatus in the permutation basis are vectors with natural components
\Mathh{
\barket{n} = \Vthree{n_1}{\vdots}{n_{\wSN}} \text{~and~}
\barket{m} = \Vthree{m_1}{\vdots}{m_{\wSN}}.}
According to the Born rule, the probability of detecting the state \Math{\barket{n}} by an instrument configured to the state \Math{\barket{m}} is 
\begin{equation*}
    \ProbBorn{m}{n}=\frac{\vect{\sum_i{m_i}n_i}^2}{\sum_i{m_i}^2\sum_i{n_i}^2}.
\end{equation*}
It is clear that the value of this expression on the natural vectors  \Math{\barket{n}} and \Math{\barket{m}} is a non-negative rational number.
Destructive quantum interference among  nonzero natural vectors in the permutation space is possible only if for any \Math{i} either \Math{m_i=0} or \Math{n_i=0}.
In the proper \emph{invariant subspaces} of the permutation representation  the destructive interference is possible also for vectors in which all components are strictly  greater than zero natural numbers.
Let us first consider a simple example.
\subsection{Examples of Finite Quantum Systems}
\subsubsection{A detailed example: the smallest noncommutative group \Math{\SymG{3}}}
We take \Math{\SymG{3}} as an example because of the ease of analysis.
However, this group has important applications in physics.
In particular, it is used to describe the so-called \emph{tribimaximal mixing} in neutrino oscillations \cite{HPS02,HS03} (for more details see Appendix \hyperref[Enappflavor]{E}).
The group consists of six elements that can be represented by the following permutations of three objects
\begin{equation*}
\wg_1=\vect{}\!,~\wg_2=\vect{2,3}\!,~\wg_3=\vect{1,3}\!,~\wg_4=\vect{1,2}\!,
    ~\wg_5=\vect{1,2,3}\!,~\wg_6=\vect{1,3,2}.
\end{equation*}
The exponent of \Math{\SymG{3}} is equal to six, since the orders of its elements are 2 and 3.
The group can be generated by different pairs of its elements.
For example, as a generating set, one can take the pair \Math{\set{\wg_2, \wg_6}}.
The group \Math{\SymG{3}} splits into three conjugacy classes:
\begin{equation*}
    \class{1}=\set{\wg_1},~~\class{2}=\set{\wg_2,~\wg_3,~\wg_4},
    ~~\class{3}=\set{\wg_5,~\wg_6}.
\end{equation*}
The character table (see Appendix \hyperref[Enirreps]{D}) of \Math{\SymG{3}} has the form
\begin{center}
    \text{\begin{tabular}{c|crr}
    &\Math{\class{1}}&\Math{\class{2}}&\Math{\class{3}}\\\hline
    \Math{\chi_1}&1&1&1\\
    \Math{\chi_2}&1&-1&1\\
    \Math{\chi_3}&2&0&-1
    \end{tabular}\enspace.}
\end{center}
Following the tradition adopted in the physics literature, we will denote \emph{irreducible} 
repre\-sentations by their dimensions in bold font.
Thus, we have three irreducible representations \Math{\IrrRep{1}, \IrrRep{1'}} and 
\Math{\IrrRep{2}}. The only faithful of them is the representation \Math{\IrrRep{2}}.
To denote \emph{permutation} representations that play an important role in our approach, 
we will also use their dimen\-sions in bold, but with additional underscore.
\par
The generators \Math{\set{\wg_2, \wg_6}} have the following matrix form
\begin{equation*}
    P_2=\Mthree{1}{~\cdot}{~\cdot}{\cdot}{~\cdot}{~1}{\cdot}{~1}{~\cdot}
    \text{~~and~}
    ~P_6=\Mthree{\cdot}{~\cdot}{~1}{1}{~\cdot}{~\cdot}{\cdot}{~1}{~\cdot}.
\end{equation*}
These matrices have, respectively, the following sets of eigenvalues:
\Math{\vect{1, 1, -1}} and \Math{\vect{1, \runi{3}, \runi{3}^2}},
where \Math{\runi{3}} is a third primitive root of unity defined by the cyclotomic polynomial \Math{\Phi_3\vect{r} = 1+r+r^2}.
\par
As noted above, any permutation representation has one-dimensional invariant sub\-space with the basis vector \Math{\vect{1,\ldots,1}^\mathrm{T}}.
Thus, the only possible decomposition of the permutation representation is \Math{\PermRep{3}\cong\IrrRep{1}\oplus\IrrRep{2}},
or in the explicit matrix form
\begin{equation}
    \widetilde{U}_j=\Mtwo{\IrrRep{1}}{0}{0}{U_j},~~j =1,\ldots,6,
    \label{EnS3permqEn}
\end{equation}
where the matrices \Math{U_j} are elements of the faithful irreducible representation \Math{\IrrRep{2}}. 
\par
To construct  decomposition \eqref{EnS3permqEn} one should determine the matrices \Math{U_j} and \Math{\transmatr} such that
\Math{\widetilde{U}_j=\transmatr^{-1}P_j\transmatr}. 
We can always assume that all these matrices are unitary.
It is clear that to construct the matrices  \Math{U_j} and \Math{\transmatr} it is sufficient to use only the permutation matrices of generators.%
\footnote{\label{EnMeatAxePage}
Note that the problem of constructing common invariant subspaces for several matrices, or, in a more general sense, 
the problem of splitting modules over associative algebras into irreducible submodules is algorithmically rather nontrivial.
A universal enough algorithm --- called \textbf{\emph{MeatAxe}} \cite{Holt} and implemented in the computer algebra systems \GAP and \Magma\! --- is developed only for algebras over finite fields.
However, in the case of group algebras, at least for small groups, necessary constructions can be carried out, as a rule, relatively easy by the
``trial and error'' method over the fields of zero characteristic,
in particular, over Abelian number fields.}
Decomposition \eqref{EnS3permqEn}  can be constructed in different ways.
Clearly, we can always make diagonal one of the matices \Math{U_j}. 
\par
If we start with diagonalization of \Math{P_6}, then we
obtain the following set of matrices \Math{U_j}:
\begin{align}
        U_1=\Mtwo{1}{0}{0}{1},~U_2=\Mtwo{0}{\runi{3}^2}{\runi{3}}{0},
        ~U_3=\Mtwo{0}{\runi{3}}{\runi{3}^2}{0},\nonumber\\
        \label{Enumatrices}\\[-12pt]
        ~U_4=\Mtwo{0}{1}{1}{0},
        ~U_5=\Mtwo{\runi{3}^2}{0}{0}{\runi{3}},
        ~U_6=\Mtwo{\runi{3}}{0}{0}{\runi{3}^2}.\nonumber
\end{align}
The transformation matrix (up to inessential arbitrariness in choosing its elements) takes the form
\begin{equation}
\transmatr=\frac{1}{\sqrt{3}}
    \Mthree{1}{1}{\runi{3}^2}
     {1}{\runi{3}^2}{1}
     {1}{\runi{3}}{\runi{3}},~~~~
\transmatr^{-1}=\frac{1}{\sqrt{3}}
    \Mthree{1}{1}{1}
     {1}{\runi{3}}{\runi{3}^2}
     {\runi{3}}{1}{\runi{3}^2}.
\label{EntransS3monomial}
\end{equation}
\par
The minimal Abelian number field containing all elements of matrices \eqref{Enumatrices} and \eqref{EntransS3monomial} is the cyclotomic field \Math{\Q_{12}}.
Thus, we can rewrite, say, matrix  \eqref{EntransS3monomial} in terms of elements from \Math{\Q_{12}}:    
\begin{equation}
\transmatr=\frac{1}{3}
    \Mthree{2\runi{12}+\runi{12}^{9}}{2\runi{12}+\runi{12}^{9}}
    {\runi{12}^7+\runi{12}^9}
     {2\runi{12}+\runi{12}^{9}}{\runi{12}^7+\runi{12}^9}
     {2\runi{12}+\runi{12}^{9}}
     {2\runi{12}+\runi{12}^{9}}{2\runi{12}^3+\runi{12}^7}{2\runi{12}^3+\runi{12}^7},
     \label{EntransS3monomialcyclo}
\end{equation}
where \Math{\runi{12}} is a primitive 12th root of unity, i.e., an arbitrary (abstract) solution of the equation \Math{\Phi_{12}\vect{r} \equiv 1-r^2+r^4 =0}.
In \eqref{EntransS3monomialcyclo} we use the ``natural'' representation of cyclotomic integers.
Reduction \emph{modulo} the cyclotomic polynomial \Math{\Phi_{12}\vect{r}} gives the ``integer'' representation in the power basis, i.e., representation allowing negative coefficients, but with minimal powers of \Math{\runi{12}}:
 \Math{\runi{12}}:
\begin{equation*}
\transmatr=\frac{1}{3}
    \Mthree{2\runi{12}-\runi{12}^3}{2\runi{12}-\runi{12}^3}
    {-\runi{12}-\runi{12}^3}
     {2\runi{12}-\runi{12}^3}{-\runi{12}-\runi{12}^3}
     {2\runi{12}-\runi{12}^3}
     {2\runi{12}-\runi{12}^3}{-\runi{12}+2\runi{12}^3}{-\runi{12}+2\runi{12}^3}.
\end{equation*}
\par
Diagonalization of \Math{P_2} leads to another form of decomposition  \eqref{EnS3permqEn}.
In this case, the matrices of generators are
\begin{equation}
U'_2=\Mtwo{1}{0}{0}{-1},
~~~~U'_6=\Mtwo{-\frac{1}{2}}{\frac{\sqrt{3}}{2}}{-\frac{\sqrt{3}}{2}}{-\frac{1}{2}}.
\label{EnUS3}
\end{equation}
The transformation matrix takes the form
\begin{equation}
\transmatrprim
=\Mthree{\frac{1}{\sqrt{3}}}{~\,\,\sqrt{\frac{2}{3}}}{~0}
          {\frac{1}{\sqrt{3}}}{\,-\!\frac{1}{\sqrt{6}}}{-\!\frac{1}{\sqrt{2}}}
          {\frac{1}{\sqrt{3}}}{\,-\!\frac{1}{\sqrt{6}}}{~\,\,\frac{1}{\sqrt{2}}},
~~~~
\transmatrprim^{-1}
=\Mthree{\frac{1}{\sqrt{3}}}{~\,\frac{1}{\sqrt{3}}}{~\frac{1}{\sqrt{3}}}
        {\sqrt{\frac{2}{3}}}{\,-\!\frac{1}{\sqrt{6}}}{-\!\frac{1}{\sqrt{6}}}
        {~0}{\,-\!\frac{1}{\sqrt{2}}}{~\,\,\frac{1}{\sqrt{2}}}.
\label{EntransS3sqrt}
\end{equation}
This matrix is used in particle physics in the phenomenology of neutrino oscillations, where it is
known as the \emph{tribimaximal mixing}  matrix or the \emph{Harrison–Perkins–Scott} matrix.
\par
The minimal Abelian number field \Math{\NF} containing all elements of matrices \eqref{EnUS3} and \eqref{EntransS3sqrt} is the subfield of the cyclotomic field \Math{\Q_{24}} fixed by the Galois automorphism \Math{\runi{24}\rightarrow\runi{24}^{23}}, where \Math{\runi{24}} is a primitive \Math{24}th root of unity. The corresponding cyclotomic polynomial is \Math{\Phi_{24}\vect{r}=1-r^4+r^8}. 
In terms of cyclotomic numbers, matrices \eqref{EnUS3} and \eqref{EntransS3sqrt} take the form
\Mathh{U'_6=\frac{1}{2}\Mtwo{-1}{2\runi{24}^2-\runi{24}^6}{-2\runi{24}^2+\runi{24}^6}{-1}}
and
\Mathh{\transmatrprim
=\frac{1}{6}\Mthree{4\runi{24}^2-2\runi{24}^6}{2\runi{24}+2\runi{24}^3
+2\runi{24}^5-4\runi{24}^7}{0}{4\runi{24}^2-2\runi{24}^6}
{-\runi{24}-\runi{24}^3-\runi{24}^5+2\runi{24}^7}
{-6\runi{24}+6\runi{24}^5}
{4\runi{24}^2-2\runi{24}^6}{-\runi{24}-\runi{24}^3-\runi{24}^5+2\runi{24}^7}
{6\runi{24}-6\runi{24}^5}.}
\par
In fact, information about the ``quantum behavior'' is encoded in the transformation matrices, such as \eqref{EntransS3monomial} or \eqref{EntransS3sqrt}. 
\par
Let \Math{\barket{n} = \Vthree{n_1}{n_2}{n_3}} and \Math{\barket{m} = \Vthree{m_1}{m_2}{m_3}} be vectors of states of a system and apparatus in the ``permutation'' basis.
Using transformation matrix \eqref{EntransS3monomial} we can write these vectors in the ``quantum'' basis:
\begin{align*}
    \barket{\widetilde{\psi}}=\transmatr^{-1}\barket{n}
    &=\frac{1}{\sqrt{3}}\Vthree{n_1+n_2+n_3}
    {n_1+n_2\runi{3}+n_3\runi{3}^2}{n_1\runi{3}+n_2+n_3\runi{3}^2},\\
    \barket{\widetilde{\phi}}=\transmatr^{-1}\barket{m}
    &=\frac{1}{\sqrt{3}}\Vthree{m_1+m_2+m_3}
    {m_1+m_2\runi{3}+m_3\runi{3}^2}{m_1\runi{3}+m_2+m_3\runi{3}^2}.
\end{align*}
Projections of the state vectors into the two-dimensional invariant subspace are
\begin{equation*}
    \barket{\psi} = \frac{1}{\sqrt{3}}\Vtwo{n_1+n_2\runi{3}+n_3\runi{3}^2}
    {n_1\runi{3}+n_2+n_3\runi{3}^2},~~~~
    \barket{\phi} = \frac{1}{\sqrt{3}}\Vtwo{m_1+m_2\runi{3}+m_3\runi{3}^2}
    {m_1\runi{3}+m_2+m_3\runi{3}^2}.
\end{equation*}
The components of Born's probability  \eqref{EnBornEn} have the form
\begin{equation}
    \inner{\psi\!}{\!\psi}=\invarQ{3}{n}{n}-\frac{1}{3}\invarL{3}{n}^2,
\label{EnBorn2den1En}
\end{equation}
\begin{equation}
    \inner{\phi\!}{\!\phi}=\invarQ{3}{m}{m}-\frac{1}{3}\invarL{3}{m}^2,
\label{EnBorn2den2En}
\end{equation}
\begin{equation}
\cabs{\inner{\phi\!}{\!\psi}}^2=
\vect{\invarQ{3}{m}{n}-\frac{1}{3}\invarL{3}{m}\invarL{3}{n}}^2,
\label{EnBorn2numEn}
\end{equation}
where \Math{\invarL{\wSN}{n}=\sum\limits_{i=1}^{\wSN}n_i} and
\Math{\invarQ{\wSN}{m}{n}=\sum\limits_{i=1}^{\wSN}m_in_i}
are linear and quadratic permutation invariants, which are common to all permutation groups. 
\par
Let us note that
\begin{enumerate}
    \item 
Expressions \eqref{EnBorn2den1En}--\eqref{EnBorn2numEn} are combinations of the
\emph{permutation invariants}.
    \item
\emph{Destructive quantum interference} ---
i.e., vanishing of Born's probability --- is deter\-mined by the equation 
\Mathh{3\vect{m_1n_1+m_2n_2+m_3n_3}-\vect{m_1+m_2+m_3}\vect{n_1+n_2+n_3}=0.}
This equation has infinitely many ``natural'' solutions with all non-zero components, for example,
\Mathh{{\barket{n} = \Vthree{1}{1}{2},~~\barket{m} = \Vthree{1}{3}{2}}.}
\par
Thus, by simple transition to an invariant subspace we obtain essential features of the quantum behavior
from the ``permutation dynamics'' and ``natural'' interpretation \eqref{Ennatvect} of quantum amplitudes.
\end{enumerate}
\par
This example can be slightly generalized.
Any permutation representation contains the \Math{\vect{\wSN-1}}-dimensional invariant subspace.
The scalar product in this subspace in terms of permutation invariants takes the form 
\Mathh{\inner{\phi\!}{\!\psi}=
\invarQ{\wSN}{m}{n}-\frac{1}{\wSN}\invarL{\wSN}{m}\invarL{\wSN}{n}.}
The identity
\Mathh{\displaystyle\invarQ{\wSN}{n}{n}\,-\,\frac{1}{\wSN}\invarL{\wSN}{n}^2~\equiv~
\frac{1}{\wSN^2}\sum\limits_{i=1}^\wSN\vect{\invarL{\wSN}{n}-\wSN{}n_i}^2}
shows explicitly that
\Math{\inner{\psi\!}{\!\psi}>0} for \Math{\barket{n}} with different%
\footnote{The vectors with equal components belong to the one-dimensional
complement of the \Math{\vect{\wSN-1}}-dimensional subspace.}
components \Math{n_i}.
This scalar product takes only rational values on the natural vectors  
\Math{\barket{n}} and \Math{\barket{m}}. 
\subsubsection{Icosahedral group \Math{\AltG{5}}}
\begin{figure}[!h]
\centering
\includegraphics[width=0.62\textwidth]{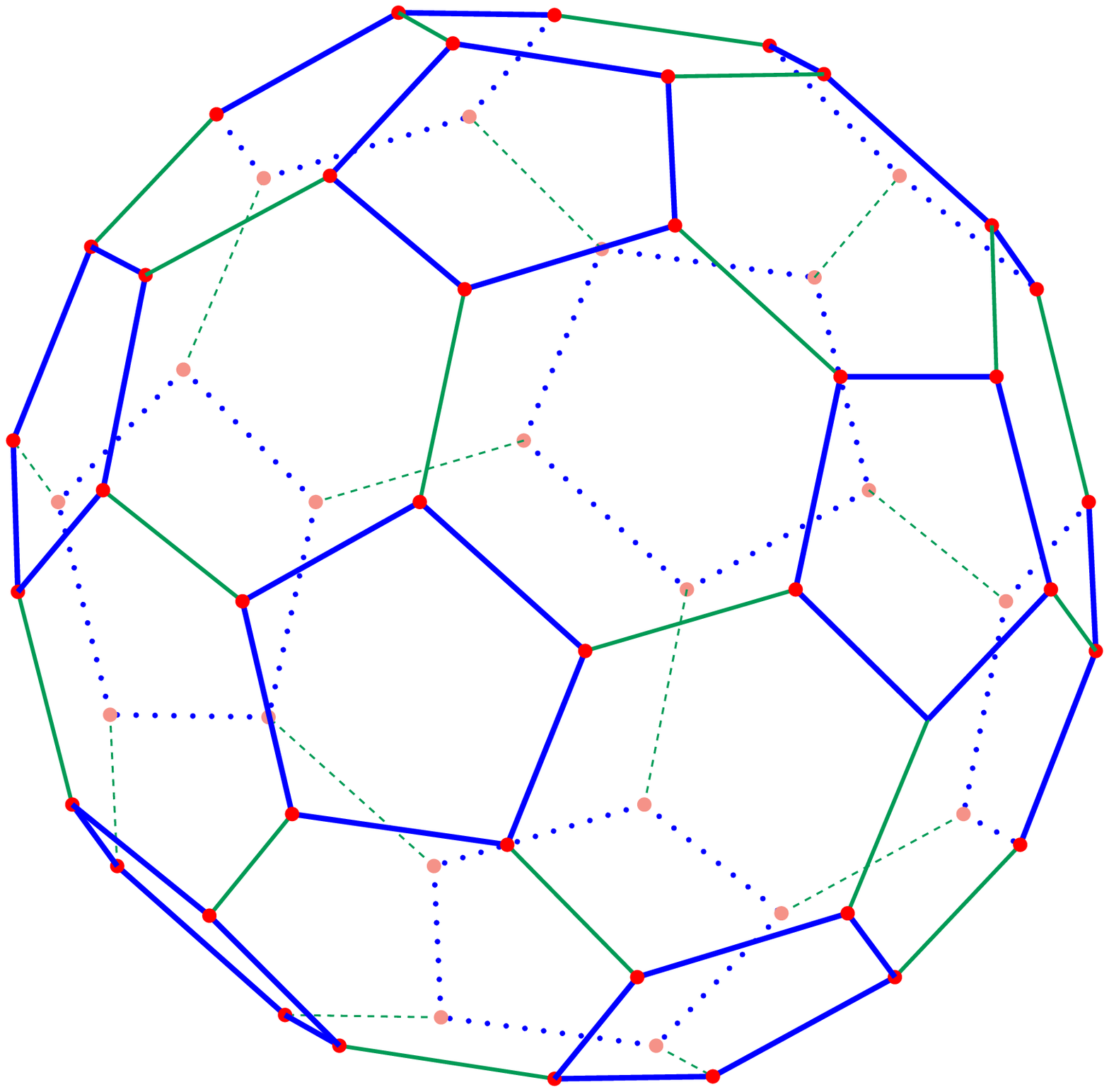}
\caption{Cayley graph of \Math{\AltG{5}}. 
Pentagons, hexagons and links between adjacent pentagons correspond to the
relators \Math{a^5}, \Math{\vect{ab}^3} and  \Math{b^2} in presentation \eqref{Enpresent}.}
\label{Enbucky}
\end{figure}

The alternating group \Math{\AltG{5}} --- also the group of orientation-preserving symmetries of an icosahedron --- is the smallest simple noncommutative group.
It consists of \Math{60} elements and its exponent is \Math{30}.
The group is so important in mathematics and applications, that F. Klein dedicated a whole book to it  \cite{Klein}.
In the physics literature, the icosahedral group is often denoted by \Math{\Sigma\!\vect{60}}.
It is interesting to note that \Math{\AltG{5}} has a ``physical incarnation'': the fullerene \Math{C_{60}} molecule has the structure of a Cayley graph of the group (see Fig. \ref{Enbucky}).
This is clear from the following \emph{presentation} of \Math{\AltG{5}} by two generators with three \emph{relators}:%
\footnote{Relator \Math{R} is an abbrevation for the relation \Math{R=\id}.} 
\begin{equation}
	\AltG{5}\cong\left\langle{}a, b\mid{}a^5,
	b^2, \vect{ab}^3\right\rangle.
	\label{Enpresent}
\end{equation}
\par
The group \Math{\AltG{5}} splits into five conjugacy classes \Math{\class{1},} \Math{\class{15},} \Math{\class{20},} \Math{\class{12},} \Math{\class{12'}}.
To distinguish these classes we use their sizes as subscripts.
The character table has the form
\begin{center}
  \begin{tabular}{c|c|c|c|c|c}
&\Math{\class{1}}&\Math{\class{15}}&\Math{\class{20}}&\Math{\class{12}}&\Math{\class{12'}}
\\\hline  
\Math{\chi_{1}}&\Math{1}&\Math{1}&\Math{1}&\Math{1}&\Math{1}
\\\hline  
\Math{\chi_{3}}&\Math{3}&\Math{-1}&\Math{0}&\Math{\phi}&\Math{1-\phi}
\\\hline  
\Math{\chi_{3'}}&\Math{3}&\Math{-1}&\Math{0}&\Math{1-\phi}&\Math{\phi}
\\\hline  
\Math{\chi_{4}}&\Math{4}&\Math{0}&\Math{1}&\Math{-1}&\Math{-1}
\\\hline  
\Math{\chi_{5}}&\Math{5}&\Math{1}&\Math{-1}&\Math{0}&\Math{0}
	\end{tabular}\enspace.
\end{center}
Here \Math{\phi=\frac{1+\sqrt{5}}{2}} is the ``\alert{golden ratio}''.
Note that \Math{\phi} and \Math{1-\phi} are cyclotomic integers (or ``\emph{cyclotomic naturals}''): 
\Math{\phi=-\runi{5}^2-\runi{5}^3~\equiv~1+\runi{5}+\runi{5}^4} ~and~ 
\Math{1-\phi=-\runi{5}-\runi{5}^4~\equiv~1+\runi{5}^2+\runi{5}^3},
where \Math{\runi{5}} is a fifth primitive root of unity.
The character table shows that \Math{\AltG{5}} has five irreducible representations: the trivial \Math{\IrrRep{1}} and four faithful \Math{\IrrRep{3}, \IrrRep{3'}, \IrrRep{4}, \IrrRep{5}}.
\par
As for permutations, the group has three \emph{primitive} actions on sets of \Math{5}, \Math{6} and \Math{10} elements.
The respective permutation representations have the following decompositions into irreducible components:
\Mathh{\PermRep{5}\cong\IrrRep{1}\oplus\IrrRep{4},~~
 \PermRep{6}\cong\IrrRep{1}\oplus\IrrRep{5},~~
 \PermRep{10}\cong\IrrRep{1}\oplus\IrrRep{4}\oplus\IrrRep{5}.}
Recall that a transitive action of a group on a set is called \emph{imprimitive} \cite{Wielandt}, if there is a \emph{nontrivial partition} of the set, invariant under the action of the group.
By definition, the \emph{trivial partitions} are decompositions into singleton blocks and into two blocks: the whole set and complementary to it empty block.
A nontrivial invariant partition is called an \emph{imprimitivity system} or a \emph{block system}.
An action is called \emph{primitive}, if only  trivial  partitions are invariant. 
Primitive actions are considered to be the most fundamental among all permutation actions.
\begin{figure}[!h]
\centering
\includegraphics[width=0.65\textwidth]{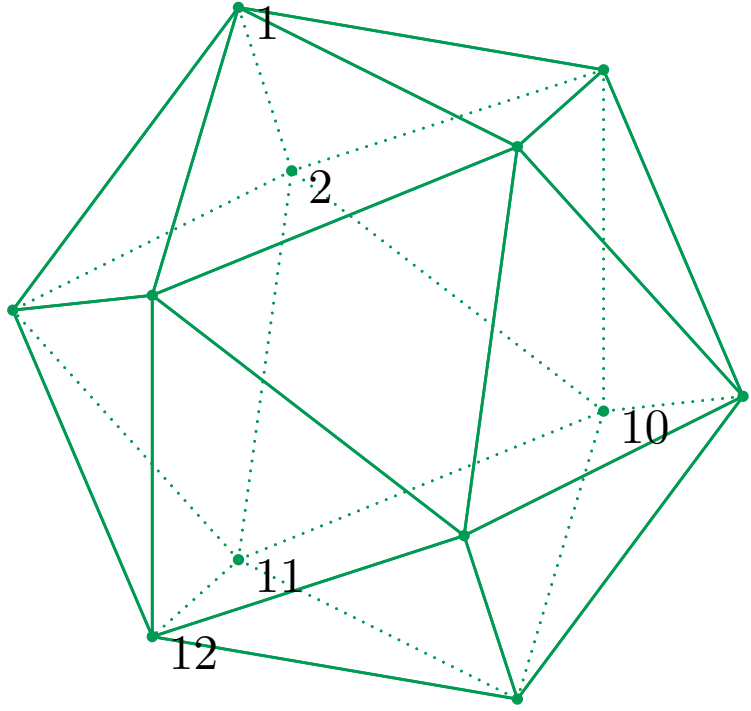}
\caption{Icosahedron. Invariant blocks are pairs of opposite vertices.}
\label{Enico}
\end{figure}
\par
Consider the action of \Math{\AltG{5}} on the vertices \Math{\wS_{12}} of an icosahedron.
This action is transitive, but imprimitive with the following block system 
\Mathh{\set{\mid{}B_1\mid\cdots\mid{}B_i\mid\cdots\mid{}B_6\mid}\equiv
\set{\mid1,7\mid\cdots\mid{}i,i+6\mid\cdots\mid6,12\mid},}
assuming the vertex numbering shown in Fig. \ref{Enico}.
Each block \Math{B_i} consists of a pair of opposite vertices of the icosahedron.
The group \Math{\AltG{5}} transitively permutes the blocks as single objects.
We will denote the mapping between opposite vertices of blocks (``complementarity'') by the symbol \Math{\Oppbare}, i.e., if \Math{B_i=\set{p,q}}, then \Math{q=\Opp{p}} and  \Math{p=\Opp{q}}.
For the numbering used in Fig. \ref{Enico}, the complementarity can be expressed by the formula \Math{\Opp{p}\equiv1+\vect{p+5\mod12}}. 
\par
Permutation representation of the action of \Math{\AltG{5}} on icosahedron vertices has the following decomposition into irreducible components:
\begin{equation}
	\PermRep{12}\cong\IrrRep{1}\oplus\IrrRep{3}\oplus\IrrRep{3'}\oplus\IrrRep{5}
	\text{~~or~~}\transmatr^{-1}\vect{\PermRep{12}}\transmatr
	=\IrrRep{1}\oplus\IrrRep{3}\oplus\IrrRep{3'}\oplus\IrrRep{5}.
	\label{Endecoico}
\end{equation}
With notations
\Mathh{\displaystyle\alpha=\frac{\phi}{4}\sqrt{10-2\sqrt{5}},~
\beta=\frac{\sqrt{5}\sqrt{10-2\sqrt{5}}}{20},~
\gamma=\frac{\sqrt{3}}{8}\vect{1-\frac{\sqrt{5}}{3}},~
\delta=-\frac{\sqrt{3}}{8}\vect{1+\frac{\sqrt{5}}{3}},}
the transformation matrix \Math{\transmatr} in \eqref{Endecoico} can be written, in particular, as
\begin{equation}
\transmatr=
\bmat
\frac{\sqrt{3}}{6}&\alpha&\beta&0&\alpha&\beta&0&\frac{1}{4}&-\frac{1}{2}&0&0&\frac{\sqrt{15}}{12}
\\
\frac{\sqrt{3}}{6}&0&\alpha&\beta&-\beta&0&\alpha&-\frac{\phi}{4}&0&-\frac{1}{2}&0&\gamma
\\
\frac{\sqrt{3}}{6}&\beta&0&\alpha&0&-\alpha&-\beta&\frac{\phi-1}{4}&0&0&-\frac{1}{2}&\delta
\\
\frac{\sqrt{3}}{6}&0&\alpha&-\beta&-\beta&0&-\alpha&-\frac{\phi}{4}&0&\frac{1}{2}&0&\gamma
\\
\frac{\sqrt{3}}{6}&-\beta&0&\alpha&0&\alpha&-\beta&\frac{\phi-1}{4}&0&0&\frac{1}{2}&\delta
\\
\frac{\sqrt{3}}{6}&\alpha&-\beta&0&-\alpha&\beta&0&\frac{1}{4}&\frac{1}{2}&0&0&\frac{\sqrt{15}}{12}
\\
\frac{\sqrt{3}}{6}&0&-\alpha&\beta&\beta&0&\alpha&-\frac{\phi}{4}&0&\frac{1}{2}&0&\gamma
\\
\frac{\sqrt{3}}{6}&\beta&0&-\alpha&0&-\alpha&\beta&\frac{\phi-1}{4}&0&0&\frac{1}{2}&\delta
\\
\frac{\sqrt{3}}{6}&-\alpha&\beta&0&\alpha&-\beta&0&\frac{1}{4}&\frac{1}{2}&0&0&\frac{\sqrt{15}}{12}
\\
\frac{\sqrt{3}}{6}&-\alpha&-\beta&0&-\alpha&-\beta&0&\frac{1}{4}&-\frac{1}{2}&0&0&\frac{\sqrt{15}}{12}
\\
\frac{\sqrt{3}}{6}&0&-\alpha&-\beta&\beta&0&-\alpha&-\frac{\phi}{4}&0&-\frac{1}{2}&0&\gamma 
\\
\frac{\sqrt{3}}{6}&-\beta&0&-\alpha&0&\alpha&\beta&\frac{\phi-1}{4}&0&0&-\frac{1}{2}&\delta
\emat.\label{EnTA5}
\end{equation}
Note that the standard computer algebra systems, such as \emph{Maple} or \emph{Mathematica}, can not handle matrices of such a type due to limited abilities in simplifying complicated expressions with irrationalities, especially with nested roots.
However, if the matrix elemens are expressed in terms of suitable Abelian number field, then the problem of simplification is reduced to simple computations with univariate polynomials \emph{modulo} a cyclotomic poly\-nomial.
\par
In this case, the minimal field \Math{\NF} is a subfield of the cyclotomic field \Math{\Q_{60}}.
\Math{\NF} is fixed in \Math{\Q_{60}} by the Galois automorphism \Math{\runi{}\rightarrow\runi{}^{59}}, where \Math{\runi{}} is a primitive \Math{60}th root of unity.
The corresponding cyclotomic polynomial is \Math{\Phi_{60}\vect{r}=1+r^2-r^6-r^8-r^{10}+r^{14}+r^{16}}.
In the language of cyclotomic numbers, the irrational elements of matrix  \eqref{EnTA5} have the form:
\begin{align*}
\frac{\sqrt{3}}{6}&=\frac{1}{6}\vect{2\runi{}^5-\runi{}^{15}},
\\
\frac{\phi}{4}\sqrt{10-2\sqrt{5}}&=
\frac{1}{2}\vect{\runi{}+\runi{}^3-\runi{}^9-\runi{}^{11}+\runi{}^{15}},
\\
\frac{\sqrt{5}\sqrt{10-2\sqrt{5}}}{20}&=
\frac{1}{2}\vect{\runi{}+4\runi{}^3-3\runi{}^9-\runi{}^{11}+2\runi{}^{15}},
\\
-\frac{\phi}{4}&=\frac{1}{4}\vect{-\runi{}^4-\runi{}^6+\runi{}^{14}},
\\
\frac{\phi-1}{4}&=\frac{1}{4}\vect{-1+\runi{}^4+\runi{}^6-\runi{}^{14}},
\\
\frac{\sqrt{15}}{12}&=\frac{1}{12}\vect{-2\runi{}+2\runi{}^5+4\runi{}^7
+2\runi{}^9+2\runi{}^{11}-4\runi{}^{13}-3\runi{}^{15}},
\\
\frac{\sqrt{3}}{8}\vect{1-\frac{\sqrt{5}}{3}}&=
\frac{1}{12}\vect{\runi{}+2\runi{}^5-2\runi{}^7-\runi{}^9
-\runi{}^{11}+2\runi{}^{13}},
\\
-\frac{\sqrt{3}}{8}\vect{1+\frac{\sqrt{5}}{3}}&=
\frac{1}{12}\vect{\runi{}-4\runi{}^5-2\runi{}^7-\runi{}^9
-\runi{}^{11}+2\runi{}^{13}+3\runi{}^{15}}.
\end{align*}
\par
Scalar products in the invariant subspaces of representation \eqref{Endecoico} in terms of permuta\-tion invariants take the form:
\begin{align}
\displaystyle\inner{\Phi_{\IrrRep{1}}}{\Psi_{\IrrRep{1}}}&=
\frac{1}{12}\invarL{12}{m}\invarL{12}{n},\nonumber
\\
\displaystyle\inner{\Phi_{\IrrRep{3}}}{\Psi_{\IrrRep{3}}}&=
\frac{1}{20}\vect{5\invarQ{12}{m}{n}-5\invar{A}{m}{n}
+\sqrt{5}\vect{\invar{B}{m}{n}-\invar{C}{m}{n}}},\label{Eninn3}
\\
\displaystyle\inner{\Phi_{\IrrRep{3'}}}{\Psi_{\IrrRep{3'}}}&=
\frac{1}{20}\vect{5\invarQ{12}{m}{n}-5\invar{A}{m}{n}
-\sqrt{5}\vect{\invar{B}{m}{n}-\invar{C}{m}{n}}},\label{Eninn3'}
\\
\displaystyle\inner{\Phi_{\IrrRep{5}}}{\Psi_{\IrrRep{5}}}&=
\frac{1}{12}\vect{5\invarQ{12}{m}{n}+5\invar{A}{m}{n}
-\invar{B}{m}{n}-\invar{C}{m}{n}},\nonumber
\end{align}
where
\begin{align}
\displaystyle\invar{A}{m}{n}&
=\invar{A}{n}{m}=\sum\limits_{k=1}^{12}m_kn_{\Opp{k}},\label{EnAmn}\\
\displaystyle\invar{B}{m}{n}&
=\invar{B}{n}{m}=\sum\limits_{k=1}^{12}m_k\!\!\sum\limits_{i\in\Nei{k}}\!\!n_{i},\label{EnBmn} 
\end{align}
\begin{align}
\displaystyle\invar{C}{m}{n}&
=\invar{C}{n}{m}=\sum\limits_{k=1}^{12}m_k\!\!\!\sum\limits_{i\in\Nei{\Opp{k}}}\!\!\!\!\!n_{i}.
\label{EnCmn}
\end{align}
In \eqref{EnBmn} and \eqref{EnCmn}, \Math{\Nei{k}} denotes the ``neighborhood'' of \Math{k}th vertex of the icosahedron, i.e., the set of vertices adjacent to \Math{k}.
For example, \Math{\Nei{1}=\set{2,3,4,5,6}} in Fig. \ref{Enico}.  
Thus, the inner sum in \eqref{EnBmn} is the sum of data on the neighborhood of the vertex \Math{k}, and a similar sum in \eqref{EnCmn} is the sum of data on the neighborhood of the vertex opposite to \Math{k}.
\par
Quadratic invariants \eqref{EnAmn}--\eqref{EnCmn} are not independent.
They are related by the identity
\begin{equation}
	\displaystyle\invar{A}{m}{n}+\invar{B}{m}{n}+\invar{C}{m}{n}+\invarQ{12}{m}{n}
	=\invarL{12}{m}\invarL{12}{n}.
\label{EnA5formidentity}	
\end{equation}
\par
Scalar products \eqref{Eninn3} and \eqref{Eninn3'}, when considered separately, lead to the above-mentioned problem with probabilities.
Namely, the Born probabilities for subrepresentations \Math{\IrrRep{3}} and \Math{\IrrRep{3'}} contain irrationalities that contradicts the frequency interpretation of probability for finite sets.
Obviously, this is a consequence of the imprimitivity: you can not move an icosahedron vertex without simultaneous movement of the opposite vertex.
To resolve the contradiction, mutually conjugate subrepresentations  \Math{\IrrRep{3}} and \Math{\IrrRep{3'}} must be considered together.
Scalar product in the subrepresentation \Math{\IrrRep{3}\oplus\IrrRep{3'}} 
\begin{equation*}
\displaystyle\inner{\Phi_{\IrrRep{3\oplus3'}}}{\Psi_{\IrrRep{3\oplus3'}}}=
\frac{1}{2}\vect{\invarQ{12}{m}{n}-\invar{A}{m}{n}}	
\end{equation*}
always gives rational Born's probabilities for vectors with natural ``occupation numbers''.
\section{Permutation Invariant Bilinear Forms}
To calculate quantum interferences of natural amplitudes in accordance with the Born rule it is sufficient to know expressions for scalar products in invariant subspaces of the permutation representation.
Above we have already obtained these expressions by straightforward calculation using the matrix of transition from the ``permutation'' to ``quan\-tum'' basis.
As was noted in the footnote on page \pageref{EnMeatAxePage} there is no satisfactory algorithm for splitting a module over an associative algebra into irreducible submodules (and, hence, for computing  corresponding transition matrix) over important to us systems of numbers.
However, one can try to get around these difficulties, if we can get the expressions for the scalar products directly, i.e. without the use of transition matrices.
Let us describe here one of the possible approaches.
\subsection{Basis of Permutation Invariant Forms}
For a start we construct the set of all invariant bilinear forms in the permutation basis.
Suppose that the group \Math{\wG\equiv\wG\vect{\wS}}, that acts on the set \Math{\wS=\set{\ws_1,\ldots,\ws_{\wSN}}}, is generated by \Math{K} elements \Math{g_1,\ldots,g_k,\ldots,g_K}.
Clearly, it suffices to verify invariance of the form with respect to these elements.
For convenience, the set \Math{\wS} is identified canonically with the set of indices of its elements: \Math{\wS=\onetonset{\wSN}=\set{1,\ldots,{\wSN}}}.
Denote the matrices associated to the generators by the permutation representation by \Math{P_k=\regrep\vect{g_k}}. Since \Math{P_k} is unitary, \Math{\transpose{P_k\!}=\regrep\vect{g_k^{-1}}}.
The condition of invariance of the bilinear form \Math{A=\Mone{a_{ij}}} under 
the group \Math{\wG} can be written as a system of matrix equations 
\begin{equation*}
A = P_kA\transpose{P_k\!},~~1\leq{}k\leq{}K.
\end{equation*}
It is easy to check that in terms of components these equations have the form
\begin{equation}
a_{ij} = a_{ig_kjg_k}.
\label{Eninvaracomp}
\end{equation}
Thus, the basis of all invariant bilinear forms is in one-to-one correspondence with the set of orbits \Math{\Delta_1,\ldots,\Delta_\baseformN} of the action of \Math{\wG} on the Cartesian product \Math{\wS\times\wS}. 
Orbits of the group on the product \Math{\wS\times\wS} are called \emph{orbitals}%
\footnote{Orbitals are special case of \emph{coherent configurations} (see e.g. \cite{Cameron}, Chapter 3). Namely, they are so-called \emph{Schurian coherent configurations}.} and play an important role in the theory of permutation groups and their representations \cite{Cameron,Dixon}.
For a transitive group, there is an orbital consisting of all pairs of equal indices \Math{\vect{i,i}}.
Such an orbital is called \emph{trivial} or \emph{diagonal}.
The number of orbitals \Math{\baseformN} is called the \emph{rank} of the group \Math{\wG}.
Each orbital \Math{\Delta_r} can be associated with
\begin{enumerate}
	\item 
a directed graph, whose vertices are the elements of \Math{\wS} and edges are pairs  \Math{\vect{i,j}\in\Delta_r};	
	\item 
an \Math{\wSN\times\wSN} matrix \Math{A\vect{\Delta_r}}, which is constructed according to the rule
\Mathh{
A\vect{\Delta_r}_{ij} =
\begin{cases}
1, & \text{if~} \vect{i,j}\in\Delta_r,\\
0, & \text{if~} \vect{i,j}\notin\Delta_r.
\end{cases}
}
\end{enumerate}
\par
Properties of the graphs of orbitals reflect important properties of the groups.
For example, a transitive permutation group is primitive if and only if the graphs of all its non-trivial orbitals are connected.
\par
The matrices of the orbitals form a basis of the \emph{centralizer ring} of the permutation representation \Math{\regrep} of \Math{\wG}.
This ring plays an important role in the theory of group representations.
We will denote it by \Math{\ZR\equiv\ZR\vect{\wG\vect{\wS}}}.
The multiplication table of basis elements of the centralizer ring \Math{\ZR} has the form
\begin{equation}
	A\vect{\Delta_p}A\vect{\Delta_q}=\sum\limits_{r=1}^{\baseformN}\alpha^{r}_{pq}A\vect{\Delta_r},
	\label{Enring}
\end{equation}
where the structure constants \Math{\alpha^{r}_{pq}} are natural numbers such that \Math{0\leq{}\alpha^{r}_{pq}\leq{}\wSN}. 
\par
We will call the matrices of orbitals \Math{\baseform{r}\equiv{A}\vect{\Delta_r}}  \emph{basic forms}, since any permutation invariant bilinear form can be presented by their linear combination.
\par
Algorithm for computing the basic forms is reduced to construction of orbitals in accordance with \eqref{Eninvaracomp}.
In a few words, it scans the elements of the set \Math{\wS\times\wS} in some, say, lexicographic, order and distributes these elements over the equivalence classes.
The output of the algorithm is a complete basis \Math{\baseform{1},\ldots,\baseform{\baseformN}} of permutation invariant bilinear forms.
The algorithm is quite simple. Our implementation is just a few lines in C.
\par
The  identity 
\begin{equation*}
	\baseform{1}+\baseform{2}+\cdots+\baseform{\baseformN}=\transpose{L}L=
	\matones_{\wSN}\equiv
	\bmat
	1&1&\cdots&1\\
	1&1&\cdots&1\\
	\vdots&\vdots&\vdots&\vdots\\
	1&1&\cdots&1
	\emat
\end{equation*}
follows directly from the construction.
Here \Math{L} is a covector of the form \Math{\overbrace{\vect{1,1,\ldots,1}}^{\wSN}},
and \Math{\matones_{\wSN}} is the \Math{\wSN\times\wSN} all-1  matrix.
Formula  \eqref{EnA5formidentity} is a special case of this identity.
\par
Let us illustrate the algorithm with the example of the group \Math{\CyclG{3}} acting on the set \Math{\wS=\set{1,2,3}}.
The group is generated by a single element, for example, \Math{g_1=\vect{1,2,3}}.
We need to distribute the set of pairs of indices
\begin{equation}
\begin{matrix}
\vect{1,1}&\vect{1,2}&\vect{1,3}\\
\vect{2,1}&\vect{2,2}&\vect{2,3}\\
\vect{3,1}&\vect{3,2}&\vect{3,3}
\end{matrix}
\label{En(i,j)tab}	       
\end{equation}
over the equivalence classes in accordance with \eqref{Eninvaracomp}.
If we start from the top left corner of table \eqref{En(i,j)tab} and will look for untreated pairs in the lexicographic order, the orbitals will be constructed in the following order
\Mathh{\Delta_1=\set{\vect{1,1},\vect{2,2},\vect{3,3}}, 
\Delta_2=\set{\vect{1,2},\vect{2,3},\vect{3,1}}, 
\Delta_3=\set{\vect{1,3},\vect{2,1},\vect{3,2}}.}
The corresponding complete basis of invariant forms has the form
\begin{equation*}
\baseform{1} =
\Mthree{1}{\cdot}{\cdot}
	     {\cdot}{1}{\cdot}
	     {\cdot}{\cdot}{1},~~
\baseform{2} =
\Mthree{\cdot}{1}{\cdot}
	     {\cdot}{\cdot}{1}
	     {1}{\cdot}{\cdot},~~
\baseform{3} =
\Mthree{\cdot}{\cdot}{1}
	     {1}{\cdot}{\cdot}
	     {\cdot}{1}{\cdot}.
\end{equation*}
Note that in general, if we  start the algorithm with the pair \Math{\vect{1,1}} and \Math{\wG} acts transitively on \Math{\wS}, then the first basis form will always be
the identity matrix \Math{\baseform{1} = \idmat_{\wSN}}, corresponding to the trivial orbital.
\par
By introducing commutators \Math{\comm{\baseform{p}}{\baseform{q}}=\baseform{p}\baseform{q}-\baseform{q}\baseform{p}} we can construct a Lie algebra associated with the centralizer ring. 
The commutator table of this Lie algebra has the form
\begin{equation}
	\comm{\baseform{p}}{\baseform{q}}=\sum\limits_{r=1}^{\baseformN}\gamma^{r}_{pq}\baseform{r},
\label{EnLie}
\end{equation}
where the structure constants \Math{\gamma^{r}_{pq}=\alpha^{r}_{pq}-\alpha^{r}_{qp}} are integers from the
interval  \Math{\left[-\wSN,\wSN\right]}; \Math{\alpha^{r}_{pq}} are the structure constants from \eqref{Enring}.
The Lie algebra is more convenient for our further considerations.
\subsection{Permutation Invariant Forms and Decomposition into Irredu\-cible Components}
Permutation invariant forms, despite the ease of their calculation, contain essential information about the structure of  permutation representation.
\par
Consider the decomposition of permutation representation into irreducible com\-ponents using the transformation matrix \Math{\transmatr}.
This decomposition for a transitive groups has the form
\begin{equation}
    \transmatr^{-1}\regrep(g)\transmatr=
    \bmat
       \IrrRep{1} &&&&&
    \\[5pt]
    &
    \hspace*{-8pt}
    \idmat_{m_2}\otimes\repq_2(g)
    &&&&
    \\
    &&
    \hspace*{-45pt}
    \ddots&&&\\
    &&&
    \hspace*{-38pt}
    \idmat_{m_{k}}\otimes\repq_{k}(g)&&\\
    &&&&
    \hspace*{-40pt}
    \ddots&\\
    &&&&&
    \hspace*{-45pt}
    \idmat_{m_{\irrepsN}}\otimes\repq_{\irrepsN}(g)\\
    \emat,~~~~g\in\wG\enspace.
		\label{Enrepdecomp}
\end{equation}
Here \Math{\otimes} denotes the \emph{Kronecker product} of matrices; \Math{\irrepsN} is the total number of different irreducible representations \Math{\repq_k} \Math{\vect{\repq_1\equiv\IrrRep{1}}} of the group \Math{\wG}, which are contained in the permuta\-tion representation \Math{\regrep}; \Math{m_k} is 
multiplicity of the subrepresentations \Math{\repq_k} in the representation \Math{\regrep}; \Math{\idmat_m} is an \Math{m\times{}m} identity matrix.
\par
The most general permutation invariant bilinear form can be written as a linear com\-bination of the basic forms
\begin{equation}A=\baseformcoegen{1}\baseform{1}+\baseformcoegen{2}\baseform{2}
+\cdots+\baseformcoegen{\baseformN}\baseform{\baseformN},
\label{EngenAform}	
\end{equation}
where the coefficients \Math{\baseformcoegen{i}} are elements of some Abelian number field \Math{\NF},
which is defined concretely in the computations to be described below.
\par
It is easy to show (see \cite{Cameron,Wielandt}) that in a basis, which splits the permutation representation into irreducible components, matrix \eqref{EngenAform} takes the form
\begin{equation}
\transmatr^{-1}A\transmatr=
\bmat
\bornform{1}&&&&&\\[3pt]
&\hspace*{-15pt}\bornform{2}\otimes\idmat_{\dimirr_2}&&&&\\
&&\hspace*{-15pt}\ddots&&&\\
&&&\hspace*{-15pt}\bornform{k}\otimes\idmat_{\dimirr_k}&&\\
&&&&\hspace*{-15pt}\ddots&\\
&&&&&\hspace*{-23pt}\bornform{\irrepsN}\otimes\idmat_{\dimirr_{\irrepsN}}
\emat~.
\label{EngenAformdecomp}	
\end{equation}
Here \Math{\bornform{k}} is an \Math{m_k\times{}m_k} matrix, whose elements are linear
combinations of the coefficients  \Math{\baseformcoegen{i}} from \eqref{EngenAform},
\Math{m_k} is a multiplicity of the irreducible component \Math{\repq_k};
\Math{\idmat_{\dimirr_k}} is a \Math{\dimirr_k\times\dimirr_k} identity matrix,
where \Math{\dimirr_k} is the dimension of \Math{\repq_k}.
The structure of matrix \eqref{EngenAformdecomp} implies that the rank \Math{\rankG} of the group (i.e., dimension of the centralizer ring) is equal to the sum of squares of the multiplicities: \Math{\rankG=1+m_2^2+\cdots+m_{\irrepsN}^2}.
\par 
Let us consider the determinant \Math{\det\vect{A}}.
In terms of the variables \Math{\baseformcoegen{1},\ldots,\baseformcoegen{\baseformN}} this determinant is a homogeneous polynomial of the  degree \Math{\wSN}.
Since the determinant of a form does not depend on the choice of basis, and the determinant of a block diagonal matrix is the product of determinants of its blocks, from decomposition \eqref{EngenAformdecomp} it follows
\begin{equation*}
\det\vect{A}=\det\vect{\bornform{1}}\det\vect{\bornform{2}}^{\dimirr_2}
\cdots\det\vect{\bornform{k}}^{\dimirr_k}\cdots
\det\vect{\bornform{\irrepsN}}^{\dimirr_{\irrepsN}}.
\end{equation*}
Here we use the identity 
\Math{\det\vect{X\otimes{}Y} = \det\vect{X}^m\det\vect{Y}^n} 
for the Kronecker product of \Math{n\times{}n} matrix \Math{X}
and \Math{m\times{}m} matrix \Math{Y}.
\par
It is clear that \Math{\det\vect{\bornform{k}}} is a homogeneous polynomial of the degree \Math{m_k} in the variables 
\Math{\baseformcoegen{1},\ldots,\baseformcoegen{\baseformN}}:
\Math{~\det\vect{\bornform{k}}=\efacdetA_k\vect{\baseformcoegen{1},\ldots,\baseformcoegen{\baseformN}}.}
Thus, we have the following
\par
\textbf{Proposition}.
\emph{The determinant of linear combination \eqref{EngenAform} has the following decompo\-sition into factors over a certain ring of cyclotomic integers
\begin{equation}
	\det{\sum\limits_{i=1}^{\baseformN}\baseformcoegen{i}\baseform{i}}
	=\prod\limits_{k=1}^{\irrepsN}\efacdetA_k\vect{\baseformcoegen{1},\ldots,
	\baseformcoegen{\baseformN}}^{\dimirr_k},
	~~\deg{\efacdetA_k\vect{\baseformcoegen{1},\ldots,
	\baseformcoegen{\baseformN}}} = m_k,
\label{Enhypodet}	
\end{equation}
where \Math{\efacdetA_k} is an irreducible polynomial corresponding to the irreducible component \Math{\repq_k} in the representation \Math{\regrep}.}
\par
This proposition 
is very similar to the \emph{Frobenius determinant theorem},%
\footnote{This theorem, discovered by Dedekind and proved by Frobenius, initiated the development of the group representation theory.
The history of the subject is presented, in particular, in \cite{Lam}.
}
which can be obtained by applying to decomposition \eqref{Enrepdecomp} the following modification of the above procedure.
Let \Math{\wGN} be the size of the group \Math{\wG}, i.e., \Math{\wG=\set{\wg_1,\ldots,\wg_{\wGN}}}.
Let us introduce the linear combination
\begin{equation}
	\regrep = x_1\regrep\vect{\wg_1}+x_2\regrep\vect{\wg_2}+\cdots+x_{\wGN}\regrep\vect{\wg_{\wGN}},
	\label{Engengalg}
\end{equation}
where \Math{x_1,\ldots,x_{\wGN}} are commuting indeterminates.
It follows from the structure of decomposition \eqref{Enrepdecomp} that
\begin{equation}
	\det{\sum\limits_{i=1}^{\wGN}x_i\regrep\vect{\wg_i}}
	=\prod\limits_{k=1}^{N_{\chi}}F_k\vect{x_1,\ldots,x_{\wGN}}^{m_k},
	~~\deg{F_k\vect{x_1,\ldots,x_{\wGN}}} = \dimirr_k. 
\label{Engalgdet}	
\end{equation}
The Frobenius determinant theorem is a special case of \eqref{Engalgdet}, when \Math{\regrep} is the \emph{regular representation} of \Math{\wG}.
In this case \Math{\dimirr_k=m_k} and \Math{\irrepsN=N_{\chi}}, where \Math{N_{\chi}} is the number of irreducible characters (~=~number of conjugacy classes) of the group \Math{\wG}.
We see some kind of duality between factorizations \eqref{Enhypodet} and \eqref{Engalgdet}: \Math{\efacdetA_k\vect{\baseformcoegen{1},\ldots,\baseformcoegen{\baseformN}}} is raised to the \emph{power} \Math{\dimirr_k} in \eqref{Enhypodet}
and \Math{\dimirr_k} is the \emph{degree} of \Math{F_k\vect{x_1,\ldots,x_{\wGN}}} in \eqref{Engalgdet}, whereas \Math{F_k\vect{x_1,\ldots,x_{\wGN}}} is raised to the \emph{power} \Math{m_k}  in \eqref{Engalgdet} and \Math{m_k} is the \emph{degree} of  \Math{\efacdetA_k\vect{\baseformcoegen{1},\ldots,\baseformcoegen{\baseformN}}} in \eqref{Enhypodet}.
\par
Now let us compare \eqref{Enhypodet} and \eqref{Engalgdet} from a computational point of view.
An advantage of \eqref{Enhypodet} is that the number of indeterminates \Math{\baseformcoegen{1},\ldots,\baseformcoegen{\rankG}} is usually much smaller than the number of \Math{x_1,\ldots,x_{\wGN}}:
\Math{\vect{\rankG=\sum\limits_{k=1}^{\irrepsN}m_k^2}\leq\vect{\wGN=\sum\limits_{k=1}^{N_{\chi}}\dimirr_k^2}}
since \Math{\irrepsN\leq{}N_{\chi}} and \Math{m_k\leq{}\dimirr_k}. For example, for the action of the icosahedral group \Math{\AltG{5}} on the icosahedron vertices \Math{\rankG=4} whereas \Math{\wGN=60}.
\par
Moreover, the polynomials \Math{F_k\vect{x_1,\ldots,x_{\wGN}}} in  \eqref{Engalgdet} are nonlinear for nontrivial dimensions of irreducible components and it is not clear whether there is a simple direct way to handle this.
In the case of factorization \eqref{Enhypodet} we can reduce nonlinear polynomials in \Math{\rankG} variables to linear polynomials in \Math{\sum\limits_{k=1}^{\irrepsN}m_k} variables via an approach  based on 
the ``coarsening'' of respective coherent configurations. This approach is described below.
\par
From \eqref{Enhypodet} the idea of an algorithm to compute the invariant forms in irreducible subspaces of the permutation representation follows.
\par 
We must first calculate the polynomial  \Math{\det\vect{A}}.
This is a relatively simple task.
The standard algorithms have the cubic (or slightly less) complexity in the matrix dimension. 
Then, the polynomial \Math{\det\!\vect{A}} must be decomposed into a maximum number of irreducible factors
over the ring of cyclotomic integers, whose conductor is a certain divisor of the group exponent.
As is well known, the practical algorithms to factor polynomials are polynomial-time algorithms of the Las Vegas type.%
\footnote{
A Las Vegas algorithm \cite{Babai} is a randomized algorithm that either gives a correct result, or reports failure (in contrast to Monte Carlo algorithms where erroneous outputs are possible) with the probability \Math{\leq1/2}.
The probability of failure is reduced to \Math{\leq2^{-t}} after \Math{t} runs of the algorithm.} 
\par
Thus, constructing decomposition \eqref{Enhypodet} is an algorithmically realizable task.
Solving it, we obtain complete information about the dimensions and multiplicities of all irreducible subrepresentations.
This gives a concretization of the structure of decomposition of the permutation representation into irreducible components.
\par
The next natural step is to try to compute explicitly invariant scalar products
\Math{\bornform{k}} in the irreducible subspaces of the permutation representation.
To do this, we exclude from consideration the factor
\Math{\efacdetA_k\vect{\baseformcoegen{1},\ldots,\baseformcoegen{\baseformN}}}, 
related to the component \Math{\bornform{k}}, and equate to zero the other factors. 
That is, we write the system of equations 
\begin{equation}
\efacdetA_1
 =
\cdots = 
\widehat{\efacdetA_k} =
\cdots = 
\efacdetA_{\irrepsN}
 = 0.
\label{EnEkEquations}	
\end{equation}
\subsubsection{The case of multiplicity-free representation.}
\label{Ensubsectmult1}
If all the multiplicities \Math{m_i=1} (under this condition
\Math{\rankG=\irrepsN} and the centralizer ring \Math{\ZR} is commutative), then
all the polynomials \Math{\efacdetA_i} are linear. 
In this case, the computation of the scalar products in the invariant subspaces can easily be completed, as is reduced to solution of a system of linear equations. 
\par
As an example, consider the group \Math{\SL{2,3}} defined as the group of special linear trans\-formations of two-dimensional space over the field of three elements \Math{\F_3}.
This group is used in particle physics, where it is often denoted by \Math{\mathrm{T}'} since it is a double cover of the orientation preserving symmetry group \Math{\mathrm{T}\cong\AltG{4}} of a tetrahedron. 
We will consider its faithful permutation action of degree 8, which can be generated, for example, by the following two permutations
\Mathh{g_1 = \vect{1,5,3,2,6,4}\vect{7,8}\text{~~and~~}g_2 = \vect{1,3,7,2,4,8}\vect{5,6}.}
The 8-dimensional permutation representation will be denoted by \Math{\PermRep{8}}.
\par
The following four matrices --- obtained via constructing orbitals --- form a basis of the ring \Math{\ZR\equiv\ZR\vect{\SL{2,3}\vect{\onetonset{8}}}} of permutation invariant forms
\begin{equation*}
		\baseform{1} = 
	\bmat
	1&\cdot&\cdot&\cdot&\cdot&\cdot&\cdot&\cdot\\
	\cdot&1&\cdot&\cdot&\cdot&\cdot&\cdot&\cdot\\
	\cdot&\cdot&1&\cdot&\cdot&\cdot&\cdot&\cdot\\
	\cdot&\cdot&\cdot&1&\cdot&\cdot&\cdot&\cdot\\
	\cdot&\cdot&\cdot&\cdot&1&\cdot&\cdot&\cdot\\
	\cdot&\cdot&\cdot&\cdot&\cdot&1&\cdot&\cdot\\
	\cdot&\cdot&\cdot&\cdot&\cdot&\cdot&1&\cdot\\
	\cdot&\cdot&\cdot&\cdot&\cdot&\cdot&\cdot&1
	\emat,~~ 
		\baseform{2} =
	\bmat
	\cdot&1&\cdot&\cdot&\cdot&\cdot&\cdot&\cdot\\
	1&\cdot&\cdot&\cdot&\cdot&\cdot&\cdot&\cdot\\
	\cdot&\cdot&\cdot&1&\cdot&\cdot&\cdot&\cdot\\
	\cdot&\cdot&1&\cdot&\cdot&\cdot&\cdot&\cdot\\
	\cdot&\cdot&\cdot&\cdot&\cdot&1&\cdot&\cdot\\
	\cdot&\cdot&\cdot&\cdot&1&\cdot&\cdot&\cdot\\
	\cdot&\cdot&\cdot&\cdot&\cdot&\cdot&\cdot&1\\
	\cdot&\cdot&\cdot&\cdot&\cdot&\cdot&1&\cdot
	\emat,
\end{equation*}
\begin{equation*}
		\baseform{3} = 
	\bmat
	\cdot&\cdot&1&\cdot&1&\cdot&1&\cdot\\
	\cdot&\cdot&\cdot&1&\cdot&1&\cdot&1\\
	\cdot&1&\cdot&\cdot&\cdot&1&1&\cdot\\
	1&\cdot&\cdot&\cdot&1&\cdot&\cdot&1\\
	\cdot&1&1&\cdot&\cdot&\cdot&\cdot&1\\
	1&\cdot&\cdot&1&\cdot&\cdot&1&\cdot\\
	\cdot&1&\cdot&1&1&\cdot&\cdot&\cdot\\
	1&\cdot&1&\cdot&\cdot&1&\cdot&\cdot
	\emat,~~
		\baseform{4} =
	\bmat
	\cdot&\cdot&\cdot&1&\cdot&1&\cdot&1\\
	\cdot&\cdot&1&\cdot&1&\cdot&1&\cdot\\
	1&\cdot&\cdot&\cdot&1&\cdot&\cdot&1\\
	\cdot&1&\cdot&\cdot&\cdot&1&1&\cdot\\
	1&\cdot&\cdot&1&\cdot&\cdot&1&\cdot\\
	\cdot&1&1&\cdot&\cdot&\cdot&\cdot&1\\
	1&\cdot&1&\cdot&\cdot&1&\cdot&\cdot\\
	\cdot&1&\cdot&1&1&\cdot&\cdot&\cdot
	\emat.
\end{equation*}
The exponent of \Math{\SL{2,3}} is \Math{12}. However the ring of cyclotomic integers \Math{\N_3} is sufficient for the ``\emph{absolute factorization}'' of determinant of the linear combination 
\Math
{A=\baseformcoegen{1}\baseform{1}+\baseformcoegen{2}\baseform{2}
+\baseformcoegen{3}\baseform{3}+\baseformcoegen{4}\baseform{4}}:
\begin{align}
\det{A} =& 	
  \vect{\baseformcoegen{1}+\baseformcoegen{2}
	     +3\baseformcoegen{3}+3\baseformcoegen{4}}
	    \nonumber\\
	&\set{\baseformcoegen{1}-\baseformcoegen{2}
	     +\vect{1+2\runi{}}\baseformcoegen{3}
	     -\vect{1+2\runi{}}\baseformcoegen{4}}^2
	     \nonumber\\
	&\set{\baseformcoegen{1}-\baseformcoegen{2}
	     -\vect{1+2\runi{}}\baseformcoegen{3}
	     +\vect{1+2\runi{}}\baseformcoegen{4}}^2
	     \label{EndecoSL23}\\
	&\vect{\baseformcoegen{1}+\baseformcoegen{2}
	     -\baseformcoegen{3}-\baseformcoegen{4}}^3,\nonumber
\end{align}
where \Math{\runi{}} is a third primitive root of unity.
From \eqref{EndecoSL23} the structure of decomposition of the 
representation \Math{\PermRep{8}} into irreducible components is seen immediately:
\Mathh{\PermRep{8}=
\IrrRep{1}\oplus\IrrRep{2}\oplus\IrrRep{2'}\oplus\IrrRep{3}.}
The linear factors as they are ordered in \eqref{EndecoSL23} correspond to the representations \Math{k = \IrrRep{1}, \IrrRep{2}, \IrrRep{2'}, \IrrRep{3}}, respectively.
Excluding the linear factor corresponding to the representation \Math{k} and equating to zero the other factors, we obtain a systems of three linear equations. 
After considering all irreducible components we come to the set of four such systems.
\par
Consider for example the subrepresentation \Math{\IrrRep{2}}.
For this component the system of equations \eqref{EnEkEquations} takes the form
\begin{align}
\baseformcoegen{1}+\baseformcoegen{2}+3\baseformcoegen{3}+3\baseformcoegen{4}&=0,
\label{Eneqex1}\\
\baseformcoegen{1}-\baseformcoegen{2}-\vect{1+2\runi{}}\baseformcoegen{3}
+\vect{1+2\runi{}}\baseformcoegen{4}&=0,\\
\baseformcoegen{1}+\baseformcoegen{2}-\baseformcoegen{3}-\baseformcoegen{4}&=0.
\label{Eneqex3}	
\end{align}
\par
The linear systems of this type consist of \Math{\rankG-1} equations, but contain \Math{\rankG} variables.
Since the bilinear form \eqref{EngenAform} describes a non-degenerate scalar product, the coefficient \Math{\baseformcoegen{1}} at the diagonal basis form can not vanish.
Thus, one can treat \Math{\baseformcoegen{1}} as a parameter and solve the system of equations for the remaining variables.
In principle, the coefficient \Math{\baseformcoegen{1}} can be taken to be
an arbitrary (nonzero) parameter.
For convenience, we always assume that \Math{\baseformcoegen{1} = 1}.
In fact, it is reasonable to set \Math{\baseformcoegen{1} = 1} before factoring the polynomial \Math{\det\vect{A}},
reducing thereby the number of variables.
\par 
Solving linear system (\ref{Eneqex1}---\ref{Eneqex3}) with \Math{\baseformcoegen{1} = 1}, we obtain
\Mathh{\baseformcoegen{2}=-1,~\baseformcoegen{3}=-\frac{1+2\runi{}}{3},
~\baseformcoegen{4}=\frac{1+2\runi{}}{3}.}
For the scalar product in the two-dimensional space of the subrepresentation \Math{\IrrRep{2}} we have the following expression
 \Mathh{\bornform{\IrrRep{2}} = C_{\IrrRep{2}}\vect{\baseform{1}-\baseform{2}-\frac{1+2\runi{}}{3}\baseform{3}+\frac{1+2\runi{}}{3}\baseform{4}}.}
Here \Math{C_{\IrrRep{2}}} is an arbitrary normalizing coefficient --- Born's
probability does not depend on its value. However, it is reasonable to choose the value 
\begin{equation}
	C_{\IrrRep{k}} = \dimirr_k/\wSN
	\label{Encoea1}
\end{equation}
for each irreducible component. 
Under such a normalization the sum of scalar products in invariant subspaces will be equal to the standard scalar product in the space of permutation representation. 
\par
Applying the procedure to all irreducible components we come to the following set of scalar products in invariant subspaces
\begin{align*}
	\bornform{\IrrRep{1}} =& \frac{1}{8}\vect{\baseform{1}+\baseform{2}+\baseform{3}+\baseform{4}}
\equiv\frac{1}{8}\matones_8,\\
	\bornform{\IrrRep{2}} =& \frac{1}{4}\vect{\baseform{1}-\baseform{2}-\frac{1+2\runi{}}{3}\baseform{3}+\frac{1+2\runi{}}{3}\baseform{4}},
	\\
	\bornform{\IrrRep{2'}} =& \frac{1}{4}\vect{\baseform{1}-\baseform{2}+\frac{1+2\runi{}}{3}\baseform{3}-\frac{1+2\runi{}}{3}\baseform{4}},
	\\
	\bornform{\IrrRep{3}} =& \frac{3}{8}\vect{\baseform{1}+\baseform{2}-\frac{1}{3}\baseform{3}
	-\frac{1}{3}\baseform{4}}.
\end{align*}
We see that  normalization \eqref{Encoea1} ensures the identity
\Math{\bornform{\IrrRep{1}}+\bornform{\IrrRep{2}}+\bornform{\IrrRep{2'}}
+\bornform{\IrrRep{3}} = \baseform{1}\equiv\idmat_8.}
Note that in all such tasks \Math{\bornform{\IrrRep{1}}} does not require calculation, since the inner product in the subspace of the trivial representation always has the form \Math{\displaystyle\frac{1}{\wSN}\matones_\wSN.}
\subsubsection{Nontrivial multiplicities of subrepresentations.}
In the case of nontrivial multiplicities of subrepresentations the situation becomes more complicated because of the nonlinearity of equations \eqref{EnEkEquations}	.
The question of whether it is possible to develop a general algorithm for the case of multiple subrepresentations requires a deeper additional study.
\par
The source of nonlinearity is an excess in the number of parameters, the definite values  of which are inessential for computing the Born probabilities.
As one can see from the structure of decomposition \eqref{EngenAformdecomp}, any block of multiple components \Math{\bornform{k}\otimes\idmat_{\dimirr_k}} contains \Math{m_k\times{}m_k} such parameters. Our convention \eqref{Encoea1} fixes
\Math{m_k} diagonal elements of the matrix \Math{\bornform{k}}. So we need to fix somehow the remaining \Math{m_k^2-m_k} parameters.
An examination of the structure of the centralizer ring may help to do this.
Recall that if the centralizer ring is commutative, then all the multiplicities are equal to one.
So, the natural idea of how to remove the extra degrees of freedom is to find a suitable \emph{commutative} subalgebra of the dimension \Math{1+m_2+\cdots+m_{\irrepsN}} in the Lie algebra of the dimension \Math{\rankG=1+m_2^2+\cdots+m_{\irrepsN}^2} defined by \eqref{EnLie}.
\par
We sketch here a simple combinatorial approach to finding such a subalgebra.
It works well in all the examples of transitive permutation groups that we have considered.
Though currently we have no proof of the universality of this approach, it can be used at least as an efficient preliminary step in solving the problems with multiple subrepresentations.
\par
Basis matrices derived from the orbitals are (0,1)-matrices with disjoint sets of unit
entries. A result of addition of such matrices belongs to the same type of matrices.
This corresponds to the \emph{coarsening} of the respective coherent configuration.
So, the natural idea is to sum up some of the matrices \Math{\baseform{i}} in order to replace the basis
\Math{\baseform{1},\ldots,\baseform{\baseformN}}  by a smaller set of mutually commuting (0,1)-matrices. 
\par
As an illustrative example, let us consider the Coxeter group of type \Math{A_2}, which is also the Weyl group of the Lie group, say, \Math{\SU{3}}. We will consider the natural action of \Math{A_2} on its root system \Math{\wS = \set{1,2,\cdots,6}}.
The vectors of this root system form a two-dimensional regular hexagon.
The generators of the action are, for example, \Math{g_1=(1,4)(2,3)(5,6)} and
\Math{g_2=(1,3)(2,5)(4,6)}.
\par
Computation of orbitals gives the following basis of the centralizer ring
\begin{align*}
		&	\baseform{1} = 
		\bmat
		1&\cdot&\cdot&\cdot&\cdot&\cdot\\
		\cdot&1&\cdot&\cdot&\cdot&\cdot\\
		\cdot&\cdot&1&\cdot&\cdot&\cdot\\
		\cdot&\cdot&\cdot&1&\cdot&\cdot\\
		\cdot&\cdot&\cdot&\cdot&1&\cdot\\
		\cdot&\cdot&\cdot&\cdot&\cdot&1
		\emat, 
			\baseform{2} =
		\bmat
		\cdot&1&\cdot&\cdot&\cdot&\cdot\\
		\cdot&\cdot&\cdot&\cdot&\cdot&1\\
		\cdot&\cdot&\cdot&\cdot&1&\cdot\\
		\cdot&\cdot&1&\cdot&\cdot&\cdot\\
		\cdot&\cdot&\cdot&1&\cdot&\cdot\\
		1&\cdot&\cdot&\cdot&\cdot&\cdot
		\emat,
			\baseform{3} =
		\bmat
		\cdot&\cdot&1&\cdot&\cdot&\cdot\\
		\cdot&\cdot&\cdot&1&\cdot&\cdot\\
		1&\cdot&\cdot&\cdot&\cdot&\cdot\\
		\cdot&1&\cdot&\cdot&\cdot&\cdot\\
		\cdot&\cdot&\cdot&\cdot&\cdot&1\\
		\cdot&\cdot&\cdot&\cdot&1&\cdot
		\emat,
	\end{align*}
	\begin{align*}
	&
			\baseform{4} = 
		\bmat
		\cdot&\cdot&\cdot&1&\cdot&\cdot\\
		\cdot&\cdot&\cdot&\cdot&1&\cdot\\
		\cdot&\cdot&\cdot&\cdot&\cdot&1\\
		1&\cdot&\cdot&\cdot&\cdot&\cdot\\
		\cdot&1&\cdot&\cdot&\cdot&\cdot\\
		\cdot&\cdot&1&\cdot&\cdot&\cdot
		\emat, 
			\baseform{5} =
		\bmat
		\cdot&\cdot&\cdot&\cdot&1&\cdot\\
		\cdot&\cdot&1&\cdot&\cdot&\cdot\\
		\cdot&1&\cdot&\cdot&\cdot&\cdot\\
		\cdot&\cdot&\cdot&\cdot&\cdot&1\\
		1&\cdot&\cdot&\cdot&\cdot&\cdot\\
		\cdot&\cdot&\cdot&1&\cdot&\cdot
		\emat,
			\baseform{6} =
		\bmat
		\cdot&\cdot&\cdot&\cdot&\cdot&1\\
		1&\cdot&\cdot&\cdot&\cdot&\cdot\\
		\cdot&\cdot&\cdot&1&\cdot&\cdot\\
		\cdot&\cdot&\cdot&\cdot&1&\cdot\\
		\cdot&\cdot&1&\cdot&\cdot&\cdot\\
	\cdot&1&\cdot&\cdot&\cdot&\cdot
	\emat.\nonumber
\end{align*}
Applying absolute factorization%
\footnote{The exponent of the group \Math{A_2\cong\SymG{3}} is 6. Due to the ring isomorphism \Math{\N_6\cong\N_3}, it is sufficient to factorize over the ring of cyclotomic integers \Math{\N_3}.}
to determinant of the linear combination 
\Mathh
{A=\baseformcoegen{1}\baseform{1}+\baseformcoegen{2}\baseform{2}
+\baseformcoegen{3}\baseform{3}+\baseformcoegen{4}\baseform{4}+\baseformcoegen{5}\baseform{5}+\baseformcoegen{6}\baseform{6},}
we obtain the following decomposition
\begin{align*}
\det{A} =& 	
  \vect{\baseformcoegen{1}+\baseformcoegen{2}
	     -\baseformcoegen{3}-\baseformcoegen{4}
	     -\baseformcoegen{5}+\baseformcoegen{6}}
	    \nonumber\\
  &\vect{\baseformcoegen{1}+\baseformcoegen{2}
	     +\baseformcoegen{3}+\baseformcoegen{4}
	     +\baseformcoegen{5}+\baseformcoegen{6}}
	    \nonumber\\
	&\left\{\baseformcoegen{1}^2+\baseformcoegen{2}^2-\baseformcoegen{3}^2
	-\baseformcoegen{4}^2-\baseformcoegen{5}^2+\baseformcoegen{6}^2\right.
	\\
	&\left.~-\baseformcoegen{1}\baseformcoegen{2}
	-\baseformcoegen{1}\baseformcoegen{6}
	-\baseformcoegen{2}\baseformcoegen{6}
	+\baseformcoegen{3}\baseformcoegen{4}+\baseformcoegen{3}\baseformcoegen{5}
	+\baseformcoegen{4}\baseformcoegen{5}\right\}^2.\nonumber
\end{align*}
The structure of permutation action of the Weyl group \Math{A_2} on its roots 
in terms of irreducible components
follows immediately from this decomposition: 
\Math{\PermRep{6}=
\IrrRep{1}\oplus\IrrRep{1'}\oplus\vect{\IrrRep{2}\oplus\IrrRep{2}}.}
Here we have four irreducible components and six-dimensional centralizer ring. 
Thus, we need to eliminate the two extra degrees of freedom.
\par
The Lie algebra commutators are
\begin{align}
&
\ZRcomm{1}{2}=\ZRcomm{1}{3}=\ZRcomm{1}{4}=\ZRcomm{1}{5}=\ZRcomm{1}{6}=\ZRcomm{2}{6}=0,
\nonumber\\[3pt]
&
\ZRcomm{2}{3}=\ZRcomm{3}{6}=\baseform{4}-\baseform{5},
\label{EnA2com23}\\
&
\ZRcomm{2}{4}=\ZRcomm{4}{6}=-\baseform{3}+\baseform{5},
\\
&
\ZRcomm{2}{5}=\ZRcomm{5}{6}=\baseform{3}-\baseform{4},
\label{EnA2com56}\\[3pt]
&
\ZRcomm{3}{4}=-\ZRcomm{3}{5}=\ZRcomm{4}{5}=-\baseform{2}+\baseform{6}.\nonumber
\end{align}
\eqref{EnA2com23}-\eqref{EnA2com56}  imply, that \Math{\ZRcomm{2}{X}=\ZRcomm{6}{X}=0}, 
where \Math{\baseform{X}=\baseform{3}+\baseform{4}+\baseform{5}}. It is easy to check 
that the matrices \Math{\baseform{1}, \baseform{2}, \baseform{6}, \baseform{X}} form 
a basis of four-dimensional commutative Lie algebra.
Now the determinant of the combination
\Mathh
{A'=\baseformcoegen{1}\baseform{1}+\baseformcoegen{2}\baseform{2}
+\baseformcoegen{6}\baseform{6}+\baseformcoegen{X}\baseform{X}}
 is factorizable to the linear factors over the cyclotomic integer ring \Math{\N_3}:
\begin{align*}
\det{A'} =& 	
  \vect{\baseformcoegen{1}+\baseformcoegen{2}
	     +\baseformcoegen{6}-3\baseformcoegen{X}}
	    \nonumber\\
  &\vect{\baseformcoegen{1}+\baseformcoegen{2}
	     +\baseformcoegen{6}+3\baseformcoegen{X}}
	    \\
	&\set{\baseformcoegen{1}+\runi{}\baseformcoegen{2}
	     -\vect{1+\runi{}}\baseformcoegen{6}}^2\nonumber\\
&\set{\baseformcoegen{1}-\vect{1+\runi{}}\baseformcoegen{2}
	     +\runi{}\baseformcoegen{6}}^2,\nonumber
\end{align*}
where \Math{\runi{}} is a third primitive root of unity.
After the same manipulations as in Subsection \ref{Ensubsectmult1}, we come to the following set of scalar product forms in the invariant subspaces
\par
\begin{align*}
	\bornform{\IrrRep{1}} =& \frac{1}{6}\vect{\baseform{1}+\baseform{2}+\baseform{6}+\baseform{X}}
\equiv\frac{1}{6}\matones_6,\\
	\bornform{\IrrRep{1'}} =& \frac{1}{6}\vect{\baseform{1}+\baseform{2}+\baseform{6}-\baseform{X}},\\
	\bornform{\IrrRep{2}} =& 
\frac{1}{3}\set{\baseform{1}-\vect{1+
	\runi{}}\baseform{2}+\runi{}\baseform{6}},
	\\
	\bornformequi{\IrrRep{2}} =& \frac{1}{3}\set{\baseform{1}+\runi{}\baseform{2}-\vect{1+
	\runi{}}\baseform{6}}.
\end{align*}
Here \Math{\bornform{\IrrRep{2}}} and \Math{\bornformequi{\IrrRep{2}}} are different coordinate presentations of the same form, associated with the irreducible representation  \Math{\IrrRep{2}}.
\section*{Conclusions}
This review is based on the idea that any problem that has a meaningful empirical content, can be formulated in a constructive, or, to put it more definitely, finite terms.
Exclusion of actual infinities from the description of physical reality, freeing from many technical difficulties, allows us to concentrate on the substantial aspects of physical problems.
In addition, constructiveness is a necessary requirement for the very possibility of creating computer models of physical systems.
We also followed the principle of economical introduction of new elements in descriptions, i.e. mathematical concepts were introduced only when they were really needed (of course, up to our understanding of the problems).
The uncontrolled introduction of new mathematical structures (or extension of old ones simply because of the possibility to generalize them) complicates separation of substantive elements of description from artifacts.
\par
Adhering to these positions, we examined classical and quantum dynamical systems with non-trivial symmetry properties.
Special attention was paid to one of the central principles in physics --- the gauge invariance --- which we reformulated in a ``finite'' form.
We have considered deterministic dynamical systems, and discussed the group nature of moving ``soliton-like'' structures that are often observed in the dynamics of such systems. 
We also discussed how the unitarity, which is inherent in the fundamental laws of nature and underlies the quantum  behavior, may arise in deterministic dynamics.
\par 
A constructive analysis of the quantum-mechanical behavior leads to the following conclusions:
\begin{enumerate}
\item
Quantum mechanics is essentially an \emph{a priori mathematical scheme} based on the fundamental impossibility to trace identity of indistinguishable objects during the evolution of a collection of such objects.
In fact, this is a part of combinatorics, which can be called ``the calculus of indistinguishables''.
\item
Any quantum-mechanical problem can be reduced to \emph{permutations}.
\item
\emph{Quantum interferences} are phenomena observed in invariant subspaces of permutation representations and expressed in terms of \emph{permutation invariants}.
\item
Natural interpretation of indistinguishable objects, on which a symmetry group acts by permutations, as ``\emph{particles}'', and \emph{quantum amplitudes} (``\emph{waves}'') as vectors of multiplicities representing states of ensembles of these objects, leads to \emph{rational} quantum probabilities.
This is consistent with the \emph{frequency interpretation} of proba\-bility for finite sets.
\end{enumerate}
The idea of natural quantum amplitudes looks very attractive.
This idea leads to a simple and self-consistent picture of quantum behavior.
In particular, it allows us to ``deduce'' the complex numbers which are postulated in  standard quantum mechanics.
However, it requires verification.
If the idea is correct, then the quantum phenomena in different invariant subspaces are different manifestations --- visible in appropriate  ``\emph{observational (experimental) setups}'' --- of a single process of permutations of the same collection of objects.
Data relating to different invariant subspaces of permutation representation require interpretation.
For example, the trivial one-dimensional subrepresentation of any permutation representation can be interpreted as a ``\emph{counter of particles}'':
the permutation invariant \Math{\invarL{\wSN}{n}}, corresponding to this sub\-representation, is the total number of particles in the ensemble.
Interpretation of data relating to other invariant subspaces requires further study.
\section*{Acknowledgments}
The author would like to thank  S.A.~Abramov, Yu.A.~Blinkov, S.I.~ Vinitsky, V.P.~Gerdt, V.V.~Ivanov, V.S.~Melezhik, and A.A.~Mikhalev for their participation in discussions of the results presented in the review. 
The work was supported in part by the Russian Foundation for Basic Research (project no. 13-01-00668) and the Ministry of Education and Science of the Russian Federation (project no. 3802.2012.2).

\section*{A~~~Combining Spatial and Internal Symmetries}
\addcontentsline{toc}{section}{A~~Combining Spatial and Internal Symmetries}
\label{Ensymmetryunification}
In general, the group of symmetries of the set of states \Math{\wS} can be an arbitrary subgroup of the symmetric group \Math{\Perm{\wS}}.
However, if we wish the symmetry group \Math{\wG} be constructed from the groups of local \Math{\sG} and spatial \Math{\iG} symmetries, then we must equip the Cartesian product \Math{\iGX\otimes\sG} with a group structure.
Here \Math{\iGX} is the set of functions on the space \Math{\X} with values in the group of internal symmetries \Math{\iG}.
Thus, each element \Math{u\in\wG} is a pair \Math{u = \welem{\alpha(x)}{a}}, where \Math{\alpha(x)\in\iGX} and \Math{a\in\sG}.
It is natural to assume that the result of applying a spatial symmetry to a function from \Math{\iGX} is also an element of \Math{\iGX}.
This means that the group \Math{\iGX} must be a normal subgroup of the desired group  \Math{\wG}, i.e., \Math{\wG} is an \emph{extension} of the group \Math{\sG} by the group \Math{\iGX}.
If we further assume that the symmetries of the space are embedded in the group \Math{\wG} as a subgroup, then we get a particular type of extensions referred to as \emph{split extensions}.
\par
The most important example of split extension is the \emph{wreath product} introduced on page~\pageref{Enwreathpage}.
This construction is adequate to the gauge invariance principle.
In physical theories the independence between the spatial and internal symmetries is usually assumed.
This leads to a simpler version of split extension, namely, to the \emph{direct product} \Math{\wG\cong\iGX\times\sG}. 
Action and group multiplication for the direct product take the form
\begin{align}
	\sigma(x)\welem{\alpha\vect{x}}{a}
	&=
	\sigma\vect{x}\alpha\vect{x},\nonumber\\[-7pt]
	&\label{Endir}\\[-7pt]
	\welem{\alpha\vect{x}}{a}\wmult\welem{\beta\vect{x}}{b}
	&=
	\welem{\alpha\vect{x}\beta\vect{x}}{ab}.\nonumber
\end{align}
The following statement generalizes both of these constructs:\\
\emph{Any antihomomorphism
\Math{\mu: \sG \rightarrow \sG} (\emph{the term ``antihomomorphism'' means, 
that \Math{\mu(a)\mu(b)=\mu(ba)}}) of the group of spatial symmetries specifies an equivalence class of split group extensions:
\begin{equation*}
	\id\rightarrow\iGX\rightarrow\wG\rightarrow\sG\rightarrow\id.
\end{equation*}
The equivalence is described by an arbitrary function \Math{\kappa: \sG \rightarrow \sG.}
The basic operations --- action on \Math{\lSX}, group multiplication, and taking an inverse element --- 
can be written explicitly:}
\begin{align*}
		\sigma(x)\welem{\alpha\vect{x}}{a}
		&=\sigma\vect{x\mu(a)}\alpha\vect{x\kappa(a)}\\[3pt]
		\welem{\alpha\vect{x}}{a}\wmult\welem{\beta\vect{x}}{b}
		&=
	\welem{\alpha\vect{x\kappa(ab)^{-1}\mu(b)\kappa(a)}\beta\vect{x\kappa(ab)^{-1}
	\kappa(b)}}{ab}, \\[3pt]
		\welem{\alpha(x)}{a}^{-1}
		&=
		\welem{\alpha\vect{x\kappa\vect{a^{-1}}^{-1}\mu(a)^{-1}\kappa(a)}^{-1}}
		{a^{-1}}.
\end{align*}
\par
This statement is obtained by specializing the general construction of split extensions of a group \Math{F} by a group \Math{H} (see, for example, \cite{Kirillov}) to the case when  \Math{H} is a group of \Math{\iG}-valued functions on \Math{\X}, while \Math{F} acts on the arguments of these functions.
The equivalence of extensions with the same antihomomorphism \Math{\mu} but different functions \Math{\kappa} is expressed by the following commutative diagram:
\begin{equation*}
\begin{diagram}
\node{\id}
\arrow[1]{e}
\node{\iGX}
\arrow[1]{e}
\arrow[1]{s,=}
\node{\wG}
\arrow[1]{e}
\arrow[1]{s,l}{K}
\node{\sG}
\arrow[1]{e}
\arrow[1]{s,=}
\node{\id}\\
\node{\id}
\arrow[1]{e}
\node{\iGX}
\arrow[1]{e}
\node{~\wG'}
\arrow[1]{e}
\node{\sG}
\arrow[1]{e}
\node{\id}
\end{diagram},
\end{equation*}
where the mapping \Math{K} has the form:~ 
\Math{K: \welem{\alpha(x)}{a}\mapsto\welem{\alpha\vect{x\kappa(a)}}{a}.}
\par
The standard wreath and direct products are obtained from this general construction by choosing the antihomomorphisms \Math{\mu(a)=a^{-1}} and \Math{\mu(a)=\id}, respectively.
As for an arbitrary function \Math{\kappa,} in the mathematical literature is commonly  used \Math{\kappa(a)=a^{-1}} for the wreath product and \Math{\kappa(a)=\id} for the direct product.
Just these functions are used in (\ref{Enwreathaction}--\ref{Enwreathinverse}) and \eqref{Endir}.
\section*{B~~Structural Analysis of Discrete Relations}
\addcontentsline{toc}{section}{B~~Structural Analysis of Discrete Relations}
\label{Endiscreterelations}
The methods of compatibility analysis, such as the Gr?bner basis computation or reduction to involutive form, are widely used to study systems of polynomial and differential equations.
In this appendix we develop similar techniques for discrete systems \cite{Kornyak05,Kornyak06a}.
To illustrate the potential of our approach we present its applications to the study of cellular automata.
\par
Consider the Cartesian product  \Math{\Sigma^n=\Sigma_1\times{}\Sigma_2\times\cdots\times{}\Sigma_n}, i.e., the set of ordered \Math{n}-tuples \Math{\vect{\sigma_1,\sigma_2,\ldots,\sigma_n}},  where \Math{\sigma_i\in{}\Sigma_i} for each \Math{i}. 
By definition, \Math{n}-ary relation is any subset of the \Math{n}-dimensional hyperparallelepiped \Math{\Sigma^n}.
We assume that each \Math{\Sigma_i} is a finite sets of \Math{q_i = \cabs{\Sigma_i}} elements that we shall call \emph{states}.
\par
We can interpret \Math{n} dimensions of the hyperparallelepiped \Math{\Sigma^n} as a set of points \Math{X=\vect{x_1,x_2,\ldots,x_n}}. 
This --- initially amorphous --- set can be turned into a ``space'' by providing \Math{X} with an additional structure that says how ``close'' to each other are different points. 
The most suitable \emph{constructive} mathematical structure for this purpose is an abstract simplicial complex.
\par
The conventional concept of space implies the homogeneity of points of space.
This means that there exists a symmetry group acting transitively on \Math{X}, i.e., the group provides the possibility to ``move'' any point into any other.
The homogeneity is only possible if all the \Math{\Sigma_i} are equivalent.
Let \Math{\Sigma} denote the equivalence class.
We can represent \Math{\Sigma} canonically in the form \Math{\Sigma=\vect{0,\ldots,q-1}}, \Math{q = \cabs{\Sigma}}.
\par
If the number of states is a prime power, \Math{q=p^m}, we can additionally equip the set \Math{\Sigma} with the structure of the Galois field \Math{\F_q}.
Using the functional completeness of polynomials over finite fields \cite{Lidl},
we can represent any \Math{k}-ary relation on \Math{\Sigma} as a set of zeros of some polynomial from the ring \Math{\F_q\ordset{x_1,x_2,\ldots,x_n}}. 
Thus, a set of relations can be implemented as a system of polynomial equations.
Though such a representation is not necessary (and does not work, if \Math{\Sigma_i} are different sets or \Math{q} is not a prime power), 
it is useful due to our habit to employ polynomials wherever possible and possibility to exploit advanced tools of polynomial algebra, e.g.  Gr\"{o}bner bases.
\subsection*{B.1~~Basic Definitions and Constructions} 
\addcontentsline{toc}{subsection}{B.1~~Basic Definitions and Constructions}
Aside from simplices, which are marked out subsets of the set \Math{X}, we will consider \emph{arbit\-rary} subsets of \Math{X}.
For brevity, the sets containing \Math{k} points will be called \Math{k}-sets. 
When dealing with systems of relations defined on different sets of points, we need to establish a correspondence between the points and dimensions of the hypercube \Math{\Sigma^k}. 
To do this we will use exponential notation.
The symbol \Math{\Sigma^{\set{x_i}}} fixes \Math{\Sigma} as the set of values of the point \Math{x_i}.
For \Math{k}-set \Math{\delta=\set{x_{i_1},\ldots,x_{i_k}}}, we introduce the notation \Math{\Sigma^\delta=\Sigma^{\set{x_{i_1}}}\times\cdots\times\Sigma^{\set{x_{i_k}}}}.
The set \Math{\delta} is called the \emph{domain} of relation \Math{R^\delta}, if \Math{R^\delta\subseteq\Sigma^\delta}.
The whole hypercube \Math{\Sigma^\delta} will be called a \emph{trivial relation}.
Note that the relations are very economically represented by the bit strings in the computer memory, and the manipulations with them are efficiently implemented by the bit operations from the basic set of processor instructions.
\subsubsection*{B.1.1~~Relations} 
\addcontentsline{toc}{subsubsection}{B.1.1~~~Relations}
Thus, we have
\par
\textbf{Definition 1} (relation).
A \emph{relation} \Math{R^\delta} on a set of points \Math{\delta=\set{x_{i_1},\ldots,x_{i_k}}} is any subset of the hypercube \Math{\Sigma^\delta}, i.e., \Math{R^\delta\subseteq\Sigma^\delta}.\\
The relation  \Math{R^\delta} can be identified with a Boolean-valued function \Math{R^\delta:} \Math{\Sigma^\delta\rightarrow\set{0,1}},  which is called a \emph{characteristic} or \emph{indicator function}.
\par
An important special case of relations:
\par
\textbf{Definition 2} (functional relation). 
A relation \Math{R^\delta} on a set of points \Math{\delta=\set{x_{i_1},\ldots,x_{i_k}}} is called \emph{functional}, if there exists a position \Math{p\in\vect{1,\ldots,k}}, such that for any\\ 
\Math{\sigma_{i_1},\ldots,\sigma_{i_{p-1}},\sigma_{i_{p+1}},\ldots,\sigma_{i_k},\varsigma,\tau\in\Sigma},\\ from
\Math{\vect{\sigma_{i_1},\ldots,\sigma_{i_{p-1}},\varsigma,\sigma_{i_{p+1}},\ldots,\sigma_{i_k}}\in{}R^\delta}
and \Math{\vect{\sigma_{i_1},\ldots,\sigma_{i_{p-1}}, \tau,\sigma_{i_{p+1}},\ldots,\sigma_{i_k}}\in{}R^\delta}\\ it follows that \Math{\varsigma=\tau}.
\\
If points of space are treated as variables, then, the functional relation can be written as the function
\Mathh{x_{i_p}=F\vect{x_{i_1},\ldots,x_{i_{p-1}},x_{i_{p+1}},\ldots,x_{i_k}},
\text{~where~}F: \Sigma^{\delta\setminus\set{x_{i_p}}}\rightarrow\Sigma\enspace.} 
This explains the name of this type of relations.
\par
We will need to be able to extend relations from sets of points to their supersets:
\par
\textbf{Definition 3} (extension of relation). For a given set of points \Math{\delta},  its subset \Math{\tau\subseteq\delta} and relation \Math{R^\tau} on the subset \Math{\tau} we define \emph{extension} of \Math{R^\tau} as the relation 
\Mathh{R^\delta=R^\tau\times\Sigma^{\delta\setminus\tau}.}
This definition, in particular, allows to extend the relations \Math{R^{\delta_{1}},\ldots,R^{\delta_{m}}}, that are defined on different domains, to the common domain, i.e., to the union \Math{\delta_{i_1}\cup\cdots\cup\delta_{i_m}}.
\par
(Logical) consequences of relations are defined in a natural way:
\par
\textbf{Definition 4} (consequence of relation). A relation \Math{Q^\delta} is called a \emph{consequence} of the relation \Math{R^\delta}, if \Math{R^\delta\subseteq{}Q^\delta\subseteq{}\Sigma^\delta}, i.e., \Math{Q^\delta} is an arbitrary \emph{superset} of the set \Math{R^\delta}.
\par
A relation \Math{R^\delta} has many different consequences: their total number is obviously equal to \Math{2^{\cabs{\Sigma^\delta}-\cabs{R^\delta}}}.
\par
While analyzing the structure of a relation it is important to single out the consequences that can be reduced to relations on smaller sets of points:
\par
\textbf{Definition 5} (proper consequence). A \emph{nontrivial} relation \Math{Q^\tau} is called a \emph{proper con\-sequence} of a relation \Math{R^\delta}, if \Math{\tau} is a \emph{proper subset} of the domain \Math{\delta} (i.e., \Math{\tau\subset\delta}) and the relation \Math{Q^\tau\times\Sigma^{\delta\setminus\tau}} is a consequence of \Math{R^\delta}.
\par
A relation that has no proper consequences will be referred to as a \emph{prime relation}.
\subsubsection*{B.1.2~~Compatibility of systems of relations} 
\addcontentsline{toc}{subsubsection}{B.1.2~~~Compatibility of systems of relations}
Compatibility of several relations can naturally be defined by the intersection of their extensions to the common domain:
\par
\textbf{Definition 6} (base relation). The \emph{base relation} of a system of relations  \Math{R^{\delta_1},\ldots,R^{\delta_m}} is the relation
\begin{equation}
	R^\delta=\bigcap\limits_{i=1}^m{}R^{\delta_i}\times\Sigma^{\delta\setminus\delta_i},
	\text{~where~} \delta=\bigcup\limits_{i=1}{\delta_i}\enspace.
	\label{Enbasisrelation}
\end{equation}
In the polynomial case, when \Math{q=p^n}, a standard tool for the compatibility analysis is the Gr\"{o}bner bases method.
Let us give two remarks about relationship between the Gr\"{o}bner bases and our definitions:
\begin{itemize}
	\item
The compatibility condition determined by the base relation can always be represented by a \emph{single} polynomial, unlike the Gr\"{o}bner basis, which is normally a system of several polynomials.
	\item
Any possible Gr\"{o}bner basis of relations \Math{R^{\delta_1},\ldots,R^{\delta_m}}, represented by polynomials, is some combination of \emph{consequences} of the base relation.	
\end{itemize}
\subsubsection*{B.1.3~~Decomposition of relations} 
\addcontentsline{toc}{subsubsection}{B.1.3~~~Decomposition of relations}
If a relation has proper consequences, we can try to express it in terms of these consequences, i.e., in terms of relations on smaller sets of points.
To this end, we introduce
\par
\textbf{Definition 7} (canonical decomposition). The \emph{canonical decomposition} of a relation \Math{R^\delta} with proper consequences \Math{Q^{\delta_1},\ldots,Q^{\delta_m}} is representation of \Math{R^\delta} in the form
\begin{equation}
	R^\delta=P^\delta\bigcap\vect{\bigcap\limits_{i=1}^m{}Q^{\delta_i}
	\times\Sigma^{\delta\setminus\delta_i}}\enspace,
	\label{Encanondeco}
\end{equation}
where the factor \Math{P^\delta} is defined by the following
\par
\textbf{Definition 8} (principal factor). The \emph{principal factor} of a relation \Math{R^\delta} with proper consequences \Math{Q^{\delta_1},\ldots,Q^{\delta_m}} is the relation
\begin{equation*}
	P^\delta=R^\delta\bigcup\vect{\Sigma^{\delta}\setminus\bigcap\limits_{i=1}^m{}Q^{\delta_i}\times
	\Sigma^{\delta\setminus\delta_i}}\enspace.
\end{equation*}
\par
The principal factor is the maximally ``free'' — i.e., closest to the trivial — relation that, in combination with the proper consequences, makes it possible to restore the original relation.
\par
If the principal factor in the canonical decomposition is trivial, the relation can be completely reduced to relations on smaller sets of points.
\par
\textbf{Definition 9} (reducible relation). A relation \Math{R^\delta}  is called \emph{reducible} if it can be represented as
\begin{equation}
	R^\delta=\bigcap\limits_{i=1}^m{}Q^{\delta_i}\times\Sigma^{\delta\setminus\delta_i}\enspace,
	\label{Enreducibledeco}
\end{equation}
where \Math{\delta_i} are proper subsets of \Math{\delta}.
\par
This definition makes it possible to impose a \emph{``topology''} --- i.e., the structure of an abstract simplicial complex with the associated theories of homologies, cohomologies, etc. --- on an \emph{arbitrary} \Math{n}-ary relation \Math{R\subseteq\Sigma^n}. 
To do this, you need to
\begin{enumerate}
	\item
name the dimensions of the hypercube \Math{\Sigma^n} as the ``points''	\Math{x_1,\ldots,x_n\in{}X};	
	\item
decompose the relation \Math{R} (which can now be denoted by \Math{R^X}) into \emph{irreducible} components;
	\item
declare the domains of the irreducible components of the relation \Math{R^X} as \emph{maximal simplices} of a simplicial complex.	
\end{enumerate}
\subsection*{B.2~~Application to Study of Cellular Automata} 
\addcontentsline{toc}{subsection}{B.2~~Application to Study of Cellular Automata}
\subsubsection*{B.2.1~~Conway's Game of Life} 
\addcontentsline{toc}{subsubsection}{B.2.1~~~Conway's Game of Life}
Conway's automaton Life belongs to the family of binary (i.e., \Math{\Sigma=\set{0,1}; q=2}) cellular automata on a two-dimensional square lattice with local rules set on the \Math{3\times3} Moore neighborhood.
A square lattice with such a neighborhood can be represented by an 8-valent (8-regular) graph. 
The Life automaton rule is described as follows: a cell (a graph vertex) is ``born'' if it has exactly three ``alive'' neighbors,  ``survives'' if it has two to three ``alive'' neighbors, and ``dies'' in all other cases.
Symbolically, this rule can be written in the form of a ``birth''/``survival'' list as B3/S23.
Other examples of automata from this family are HighLife (the rule B36/S23) and Day\&Night (the rule B3678/S34678).
\par
In order to generalize this type of rules we define a \emph{\Math{k}-valent Life rule} as a \emph{binary} rule  on the neighborhood of a \Math{k}-valent graph. 
We will denote the central vertex of the neighborhood by \Math{x_{k+1}}, and the vertices adjacent to \Math{x_{k+1}} by, respectively, \Math{x_{1},\ldots,x_{k}}.
The rule is defined by two \emph{arbitrary} subsets \Math{B} and \Math{S} of the set  \Math{\set{0,1,\ldots,k}}. 
The subsets \Math{B} and \Math{S} contain, respectively, conditions for the forms \Math{0\rightarrow1} (``birth'') and \Math{1\rightarrow1} (``survival'') of one step transition in time \Math{x_{k+1}\rightarrow{}x'_{k+1}}. 
Since the number of subsets of any finite set \Math{A} is \Math{2^{\cabs{A}}} and {different} pairs \Math{B/S}  define different rules, the total number of different rules is equal to
\begin{equation}
	N_{B/S,k} = 2^{k+1}\times2^{k+1}=2^{2k+2}\enspace.\label{Enliferule}
\end{equation}
\par
Let us introduce another definition: a \emph{\Math{k}-symmetric \Math{q}-ary rule} is defined as a rule on a \Math{k}-valent neighborhood, symmetric with respect to all permutations of \Math{k} external vertices of the neighborhood.
It is easy to count the total number of all distinct \Math{k}-symmetric \Math{q}-ary rules:
\begin{equation}
	N_{q,\SymG{k}}=q^{\binom{k+q-1}{q-1}q}\enspace.\label{Ensymrule} 
\end{equation}
We see that for \Math{q=2} the numbers in \eqref{Enliferule} and \eqref{Ensymrule} coincide: \Math{N_{B/S,k}=N_{2,\SymG{k}}}. 
Thus, any \Math{k}-symmetric binary rule can be written in the form of a ``birth''/``survival'' list.
\par
The local relation of the Life automaton, which will be denoted by \Math{R^\delta_{\text{Life}}}, is defined on the 10-set \Math{\delta=\set{x_1,\ldots,x_{10}}}:
\begin{center}
\includegraphics[width=0.5\textwidth]{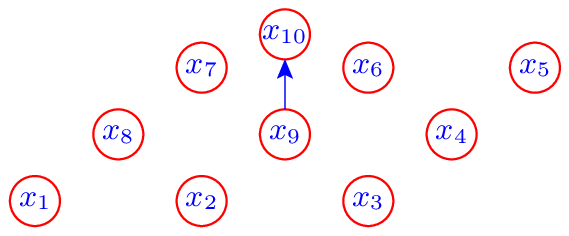}\enspace,
\end{center}
where the point \Math{x_{10}\equiv{}x'_9} is identified with the ``point \Math{x_9} at the next moment of time''.
The definition of the Life automaton rule implies that a point \Math{\vect{x_1,\ldots,x_{10}}\in\Sigma^\delta} of the 10-dimensional hypercube belongs to the relation \Math{R^\delta_{\text{Life}}}, i.e., \Math{\vect{x_1,\ldots,x_{10}}\in{}R^\delta_{\text{Life}}}, at the following conditions
\begin{enumerate}
	\item
\Math{\vect{\sum_{i=1}^8x_i=3}\wedge\vect{x_{10}=1}\enspace,}	 
	\item 
\Math{\vect{\sum_{i=1}^8x_i=2}\wedge\vect{x_{9}=x_{10}}\enspace,}	 
	\item
\Math{x_{10}=0}, if none of the above conditions holds.	 
\end{enumerate}
The number of elements of the relation is \Math{\cabs{R^\delta_{\text{Life}}}=512.}
The relation \Math{R^\delta_{\text{Life}}}, as in the case of any cellular automaton, is \emph{functional}: a state of \Math{x_{10}} is uniquely determined by the states of other points.
If the set \Math{\Sigma=\set{0,1}} is additionally endowed with the structure of the field \Math{\F_2}, then the relation \Math{R^\delta_{\text{Life}}} can be written as a polynomial from the ring \Math{\F_2\ordset{x_1,\ldots,x_{10}}}. 
This allows to study the relation both by our method of structural analysis and by polynomial methods. 
Let us compare our approach with the method of Gr\"{o}bner bases.
\par
Polynomial representation for the relation \Math{R^\delta_{\text{Life}}} has the form 
\begin{equation}
P_{\text{Life}}= x_{10}
+x_9
\vect{
\esp_7
+\esp_6
+\esp_3
+\esp_2
}
+\esp_7
+\esp_3,
\label{Enpolylife}
\end{equation}
where \Math{\esp_k\equiv\esymm{k}{x_1,\ldots,x_8}} is the \Math{k}th \emph{elementary symmetric polynomial} defined for \Math{n} variables \Math{x_1,\ldots,x_{n}} by the formula
\Mathh{\esymm{k}{x_1,\ldots,x_{n}} =
\sum\limits_{1\leq{}i_1<i_2<\cdots<i_{k}\leq{}n}x_{i_1}x_{i_2}\cdots x_{i_{k}}.}
Further we will use the following notation
\Mathh{\esp_k\equiv\esymm{k}{x_1,\ldots,x_{8}},~\esp_k^i\equiv\esymm{k}{x_1,\ldots,\widehat{x_i},\ldots,x_{8}},~\esp_k^{ij}\equiv\esymm{k}{x_1,\ldots,\widehat{x_i},\ldots,\widehat{x_j},\ldots,x_{8}}.}
\par 
Applying our program for structural analysis of discrete relations to \Math{R^\delta_{\text{Life}}}, we find that the relation is \emph{reducible} and its canonical decomposition is of the form
\begin{equation}
R^\delta_{\text{Life}} =
R_2^{\delta\setminus\set{x_9}}\bigcap
\vect{\bigcap\limits_{k=1}^7
R_1^{\delta\setminus\set{x_{i_k}}}}\enspace,
\label{Enlifedecomp}
\end{equation}
where \Math{\vect{i_1,\ldots,i_7}} is an arbitrary 7-element subset of the set \Math{\vect{1,\ldots,8}}.
For brevity we have dropped in \eqref{Enlifedecomp} the trivial factors \Math{\Sigma^{\set{x_{i_k}}}}, that are present in general formula \eqref{Enreducibledeco}.
\par
Eight relations \Math{R_1^{\delta\setminus\set{x_{i}}}}, \Math{1\leq{}i\leq8} --- to construct decomposition \eqref{Enlifedecomp} it suffices to take any seven of them --- have the following polynomial form
\Mathh{x_9x_{10}
\vect{\esp^i_6+\esp^i_5+\esp^i_2+\esp^i_1}
+x_{10}\vect{\esp^i_6+\esp^i_2+1}
+x_9\vect{\esp^i_7+\esp^i_6+\esp^i_3+\esp^i_2}=0.}
The polynomial form of the relation \Math{R_2^{\delta\setminus\set{x_9}}} is of the form
\Mathh{x_{10}\vect{\esp_7+\esp_6+\esp_3+\esp_2+1}+\esp_7+\esp_3=0.}
The relations \Math{R_1^{\delta\setminus\set{x_{i}}}} and \Math{R_2^{\delta\setminus\set{x_9}}} are \emph{irreducible} but \emph{not prime}, and can be decomposed in accordance with formula \eqref{Encanondeco}.
Continuing the iterations of decompositions, we eventually come to the maximally simplified system of relations, which is equivalent to the original relation \Math{R^\delta_{\text{Life}}}.
Let us present this simplified system in the polynomial form:
\begin{align}
	x_9x_{10}\vect{\esp^i_2+\esp^i_1}
	+x_{10}\vect{\esp^i_2+1}
	+x_9\vect{\esp^i_7+\esp^i_6+\esp^i_3+\esp^i_2}&=0,
	\label{Enpoly1red}
	\\
	x_{10}\vect{\esp_3+\esp_2+1}+\esp_7+\esp_3&=0,
	\label{Enpoly2red}
	\\
	\vect{x_9x_{10}+x_{10}}\vect{\esp^{ij}_3+\esp^{ij}_2+\esp^{ij}_1+1}&=0,
	\label{Enpoly11red}
	\\
	x_{10}\vect{\esp^{i}_3+\esp^{i}_2+\esp^{i}_1+1}&=0,
	\label{Enpoly12red}
	\\
	x_{10}x_{i_1}x_{i_2}x_{i_3}x_{i_4}&=0.
	\label{Enpoly0123red}
\end{align}
The simplest ones of these relations, namely \eqref{Enpoly0123red}, can be easily interpreted:
if the state of point \Math{x_{10}} is 1, then at least one of any of the four points surrounding \Math{x_{9}}, must be in the state 0.
\par
The above analysis of the relation \Math{R^\delta_{\text{Life}}} is performed in less than a second on an average laptop.
For computing the Gr\"{o}bner basis, ten polynomials 
\Mathh{x_i^2+x_i,~~i=1,\ldots,10\enspace}
must be added to polynomial \eqref{Enpolylife}, in accordance with the identity \Math{x^q=x}, which is hold for any element \Math{x} of any finite field  \Math{\F_q}. 
The calculation using the standard \textbf{Maple} function takes approximately one hour.
The result depends on the ordering of the variables that must be set when using the method of Gr\"{o}bner bases:
\begin{itemize}
	\item 
pure lexicographic order with the variable ordering \Math{x_{10}\succ x_9\succ\cdots\succ x_1} does not provide any new information leaving original polynomial \eqref{Enpolylife} intact;	
	\item
pure lexicographic order with the variable ordering \Math{x_1\succ x_2\succ\cdots\succ x_{10}} and  degree-reverse-lexicographic order reproduce relations \eqref{Enpoly1red}---\eqref{Enpoly0123red}
(up to several additional polynomial reductions violating the symmetry of polynomials).
\end{itemize}
\subsubsection*{B.2.2~~Elementary cellular automata} 
\addcontentsline{toc}{subsubsection}{B.2.2~~~Elementary cellular automata}
The simplest binary one-dimensional cellular automata were studied by S. Wolfram, who named them \emph{elementary cellular automata} \cite{Wolfram}.
They are much easier for study than Conway’s Life, so we may pay more attention to topological aspects of the structural analysis.
\par
Local rules of the elementary cellular automata are defined on the 4-set \Math{\delta=\set{p,q,r,s}}, which can be pictured by the diagram
\begin{center}
\includegraphics[width=0.15\textwidth]{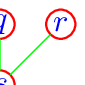}.
\end{center}
The local rule is a binary function of the form \Math{s=f(p,q,r)}.
The total number of such functions is \Math{2^{2^3}=256}.
Therefore, each rule can be indexed by an 8-bit binary number.
\par
Our calculations show that the 256 relations that corresponds to these rules are divided into 118 \emph{reducible} and 138 \emph{irreducible} relations.
Only two of the irreducible relations are \emph{prime}, namely the rules 105 and 150 in the Wolfram numbering.%
\footnote{Bit string of values of a function \Math{s=f(p,q,r)} in lexicographically ascending order of combinations of values of the variables \Math{p,q,r} is interpreted as binary digits in descending order of the rule ordinal number.}
The prime rules 105 and 150 are represented by the linear polynomials \Math{s=p+q+r+1} and \Math{s=p+q+r}, respectively.
\par
Space-time of elementary automata is a lattice with integer coordinates \Math{(x,t)\in\Z\times\Z}.
Of course, we can without any problem replace \Math{\Z\times\Z} by the finite lattice \Math{\Z_m\times\ordset{\tin,1,\ldots,\tfin}}, if we take sufficiently large \Math{m} and \Math{\tfin}.
State of a lattice will be represented by a function \Math{u(x,t)} with values in \Math{\Sigma=\set{0,1}}.
If we represent the lattice as a graph, then in the general case, we assume that the lattice points are connected as is shown in the picture
\begin{center}
\includegraphics[width=0.3\textwidth]{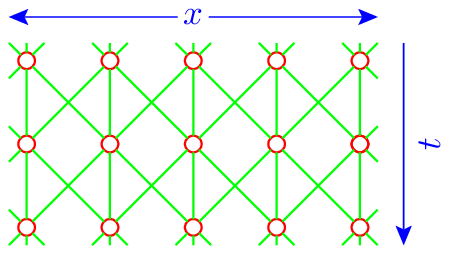}.
\end{center}
The absence of horizontal ties expresses the independence of states of the ``space-like'' points in cellular automata.
\paragraph{Reducible automata.}
The analysis shows that some automata with reducible local relations can be represented as unions of automata defined on disconnected (not intersecting) subcomplexes:
\par
\begin{itemize}
	\item 
Two automata 0 and 255 are determined by the unary relations \Math{s=0} and \Math{s=1} on the disconnected set of points:
\begin{center}
\includegraphics[width=0.2\textwidth]{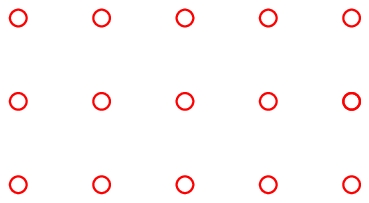}.
\end{center}
Note that in mathematics an unary relation is usually called a \emph{property}.
	\item 
Six automata 15, 51, 85, 170, 204 and 240 are actually disjoint collections of spacially zero-dimensional automata, i.e., single points evolving in time.
For example, consider the automaton 15. Its local relation, which is defined on the set
\raisebox{-0.02\textwidth}{\includegraphics[width=0.1\textwidth]{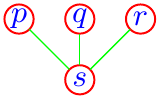}},
can be represented by the bit string  0101010110101010.
This relation is reduced to a relation on the face 
\raisebox{-0.02\textwidth}{\includegraphics[width=0.062\textwidth]{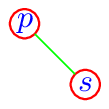}}
with the bit string 0110.
The space-time lattice splits as follows: 
\begin{center}
\includegraphics[width=0.24\textwidth]{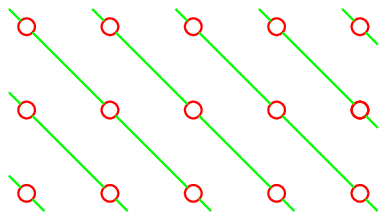}.
\end{center}
The bit string 0110 means that the points \Math{p} and \Math{s} are always in opposite states, and we can write immediately the general solution for the automaton 15:
\Mathh{
u(x,t) = a(x-t)+t\mod 2,
}
where \Math{u(x,0)\equiv{}a(x)} is an arbitrary initial condition.
	\item
Each of the ten automata 5, 10, 80, 90, 95, 160, 165, 175, 245, 250 is decomposed into two identical automata.
For example, consider the rule 90.
This automaton is distinguished as producing a fractal of topological dimension 1 and Hausdorff dimension \Math{\ln3/\ln2\approx1.58}, which is known as the ``Sierpinski sieve'' (or ``Sierpinski gasket'' or ``Sierpinski triangle'').
The local relation of the rule 90 on the set 
\raisebox{-0.02\textwidth}{\includegraphics[width=0.1\textwidth]{ECA-Set-pqrs-Color}}
is represented by the bit string 1010010101011010.
This relation, in accordance with decomposition \eqref{Enreducibledeco}, is reduced to a relation with the bit string 
\begin{equation}
10010110 \text{~~~~on the face~~
\raisebox{-0.02\textwidth}{\includegraphics[width=0.12\textwidth]{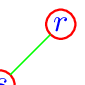}}}.
\label{Enbt90}
\end{equation}
It is seen from the structure of this face that the space-time lattice splits into two identical independent complexes:
\begin{center}
\includegraphics[width=0.8\textwidth]{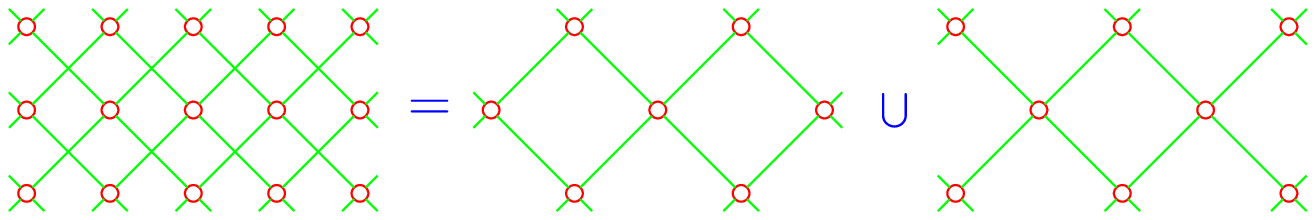}\enspace.
\end{center}
To find a general solution of the automaton 90 it is convenient to use the polynomial form \Math{s+p+r=0} of relation \eqref{Enbt90}. 
With this linear expression, the general solution is easily constructed:
\Mathh{
u(x,t)=\sum\limits_{k=0}^t \binom{t}{k}a(x-t+2k)\mod 2,\qquad{} u(x,0)\equiv{}a(x).
}
\end{itemize}
\paragraph{Using proper consequences.}
Proper consequences --- even if they are not functional --- contain useful information about the global behavior of a cellular automaton.
\par
For example, 64 automata%
\footnote{The full list of such automata in the Wolfram numbering is
  2,   4,   8,  10,  16,  32,  34,  40,
 42,  48,  64,  72,  76,  80,  96, 112,
128, 130, 132, 136, 138, 140, 144, 160,
162, 168, 171, 174--176, 186, 187,
190--192, 196, 200, 205, 206, 208,
220, 222--224, 234--239,
241--254. 
The automata in this list have both reducible and irreducible local relations.
}
have proper consequences with the bit string 
\begin{equation}
1101
\label{Enfiniterod}
\end{equation}
on, at least, one of the faces
\begin{equation}
\text{\includegraphics[width=0.09\textwidth]{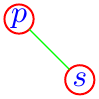}
~~~~
\includegraphics[width=0.035\textwidth]{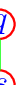}
~~~~
\includegraphics[width=0.09\textwidth]{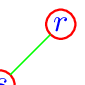}}.
\label{Enfiniterodsets}
\end{equation}
Relation \eqref{Enfiniterod} has on faces \eqref{Enfiniterodsets}, respectively, the following polynomial forms
\Mathh{ps+s=0,~~qs+s=0,~~rs+s=0.}
\par
\begin{figure}[!h]
\begin{center}
\includegraphics[width=0.5\textwidth]{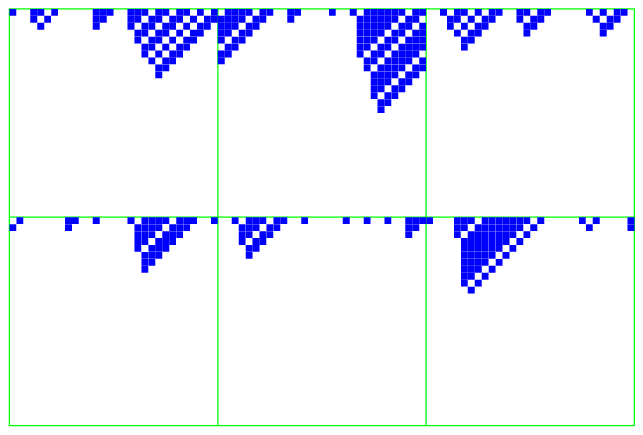}
\caption{Rule 168. Several evolutions with random initial conditions.\label{Enfigu}}
\end{center}
\end{figure}

Relation \eqref{Enfiniterod} is \emph{not functional}, and therefore cannot describe any deterministic evolution.
Nevertheless, it imposes severe restrictions on the behavior of respective automata. 
The features of behavior resulting from relation \eqref{Enfiniterod} are clearly seen in many illustrations to computational results presented in the atlas \cite{site}.
A typical picture from this atlas, showing several evolutions of the automaton  168, is reproduced in Figure \ref{Enfigu}.
In the figure, zeros and ones are represented by empty and filled square cells, respectively.
Note that the authors of the calculations used periodicity in the spatial dimension: \Math{x\in\Z_{30}}.
\par
The polynomial form of local relation of the automaton 168 is \Math{pqr+qr+pr+s=0}.
This relation has the proper consequence \Math{rs+s=0}, which means that if \Math{r} is in state 1, \Math{s} may be in either of two states 0 or 1, but if the state of \Math{r} is 0,  than \Math{s} \emph{must} be in state 0.
Symbolically:
\begin{align*}
	r=1\Rightarrow{}&s=0\vee{}s=1,\\
	r=0\Rightarrow{}&s=0.
\end{align*}
Figure \ref{Enfigu} shows that any evolution consists of left down diagonals without gaps and of finite length.
Each diagonal begins with several consecutive units, but after the first occurrence of zero all subsequent points of the diagonal can only be zero.
\paragraph{Canonical decomposition vs Gr\"{o}bner basis.}
It is clear that our canonical decom\-position \eqref{Encanondeco} is a more general method of studying discrete relations than the Gr\"{o}bner bases.
However, in the polynomial case, both methods can be compared.
As an example, let us give the Gr\"{o}bner bases and canonical decompositions for two elementary cellular automata.
The Gr\"{o}bner bases were computed in the total degree and reverse lexicographic order of monomials.
The trivial polynomial \Math{p^2+p,} \Math{q^2+q,} \Math{r^2+r} and \Math{s^2+s} are excluded from the descriptions of Gr\"{o}bner bases.
\begin{itemize}
\item 
\textbf{Automaton 30} is remarkable by its chaotic behavior. It is even used as a random number generator in \emph{\textbf{Mathematica}}.
 \par
Relation: \Math{1001010101101010} ~or~  \Math{qr+s+r+q+p=0}.
\par
\textbf{Canonical decomposition:}
\par
Proper consequences:
\par
\begin{tabular}{lcc}
face&
\includegraphics[width=0.09\textwidth]{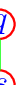}
&
\includegraphics[width=0.14\textwidth]{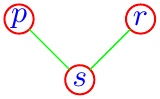}
\\[10pt]
bit string&11011110 & 11011110\\[10pt]
polynomial \hspace*{30pt}&\hspace*{20pt}\Math{qs+pq+q}\hspace*{20pt} & \Math{rs+pr+r}.\\[10pt]
\end{tabular}
\par
Principal factor: \Math{1011111101111111} ~or~ \Math{qrs+pqr+rs+qs+pr+pq+s+p=0.}
\par
\textbf{Gr\"{o}bner basis:}
\Math{\set{qr+s+r+q+p,~qs+pq+q,~rs+pr+r}.}\\[5pt]
Thus, for the automaton 30 the set of polynomials of the canonical decomposition coincides with the Gr\"{o}bner basis (up to the obvious polynomial substitutions into the principal factor).
\item 
\textbf{Automaton 110} is --- like a Turing machine --- \emph{universal}, i.e., it can be used to implement any computational process, in particular, to reproduce the behavior of any other cellular automaton.
\par
Relation: \Math{1100000100111110} ~or~ \Math{pqr+qr+s+r+q=0.}
\par
\textbf{Canonical decomposition:}
\par
Proper consequences:
\par
\begin{tabular}{lccc}
face&
\includegraphics[width=0.09\textwidth]{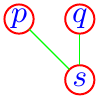}
&
\includegraphics[width=0.14\textwidth]{ECA-Set-prs-Color}
&
\includegraphics[width=0.09\textwidth]{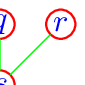}
\\[10pt]
bit string&11011111 & 11011111 &10010111\\[10pt]
polynomial\hspace*{30pt}&
\Math{pqs+qs+pq+q} &
\Math{prs+rs+pr+r} & \Math{qrs+s+r+q}.\\[10pt]
\end{tabular}
\par
Principal factor:
\Math{1111111111111110} ~or~ \Math{pqrs=0.}
\par
\textbf{Gr\"{o}bner basis:}
\Mathh{\set{prs+rs+pr+r,~qs+rs+r+q,~qr+rs+s+q,~pr+pq+ps}.}
In this case, the Gr\"{o}bner basis polynomials do not coincide with polynomials of the canonical decomposition.
The system of relations corresponding to the Gr\"{o}bner basis has the form
\begin{eqnarray*}
R_1^{\set{p,r,s}}&=&11011111=\vect{prs+rs+pr+r=0},
\\
R_2^{\set{q,r,s}}&=&10011111=\vect{qs+rs+r+q=0},
\\
R_3^{\set{q,r,s}}&=&10110111=\vect{qr+rs+s+q=0},
\\
R_4^{\set{p,q,r,s}}&=&1110101110111110=\vect{pr+pq+ps=0}.
\end{eqnarray*}
\end{itemize}
In general, we can mention the following distinctions between our approach and the method of Gr\"{o}bner bases.
\begin{itemize}
	\item 
In contrast to the Gr\"{o}bner basis, the base relation, defined in \eqref{Enbasisrelation} as an intersection of conditions, is consistent with a natural for logic and set theory notion of compatibility.	
	\item
Unlike the canonical decomposition, a Gr\"{o}bner basis may look outside the polynomial context as a set of accidental supersets.
	\item
However, there is a certain analogy between  Gr\"{o}bner bases and canonical decompo\-sitions. 
In fact, we observed the coincidence of respective sets of polynomials in about half of our calculations.
	\item
The canonical decomposition is more efficient in the problems, in which polynomials may have arbitrary  degree ---  the given above calculation with the Conway automaton illustrates this fact.	
  \item
The Gr\"{o}bner bases method outperforms the canonical decomposition in problems with small degree polynomials in a large number \Math{n} of variables. 
The number of polynomials of limited degree depends polynomially on \Math{n}, whereas the canonical decomposition algorithm  scans exponentially large number \Math{q^{\textstyle{n}}} of points of a hypercube \Math{\Sigma^n}.
\end{itemize}
\section*{C~~Local Quantum Models on Regular Graphs}
\addcontentsline{toc}{section}{C~~Local Quantum Models on Regular Graphs}
\label{Enquantumonregulargraphs}
In standard quantum mechanics, the amplitude along the path is proportional to the exponential of the action: \Math{A\propto\exp\vect{\frac{i}{\hbar}S}}.
It is clear that this expression never vanishes.
Therefore, in the standard formulation quantum transitions between any states of a system are possible in principle.
Here we consider more convenient to study models with restricted sets of possible transitions.
Namely, in our models \cite{Kornyak09c,Kornyak10}, which are defined on regular graphs, the transitions are possible only within the neighborhoods of vertices of graphs. 
\subsection*{C.1~~General Definition of Quantum Model on Graph} 
\addcontentsline{toc}{subsection}{C.1~~General Definition of Quantum Model on Graph}
Our definition of \emph{local quantum model on \Math{k}-valent graph} uncludes the following:
\begin{enumerate}
\item
\emph{Space} \Math{\X=\set{\x_1,\ldots, \x_\XN}}  is a \Math{k}-valent graph.
 The vertices of the graph, which are adjucent to a given vertex \Math{\x_i}, will be denoted by  \Math{\x_{1,i},\ldots,\x_{k,i}}.
\item
\emph{Scheme of local transitions} \Math{E_i=\set{e_{0,i}, e_{1,i},\cdots,e_{k,i}}} is the set of edges adjucent to a vertex \Math{\x_i}, supplemented by the edge \Math{e_{0,i}=\vect{\x_i\rightarrow{}\x_i}\equiv\vect{\x_i\rightarrow{}\x_{0,i}}}. 
Each edge \Math{e_{m,i}=\vect{\x_i\rightarrow{}\x_{m,i}}} symbolizes the one-time-step transition from the vertex \Math{\x_i} to the vertex \Math{\x_{m,i}}.
\item
We assume that the space symmetry group \Math{\sG=\mathrm{Aut}\vect{\X}} acts transitively on the set of schemes of local transitions \Math{\set{E_1,\cdots,E_{\XN}}}.
\item
We define the \emph{local space symmetry group} as a maximal subgroup of \Math{\sG} that fixes a vertex  \Math{\x_i}: \Math{\sGloc=\mathrm{Stab}_{\sG}\vect{\x_i}\leq\sG}.
By the transitivity of the action of \Math{\sG} on the set of neighborhoods, the group \Math{\sGloc} does not depend on the choice of the vertex.
\item
\Math{\Omega_i=\set{\omega_{0,i},\omega_{1,i},\cdots,\omega_{h,i}}}	denotes the \emph{set of orbits} of \Math{\sGloc} on \Math{E_i}.
It is clear that these sets are isomorphic for all vertices \Math{\x_i}.
\item
We assume that each vertex has internal degrees of freedom, i.e., the set of local states \Math{\lS} on which the group of internal symmetries  \Math{\iG} acts.
\item
\emph{Evolution rule} \Math{R} is a function on \Math{E_i} with values in some representation \Math{\Rep{\iG}} of the internal symmetry group.
The evolution rule prescribes \Math{\Rep{\iG}}-valued weights to one-step-in-time transitions between vertices of the neighborhood of a vertex \Math{\x_i}. 
For symmetry reasons, the rule \Math{R} must be a function on orbits from the set \Math{\Omega_i}, i.e., \Math{R\vect{e_{m,i}g}=R\vect{e_{m,i}}}, if \Math{g\in\sGloc}.
\end{enumerate}
To illustrate these constructions consider the local quantum model on the graph of the fullerene \Math{C_{60}}.
The graph is shown in Figures \ref{EnNanoCarbons} and \ref{Enbucky}. 
In this case, the symmetry group of the space  \Math{X=\set{\x_1,\cdots,\x_{60}}} is \Math{\sG=\mathrm{Aut}\vect{\X}=\AltG{5}\times\CyclG{2}}.
The neighborhood of a vertex \Math{\x_i} is arranged as follows
\begin{center}
\includegraphics[width=0.17\textwidth]{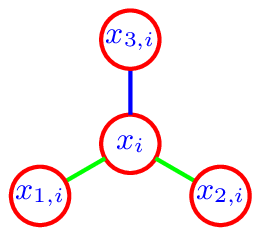}.
\end{center}
The scheme of local transitions has the form \Math{E_i=\set{e_{0,i}, e_{1,i}, e_{2,i}, e_{3,i}}}, 
where
\begin{align*}
		 e_{0,i}&=\vect{\x_i\rightarrow{}\x_i},\\
		 e_{1,i}&=\vect{\x_i\rightarrow{}\x_{1,i}},\\
		 e_{2,i}&=\vect{\x_i\rightarrow{}\x_{2,i}},\\
		 e_{3,i}&=\vect{\x_i\rightarrow{}\x_{3,i}}.
\end{align*}
Note that the edges of the neighborhood are not equivalent:
the edges \Math{e_{1,i}}  and \Math{e_{2,i}} belong to the pentagon adjacent to the vertex \Math{\x_i}, whereas the edge \Math{e_{3,i}} separates two hexagons;
or, in terms of chemistry, the edge \Math{e_{3,i}} corresponds to the double bond in the carbon molecule \Math{C_{60}}, whereas the other edges represent single bonds.
Thus the group of local symmetries of space, i.e., stabilizer of a vertex \Math{\x_i}, is \Math{\sGloc=\mathrm{Stab}_{\sG}\vect{\x_i}=\CyclG{2}}.
The set of orbits of \Math{\sGloc} on \Math{E_i} consists of three elements (subsets):
 \Mathh{\Omega_i=\set{\omega_{0,i}=\set{e_{0,i}}, \omega_{1,i}=\set{e_{1,i}, e_{2,i}}, \omega_{2,i}=\set{e_{3,i}}},}
i.e., the stabilizer  interchanges the edges  \Math{\vect{\x_i\rightarrow{}\x_{1,i}}} and
\Math{\vect{\x_i\rightarrow{}\x_{2,i}}},  but leaves fixed the edges \Math{\vect{\x_i\rightarrow{}\x_i}}
and
 \Math{\vect{\x_i\rightarrow{}\x_{3,i}}}. 
\par
In view of these symmetries the evolution rule takes the form
\begin{align*}
&R\vect{\x_i\rightarrow{}\x_i}~~=\Rep{\alpha_0},\\
R\vect{\x_i\rightarrow{}\x_{1,i}}=
&R\vect{\x_i\rightarrow{}\x_{2,i}}=\Rep{\alpha_1},\\
&R\vect{\x_i\rightarrow{}\x_{3,i}}=\Rep{\alpha_2},	
\end{align*}
where \Math{\alpha_0,\alpha_1,\alpha_2\in\iG}. 
\subsection*{C.2~~Illustration: Discrete Model of Free Quantum Particle} 
\addcontentsline{toc}{subsection}{C.2~~Illustration: Discrete Model of Free Quantum Particle}
Consider the quantization of a free particle moving in one dimension.
Such a particle is described by the Lagrangian \Math{L = \frac{\textstyle{m\dot{x}^2}}{\textstyle{2}}.}
Assuming that there are quantum transitions only to the nearest points of the discretized space, we arrive at the following rule of transitions in one step in time:
\begin{center}
\includegraphics[width=0.20\textwidth]{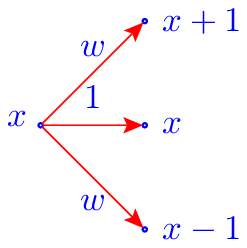}\hspace*{20pt}
\raisebox{34pt}{\begin{tabular}{l}
\Math{e^{\textstyle{\frac{i}{\hbar}\frac{m\left\{(x+1)-x\right\}^2}{2}}}
=e^{\textstyle{i\frac{m}{2\hbar}}}}\\[10pt]
\Math{e^{\textstyle{\frac{i}{\hbar}\frac{m\vect{x-x}^2}{2}}}\hspace*{17pt}=~\id}\\[10pt]
\Math{e^{\textstyle{\frac{i}{\hbar}\frac{m\left\{(x-1)-x\right\}^2}{2}}}=e^{\textstyle{i\frac{m}{2\hbar}}},}
\end{tabular}}
\end{center}
i.e., the evolution rule \Math{R} is a function with values in the unitary representation \Math{\UG{1}} of a circle:
\begin{align*}
		R\vect{x\rightarrow{}x}&=1~~\in~~\UG{1},\\
		R\vect{x\rightarrow{}x-1}=R\vect{x\rightarrow{}x+1}
		&=w=e^{\textstyle{i\frac{m}{2\hbar}}}~~\in~~\UG{1}.
\end{align*}
Suppose further that \Math{w} is an element of a one-dimensional representation of a finite group \Math{\iG}.
One-dimensionality of representation implies that without loss of generality, we can assume the group is cyclic \Math{\iG=\Z_M},
and \Math{w} is a primitive \Math{M}th root of unity.
Rearranging the \emph{multinomial coefficients} --- \emph{trinomial} in this particular case --- it is easy to write the sum of amplitudes over all possible paths between the space-time points \Math{\vect{0,0}} and \Math{\vect{x,t}}
\begin{equation*}
A_x^t\vect{w}=\sum\limits_{\tau=0}^t\frac{\tau!}{\vect{\frac{\tau-x}{2}}!\vect{\frac{\tau+x}{2}}!}
	\times
	\frac{t!}{\tau!\vect{t-\tau}!}~w^{\tau}.
\end{equation*}
Note that \Math{x} must lie within \Math{\pm{}t}: \Math{x\in\left[-t,t\right].}
\par
The model can be complicated by adding, for example, gauge connections acting along the paths, imposing restrictions on possible paths, such as the screen in the ``double-slit experiment'', etc.
In any case, the amplitude \Math{A\vect{w}} will be a polynomial in \Math{w}.
In order to reproduce, say, destructive interference --- one of the striking manifestations of quantum behavior --- it is necessary to solve the system of polynomial equations \Math{A\vect{w}=\Phi_M\vect{w}=0}, 
where \Math{\Phi_M\vect{w}} is \Math{M}th cyclotomic polynomial.
In this model, the smallest group, in which destructive interference appears, is \Math{\CyclG{4}}.
\begin{figure}[!h]
\centering
\includegraphics[width=0.95\textwidth]{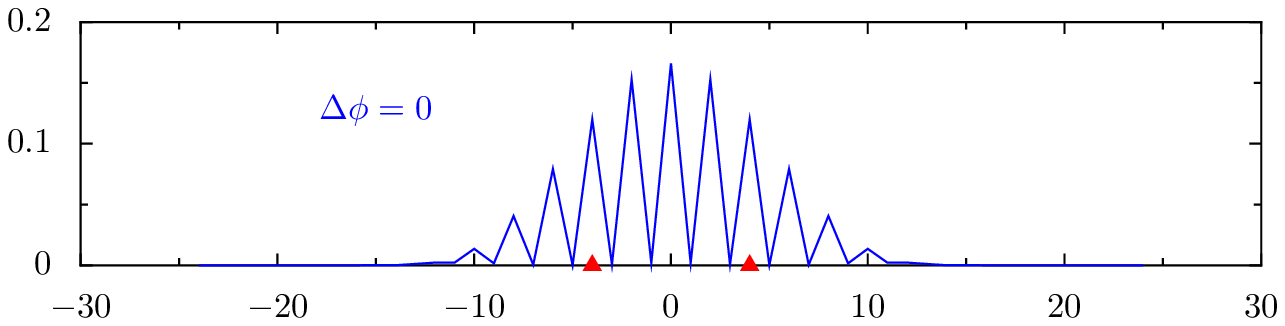}\\
\includegraphics[width=0.95\textwidth]{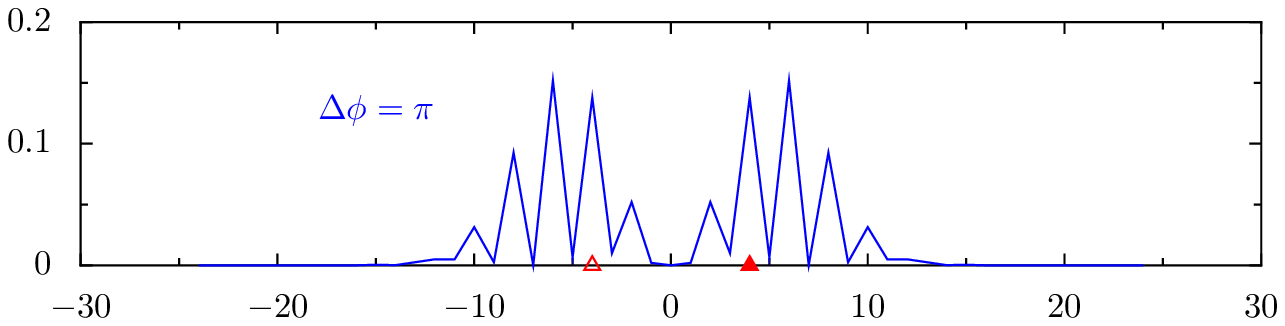}
\caption{Group \Math{\CyclG{4}}. Interference from two sources located at positions -4 and 4. 
Number of time steps \Math{T=20}. \Math{\Delta\phi=\phi_4-\phi_{-4}} is a phase shift between the sources.}
	\label{EnInterf2}
\end{figure}
\par
Figure  \ref{EnInterf2} shows the interference patterns --- the normalized squared amplitudes (``probabilities'') --- from two sources placed at points \Math{x=-4} and \Math{x=4} for 20 time steps. 
The top and bottom panels show interference in the cases, when the source phases are identical (\Math{\Delta\phi=0}) and opposite (\Math{\Delta\phi=\pi}), respectively.
\section*{D~~Linear Representations of Finite Groups}
\addcontentsline{toc}{section}{D~~Linear Representations of Finite Groups}
\label{Enirreps}
Any linear representation of a finite group is equivalent to a unitary representation, since an invariant inner product can be constructed from an arbitrary inner product by means of ``averaging over the group''.
For example, we can start from the standard inner product 
\begin{equation}
    \innerstandard{\phi}{\psi}\equiv\sum\limits_{i=1}^{\adimH}\cconj{\phi^i}\psi^i
\label{EninnerstdEn}
\end{equation}
in the \Math{\adimH}-dimensional Hilbert space \Math{\Hspace}.
Averaging \eqref{EninnerstdEn} over the group we come to the \emph{invariant scalar product}
\begin{equation*}
\inner{\phi}{\psi}\equiv\frac{\textstyle{1}}{\textstyle{\cabs{G}}}\sum\limits_{g\in{}G}
\!\innerstandard{U\!\vect{g}\phi}{U\vect{g}\psi},
\end{equation*}
which ensures unitarity of the representation \Math{U} of the group \Math{G} in the space \Math{\Hspace}.
\par
An important transformation of the group elements is the conjugation: \Math{a^{-1}ga\rightarrow{}g',} \Math{g, g'\in\wG,} \Math{a\in\SymG{\wG}} --- it is an analog of a change in the reference frame in physics.
Conjugation by an element of the group itself, i.e., if \Math{a\in\wG}, is called an \emph{inner automorphism}.
The equivalence classes with respect to inner automorphisms are called \emph{conjugacy classes}.
The decomposition of a group into conjugacy classes
\Math{~\wG=\class{1}\sqcup\class{2}\sqcup\cdots\sqcup\class{\classN}~}
is a starting point in studying its representations.
\par
The group multiplication induces a \emph{multiplication on classes}.
The product of classes \Math{\class{i}} and \Math{\class{j}} is the \emph{multiset} of all possible products \Math{ab,~a\in\class{i},~b\in\class{j}}, decomposed into classes.
Obviously, this multiplication is commutative since \Math{ab} and \Math{ba} belong to the same class: \Math{ab\sim{}a^{-1}\vect{ab}a=ba}. 
Thus, we have the following multiplication table for classes
\begin{equation*}
    \class{i}\class{j}= \class{j}\class{i} = \sum\limits_{k=1}^{\classN}c_{ijk}\class{k}.
\end{equation*}
\emph{Natural integer} \Math{c_{ijk}} --- multiplicities of classes in multisets --- are called \emph{class coefficients}.
Obtained in this way algebra of classes contains all the information needed to construct representations of the group.
\par
Here is a short list of the basic properties of linear representations of finite groups:
\begin{enumerate}
    \item
Any irreducible representation is contained in the regular representation.
More concretely, there is a transformation matrix \Math{\transmatr} that reduces simultaneously all  matrices of the regular representation to the form
\begin{equation}
    \transmatr^{-1}\regrep(g)\transmatr=
    \bmat
            \repirr_1(g) &&&
    \\[5pt]
    &
    \hspace*{-27pt}d_2\left\{
    \begin{matrix}
    \repirr_2(g)&&\\
    &\hspace*{-10pt}\ddots&\\
    &&\hspace*{-7pt}\repirr_2(g)
    \end{matrix}
    \right.
    &&
    \\
    &&\hspace*{-10pt}\ddots&\\
    &&&
    \hspace*{-25pt}d_{\classN}\left\{
    \begin{matrix}
    \repirr_{\classN}(g)&&\\
    &\hspace*{-10pt}\ddots&\\
    &&\hspace*{-7pt}\repirr_{\classN}(g)
    \end{matrix}
    \right.
    \emat,
\label{EnregrepdecompEn}
\end{equation}
and any irreducible representation is one of \Math{\repirr_j}.
The number of nonequivalent irreducible representations is equal to the number of conjugacy classes.
The number \Math{d_j}  is simultaneously the dimension of the irreducible component 
\Math{\repirr_j} and the multiplicity of its occurence in the regular representation.
From \eqref{EnregrepdecompEn} one can see that the relation
\Math{d^2_1+d^2_2+\cdots+d^2_{\classN}=\cabs{\wG}=\wGN} holds for the dimensions of irreducible representations. 
Moreover, the dimensions of irreducible representations are divisors of the order of the group: \Math{d_j\mid\wGN.}
    \item
Any irreducible representation \Math{\repirr_j}  is unambiguously determined by its \emph{character} \Math{\chi_j}, i.e., by the set of traces of representation matrices \Math{\chi_j\vect{g}=\mathrm{Tr}\repirr_j\vect{g}}.
This set is in fact a function on conjugacy classes,
since, by virtue of standard properties of traces of matrices, we have \Math{\chi_j\vect{g}=\chi_j\vect{a^{-1}ga}}.
it is obvious that \Math{\chi_j\vect{\id}=d_j}.
    \item
A compact way of enumerating all the irreducible representations is the \emph{character table}.
The columns of this table are numbered by the conjugacy classes, and the rows contain values of the characters of non-equivalent representations.
\begin{table}[h]
\begin{center}
\begin{tabular}{c|cccc}
&\Math{\class{1}}&\Math{\class{2}}&\Math{\cdots}&\Math{\class{\classN}}\\\hline
\Math{\chi_1}&1&1&\Math{\cdots}&1\\
\Math{\chi_2}&\Math{\chi_2\vect{\class{1}}=d_2}&\Math{\chi_2\vect{\class{2}}}&
\Math{\cdots}&\Math{\chi_2\vect{\class{\classN}}}\\
\Math{\vdots}&\Math{\vdots}&\Math{\vdots}&&\Math{\vdots}\\
\Math{\chi_{\classN}}&\Math{\chi_{\classN}\vect{\class{1}}=d_{\classN}}&
\Math{\chi_{\classN}\vect{\class{2}}}&\Math{\cdots}&\Math{\chi_{\classN}\vect{\class{\classN}}}
\end{tabular}.
\end{center}
\caption{Character table.}
\label{Enchartab}
\end{table}
According to the tradition, the first column corresponds to the class of the \emph{identity} element, and the first row contains the \emph{trivial} representation.
Both the rows and the columns are pairwise \emph{orthogonal}.
Important information about the group and its representations can be read directly from the character table.
For example, an irreducible representation is \emph{faithful} if and only if the value of the character in the first column (that is, the dimension) is not repeated anywhere in the respective row. 
Tables of characters determine groups almost completely.
They do not distinguish only isoclinic groups \cite{atlas}.
Groups are called \emph{isoclinic} if their quotients by their centers are isomorphic.
Isoclinic groups have identical tables of characters, since the characters ``ignore'' centers.
An example of isoclinic groups is given by the eight-element dihedral group \Math{\DihG{8}=\set{\text{\textit{symmetries of a square}}}} and the quaternion group \Math{\QuatG{8}=\set{\pm1,\pm{}i,\pm{}j,\pm{}k}},  where \Math{i,j,k}  are quaternionic imaginary
units.
\end{enumerate}
\section*{E~~Finite Symmetry Groups and Phenomenology of Ele\-mentary Particles}
\addcontentsline{toc}{section}{E~~Finite Symmetry Groups and Phenomenology of Elementary Particles}
\label{Enappflavor}
Currently, all experimental data \cite{RevPartPhys} on fundamental particles are consistent with the Standard Model, which is a gauge theory whose group \Math{\iG} of internal (gauge) symmetries is the direct product of the groups \Math{\UG{1}}, \Math{\SU{2}} and \Math{\SU{3}}.
In the context of Grand Unified Theories, it is assumed that \Math{\iG} is a subgroup of some larger (presumably simple) group.
With respect to the space-time symmetries, the elementary particles are divided into two classes: \emph{bosons}, which are responsible for physical forces (roughly speaking, they correspond to elements of the gauge group), and \emph{fermions}, which are interpreted as particles of matter.
The fermions of the Standard Model form three \emph{generations} of \emph{quarks} and \emph{leptons}, which are presented in Table \ref{Ensmfermions}.
\begin{table}[h]
\begin{center}
\begin{tabular}{l|c|c|c}
generations&1&2&3
\\\hline
up quarks&\Math{u}&\Math{c}&\Math{t}
\\
down quarks&\Math{d}&\Math{s}&\Math{b}
\\\hline
charged leptons&\Math{e^-}&\Math{\mu^-}&\Math{\tau^-}
\\
neutrinos&\Math{\nu_e}&\Math{\nu_\mu}&\Math{\nu_\tau}
\end{tabular}
\end{center}
\caption{Fermions of the Standard Model (antiparticles are omitted)}
\label{Ensmfermions}
\end{table}
Particles of different generations differ from each other only by their masses and quantum property called \emph{flavor}.
Flavor-changing physical processes, such as weak quark decays or neutrino oscillations, are described by unitary \Math{3\times3} \emph{mixing matrices.} 
The experimental data allow to determine absolute values of the elements of these matrices.
\par
In the case of quarks (``in the quark sector''), the matrix of transitions between up and down quarks is known as the \emph{Cabibbo–Kobayashi–Maskawa (CKM) matrix} 
\Mathh {
V_{\text{CKM}}=\Mthree{V_{ud}}{V_{us}}{V_{ub}}
       {V_{cd}}{V_{cs}}{V_{cb}}
       {V_{td}}{V_{ts}}{V_{tb}},
}
where \Math{\cabs{V_{\alpha\beta}}^2} is the probability of transition of a quark (with flavor) \Math{\beta} to a quark \Math{\alpha} in a weak process.
The experimental data obtained so far give the following values for the matrix element moduli:
\Mathh{
\Mthree{\cabs{V_{ud}}}{\cabs{V_{us}}}{\cabs{V_{ub}}}
       {\cabs{V_{cd}}}{\cabs{V_{cs}}}{\cabs{V_{cb}}}
       {\cabs{V_{td}}}{\cabs{V_{ts}}}{\cabs{V_{tb}}}
=\Mthree{0.974}{~0.225}{~0.004}
        {0.225}{~0.974}{~0.041}
        {0.009}{~0.040}{~0.999}.
}
We used the rounding to three decimal digits --- more accurate values can be found in \cite{RevPartPhys}.
\par
In the lepton sector, weak processes are described by the \emph{Pontecorvo-Maki-Nakagawa-Sakata (PMNS) mixing matrix}
\Mathh
{
U_{\text{PMNS}}=\Mthree{U_{e1}}{U_{e2}}{U_{e3}}
       {U_{\mu1}}{U_{\mu2}}{U_{\mu3}}
       {U_{\tau1}}{U_{\tau2}}{U_{\tau3}}.
}
Here, the indices \Math{e, \mu, \tau} correspond to the neutrino flavors.
This means that in the weak processes the neutrinos \Math{\nu_e, \nu_\mu, \nu_\tau} are produced in association with \Math{e^+, \mu^+, \tau^+}  (or generate \Math{e^-, \mu^-, \tau^-}), respectively. 
The indices \Math{1, 2, 3} correspond to the eigenvalues of the mass operator, i.e., the symbols \Math{\nu_1, \nu_2, \nu_3} denote neutrinos with definite masses \Math{m_1, m_2, m_3}.
Numerous observations of solar and atmospheric neutrinos and experiments with neutrinos produced by reactors and accelerators reveal discrete (in fact, finite) symmetries that cannot be derived from the Standard Model.
The phenomenological picture with reasonable accuracy looks as follows \cite{Smirnov}:
\begin{enumerate}
    \item  
The flavors \Math{\nu_\mu} and \Math{\nu_\tau} are distributed over all three masses \Math{\nu_1, \nu_2, \nu_3} with equal weights (this is called ``bimaximal mixing''):
\Math{\cabs{U_{\mu{i}}}^2=\cabs{U_{\tau{i}}}^2,} \Math{i=1,2,3};
    \item
All three flavors are present equally in the mass eigenstate \Math{\nu_2} (``trimaximal mixing''): \Math{\cabs{U_{e{2}}}^2=\cabs{U_{\mu{2}}}^2=\cabs{U_{\tau{2}}}^2};
    \item  \Math{\nu_e} is absent in \Math{\nu_3}: \Math{\cabs{U_{\mu3}}^2=0}.
\end{enumerate}
These relations, together with the probability normalization conditions, lead to the following values for the squares of the matrix element moduli:
\begin{equation}
    \Mone{\cabs{U_{l{i}}}^2}
    =\Mthree{~\fra{2}{3}}{~~\fra{1}{3}}{~~0~}
            {\\[-8pt]~\fra{1}{6}}{~~\fra{1}{3}}{~~\fra{1}{2}~}
            {\\[-8pt]~\fra{1}{6}}{~~\fra{1}{3}}{~~\fra{1}{2}~}.
\label{Entribi2}
\end{equation}
A particular form of the unitary matrix that satisfied \eqref{Entribi2} was proposed by Harrison, Perkins, and Scott \cite{HPS02}:
\begin{equation*}
    U_{\text{TB}}
    =\Mthree{~\,\sqrt{\fra{2}{3}}}{~~\fra{1}{\sqrt{3}}}{~~0}
            {-\fra{1}{\sqrt{6}}}{~~\fra{1}{\sqrt{3}}}{~-\!\fra{1}{\sqrt{2}}}
            {-\fra{1}{\sqrt{6}}}{~~\fra{1}{\sqrt{3}}}{~~~~\,\fra{1}{\sqrt{2}}}.
\end{equation*}
This so-called \emph{tribimaximal} (TB) mixing matrix coincides, up to trivial permutation of two columns corresponding to the renaming \Math{\nu_1\rightleftarrows\nu_2} of two mass states, with transformation matrix \eqref{EntransS3sqrt}, which decomposes the three-dimensional permutation representation of the group \Math{\SymG{3}} into irreducible components. 
This means that we can identify the flavor basis with the basis of permutations of three elements, and the mass basis with the basis in which the permutation representation is decomposed into irreducible components.
In \cite{HS03}, Harrison and Scott study in detail the relation between the neutrino mass operator and the character table and class algebra of the group \Math{\SymG{3}}.
\par 
Presently,  models based on finite group of flavor symmetries are intensively developed and studied (reviews of this activity are presented in, for example, \cite{Ishimori,Ludlgen}). 
We list here some of the popular groups used in constructing such models. 
\begin{itemize}
    \item \Math{\mathsf{T}=\AltG{4}} --- tetrahedral group,
    \item   \Math{\TprimG} --- double cover of the  tetrahedral group \Math{\AltG{4}},
    \item \Math{\mathsf{O}=\SymG{4}} --- octahedral group,
    \item \Math{\mathsf{I}=\AltG{5}} --- icosahedral group,
    \item \Math{\DihG{N}} --- dihedral group (\Math{N} is even),
    \item \Math{\QuatG{N}} --- quaternion group (4 divides \Math{N}),
    \item   \Math{\Sigma\vect{2N^2}}
--- groups of this series have the structure
\Math{\vect{\CyclG{N}\times\CyclG{N}}\rtimes\CyclG{2}},
    \item   \Math{\Delta\vect{3N^2}} --- structure
\Math{\vect{\CyclG{N}\times\CyclG{N}}\rtimes\CyclG{3}},
    \item   \Math{\Sigma\vect{3N^3}}
--- structure
\Math{\vect{\CyclG{N}\times\CyclG{N}\times\CyclG{N}}\rtimes\CyclG{3}},
    \item   \Math{\Delta\vect{6N^2}} --- structure
\Math{\vect{\CyclG{N}\times\CyclG{N}}\rtimes\SymG{3}}.
\end{itemize}
\par
In the quark sector observations do not give such a clear picture as in the case of leptons.
There is a number of studies devoted to the search for finite symmetries in the quark sector and symmetries common to quarks and leptons.
For example,  in \cite{BlumHagedorn} the group \Math{\DihG{14}} was used to explain the value of the Cabibbo angle (one of the parameters of the CKM mixing matrix), but without any connection with lepton symmetries.
On the whole, the natural attempts to find discrete symmetries, that are common to leptons and quarks, are still not too successful.
However, there are some encouraging observations: for example, the \emph{quark-lepton complementarity} --- the experimental fact of the approximate equality of the sum of mixing angles for quarks and leptons to  \Math{\pi/4}.
\par
The origin of finite symmetries, associated with the fundamental particles, is not clear at the present time. 
There are various attempts at explaining it, sometimes slightly overcomplicated and artificial.
For example, these symmetries are interpreted as symmetries of manifolds that arise upon the compactification of extra dimensions in theories with higher dimensional spaces \cite{Altarelli}.
From our point of view, the idea that symmetries at the most fundamental level are finite in themselves looks more attractive.
In this approach, the unitary groups used in physical theories are some kind of reservoirs absorbing, for all intents and purposes, all possible finite groups that have faithful representations of respective dimension.
For example, the group \Math{\SU{n}} contains as subgroups all finite groups that have faithful \Math{n}-dimensional representations with unit determinant.
Moreover, because of the redundancy of the field \Math{\C}, the group \Math{\SU{n}} is not a minimal group that contains all finite groups with these properties.
\par
Most likely, such small groups as \Math{\SymG{3},~\AltG{4}}, etc., are just remnants of large combinations (extensions, direct and semidirect products) of more fundamental finite symmetries. 
It can be expected that the natural scale at which such fundamental symmetries  manifest themselves, is the scale of Grand Unification.
One could try to use the symmetries of the Standard Model for the purposeful search for fundamental symmetries.
For example, finite subgroups of the group \Math{\SU{3}}, that have faithful three-dimensional representations, were completely described. 
Their list contains several infinite series and several individual groups.
However, such an attempt would hardly be successful, since the Standard Model is a compact framework for describing phenomenology obtained at energy scales of about \Math{10^{4}} GeV,
whereas the Grand Unification scale is approximately \Math{10^{16}} GeV,  which is quite close to the Planck scale (\Math{10^{19}} GeV), and which is far beyond the experimental capabilities.
Apparently, the only practical approach is constructing models and comparing their implications with available experimental data.
In such kind of studies, methods of computational group theory \cite{Holt} may prove to be useful.
\cleardoublepage
\selectlanguage{russian}
\renewcommand{\thesection}{\arabic{section}}
\renewcommand{\figurename}{Рис.}
\renewcommand{\tablename}{Таблица}
\stepcounter{rupage}
\stepcounter{rusection}
\stepcounter{rufootnote}
\stepcounter{ruequation}
\stepcounter{rufigure}
\stepcounter{rutable}
\addtocounter{page}{1}
\begin{center}
\textbf{\Large{}Классические и квантовые дискретные динамические системы}\\[10pt]
{\Large{}Владимир В. Корняк}\\[8pt]
{\large{}Объединённый институт ядерных исследований, Дубна, Россия}
\end{center}
\def\abstractname{Аннотация}
\begin{abstract}
\noindent
Мы рассматриваем детерминистическую и квантовую динамику с конструк\-тив\-ной ``конечной'' точки 
зрения, поскольку введение континуума или других актуальных бесконечностей в физику создаёт серьёзные
концептуальные и технические трудности без какой-либо необходимости в этих понятиях для 
физики как эмпирической науки.
Особое внимание уделяется симметрийным свойствам дискретных систем.
Для согласованного описания симметрий динамических систем в различные моменты времени
и симметрий различных частей таких систем мы вводим дискретные аналоги калибровочных связностей.
Эти структуры особенно важны для описания квантового поведения.
Симметрии определяют фундаментальные свойства поведения динамических систем.
В частности, можно показать, что движущиеся солитоноподобные структуры возникают неизбежно
в детерминистической  
динамической системе, группа симметрий которой разбивает 
множество состояний на конечное число групповых орбит.
Мы показываем, что квантовое поведение динамических систем является естественным следствием 
их симметрий.
Такое поведение является результатом фундаментальной невозможности
проследить идентичность неразличимых объектов в процессе эволюции.
Доступна лишь информация об инвариантных утверждениях и величинах,
относящихся к таким объектам. 
Используя математические аргументы общего характера можно показать,
что любая квантовая динамика может быть сведена к последовательности перестановок.
Квантовые интерференции возникают в инвариантных подпространствах
перестановочных представлений групп симметрий динамических систем.
Наблюдаемые величины могут быть выражены в терминах перестановочных инвариантов.
Мы показываем также, что для описания квантовых явлений,
вместо неконструктивной числовой системы --- поля комплексных чисел,
достаточно использовать циклотомические поля --- минимальные расширения натуральных
чисел, пригодные для квантовой механики.  
Конечные группы симметрий играют центральную роль в данной статье.
Интерес к таким группам имеет дополнительную мотивацию в физике. 
Многочисленные эксперименты и наблюдения в физике элементарных частиц указывают на важную 
роль конечных групп относительно невысоких порядков в ряде фундаментальных процессов. 
Происхождение этих групп не имеет объяснения в рамках признанных в настоящее время
теорий, в частности, в Стандартной Модели.
\end{abstract}
{PACS: 03.65.Aa,~03.65.Fd,~03.65.Ta,~02.20.-a,~11.30.-j,~11.15.-q}
\section{Введение}
Существует много аргументов, как прагматического так и концептуального характера, 
в пользу того, что дискретная, или даже конечная математика более пригодна для 
описания физической реальности, чем непрерывная.
Особенно отчетливо это проявляется на фундаментальном уровне.
В частности, изучение физических процессов, пространственно-временн\'ые масштабы 
которых близки к планковским,
почти неизбежно приводит к необходимости введения дискретных структур.
В качестве примера можно привести \emph{``голографический принцип''} Г. `т Хоофта \cite{tHooft}, 
возникший при изучении термодинамики чёрных дыр.
Согласно этому принципу вся физическая информация, содержащаяся
в некоторой области пространства, полностью определяется дискретными
данными на двумерной границе этой области.
Плотность информации этих данных равна, самое большее, 
одному биту на планковскую площадь (\emph{``предел Бекенштейна''}).
С более спекулятивной точки зрения вся Вселенная представляет собой конечную 
информационную структуру на замкнутой двумерной решётке --- 
\emph{``космологическом горизонте''},
а наблюдаемые три измерения возникают лишь как результат эффективного описания 
в макроскопических масштабах и при низких энергиях.
\par 
Изучение дискретных систем важно и с прикладной точки зрения:
многие физические объекты, даже если адекватные для их описания 
пространственно-временн\'ые масштабы существенно больше фундаментальных 
--- \emph{наноструктуры}, например, ---
являются по существу дискретными, а не непрерывными, образованиями.
\par
С концептуальной точки зрения вопрос о том, 
``является ли реальный мир дискретным или непрерывным'' или даже
``конечным или бесконечным'' относится исключительно к метафизике, 
так как никакие эмпирические наблюдения или логические аргументы не 
в состоянии обосновать тот или иной выбор ---
это вопрос веры, вкуса или привычки.
Вообще, в практике идея непрерывности возникает из макроскопического опыта 
как приближённое (\emph{``осреднённое'', ``термодинамическое''}) описание больших 
совокупностей дискретных элементов. 
Типичный пример физической системы, порождающей эмпирическое представление о 
непрерывности  --- жидкость. Динамика жидкости описывается
уравнениями в частных производных типа уравнений Навье-Стокса, 
тогда как на более фундаментальном уровне, жидкость --- это совокупность молекул, 
а на ещё более фундаментальном  --- комбинация квантовых
частиц, подчиняющихся дискретным законам%
\footnote{Описание физической системы называют \emph{феноменологическим}, если 
известно или предполагается, что оно может быть, хотя бы в принципе, получено из 
некоторого более фундаментального описания с помощью приближений и упрощающих предположений.
В случае уравнений гидродинамики цепочку таких упрощений, начиная от 
квантово-механического описания, легко проследить: ВКБ-приближение \Math{\rightarrow} уравнение
Лиувилля для функции распределения \Math{\rightarrow} разложение функции распределения по порядкам
моментов \Math{\rightarrow} ``гипотезы замыкания'' (феноменологические соотношения, выражающие
``высшие'' моменты через ``низшие'' и, таким образом,  позволяющие ``оборвать'' 
бесконечную цепочку уравнений для моментов) \Math{\rightarrow} уравнения гидродинамики.}. 
Математическое понятие континуума представляет собой логическое уточнение
эмпирической идеи непрерывности.
\par
Конечно, дискретные и непрерывные \emph{математические}
теории существенно различаются и эффективность их применения в физике зависит от 
конкретных исторических обстоятельств.
Непрерывность позволяет создавать модели физических систем, основанные на
дифференциальных уравнениях, для исследования и решения которых было разработано 
большое количество разнообразных методов.
Эмпирическим эквивалентом понятия производной является предположение о том, что
малые изменения физических величин приблизительно пропорциональны малым изменениям
координат, в которых эти величины заданы.
Ясно, что такая гипотеза существенно упрощает физические модели.
Фактически, со времён Ньютона 
и до появления современных компьютеров анализ и дифференциальная геометрия были 
единственным средством математического изучения физических систем.
Высоко оценивал эвристическую силу понятия непрерывности Пуанкаре (хотя он и  
не признавал фундаментальную обоснованность этого понятия). 
В книге ``Ценность науки''
(\cite{PoincareVal}, стр. 288--289 русского перевода) Пуанкаре пишет:
``\emph{Единственный естественный предмет 
математической мысли есть целое число. Непрерывность \textellipsis, без сомнения,
изобретена нами, но изобрести ее нас вынудил внешний мир.}
\emph{Без него не было бы анализа бесконечно малых. 
Все математическое знание свелось бы к \emph{арифметике} 
или к \emph{теории подстановок}.}
\emph{Но мы, напротив, посвятили изучению 
непрерывности почти все наше время, почти все наши силы.}
\textellipsis\emph{Вам, без сомнения, скажут, что вне целого числа 
нет строгости, а следовательно, нет математической 
истины, что оно скрывается всюду и что нужно 
стараться разоблачить его покровы, хотя бы для этого 
пришлось обречь себя на нескончаемые повторения.}
\emph{Но мы не будем столь строги; мы будем 
признательны непрерывности, которая, если даже {всё} 
исходит из целого числа, одна только была способна 
извлечь из него {так много}.}'' 
\par
В настоящее время цифровые технологии почти повсеместно вытесняют аналоговые.
Благодаря компьютерам реальные возможности дискретной математики в 
приложениях существенно выросли.
Соответственно и ``дискретный'' стиль мышления становится всё более популярным.
Важным преимуществом дискретного описания является его 
концептуальная ``экономность'' в оккамовском смысле --- отсутствие ``лишних сущностей'' 
основанных на идеях актуальной бесконечности типа ``дедекиндовых сечений'', 
``последовательностей Коши'' и т. п. 
Более того, дискретная математика  содержательно 
богаче непрерывной --- непрерывность ``сглаживает'' тонкие детали строения структур.
Упрощения описания, достигаемые введением производных в математические структуры, 
могут обходиться достаточно дорого. 
Для иллюстрации этого тезиса можно сравнить списки
простых групп Ли и простых конечных групп.
Благодаря дифференцируемости и, следовательно, возможности введения алгебр Ли, классификация 
простых групп Ли --- сравнительно несложная задача, которая была решена в начале 20-го века.
Результат классификации: 4 бесконечных последовательности и 5 исключительных групп.
Классификация простых конечных групп --- исключительно трудная проблема, 
решение которой, как считается, было завершено к 2004 году.
Список простых конечных групп состоит из 18 бесконечных последовательностей 
(16 из которых содержат непосредственные аналоги всех простых групп Ли) и 26 спорадических групп.
\par
Статья начинается с общего обсуждения дискретных динамических систем.
Наиболее фундаментальными понятиями являются дискретное время и множество состояний,
эволюционирующих во времени. Пространство рассматривается как дополнительная структура,
позволяющая организовать множество состояний в виде множества функций на точках пространства
со значениями в некотором множестве, называемом множеством локальных состояний.
Мы обсуждаем симметрии пространства и локальных состояний и то, каким
образом эти симметрии могут быть объединены в полную группу симметрий системы в целом%
\footnote{Если подойти к проблеме с противоположной стороны, то пространство можно 
построить из заданного полного множества состояний как множество классов смежности 
по некоторой подгруппе полной группы, а группу симметрий пространства как эффективно действующую на 
этом множестве классов часть полной группы. Однако в данной статье мы не будем 
развивать эту тему.}.
\par
Важной особенностью систем с симметриями является наличие степеней свободы в их описании, 
связанных с произволом в выборе систем координат. Если этот выбор ``глобальный'', т.е. установлен
раз и навсегда для всей эволюции системы или (при наличии 
нетривиальной пространственной структуры) для разных частей системы, то он не может
сказаться на объективно наблюдаемых данных. Возможность ``локального'' выбора систем координат
приводит к физически наблюдаемым калибровочным эффектам.
Естественной математической схемой, подходящей для описания таких эффектов, является теория связностей 
в расслоениях.
Поэтому в данной статье мы используем подходящим образом модифицированные для дискретного случая понятия
расслоений и связностей в этих расслоениях. Для иллюстрации мы показываем как, исходя из введённых 
нами конструкций, можно построить континуальные версии абелевых (электродинамика) и неабелевых (теория
Янга-Миллса) калибровочных теорий.
\par
С помощью простой модели мы показываем, каким образом в дискретной динамике могут возникать
характерные пространственно-временн\'ые структуры, такие как, например, предельная скорость
и световые конусы.
\par
Далее мы обсуждаем особенности поведения классических детерминистических систем. 
Мы показываем, что при наличии симметрий в таких системах совершенно естественным образом возникают
сохраняющие форму движущиеся (солитоноподобные) структуры. Мы также обсуждаем идею `т Хоофта 
о том, каким образом может возникать наблюдаемая в природе обратимость фундаментальных процессов и показываем, что
в случае дискретных детерминистических систем (априори редкая) обратимость неизбежно возникает
в результате эволюции как эффективно наблюдаемое явление.
\par
Наиболее фундаментальную --- фактически определяющую --- роль симметрии играют
в феноменах квантового поведения.
Характерной чертой квантового поведения является его универсальность.
Его демонстрируют системы совершенно различной физической природы и 
масштабов: от субатомных элементарных частиц до больших молекул%
\footnote{В частности, в работе \cite{fullerinterferEn} описаны
эксперименты по наблюдению квантовых интерференций между
молекулами фуллерена \Math{C_{60}}, а недавно были опубликованы \cite{phthalointerfer}
данные об интерференции для гораздо большей молекулы --- производной фталоцианина 
с брутто-формулой \Math{C_{48}H_{26}F_{24}N_8O_8}.}.
Универсальность явления (т.е. независимость от физического субстрата) 
указывает на то, что описывающая его теория должна
основываться на некоторых априорных математических принципах.
Мы показываем, что в случае квантовой механики таким ведущим математическим принципом 
является симметрия. Наши аргументы основаны на том, что квантовое поведение
демонстрируют лишь системы, содержащие неразличимые частицы:
любое отклонение от точной идентичности частиц разрушает квантовые интерференции.
Неразличимость элементов системы означает, что они лежат на одной и той же
орбите группы симметрий этой системы.
Для систем с симметриями объективными являются только не зависящие от 
``переобозначений'' ``однородных'' элементов, т.е. \emph{инвариантные}, отношения и утверждения.
Например, никакого объективного смысла нельзя приписать электрическим потенциалам
\Math{\varphi} и \Math{\psi} или точкам пространства, обозначаемым векторами
\Math{\mathbf{a}} и \Math{\mathbf{b}}. Объективный смысл имеют лишь некоторые их комбинации,
например, комбинации, обозначаемые как
\Math{\psi-\varphi} или \Math{\mathbf{b}-\mathbf{a}} (или в более общей 
символике теории групп \Math{\varphi^{-1}\psi} и \Math{\mathbf{a^{-1}b}}).
\par
В основе квантового формализма лежат унитарные операторы в гильбертовых пространствах.
Эти операторы принадлежат общей унитарной группе, или в более строгих терминах,
\emph{унитарному представлению} группы автоморфизмов (преобразований, 
сохраняющих эрмитово скалярное произведение) гильбертова пространства.
Для придания конструктивности квантовым концепциям мы можем заменить
общую унитарную группу унитарными представлениями \emph{конечных групп}
без всякого риска исказить \emph{физическое} содержание проблемы поскольку, как отмечалось
выше, метафизическое различие между ``\emph{конечным}'' и ``\emph{бесконечным}''
не может иметь каких-либо эмпирически наблюдаемых последствий.
Более того, имеются надёжные экспериментальные данные о том, что в основе некоторых
фундаментальных процессов лежат конечные%
\footnote{Приложение \ref{appflavor} содержит краткий обзор того, 
как конечные симметрии проявляются в феноменологии физики ароматов.}
группы относительно невысоких порядков. 
\par
Используя тот факт, что \emph{любое линейное представление} конечной группы
является подпредставлением перестановочного, мы показываем,
что любая квантово-механическая проблема может быть сведена к \emph{перестановкам},
а квантовые наблюдаемые могут быть выражены в терминах \emph{перестановочных инвариантов}.
Более детальный анализ ``перестановочного'' подхода показывает, что
комплексные числа в квантово-механическом формализме должны быть заменены 
циклотомическими.
Если принять эту модификацию формализма, то
\begin{itemize}
	\item 
квантовые амплитуды приобретают простой и естественный смысл:
они являются проекциями в инвариантные подпространства векторов кратностей
(``чисел заполнения'') элементов множества, на котором группа действует перестановками;
	\item 
борновские	вероятности оказываются рациональными числами ---
в полном соответствии с естественной 
для конечных множеств 
``частотной интерпретацией'' вероятностей;
	\item
квантовые явления возникают как результат фундаментальной невозможности
проследить идентичность неразличимых объектов	в процессе эволюции
--- доступна только информация об инвариантных комбинациях таких объектов.
\end{itemize}
Некоторые темы, несколько отклоняющиеся от основного направления статьи 
(структурный анализ дискретных отношений, клеточные автоматы, локальные 
квантовые модели на графах и др.), а также информация технического характера
вынесены в приложения.
\par
Для обеспечения целостности изложения и единства обозначений с самого начала опишем
наиболее фундаментальные понятия, используемые в данной статье.
\paragraph{Натуральные числа.}
Исходной базовой системой чисел у нас является \emph{полукольцо натуральных чисел} 
\begin{equation*}
	\N=\set{0,1,2,\ldots}.
\end{equation*}
Подчеркнём, что мы включаем нуль в множество натуральных чисел. 
Натуральные числа наиболее естественны в дискретной математике, где они выступают как
счётчики элементов и комбинаций элементов дискретных систем. 
Остальные конструктивные системы чисел выводятся из множества \Math{\N} чисто математическими
средствами. Например, \emph{кольцо целых чисел} \Math{\Z = \set{\ldots,-1,0,1,\ldots}} 
мы рассматриваем, как \emph{алгебраическое расширение} полукольца \Math{\N} с помощью 
\emph{примитивного квадратного корня из единицы}%
\footnote{Такое определение, как будет видно из дальнейшего, мотивировано квантовой механикой.
Другой распространённый способ введения целых чисел --- 
это интерпретация их в виде классов эквивалентности пар натуральных чисел с отношением 
эквивалентности \Math{\mathbf{\sim}}: \Math{\vect{n,m}\mathbf{\sim}\vect{n',m'}} 
означает, что \Math{n+m'=n'+m}, где 
\Math{n,m,n',m'\in\N}.}, поле рациональных чисел \Math{\Q}, в свою очередь, интерпретируется,
как поле частных кольца \Math{\Z} и т.д.
\paragraph{Расслоения.} Для систем с симметриями, в особенности при исследовании их квантового
поведения, важную роль играет понятие расслоения. \emph{Расслоением}  по определению%
\footnote{В литературе теория расслоений обычно излагается как составная 
часть дифференциальной геометрии \cite{Kobayashi,Sulanke}. 
Однако существенные для нас элементы этой теории не требуют понятия дифференцируемости 
и без труда переносятся на дискретный и конечный случай.}
называется структура вида
\begin{equation*}
	\vect{E,X,S,G,\pi},
\end{equation*}
где
\begin{description}
	\item
	\Math{E} --- множество, называемое \emph{(тотальным) пространством расслоения}; 
	\item 
\Math{X} --- множество, называемое \emph{базой расслоения};	 
	\item 
\Math{S} --- множество, называемое \emph{типичным (модельным) слоем расслоения};	 
	\item 
\Math{G} --- группа преобразований типичного слоя \Math{S}, называемая 
\emph{структурной группой расслоения};	 
	\item
\Math{\pi} --- отображение \Math{E\rightarrow{}X}, называемое \emph{проекцией расслоения}. 
\end{description}
\par
Проекция \Math{\pi:E\rightarrow{}X} характеризуется следующим свойством.
Если \Math{x} --- точка базы, т.е. \Math{x\in{}X}, то множество 
\Math{S_x=\pi^{-1}\vect{x}}, называемое \emph{слоем над точкой} \Math{x}, изоморфно
типичному слою \Math{S}. Изоморфизм здесь означает, что существует элемент структурной
группы \Math{g_x\in{}G}, такой, что \Math{S_xg_x^{-1}=S}.
\par 
Роль структурной группы в физике заключается в том, что она даёт возможность с помощью 
калибровочных связностей%
\footnote{
Понятие связности, существенно зависящее от структуры базы расслоения, 
мы рассмотрим после более конкретного обсуждения дискретных версий базы.}
 ``сравнивать'' данные, относящиеся 
к различным моментам времени и разным точкам пространства.
\par
\emph{Сечением расслоения} \Math{\vect{E,X,S,G,\pi}} называется отображение 
\Math{\sigma: X\rightarrow{}E} такое, что \Math{\pi\circ{}\sigma=\mathrm{id}_X}.
Символически множество всех сечений можно представить как
\Math{\prod\limits_{x\in{}X}S_x}. Таким образом, для конечных 
\Math{X} и \Math{S} полное число всех возможных сечений равно 
\Math{\cabs{S}^{\cabs{X}}}.
\par
Расслоение называется
\emph{главным}, если типичным слоем является сама структурная группа. 
Символически главное расслоение можно записать в виде 
\begin{equation*}
	\vect{P,X,G,\rho},
\end{equation*}
где \Math{P} --- тотальное пространство, а \Math{\rho} --- проекция. 
Подразумевается, что структурная группа \Math{G} действует 
на слое \Math{G} правыми (или левыми) сдвигами. 
\section{Классическая и квантовая динамика}
\subsection{Основные понятия и структуры}
Наиболее фундаментальным понятием динамики является \emph{время}.
В дискретной версии время \Math{\Time} это последовательность целых чисел, 
нумерующих последовательность событий, например, 
\begin{equation}
	\Time=\Z \text{~~или~~} \Time=\ordset{\tin,1,\ldots,\tfin},
	\label{time} 
\end{equation}
где  \Math{\tfin} --- некоторое натуральное число.
\par
Мы предполагаем, что \emph{состояния} динамической системы 
образуют конечное множество
\begin{equation}
	\wS=\set{\ws_1,\ldots,\ws_{\wSN}}
	\label{ws} 
\end{equation}
некоторых элементов. Мы предполагаем
наличие нетривиальной \emph{группы симметрий} 
\begin{equation}
	\wG=\set{\wg_1,\ldots,\wg_{\wGN}}\leq\Perm{\wS},
	\label{wG}
\end{equation}
действующей на множестве состояний \Math{\wS}.
Без потери общности можно предполагать, что группа \Math{\wG} действует на \Math{\wS} 
\emph{точно} (в противном случае \Math{\wG} можно заменить фактор-группой по ядру действия)
и \emph{транзитивно} (в противном случае можно рассматривать орбиты действия по отдельности%
\footnote{В некоторых случаях, как например, при описании обсуждаемых ниже ``солитонов'' в дискретных
системах приходится рассматривать совокупности орбит, т.е. нетранзитивные действия.}).
\par
\emph{Эволюция} или \emph{траектория} динамической системы 
это последовательность состояний, индексированных временем 
\begin{equation}
	h=\cdots\rightarrow{}w_{t-1}\rightarrow{}w_{t}\rightarrow{}w_{t+1}\rightarrow\cdots,
	 ~\text{где}~w_i\in\wS.
	\label{evolution}
\end{equation}
\par
Здесь мы сталкиваемся с проблемой отождествления ``неразличимых'' объектов 
--- более формально, элементов множества \eqref{ws} лежащих на одной орбите группы 
симметрий \eqref{wG} --- в различные моменты времени%
\footnote{В приложении ``Ars Combinatoria'' к книге \cite{Weyl} Г. Вейль 
(автор идеи калибровочной инвариантности в физике) подробно 
обсуждает проблемы индивидуализации неразличимых объектов и их отождествления в процессе 
эволюции во времени.}.
Дело в том, что для регистрации и отождествления индивидуально неразличимых 
объектов используются некоторые индивидуально различимые метки (символы, числа).
Ясно, что не существует абсолютного способа сопоставить таким различимым 
меткам неразличимые элементы. Фиксировать такие элементы можно только относительно некоторой
дополнительной системы \cite{Shafarevich}, проявляющейся в разных контекстах
как \emph{``система координат''} 
или \emph{``наблюдатель''} или \emph{``измерительный прибор''}. 
\par
Рассмотрим более подробно понятия системы координат и замены координат в конечном случае.
Для начала фиксируем какой-нибудь (произвольный) \emph{порядок} на изначально неупорядоченном 
множестве \Math{\wS}, т.е. будем предполагать, что \Math{\wS} это вектор (в данном контексте слово ``вектор'' 
означает просто упорядоченное множество, используется также термин ``кортеж'') вида
\begin{equation*}
	\wSord=\vect{\ws_1,\ldots,\ws_{\wSN}}.
\end{equation*}
Мы будем называть \Math{\wSord} \emph{каноническим множеством состояний}.
Действие группы симметрий \Math{\wG} на множестве \Math{\wSord}
имеет вид%
\footnote{В статье мы будем в основном записывать действие группы справа.
Это соглашение более естественно и интуитивно. Оно широко используется в математике и, в частности,
в основных системах компьютерной алгебры, ориентированных на задачи 
теории групп, \GAP и \Magma.  
} 
\Mathh{\wSord{}g=\vect{\ws_1g,\ldots,\ws_kg,\ldots,\ws_{\wSN}g}
=\vect{\ws_{i_1},\ldots,\ws_{i_k},\ldots,\ws_{i_\wSN}}.}
Нам удобно в качестве меток элементов \Math{\ws_k\in\wSord} использовать натуральные числа \Math{k}. 
Введём функцию \Math{\coord{v}{V}}, которая по заданному 
элементу \Math{v}, один и только один раз входящему в упорядоченное множество \Math{V},
выдаёт его позицию в этом множестве, например,  
\Math{\coord{\ws_k}{\wSord} = k} и \Math{\coord{\ws_{i_k}}{\wSord{}g} = k}.
\par
Позицию элемента в упорядоченном множестве мы будем называть \emph{координатой}.
Применив функцию \Math{\coordname} к элементам множества 
\Math{\wSord} мы получим вектор
 \begin{equation}
	R=\vect{1,2,\ldots,\wSN},
	\label{modrefframe}
\end{equation}
который будем называть \emph{канонической системой координат}.
То есть, мы пометили некоторым фиксированным образом состояния из 
множества \Math{\wS} натуральными числами. Это позволит нам контролировать
индивидуальность состояний в процессе эволюции динамической системы.
Действие элемента группы \Math{g\in\wG} на состояния
в терминах координат можно записать в виде перестановки
\Mathh{\Vtwo{1~\cdots{}k\cdots~\wSN}{1g\cdots{}kg\cdots\wSN{}g}.}
Это \emph{``контравариантное''} действие переводит одно состояние в другое: 
\Math{\ws_kg\rightarrow\ws_j}. Соответствующее \emph{``ковариантное''} преобразование
системы координат имеет вид 
\begin{equation}
	S=Rg^{-1}=\vect{1g^{-1},2g^{-1},\ldots,\wSN{}g^{-1}}.
	\label{framechange}
\end{equation}
Это означает, что если некоторый элемент \Math{x} множества состояний
в канонической системе координат \Math{R} имеет координату 
\Math{x_{R}=\coord{x}{R}}, 
то его координатой в системе \Math{S} будет
\Math{x_{S}=\coord{x}{S}}.
Операционально получение координаты означает, что мы имеем возможность распознать \Math{x} среди
элементов \Math{S} и подсчитать число шагов, пройденных до этого элемента.
Эта процедура аналогична \emph{``координатизации''}  \cite{Shafarevich} точек векторного
 пространства. 
Например, координата точки вещественной прямой (одномерного пространства \Math{\R}) --- 
это ``расстояние'', которое необходимо пройти от начала координат до данной точки.
Начало координат и ``масштаб длины шага'' (метрика) зависят от выбора системы координат.
\par
Ясно, что в соответствии с \eqref{framechange} из канонической системы координат 
\eqref{modrefframe}
можно построить все остальные и что переходы между произвольными
системами координат описываются формулой \Math{S'=Sh^{-1}},  \Math{h\in\wG}.
\par
Вернёмся к эволюции \eqref{evolution}. Мы не знаем \emph{a priori}
от какого именно из объектов  \Math{a\in\wS} 
или \Math{b\in\wS} в момент времени \Math{t-1}, 
``произошёл'' некоторый объект \Math{c} 
в момент времени \Math{t}, если  \Math{a} и \Math{b}  лежат на одной групповой орбите, 
т.е. \Math{b=ag} для некоторого \Math{g\in\wG}.
Такое отождествление осуществляется введением дополнительной структуры, называемой
\emph{калибровочной связностью} или \emph{параллельным переносом}.
\par
Рассмотрим расслоение 
базой которого является время \eqref{time}, 
типичным слоем --- каноническое множество состояний \eqref{ws},
а структурной группой --- группа \eqref{wG} 
\begin{equation}
	\vect{\FBT,\Time,\wSord,\wG,\tau}.
	\label{bundleovert}
\end{equation}
Здесь \Math{\FBT} и \Math{\tau} --- тотальное пространство и проекция, соответственно.
Связность определяет изоморфизм между слоями этого расслоения над различными моментами времени:
 \Mathh{\wSord_{t_2}=\wSord_{t_1}\Partransportt\vect{t_1,t_2},~\text{где}~
  \Partransportt\vect{t_1,t_2}\in\wG.}
Важным принципом в физике является \emph{калибровочная инвариантность} 
--- возможность задавать системы координат в слоях расслоения над различными точками базы независимо друг от друга.
В данном случае этот принцип приводит к следующему \emph{правилу преобразования для связностей}
\begin{equation}
		\Partransportt\vect{t_1,t_2}
		\rightarrow{}g\vect{t_1}^{-1}\Partransportt\vect{t_1,t_2}g\vect{t_2},
		~~g\vect{t_1},g\vect{t_2}\in\wG.\label{connruleovert}
\end{equation}
Здесь \Math{g\vect{t_1}} и \Math{g\vect{t_2}} --- \emph{произвольные} элементы 
группы  \Math{\wG}, т.е. значения произвольного сечения главного \Math{\wG}-расслоения 
над \Math{\Time}.
Очевидно, что связность между произвольными моментами времени можно записать как произведение
связностей, соответствующих элементарным шагам по времени
\Mathh{\Partransportt\vect{t_1,t_2} = \Partransportt\vect{t_1,t_1+1}\cdots\Partransportt\vect{t,t+1}
\cdots\Partransportt\vect{t_2-1,t_2}.}
Таким образом, в данном случае связность можно определить с помощью функции на 
парах последовательных моментов времени, т.е. функции на рёбрах линейного графа, 
вершинами которого являются моменты времени.
\par
Траектории \eqref{evolution} динамической системы представляют собой сечения расслоения
\eqref{bundleovert}. Символически принадлежность траектории \Math{h} к множеству сечений 
можно записать следующим образом
\begin{equation}
	h\in\prod\limits_{t\in\Time}\wSord_t.
	\label{evolutiongeneral}
\end{equation}
В основе классического подхода лежит изучение индивидуальных траекторий в предположении 
фиксированного для всех моментов времени правила отождествления состояний \Math{\ws_i}.
Поэтому расслоение \eqref{bundleovert} можно заменить прямым произведением
\Math{\Time\times\wS} и интерпретировать траекторию как функцию от времени со значениями 
в множестве состояний \Math{\wS} \Mathh{h\in\wS^\Time.} 
\par
Физически существенным учёт локальной калибровочной инвариантности становится 
в ситуациях, когда рассматриваются совокупности различных индивидуальных траекторий, 
соединяющих точки расслоения. 
Особенно отчётливо это проявляется в фейнмановской формулировке квантовой механики
через интегралы по траекториям.
\subsection{Динамические системы с пространством}
В физике множество состояний \Math{\wS} обычно имеет специальную структуру множества функций 
\begin{equation}
\wS=\lSX,
\label{stateswithspace}
\end{equation}
где множество аргументов 
\begin{equation}
	\X=\set{\x_1,\ldots, \x_\XN}
	\label{space}
\end{equation}
называется  \emph{пространством},
а множество значений 
\begin{equation}
	\lS=\set{\ls_1,\ldots,\ls_\lSN}
	\label{localstate}
\end{equation}
 --- множеством \emph{локальных состояний}.
\par
Важным свойством динамических систем с пространством является возможность
введения нетривиальных \emph{калибровочных связностей} между точками пространства.
Калибровочные структуры приводят к наблюдаемым физическим последствиям:
кривизны нетривиальных связностей описывают силы в физических теориях.
Другой  темой, в которой важна пространственная структура,
является связь между спином и статистикой.
\par
Мы будем предполагать, что как пространство \eqref{space} так и локальные состояния
\eqref{localstate} обладают нетривиальными группами симметрий:
\begin{align}
	\sG&=\set{\sg_1,\ldots,\sg_\sGN}\leq\Perm{\X}~~\text{--- \emph{симметрии пространства},}
	\label{spacegroup}\\
	\iG&=\set{\ig_1,\ldots, \ig_\iGN}\leq\Perm{\lS}~~\text{--- \emph{внутренние симметрии}.}\nonumber
\end{align}
\par
Для интерпретации группы внутренних симметрий \Math{\iG} как локальной 
калибровочной группы
введём расслоение с тотальным пространством \Math{\FBX}, базой \Math{\X}, типичным слоем
\Math{\lS} и проекцией \Math{\pi}:
\begin{equation}
	\vect{\FBX,\X,\lS,\iG,\pi}.
	\label{bundleoverx}
\end{equation}
Теперь множество состояний вместо \eqref{stateswithspace} имеет вид
множества сечений этого расслоения
\begin{equation}
\wS=\prod\limits_{x\in\X}\lS_x.
\label{stateswithspacegauged}
\end{equation}
\par
Нетривиальность группы \eqref{spacegroup} симметрий пространства предполагает наличие 
не\-которой структуры у множества \eqref{space}. Для физики существенны топологические понятия,
позволяющие классифицировать точки по степени их ``близости'' друг к другу.
\par
В подходе, называемом \emph{теоретико-множественная топология},  понятие ``близости'' точек 
формализуется с помощью семейства \emph{открытых подмножеств}, удовлетворяющего определённым аксиомам.
Эти аксиомы являются абстракцией обычных свойств континуума и мало пригодны для дискретных пространств
(хотя формально и применимы к ним).
\par
Наиболее адекватной в дискретном случае топологической структурой
является \emph{абстрактный симплициальный комплекс} \cite{Spanier}, 
который задаётся некоторым набором \Math{K} подмножеств 
множества \Math{\X}. В данном контексте элементы множества 
\Math{\X} называются \emph{точками} или \emph{вершинами}. 
Элементы множества \Math{K}, называемые \emph{симплексами}%
\footnote{Интуитивная идея ``близости'' точек соответствует тому, что точки из одного симплекса 
``ближе друг к другу'',
чем точки из разных симплексов.}, 
обладают тем свойством,
что любые их подмножества также являются симплексами, т.е. 
входят в множество \Math{K}.
Кроме того, симплексами являются все одноэлементные
подмножества  множества вершин \Math{\X}. 
Подсимплексы симплексов часто называют \emph{гранями}. Очевидно, что структура
комплекса \Math{K} однозначно определяется \emph{максимальными по включению} симплексами.
\emph{Размерность} симплекса \Math{\delta\in{}K} определяется формулой 
\Math{\dim\delta=\cabs{\delta}-1}. Это определение мотивировано тем фактом, что 
\Math{k+1} точек, погружённых в общем положении в эвклидово пространство \Math{\R^{n}},
образуют вершины простейшего (лат. \emph{simplex = простой}) выпуклого \Math{k}-мерного многогранника, если \Math{n\geq{}k}. 
\emph{Размерность комплекса} \Math{K}
определяется как максимальная размерность его симплексов:
\Math{\dim{}K=\max\limits_{\delta\in{}K}\dim\delta}.
Одномерный симплициальный комплекс называется \emph{графом}, 
а составляющие его двухэлементные подмножества (одномерные симплексы) 
называются \emph{рёбрами}. 
С точки зрения абстрактной комбинаторной топологии возможность 
погружения в эвклидовы пространства не имеет особого значения. 
Существенно только, каким образом симплексы, составляющие комплекс, связаны (пересекаются) между собой. 
Однако заметим, что в топологической теории размерности имеется 
\emph{теорема Нёбелинга--Понтрягина} из которой следует, что любой \Math{k}-мерный
симплициальный комплекс может быть погружён в пространство \Math{\R^{2k+1}}. 
В частности, любой граф можно погрузить в трёхмерное пространство \Math{\R^{3}}.
\par
Если нас не интересуют, несущественные с дискретной точки зрения,
вопросы геометрической реализации комплексов в эвклидовых пространствах, 
то любые задачи с данными на произвольных комплексах, 
можно свести к задачам на графах воспользовавшись понятием 
\emph{остова симплициального комплекса}. По определению   
\Math{q}\emph{-мерным остовом} (\Math{q}\emph{-остовом}, \Math{q}\emph{-скелетом}) 
симплициального комплекса \Math{K} называется симплициальный комплекс, 
состоящий из всех \Math{p}-мерных симплексов комплекса \Math{K}, где \Math{p\leq{}q}.
Каждому симплексу комплекса \Math{K} соответствует некоторый полный 
подграф 1-остова этого комплекса. Напомним, что граф называется \emph{полным}, если
все пары его вершин являются рёбрами.
Таким образом, вместо произвольного комплекса \Math{K} мы можем рассматривать его
1-остов, т.е. граф.
\par   
Графы (мы будем называть их также \emph{решётками}) достаточны для всех, 
рассматриваемых в данной статье задач. В частности, они достаточны для введения калибровочных и
квантовых структур. Чтобы не перегружать текст, мы сохраним для обозначения пространства
символ \Math{\X}, подразумевая при этом, что помимо вершин 
\Math{\set{\x_1,\ldots, \x_\XN}} заданы также и рёбра, т.е. некоторый набор 
2-элементных подмножеств вида \Math{e_{ij}=\set{\x_i,\x_j}}. 
Группой симметрий \eqref{spacegroup} пространства \Math{\X} в данном случае будет 
группа автоморфизмов графа, т.е. \Math{\sG=\mathrm{Aut}\vect{\X}}. В принципе группа 
автоморфизмов графа с \Math{\XN} вершинами может иметь до \Math{\XN!} элементов.
Однако существует весьма эффективный алгоритм, разработанный и реализованный Б. МакКеем
\cite{BMcKay}. Этот алгоритм строит компактное множество элементов, порождающих группу 
\Math{\sG}.
Число порождающих элементов в реальных задачах обычно равно 2-м или 3-м и 
не может превосходить \Math{\XN-1} по построению алгоритма.
\par
\begin{figure}[!h]
\centering
\begin{minipage}[t]{0.3\textwidth}
\begin{center}
{\small{}
Тетраэдран \Math{C_4H_4}\\
\Math{\sG=\SymG{4}}\\
\Math{\sGloc=\SymG{3}\cong\DihG{6}}\\[10pt]
}
\includegraphics[width=0.5\textwidth]{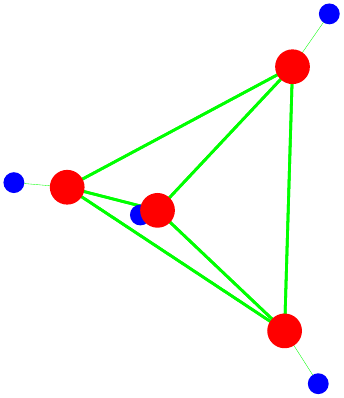}
\end{center}
\end{minipage}
~
\begin{minipage}[t]{0.3\textwidth}
\begin{center}
{\small{}
Кубан \Math{C_8H_8}\\
\Math{\sG=\SymG{4}\times\CyclG{2}}\\
\Math{\sGloc=\DihG{6}}\\[10pt]
}
\includegraphics[width=0.7\textwidth]{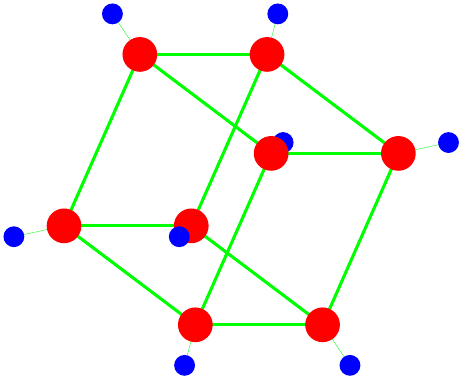}
\end{center}
\end{minipage}
~
\begin{minipage}[t]{0.3\textwidth}
\begin{center}
{\small{}
Додекаэдран \Math{C_{20}H_{20}}\\
\Math{\sG=\AltG{5}\times\CyclG{2}}\\
\Math{\sGloc=\DihG{6}}\\[5pt]
}
\includegraphics[width=0.8\textwidth]{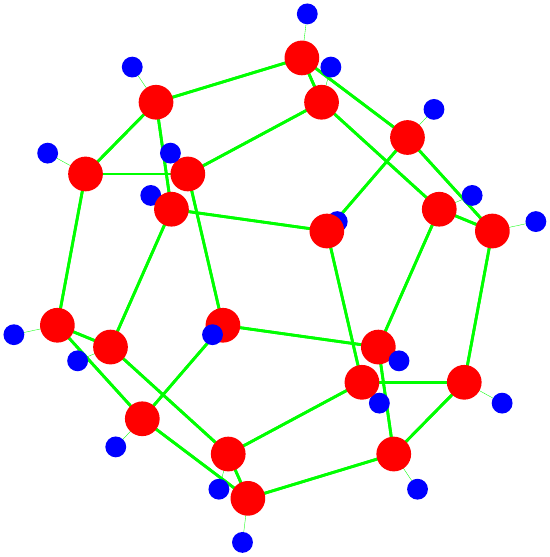}
\end{center}
\end{minipage}
\par
\vspace*{10pt}
\hspace*{10pt}
\begin{minipage}[t]{0.35\textwidth}
\begin{center}
{\small{}
Фуллерен \Math{C_{60}}\\
\Math{\sG=\AltG{5}\times\CyclG{2}}\\
\Math{\sGloc=\CyclG{2}}\\[15pt]
}
\includegraphics[width=0.9\textwidth]{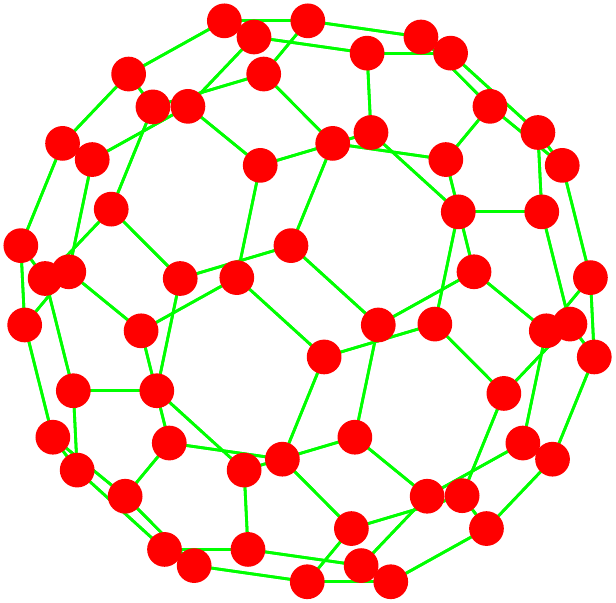}
\end{center}
\end{minipage}
\hspace{10pt}
\begin{minipage}[t]{0.5\textwidth}
\begin{center}
{\small{}
Тороидальный графен \Math{n\times{}m}\\
\Math{\sG=\DihG{n}\times\DihG{2m}}
\hspace{5pt}
\Math{\sG=\vect{\CyclG{}\times\CyclG{}}\rtimes\DihG{6}}\\
\Math{\sGloc=\CyclG{2}}
\hspace{35pt}
\Math{\sGloc=\DihG{6}}\hspace{10pt}~\\
\hspace{70pt}
\Math{n,m\rightarrow\infty}\\[3pt]
}
\includegraphics[width=0.55\textwidth]{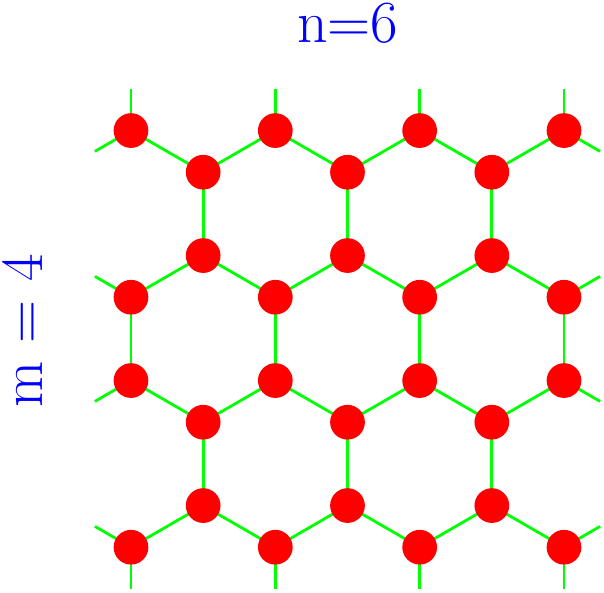}
\end{center}
\end{minipage}
\caption{Примеры угле(водо)родных наноструктур. \Math{\sG} --- группа симметрий графа структуры.
\Math{\sGloc} --- группа симметрий окрестности вершины графа.
Символы \Math{\CyclG{k},\DihG{2k},\AltG{k}} обозначают циклические, диэдральные 
и альтернирующие группы, соответственно.}
	\label{NanoCarbons}
\end{figure}

Приведём примеры решёток достаточно высокой симметрии, соответствующих реальным физическим 
объектам \cite{Kornyak08a}.
На Рис. \ref{NanoCarbons} изображены так называемые \emph{платоновы углеводороды}:
\emph{тетраэдран}, \emph{кубан} и \emph{додекаэдран}, а также \emph{фуллерен} и \emph{графен} 
--- перспективные в области нанотехнологий углеродные структуры. 
\par
Заметим, что \emph{кубан} и \emph{додекаэдран}
были искусственно синтезированы. Они достаточно устойчивы хотя и не встречаются в природе.
Тетраэдран неустойчив, но можно синтезировать его устойчивые аналоги, присоединив к 
тетраэдральному углеродному каркасу вместо атомов водорода трет-бутиловые 
\Math{\vect{CH_3}_3C} или триметилсилиловые \Math{\vect{CH_3}_3Si} радикалы.
На рисунке также приведены полные группы симметрий графов соответствующих объектов 
\Math{\sG} и их подгруппы, фиксирующие отдельные вершины \Math{\sGloc}.
Решётки, приведённые на Рис. \ref{NanoCarbons}, ввиду их невысокой сложности
интересны с методической точки зрения для построения простых моделей дискретных динамических
систем. При этом соответствующий граф интерпретируется как дискретное пространство \Math{\X}, 
группа симметрий графа \Math{\sG} --- как полная группа симметрий пространства, а подгруппа
\Math{\sGloc} --- как группа \emph{локальных симметрий пространства}.
Эта группа, описывающая симметрии окрестности вершины графа,
представляет собой аналог инфинитезимальных симметрий в физических моделях,
основанных на дифференциальных уравнениях. 
Если законы динамики дискретных моделей определяется локальными соотношениями, т.е. соотношениями,
заданными на окрестностях графов (как, например, в клеточных автоматах или описанных в Приложении
\ref{quantumonregulargraphs} квантовых моделях на регулярных графах),
то естественным требованием должна быть симметрия этих соотношений относительно группы
\Math{\sGloc}.
\par
Заметим, что графы структур на Рис. \ref{NanoCarbons} являются \emph{3-регулярными}
(\emph{3-ва\-лен\-т\-ны\-ми}). Регулярность графа аналогична в некоторых аспектах 
изотропности и однородности непрерывного
пространства. Рассуждая спекулятивно, можно было бы предположить, что пространство
на планковских масштабах может быть смоделировано каким-нибудь \Math{k}-регулярным графом.
Однако макроскопически наблюдаемая однородность и изотропность физического пространства
может быть результатом усреднения по большим ансамблям графов самых различных структур.
Тем не менее, предположение о регулярности существенно упрощает дискретные физические модели.
\par
Воспользовавшись тем, что пространство \Math{\X} имеет структуру графа, мы можем
ввести на нём связность с помощью \Math{\iG}-значной функции на рёбрах
\Math{\Partransportx\vect{\set{\x_i,\x_j}}\equiv\Partransportx\vect{\x_i,\x_j}}. 
Связность имеет очевидное свойство
\Math{\Partransportx\vect{\x_i,\x_j}=\Partransportx\vect{\x_j,\x_i}^{-1}}. 
\par
Связность \Math{\tilde{\Partransportx}\vect{\x_i,\x_j}} называется \emph{тривиальной}, 
если её можно
выразить в терминах функции на точках (вершинах графа): 
\begin{equation}
	\tilde{\Partransportx}\vect{\x_i,\x_j}=\alpha\vect{\x_i}\alpha\vect{\x_j}^{-1},~
	\alpha\vect{\x_i},\alpha\vect{\x_j}\in\iG.\label{conntriv}
\end{equation}
\par
Из локальной калибровочной инвариантности следует
аналогичное \eqref{connruleovert} \emph{правило преобразования для связностей}
\begin{equation}
		\Partransportx\vect{\x_i,\x_j}\rightarrow
		\gamma\vect{\x_i}^{-1}\Partransportx\vect{\x_i,\x_j}\gamma\vect{\x_j},~~
		\gamma\vect{\x_i},\gamma\vect{\x_j}\in\iG.\label{connruleoverx}
\end{equation}
Здесь \Math{\gamma\vect{\x_i}} и \Math{\gamma\vect{\x_j}} --- произвольные элементы
группы внутренних симметрий (можно считать, что они фиксируют локальные координаты в слоях
\Math{\lS_{\x_i}} и \Math{\lS_{\x_j}}).
\par
Предположив \emph{(одно)связность} графа \Math{\X} 
(т.е. наличие пути соединяющего любые две его вершины \Math{x_{i_1}} и \Math{x_{i_k}})
мы можем определить зависящий от пути \emph{параллельный перенос} 
\Mathh{\Partransportx\vect{x_{i_1},x_{i_2},\ldots,x_{i_k}}=
\Partransportx\vect{x_{i_1},x_{i_2}}\Partransportx\vect{x_{i_2},x_{i_3}}
\cdots \Partransportx\vect{x_{i_{k-1}},x_{i_k}},}
где каждая пара \Math{\set{\x_{i_{m}},\x_{i_{m+1}}}} является ребром.
\par
Параллельный перенос вдоль цикла графа, т.е. вдоль замкнутого пути,
\Mathh{\Partransportx\vect{x_{i_1},x_{i_2},\ldots,x_{i_k},x_{i_1}}
=\Partransportx\vect{x_{i_1},x_{i_2}}\Partransportx\vect{x_{i_2},x_{i_3}}
\cdots \Partransportx\vect{x_{i_k},x_{i_1}}}
называется \emph{голономией}. В непрерывном пространстве инфинитезимальным аналогом голономии
является \emph{кривизна связности}, которая в физических теориях описывает силовые поля.
Очевидно, что любая голономия тривиальной связности \eqref{conntriv} тривиальна, т.е.
равна единице группы: 
\Math{\tilde{\Partransportx}\vect{x_{i_1},x_{i_2},\ldots,x_{i_k},x_{i_1}} = \id}.
Это означает, в частности, что тривиальные связности не приводят к 
каким-либо физически наблюдаемым последствиям.
\par
Заметим, что в наших конструкциях время ``более фундаментально'', чем пространство.
То есть, пространство не входит в единую структуру на равных со временем, 
как это принято в релятивистски инвариантных физических теориях. 
Введение локальных времён в дискретных моделях может быть достаточно проблематичным 
и искусственным из-за отсутствия непрерывных лоренцевых симметрий.
Однако 
далее мы покажем, что такие понятия как \emph{``предельная скорость''} 
и \emph{``световой конус''} возникают в дискретной динамике вполне естественно, 
а лоренцевы симметрии могут проявляться в макроскопических масштабах как континуальные
приближения при осреднении по большим кластерам дискретных элементов.
Кроме того, общее, т.е. без локальных репараметризаций, время \Math{\Time} 
(\emph{``время жизни Вселенной в целом''}) обеспечивает единство описания динамики, независимо
от того представлены ли состояния системы в общем виде \eqref{ws} или имеют специальную 
структуру \eqref{stateswithspacegauged}, подразумевающую наличие пространства.
\par
Описание эволюции динамических систем с пространством должно учитывать эволюцию самого
пространства. Здесь также, в силу наличия пространственных симметрий \Math{\sG},
возникает естественная проблема отождествления точек пространства в разные моменты времени.
Для разрешения этой проблемы представим простра\-нство-время \Math{\XT} в виде расслоения
\Math{\sigma\!: \XT\rightarrow\Time}
с базой \Math{\Time}, типичным слоем  \Math{\X} и структурной группой \Math{\sG}
\begin{equation}
	\vect{\XT,\Time,\X,\sG,\sigma}.
	\label{xovert}
\end{equation}
Ясно, что если пространство \Math{\X} имеет
структуру графа, а время \Math{\Time} дискретно, то построенное нами пространство-время
\Math{\XT} также является графом и мы можем использовать конструкции связности и 
параллельного переноса аналогичные приведённым выше.
\par
Для того, чтобы включить эволюцию систем с пространством в общую схему, описываемую
расслоением \eqref{bundleovert}, необходимо структурную группу \Math{\wG}  этого расслоения
представить в виде комбинации пространственных и внутренних симметрий: пространственная группа 
\Math{\sG} переставляет точки расслоения
\eqref{xovert}, а группа внутренних симметрий преобразует слои в \eqref{bundleoverx}.
Более детальный анализ приводит к конструкции, называемой \emph{сплетением}  
\cite{Cameron,Dixon,Hall,Rotman}\label{wreathpage} 
\begin{equation}
	\wG=\iG\wr_\X\sG\cong\iGX\rtimes\sG\enspace.
	\label{wreath1}
\end{equation}
Последний изоморфизм означает, что сплетение имеет структуру \emph{полупрямого} произведения,
т.е. группа \Math{\iGX} функций на пространстве со значениями в группе 
внутренних симметрий является нормальным делителем группы \Math{\wG}.
\par
Из \eqref{wreath1} видно, что группа \Math{\wG} является 
частным случаем разложимого расширения группы \Math{\sG} с помощью группы \Math{\iGX}.
Приведём в явном виде операции группы \Math{\wG}, имеющей структуру \eqref{wreath1},
в терминах операций составляющих её групп \Math{\sG} и \Math{\iG}.
Действие группы \Math{\wG} на множестве состояний \Math{\wS} вида 
\eqref{stateswithspacegauged} выражается формулой
\begin{equation}
	\sigma\vect{x}\vect{\alpha\vect{x},a}=\sigma\vect{xa^{-1}}\alpha\vect{xa^{-1}},
	\label{wreathaction}
\end{equation}
где пара \Math{\vect{\alpha\vect{x},a}} --- элемент группы \Math{\wG,~x\in\X,~\sigma\vect{x}\in\wS,~a\in\sG}.
Произведение элементов \Math{\wG} имеет вид   
\begin{equation}
	\vect{\alpha\vect{x},a}\vect{\beta\vect{x},b}=\vect{\alpha\vect{x}\beta\vect{xa},ab},
\end{equation}
где \Math{\vect{\beta\vect{x},b}\in\wG}.
Для полноты приведём также формулу обратного элемента:
\begin{equation}
	\vect{\alpha\vect{x},a}^{-1}=\vect{\alpha\vect{xa^{-1}}^{-1},a^{-1}}.
	\label{wreathinverse}
\end{equation}
Заметим, что расширения групп образуют классы эквивалентности. В частности, формулы
(\ref{wreathaction}--\ref{wreathinverse}) можно записать различными эквивалентными
способами. Устройство этих эквивалентностей, а также способы объединения 
пространственных и внутренних симметрий обсуждаются в Приложении \ref{symmetryunification} 
с общей точки зрения.
\par
Поучительно посмотреть каким образом калибровочные структуры реализуются
в непрерывных физических теориях. Мы рассмотрим два примера, в которых введённые 
выше конструкции используются без предположения их дискретности. 
В обоих примерах множеством \Math{\XT} будет
4-мерное пространство Минковского с точками \Math{x=\vect{x^\nu}}, 
а множествами локальных состояний
\Math{\lS} --- комплексные линейные пространства различных размерностей.
Сечения \Math{\psi(x)} этих расслоений называются полями частиц.
\subsubsection{Электродинамика. Абелев прототип всех калибровочных теорий.} 
В данном случае группой внутренних симметрий (симметрии лагранжианов и физических наблюдаемых)
является абелева унитарная группа
\Math{\iG=\UG{1}}. Сечения главного \Math{\UG{1}}-расслоения над пространством-временем  
можно представить в виде \Math{e^{-i\alpha(x)}.}
\par
Рассмотрим параллельный перенос для двух близко расположенных точек про\-странства-времени:
\Mathh{\Partransportx\vect{x, x+\Delta{}x} = e^{-i\rho\vect{x, x+\Delta x}}.}
Фундаментальное правило преобразования связностей \eqref{connruleoverx} 
принимает вид
\Mathh{\Partransportx'\vect{x, x+\Delta{}x} 
= e^{i\alpha\vect{x}}\Partransportx\vect{x, x+\Delta{}x}e^{-i\alpha\vect{x+\Delta{}x}}.}
Далее, в соответствии с гипотезой, на которой основаны эмпирические приложения дифференциального
исчисления, мы заменяем рассматриваемые выражения линейными комбинациями разностей координат,
что приводит к приближениям:
\begin{align*}
	\Partransportx(x,x+\Delta{}x)=e^{-i\rho(x,x+\Delta{}x)}&\approx\id-i{A}(x)\Delta{}x,
	\\
	\Partransportx'(x,x+\Delta{}x)=e^{-i\rho(x,x+\Delta{}x)}&\approx\id-i{A'}(x)\Delta{}x,
	\\
	e^{-i\alpha(x+\Delta{}x)}&\approx{}e^{-i\alpha(x)}\vect{\id-i\nabla\alpha(x)\Delta{}x}.
\end{align*}
Коэффициенты \Math{{A}(x)} и \Math{{A'}(x)}, введённые в этих приближениях, 
называются \emph{1-фор\-ма\-ми связности}. 
Воспользовавшись коммутативностью группы \Math{\UG{1}}
мы получаем правило преобразования формы абелевой связности в виде
\begin{equation*}
{A'}(x)={A}(x)+\nabla\alpha(x).
\end{equation*}
В компонентах эта формула имеет вид
\begin{equation*}
{A'_\nu}(x)={A_\nu}(x)
+\frac{\textstyle{\partial\alpha(x)}}{\textstyle{\partial x^\nu}}.	
\end{equation*}
1-форма \Math{A(x)}, принимающая значения в алгебре Ли группы \Math{\UG{1}},
отождествляется с векторным потенциалом электромагнитного поля,
а её дифференциал \Math{F=\vect{F_{\nu\eta}}=\mathrm{d}A} 
--- с электрическими и магнитными силовыми полями.
Напомним, что в тензорных обозначениях в локальных координатах дифференциал \Math{F}
 формы \Math{A} имеет вид
\Mathh{F_{\nu\eta} = \frac{\textstyle{\partial{}A_\eta}}{\textstyle{\partial x^\nu}}-
\frac{\textstyle{\partial{}A_\nu}}{\textstyle{\partial x^\eta}}.}
\par
Для обеспечения калибровочной инвариантности необходимо в дифференциальных уравнениях, 
описывающих поля \Math{\psi(x)}, заменить частные производные \emph{ковариантными}:
\Mathh{\partial_\nu\rightarrow{}D_\nu=\partial_\nu-iA_\nu(x).}
\par
Чтобы завершить построение, нужно добавить уравнения, описывающие эволюцию самого калибровочного
поля \Math{{A}(x)}.
Точный вид таких уравнений не следует непосредственно из калибровочного принципа. 
Ясно лишь, что они должны быть калибровочно инвариантными. 
В данном случае простейший выбор калибровочно инвариантных уравнений приводит к
\emph{уравнениям Максвелла}:   
\begin{align}
	\mathrm{d}F&=0\hspace*{20pt}\text{\emph{первая пара}}\nonumber,\\	
	\mathrm{d}\star{}F&=0\hspace*{20pt}\text{\emph{вторая пара}}\label{m2nd}.	
\end{align}
Здесь \Math{\star} -- оператор \emph{сопряжения Ходжа} (\emph{``звезда Ходжа''}).
Заметим, что уравнение \eqref{m2nd} соответствует уравнениям Максвелла в вакууме.
В присутствие тока \Math{J} \emph{вторая пара} уравнений Максвелла принимает вид 
\Mathh{\star{}\mathrm{d}\star{}F=J.}
Заметим также, что \emph{первая пара} уравнений --- существенно априорное
утверждение, т.е. тождество, которое  выполняется для любого дифференциала внешней формы, 
каковым \Math{F} является по определению.
\subsubsection{Неабелевы калибровочные теории в непрерывном пространстве-вре\-мени.}
Если калибровочная группа \Math{\iG} является неабелевой группой Ли, 
то потребуется незначительная модификация предыдущего построения. 
Снова замена \Math{\iG}-значного параллельного переноса между двумя близко расположенными
пространственно-времен\-н\'ыми точками \Math{x} и \Math{x+\Delta{}x} линейной комбинацией
разностей этих точек приводит к 1-форме \Math{A=\vect{A_{\nu}}} со значениями
в алгебре Ли группы \Math{\iG}:
\Mathh{\Partransportx\vect{x,x+\Delta{}x}\approx\id+A_\nu(x)\Delta{}x^\nu.}
Здесь использовался тот факт, что \Math{\Partransportx\vect{x,x}=\id}.
Стандартные инфинитезимальные манипуляции с правилом преобразования связностей
\eqref{connruleoverx}
\Mathh{\gamma\vect{x}^{-1}\Partransportx\vect{x,x+\Delta{}x}\gamma\vect{x+\Delta{}x}
~\longrightarrow~
\gamma\vect{x}^{-1}\vect{\id+A_\nu\vect{x}\Delta{}x^\nu}
\vect{\gamma\vect{x}+\frac{\textstyle{\partial{}\gamma\vect{x}}}
{\textstyle{\partial{}x^\nu}}\Delta{}x^\nu}}
приводят к следующему правилу преобразования для 1-формы связности
\begin{equation*}
{A'_\nu}\vect{x}=\gamma\vect{x}^{-1}{A_\nu}\vect{x}\gamma\vect{x}+\gamma\vect{x}^{-1}
\frac{\textstyle{\partial{}\gamma\vect{x}}}{\textstyle{\partial{}x^\nu}}.	
\end{equation*}
2-форма кривизны связности
\begin{equation}
	F=\mathrm{d}A+\left[A\wedge{}A\right]\label{2form}
\end{equation}
интерпретируется как \emph{физические силовые поля}. 
1-форма \emph{тривиальной} связности имеет вид
\Mathh{\widetilde{A}_\nu\vect{x}=\gamma_0\vect{x}^{-1}
\frac{\textstyle{\partial{}\gamma_0\vect{x}}}{\textstyle{\partial{}x^\nu}}.}
Форма кривизны \eqref{2form} такой связности равна нулю. 
Тривиальная связность не порождает никаких физических сил и называется \emph{плоской} 
в дифференциальной геометрии.
\par
Существуют различные варианты динамических уравнений для калибровочных полей
\cite{Oeckl}. Наиболее важным примером является \emph{теория Янга-Миллса}, основанная на
лагранжиане
\Mathh{L_{YM}=\mathrm{Tr}\left[F\wedge\star{}F\right].}
Уравнения движения Янга-Миллса имеют вид
\begin{align}
	\mathrm{d}F+\left[A\wedge{}F\right]&=0,\label{Bianci}\\
	\mathrm{d}\star{}F+\left[A\wedge\star{}F\right]&=0.\nonumber
\end{align}
Здесь снова уравнение \eqref{Bianci} является априорным утверждением,
называемым \emph{тождеством Бьянки}. Заметим, что уравнения Максвелла являются
частным случаем уравнений  Янга-Миллса.
\par
Интересно посмотреть как выглядит лагранжиан Янга-Миллса на дискретной решётке.
Для простоты рассмотрим гиперкубическую решётку в пространстве-вре\-ме\-ни Минковского
\Math{\XT}. Вычисление с использованием инфинитезимальных аппроксимаций
показывает, что \Math{L_{YM}\sim\sum_f\sigma\vect{\gamma_f}}, где суммирование
ведётся по всем двумерным граням элементарного гиперкуба. 
Функция \Math{\sigma} имеет вид
\Mathh{
\sigma\vect{\gamma} = 2\dim\rho-\chi\vect{\rho\vect{\gamma}}
-\chi\vect{\rho^\dagger\vect{\gamma}}
\equiv2\chi\vect{\id}-\chi\vect{\rho\vect{\gamma}}
-\chi\vect{\rho^\dagger\vect{\gamma}},}
где \Math{\rho} --- фундаментальное представление калибровочной группы
(\Math{\rho^\dagger} означает дуальное представление),
 \Math{\chi} --- характер (след матрицы представления), 
\Math{\gamma_f} --- голономия группы \Math{\iG} вокруг грани \Math{f}.
\par
В теории Янга-Миллса используется оператор Ходжа, преобразующий \Math{k}-формы 
в \Math{\vect{n-k}}-формы при наличии метрики в \Math{n}-мерном пространстве \Math{\XT}.
В топологических приложениях большую роль играет так называемая \emph{BF теория}.
BF теория избегает использования метрики за счёт введения дополнительного 
динамического поля \Math{B}, представляющего собой \Math{\vect{n-2}}-форму со 
значениями в алгебре Ли калибровочной группы.
Калибровочно инвариантный лагранжиан BF теории имеет вид 
\Mathh{L_{BF}=\mathrm{Tr}\left[B\wedge{}F\right].}
\subsubsection{Возникновение пространства в дискретной динамике.}
В некоторых современных физических теориях подразумевается, что 
пространство (и даже время) являются производными понятиями, т.е. они отсутствуют 
в фундаментальной формулировке теории, но появляются как некоторые 
приближённые концепции на макроскопическом уровне.
Интерпретация пространства в таком духе возникает довольно естественно. 
Что же касается времени%
\footnote{Мотивация рассматривать время на одном уровне фундаментальности 
с пространством исходит из специальной теории относительности.}, 
то попытки вывести его из более фундаментальных понятий требуют дополнительных усилий и 
использования некоторых аргументов \textit{ad hoc}.
Проблемы вывода пространства и времени из более фундаментальных сущностей обсуждаются, 
в частности, в теории струн \cite{Seiberg}, теории каузальных множеств \cite{Markopoulou} и 
теории спиновых сетей \cite{Baez}.
\par
С помощью простой конструкции \cite{Kornyak11d} покажем, 
каким образом пространственно-временн\'ые структуры могут возникать в дискретной динамике.
Предположим, что имеется последовательность, помеченных (индексированных) временем 
событий разных типов и возможность различать эти типы. 
Более конкретно рассмотрим множество символов \Math{\Sigma=\set{\sigma_{1},\sigma_{2},\ldots,\sigma_{N}}},
обозначающих фиксируемые в восприятии некоторого ``наблюдателя'' элементы опыта, 
и предположим, что ``наблюдатель'' способен ``воспринимать'' и ``запоминать''
последовательности (потоки) символов 
\begin{equation}
	h=s_0s_1\cdots{}s_t, \text{~где~} s_i\in\Sigma, t\in\N.
 \label{path}	
\end{equation}
Понятия \emph{``причинности''} и \emph{``светового конуса''} возникают совершенно естественно. 
Невозможность преодолеть ``скорость света'' означает невозможность получить больше чем \Math{t} 
символов (``элементов восприятия'') за \Math{t} наблюдений. 
\par
\begin{figure}[!h]
\begin{center}
	\includegraphics[width=0.45\textwidth]{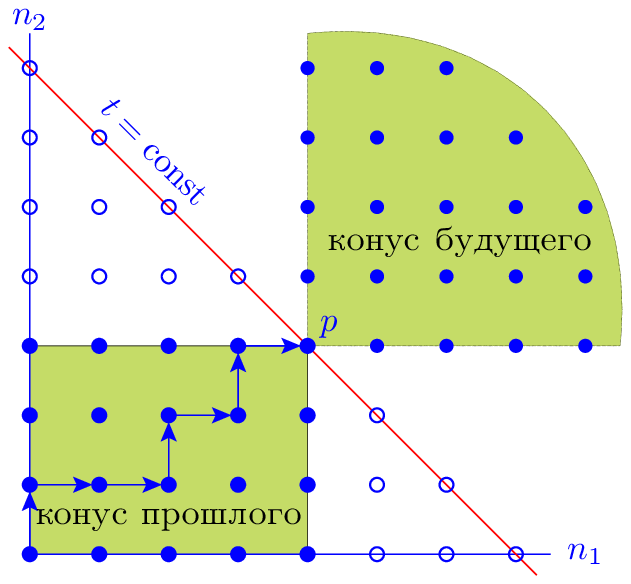}
\caption{Пространственно-временн\'ая точка \Math{p} как класс эквивалентности путей с одинаковыми
числами 
\Math{n_1, n_2, \ldots, n_{N}}\label{lightcones}.}
\end{center}
\end{figure}

В простейшем подходе счётчики событий 
различных типов можно интерпретировать как пространственные измерения, 
а  точку пространства-времени \Math{p} определить как класс эквивалентности 
последовательностей с одинаковыми кратностями вхождения каждого символа, т.е. \Math{p}
представляет собой ``коммутативный моном'' полной степени \Math{t}, описываемый
вектором из \Math{N} натуральных чисел: 
\begin{equation}
	p=\vect{n_1,\ldots,n_{N+1}}^\mathrm{T},~n_1+\cdots+n_{N}=t,
	\label{pmonom}
\end{equation}
\Math{n_i} --- кратность вхождения
символа \Math{\sigma_{i}} в последовательность \Math{h}. 
В терминах коммутативных мономов
``световой конус прошлого'' --- это множество делителей монома \Math{p}, а 
``световой конус будущего'' --- множество его кратных (см. Рис. \ref{lightcones}).
\par
Разумеется, приведённая схема имеет лишь иллюстративный характер. 
Чтобы приспособить её к описанию более реальных моделей, необходимо введение дополнительных 
структур и, возможно, другого определения понятия пространственно-временн\'ой точки,
учитывающего, в частности, симметрии между символами множества \Math{\Sigma}.
Необходимо, например, чтобы в пределе больших значений \Math{n_i}
(или, в зависимости от определения понятия точки, каких-либо других подобных 
характеристик последовательности \eqref{path}) 
в качестве континуального приближения возникали симметрии Лоренца (или Пуанкаре).
\par
Для иллюстрации этих соображений построим простую физическую модель.
Рассмотрим множество символов из двух элементов \Math{\Sigma=\set{\sigma_{1},\sigma_{2}}}
с естественным предположением, что оно симметрично относительно 
перестановки символов. Эта перестановка порождает группу \Math{\SymG{2}}. 
Допустим, что точка пространства-времени характеризуется, 
в соответствии с \eqref{pmonom}, вектором \Math{p=\Vtwo{n_1}{n_2}}. 
Представление группы \Math{\SymG{2}} в модуле, образованном такими векторами, описывается двумя матрицами 
\Math{e=\Mtwo{1}{0}{0}{1}} и \Math{r=\Mtwo{0}{1}{1}{0}}. 
Для учёта симметрии символов, необходимо выразить точку \Math{p} в терминах инвариантных подмодулей
этого представления. Для этого нужно разложить представление на неприводимые компоненты.
Переход к базису инвариантных подмодулей можно выполнить, например, с помощью матрицы 
\Math{T=\Mtwo{1}{1}{1}{-1}}. Единичная матрица \Math{e}, очевидно, не изменится, 
а матрица отражения примет вид \Math{r'=\Mtwo{1}{0}{0}{-1}}. Точка пространства-времени в 
новом базисе примет вид  \Math{p'=\Vtwo{n_1+n_2}{n_1-n_2}}, или, если ввести обозначения
\Math{t=n_1+n_2} и \Math{x=n_1-n_2}, \Math{p'=\Vtwo{t}{x}}. Переменную \Math{x}
назовём ``пространством'' (``пространственным измерением'').
Мы видим, что наличие симметрий уменьшило число ожидаемых пространственных измерений
и, кроме того, переход к инвариантным подмодулям вынудил нас ввести отрицательные числа%
\footnote{Поскольку \Math{r^2=\id}, собственные числа матрицы отражения --- квадратные корни
из единицы, один из которых \Math{\epsilon=\e^{2\pi{}i/2}} является \emph{примитивным}. 
Таким образом, мы получаем кольцо \Math{\Z} как \emph{алгебраическое расширение} 
полукольца \Math{\N} с помощью алгебраического элемента \Math{\epsilon}.}.
\par 
Теперь добавим немного физики, предположив, что последовательность символов
\Math{h=s_0s_1\cdots{}s_t,~s_i\in\set{\sigma_{1},\sigma_{2}}} соответствует
\emph{схеме испытаний Бернулли}, т.е. в достаточно длинных подпоследовательностях
символы \Math{\sigma_{1}} и \Math{\sigma_{2}} появляются с фиксированными
частотами (вероятностями) \Math{p_{1}} и \Math{p_{2}} \Math{\vect{p_{1}+p_{2}=1}}.
Тогда вероятность каждой отдельной последовательности описывается
биномиальным распределением  
\begin{equation}
P\vect{n_1,n_2}=\frac{\vect{n_1+n_2}!}{n_1!n_2!}p_1^{n_1}p_2^{n_2}.\label{bindistr}
\end{equation}
Введём величину \Math{v=p_1-p_2} и назовём её ``\emph{скоростью}''.
Скорость, очевидно, ограничена по абсолютной величине: \Math{-1\leq{}v\leq1}.
Вычислим с помощью формулы Стирлинга асимптотику выражения \eqref{bindistr} для больших 
\Math{n_1} и \Math{n_2} в окрестности стационарного значения этого выражения.
Мы ограничиваемся членами второго порядка в разложении в ряд логарифма выражения \eqref{bindistr}
в окрестности стационарного значения.
Заменив \Math{n_1} и \Math{n_2} переменными \Math{t} и \Math{x},
а вероятности --- скоростью, мы получим 
\begin{equation}
	P\vect{x,t}\approx\widetilde{P}\vect{x,t}
	=\frac{1}{\sqrt{1-v^2}}\sqrt{\frac{2}{\pi{}t}}
	\exp\set{-\frac{1}{2t}\vect{\frac{x-v t}{\sqrt{1-v^2}}}^2}\enspace.
	\label{Pxt}
\end{equation}
Заметим, что это выражение содержит ``релятивистский'' фрагмент \Math{\frac{\textstyle{x-v{}t}}{\textstyle{\sqrt{1-v^2}}}}.
Выполнив замену переменных \Math{t=T+t'} и \Math{x=vT+x'}, 
выражение \eqref{Pxt} можно переписать в виде
\begin{equation}
\widetilde{P}\vect{x,t}
=\frac{1}{\sqrt{1-v^2}}\sqrt{\frac{2}{\pi{}T}}
		\exp\set{-\frac{1}{2T}\vect{{\frac{x'-vt'}{\sqrt{1-v^2}}}}^2}+O\vect{\frac{t'}{T}}\enspace.
	\label{PxtT}
\end{equation}
Если предположить, что \Math{t'\ll{}T} (\Math{1/T} --- ``постоянная Хаббла'', а \Math{t'}
--- ``характерное время наблюдений''), то \eqref{PxtT} с высокой точностью воспроизводит
``релятивистскую инвариантность''.
Выражение \eqref{Pxt} представляет собой фундаментальное решение \emph{уравнения теплопроводности}%
\footnote{Это уравнение называется также, 
в зависимости от интерпретации функции \Math{\widetilde{P}\vect{x,t}},
\emph{уравнением диффузии} или \emph{уравнением Фоккера-Планка}.}
\begin{equation*}
	\frac{\partial{}\widetilde{P}\vect{x,t}}{\partial{}t}
	+v\frac{\partial{}\widetilde{P}\vect{x,t}}{\partial{}x} 
	=\frac{\vect{1-v^2}}{2}
	\frac{\partial^2 \widetilde{P}\vect{x,t}}{\partial{}x^2}.
\end{equation*}
В ``пределе скорости света'' \Math{\cabs{v}=1} это уравнение 
переходит в \emph{волновое уравнение}
\begin{equation*}
	\frac{\partial{}\widetilde{P}\vect{x,t}}{\partial{}t}\pm\frac{\partial{}
	\widetilde{P}\vect{x,t}}{\partial{}x}=0.
\end{equation*}
\par
Простота этой модели даёт возможность сравнить точные комбинаторные выражения с их 
континуальными приближениями. Сравнение показывает, что континуальные приближения
могут вносить серьёзные артефакты.
Рассмотрим типичную для механики задачу: найти экстремали траекторий, связывающих две
фиксированные точки пространства-времени \Math{\vect{0,0}} и \Math{\vect{X,T}}.
Нашей версией \emph{``принципа наименьшего действия''} будет поиск траекторий максимальной
вероятности. Вероятность траектории, связывающей точки \Math{\vect{0,0}} и \Math{\vect{X,T}}
и проходящей через некоторую промежуточную точку \Math{\vect{x,t}}, в соответствии с правилом
вычисления \emph{условной вероятности}, имеет вид  
\begin{align}
  P_{\vect{0,0}\rightarrow\vect{x,t}\rightarrow\vect{X,T}}
  &=\frac{P\vect{x,t}P\vect{X-x,T-t}}{P\vect{X,T}}\nonumber\\
	&=\frac{t!\vect{T-t}!\vect{\frac{T-X}{2}}!\vect{\frac{T+X}{2}}!}
	{\vect{\frac{t-x}{2}}!\vect{\frac{t+x}{2}}!
	\vect{\frac{T-t}{2}-\frac{X-x}{2}}!\vect{\frac{T-t}{2}+\frac{X-x}{2}}!T!}.
	\label{exactp}
\end{align}
Аналогичная условная вероятность для континуального приближения \eqref{Pxt} принимает вид
\begin{equation}
		\widetilde{P}_{\vect{0,0}\rightarrow\vect{x,t}\rightarrow\vect{X,T}}
		=\frac{T}{\sqrt{\frac{\pi}{2}\vect{1-v^2}tT\vect{T-t}}}
		\exp\set{-\frac{\vect{Xt-xT}^2}{2\vect{1-v^2}tT\vect{T-t}}}.
		\label{approxp}
\end{equation}
Имеются существенные различия между точным выражением \eqref{exactp} и его континуальным 
приближением \eqref{approxp}:
\begin{itemize}
	\item 
Выражение \eqref{approxp} содержит \emph{искусственную зависимость} от 	скорости 	\Math{v} 
(или, эквивалентно, от вероятностей \Math{p_1}, \Math{p_2}) в отличие от точной вероятности
\eqref{exactp}.
	\item
Легко проверить, что максимум вероятности в точном выражении \eqref{exactp} достигается не 
на единственной траектории, а на множестве различных траекторий, тогда как функционал
\eqref{approxp}	имеет единственную экстремаль, а именно, прямую линию \Math{x = \frac{X}{T}t} 
в качестве \emph{детерминистической траектории}.
Здесь мы имеем пример того, что детерминистическое поведение может 
возникнуть в результате приближения, основанного на законе больших чисел.
\end{itemize}
\subsection{Особенности детерминистической динамики.} 
Результаты этого раздела основаны на работах \cite{Kornyak07a,Kornyak07c,Kornyak08}.
\par
Отсутствие каких-либо закономерностей в поведении динамической системы означает,
что её траекторией может быть с неопределённой вероятностью любое сечение	
\eqref{evolutiongeneral} расслоения \eqref{bundleovert}.
Рассмотрим эволюции с фиксированным числом \Math{k} шагов по времени: 
\Math{h_{k,t} = \vect{\ws_{t-k},\ldots,\ws_t}.} Эти траектории являются элементами
множества \Math{H_{k,t}=\prod\limits_{s=t-k}^t\wSord_{s}}. Стандартный способ введения
закономерностей в динамику состоит в том, чтобы приписать какие-либо неотрицательные числа 
\Math{p\vect{h_{k,t}}} точкам множества \Math{H_{k,t}}.
Эти числа мы будем называть \emph{весами}%
\footnote{Отнормировав \emph{весовую функцию}
 \Math{p\vect{\cdot}} по всем элементам множества \Math{H_{k,t}} можно получить
  \emph{распределение вероятностей}.}.
Значениями функции \Math{p\vect{\cdot}} могут быть произвольные вещественные числа, 
однако, в соответствии с идеологией данной статьи, мы предпочитаем натуральные значения.
Если ограничиться для весовой функции только значениями 0 и 1, то мы получим 
\emph{характеристическую функцию} некоторого подмножества множества \Math{H_{k,t}}.
В классическом случае состояния системы во все моменты времени
\Math{\Time} фиксированно отождествлены с элементами множества \Math{\wS}.
Поэтому множество траекторий можно записать в виде 
\Math{H_{k,t}=\wS^{k+1}\equiv{}H_{k}}. 
Заметим, что мы ``потеряли'' зависимость от ``текущего момента времени'' \Math{t},
поэтому естественно предположить, что и характеристическая функция также
зависит только от длины траектории. 
Любое подмножество декартового произведения \Math{n} множеств (необязательно одинаковых)
называется \Math{n}-арным отношением%
\footnote{Более подробно дискретные отношения и их приложения 
к структурному анализу динамических систем, в частности, клеточных автоматов, 
обсуждаются в Приложении \ref{discreterelations}}. 
\Math{(k+1)}-арное дискретное отношение на множестве 
\Math{H_{k}} мы будем называть \emph{эволюционным отношением \Math{k}-го порядка}.
С помощью характеристической функции это отношение можно записать в виде уравнения
\begin{equation}
	R\vect{\ws_{t-k},\ldots,\ws_t}=0.
	\label{evrelk}
\end{equation}
Мы будем называть эволюционное отношение \Math{k}-го порядка \emph{детерминистическим},
если уравнение \eqref{evrelk} можно разрешить%
\footnote{Отношения такого типа называются \emph{функциональными}
(см. Приложение  \ref{discreterelations}).} относительно переменной \Math{\ws_t}.
Детерминистическое эволюционное уравнение \Math{k}-го порядка можно записать в виде
\begin{equation}
	\ws_t=F\vect{\ws_{t-1},\ldots,\ws_{t-k}}.
	\label{detevrelk}
\end{equation}
В приложениях обычно рассматриваются детерминистические динамические системы первого
порядка. Для любой траектории такой системы её состояние в данный момент времени 
является \emph{функцией} состояния в предыдущий момент: 
\begin{equation}
	w_{t}=F\vect{w_{t-1}}, ~w_{t},w_{t-1}\in\wS.
	\label{deterministic}
\end{equation}
Далее в этом разделе мы будем рассматривать только системы первого порядка, 
хотя,
с некоторыми техническими усложнениями, результаты, аналогичные тем, что будут изложены ниже, можно 
получить и для более общей детерминистической динамики \eqref{detevrelk}.
\par
Начнём с замечаний общего характера о связи симметрий с динамикой в детерминистическом случае.
Группа симметрий \Math{\wG} разбивает множество \Math{\wS}
на непересекающиеся орбиты конечного размера. Из функциональности отношения \eqref{deterministic}
немедленно следует, что 
\begin{itemize}
	\item 
\emph{динамические траектории} проходят \emph{групповые орбиты}	в порядке невозрастания 
размеров
орбит,
	\item \label{oncycle}
периодические траектории проходят через орбиты одного и того же размера.
\end{itemize}
\subsubsection{Солитоноподобные структуры в детерминистической динамике.} 
Обратимся к динамическим системам с пространством. Одной из характерных особенностей динамики 
таких систем является формирование движущихся в пространстве структур, сохраняющих форму. 
Покажем, что такое поведение является естественным следствием симметрий пространства.
\par
\begin{figure}[!h]
\centering
\includegraphics[width=0.65\textwidth]{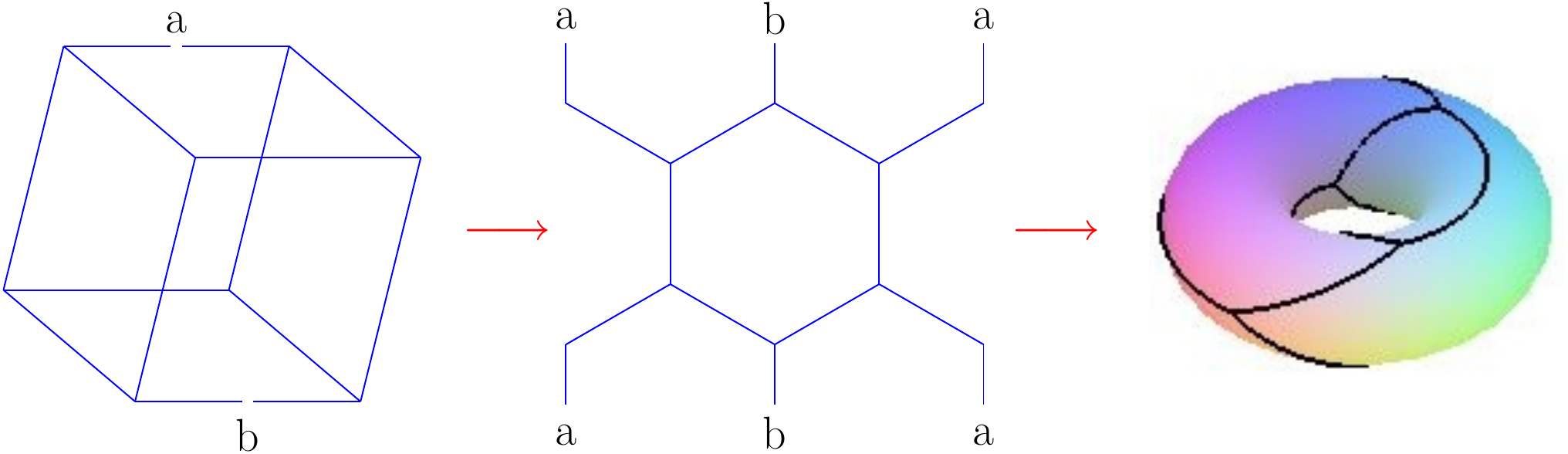} 
\caption{Граф куба образует четырёхугольную решётку на сфере (6 четырёхугольников)
и шестиугольную решётку на торе (4 шестиугольника).}
	\label{Cube-on-tor}
\end{figure}

Начнём с простого обозримого примера динамической системы. В качестве пространства \Math{\X}
возьмём куб (мы здесь, для краткости, кубом называем \emph{граф куба}). 
Заметим, что куб можно рассматривать как простейшую ``\emph{конечную модель графена}'', 
полученную замыканием гексагональной подрешётки в тор, как это показано на Рис. \ref{Cube-on-tor}.
\par
Группой симметрий куба является 48-элементная группа \Math{\sG=\SymG{4}\times\CyclG{2}}.
Если \Math{\lS=\set{0,1}} --- множество состояний  вершин куба, то полное число состояний модели
равно \Math{\cabs{\lS^{\X}}=2^8=256}. Для простоты предположим, что группа внутренних 
симметрий тривиальна \Math{\iG=\set{\id}}, тогда, в соответствии с формулой \eqref{wreath1}, 
полная группа симметрий модели фактически совпадает с группой симметрий пространства:
\Math{\wG=\iGX\rtimes\sG=\set{\id}^{\X}\rtimes\sG\cong\sG.} 
Группа \Math{\wG} расщепляет множество состояний \Math{\lS^{\X}} на 22 орбиты, размеры и числа которых
приведены в таблице
\begin{center}
\begin{tabular}{c|ccccccc}
размер орбит&1&2&4&6&8&12&24
\\\hline
число орбит&2&1&2&2&5& 4& 6
\end{tabular}\enspace.
\end{center}
\par
Рассмотрим детерминистическую динамическую систему на кубе, а именно, 
симметричный \cite{Kornyak06b,Kornyak07b} бинарный трёхвалентный клеточный автомат (см. Приложение \ref{discreterelations}) 
с правилом 86. Число 86 - это десятичное представление строки битов (подразумевается возрастающий 
порядок разрядов) из последней колонки таблицы
\ref{evol86}, 
задающей правило эволюции автомата.
\begin{table}[h] 
\begin{center}
\begin{tabular}[b]{cccc|l}
\Math{x_{1,i}}&\Math{x_{2,i}}&\Math{x_{3,i}}&\Math{x_i}&\Math{x'_i}
\\
\hline
0&0&0&0&\Math{0}\\
0&0&0&1&\Math{1}\\
1&0&0&0&\Math{1}\\
1&0&0&1&\Math{0}\\
1&1&0&0&\Math{1}\\
1&1&0&1&\Math{0}\\
1&1&1&0&\Math{1}\\
1&1&1&1&\Math{0}\\
\end{tabular}\enspace.
\end{center}
\caption{Правило эволюции автомата 86. Здесь \Math{x_i} и \Math{x'_i} --- состояния \Math{i}-й вершины куба в предыдущий и последующий 
моменты времени, соответственно; \Math{x_{1,i},x_{2,i},x_{3,i}} --- состояния вершин, смежных
с \Math{i}-й в предыдущий момент. Подразумевается симметричность правила эволюции относительно
всех перестановок вершин \Math{x_{1,i},x_{2,i},x_{3,i}}, поэтому в первых трёх столбцах таблицы
представлены симметризованные комбинации значений.}
\label{evol86}
\end{table}
\par
Это правило можно также записать
в стиле, принятом для автоматов типа Life Конвея, в терминах списков 
``\emph{рождение}''/``\emph{выживание}'' в виде B123/S0 или в виде полинома над конечным полем \Math{\F_2}
\Mathh{x'_i = x_i+\esp_1+\esp_2+\esp_3,}
где 
\Math{\esp_1 = x_{1,i}+x_{2,i}+x_{3,i},\ \esp_2 = 
x_{1,i}x_{2,i}+x_{1,i}x_{3,i}+x_{2,i}x_{3,i},\ \esp_3 = x_{1,i}x_{2,i}x_{3,i}} --- 
элементарные симметрические функции.
\par
\begin{figure}[!h]
\centering
\includegraphics[width=340pt]{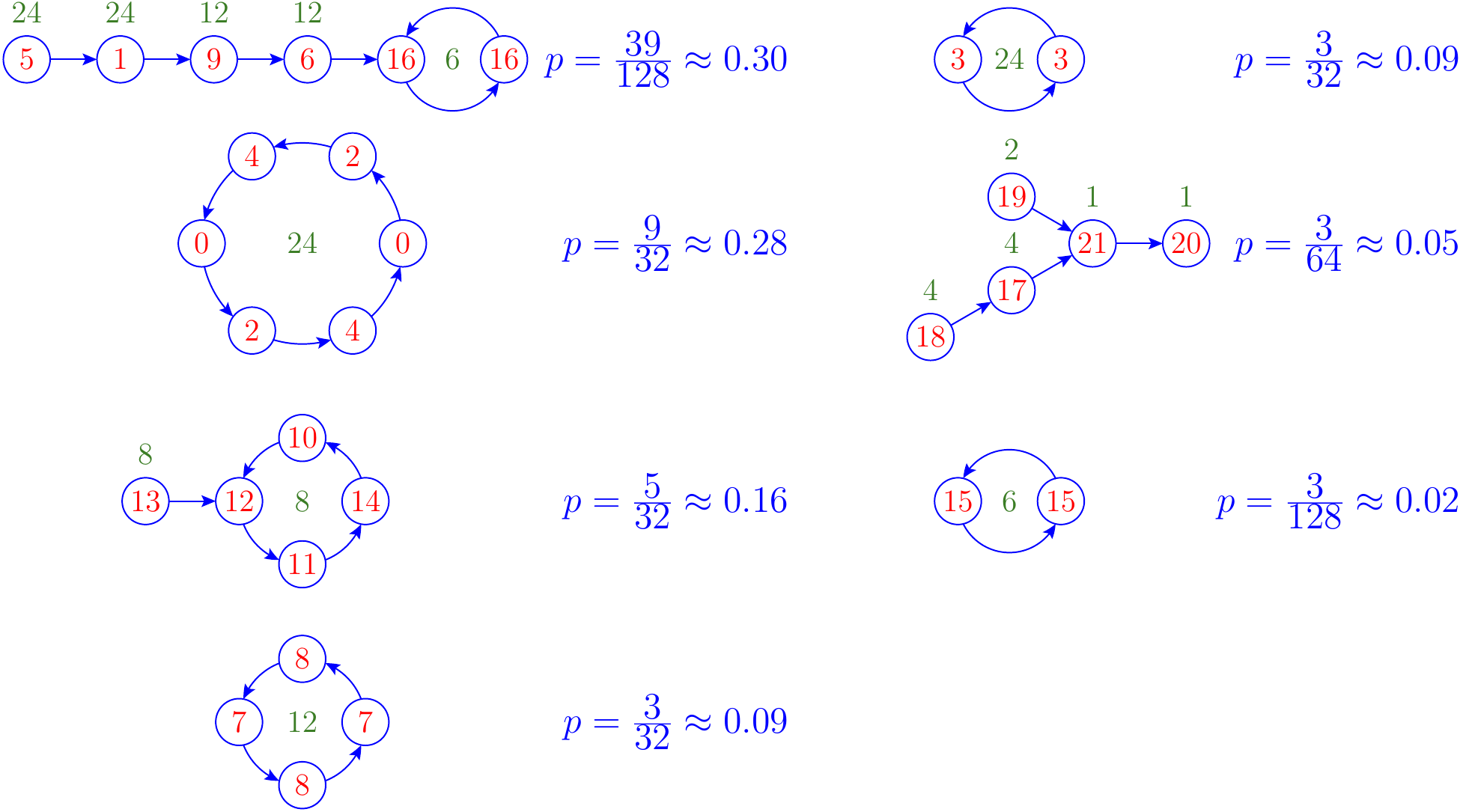}%
\caption{Классы эквивалентности траекторий для симметричного клеточного автомата на кубе с правилом эволюции 86.}
\label{PhasePortrait}
\end{figure}

Фазовый портрет автомата показан на Рис. \ref{PhasePortrait}. На этом рисунке групповые орбиты представлены
маленькими окружностями, содержащими метки орбит%
\footnote{Фактически эти метки представляют собой порядковые номера, которые компьютерная программа,
вычисляющая эволюцию модели, присваивает орбитам в процессе их появления.}.
Числа над орбитами и внутри циклов --- размеры орбит. Напомним, что циклы состоят из орбит одинакового 
размера (см. замечание на стр. \pageref{oncycle}). Рациональное число \Math{p} обозначает вес
соответствующего элемента фазового портрета. Фактически, \Math{p} --- это вероятность того, что
траектория случайно выбранного состояние окажется изолированным циклом или будет захвачена соответствующим 
аттрактором: \Math{\textstyle{}p=
\frac{\textstyle{}\text{\emph{размер бассейна}}}{\textstyle{}\text{\emph{полное число состояний}}}},
 где
\emph{размер бассейна} --- это сумма размеров орбит, входящих в данную структуру.
\par
Обобщая этот пример мы видим, что если группа симметрий 
\emph{детерминистической} динамической 
системы расщепляет множество состояний на \emph{конечное} число орбит, то \emph{любая} траектория
за конечное число шагов по времени \emph{неизбежно} перейдёт в цикл над некоторой последовательностью орбит.
Это как раз и означает формирование \emph{солитоноподобных структур}. А именно, рассмотрим эволюцию
\begin{equation}
	\sigma_{t_0}(x)\rightarrow\sigma_{t_1}(x)=A_{t_1t_0}\vect{\sigma_{t_0}(x)}.
	\label{evol}
\end{equation}
Если состояния в моменты \Math{t_0} и \Math{t_1} принадлежат \emph{одной и той же орбите}, т.е.
\Mathh{\sigma_{t_0}(x), \sigma_{t_1}(x)\in{}O_i\subseteq\lSX;}
то эволюцию \eqref{evol} можно заменить
\emph{групповым действием} 
\begin{equation}
	\sigma_{t_1}(x)=\sigma_{t_0}(x)g,~~g\in{}\wG.
	\label{move}
\end{equation}
Поскольку действие группы симметрий пространства эквивалентно ``\emph{движению}'' в пространстве,
то \eqref{move} означает, что начальное состояние (``\emph{форма}'') \Math{\sigma_{t_0}(x)} воспроизводится
после некоторого движения в пространстве.
\par
Приведем несколько примеров циклов над групповыми орбитами с 
указанием соответствующих симметрий (две из которых непрерывные):
\begin{itemize}
	\item \emph{бегущие волны} \Math{\sigma\vect{x-vt}} в математической физике --- группа Галилея;
	\item ``\emph{обобщённые когерентные состояния}'' в квантовой физике --- унитарные 
	представления компактных групп Ли;
	\item ``\emph{космические корабли}'' в клеточных автоматах --- симметрии решёток. 
\end{itemize}
\par
В качестве более детального примера рассмотрим ``\emph{глайдер}'' --- один из ``космических кораблей'' конвеевского автомата
Life.
Пространство \Math{\X} этого автомата представляет собой квадратную решётку. Мы будем считать, 
что эта решётка реализована в виде дискретного тора размера \Math{N\times{}N}. Если \Math{N\neq4},
то группа симметрий пространства \Math{\X} представляет собой полупрямое произведение двумерных трансляций 
\Math{\mathrm{T}^2=\Z_N\times\Z_N} и диэдральной группы \Math{\DihG{8}\cong\Z_4\rtimes\Z_2}:
\begin{equation*}
	\sG=\mathrm{T}^2\rtimes\DihG{8},\mbox{~~если~~} N = 3,5,6,\ldots,\infty.
\end{equation*}
В случае \Math{N=4} подгруппа трансляций \Math{\mathrm{T}^2=\Z_4\times\Z_4} уже не является нормальным
делителем и группа \Math{\sG} приобретёт несколько более сложную структуру \cite{Kornyak09c}:
\begin{equation*}
	\sG = \overbrace{
	\vect{\vect{\vect{\vect{\Z_2 \times \DihG{8}} 
	\rtimes \Z_2} \rtimes{\color{red}\Z_3}} \rtimes \Z_2}
	}^{
	\mbox{нормальное замыкание~}\textstyle\mathrm{T}^2}
	\rtimes \Z_2.
\end{equation*}
Наличие дополнительного сомножителя \Math{\CyclG{3}} в этом выражении объясняется симметрией 
диаграммы Дынкина с четырьмя вершинами  
\Mathh{D_4=~~\text{\raisebox{-0.025\textwidth}{\includegraphics[width=0.065\textwidth]{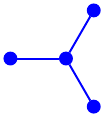}}},}
 соответствующей случаю \Math{N=4}.
\par
\begin{figure}[!h]
\centering
\includegraphics[width=0.7\textwidth]{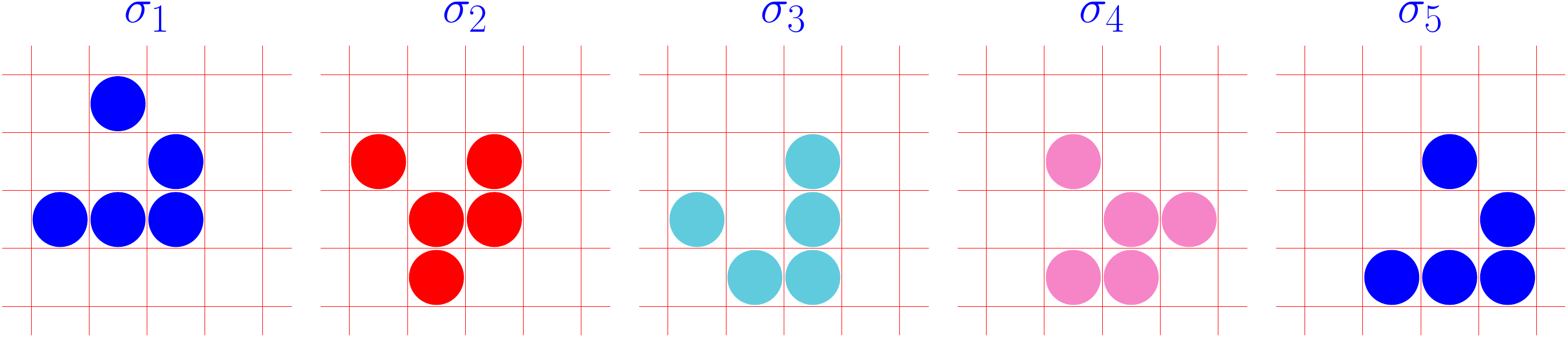}
\caption{Пример солитоноподобной структуры.
``Глайдер'' клеточного автомата Life представляет собой цикл над \emph{двумя} орбитами
группы \Math{\sG=\mathrm{T}^2\rtimes\DihG{8}}. Конфигурации \Math{\sigma_3} и  
\Math{\sigma_4} получаются из \Math{\sigma_1} и  
\Math{\sigma_2}, соответственно, одной и той же комбинацией \emph{сдвига}, 
 \emph{поворота} на 90\textdegree по часовой стрелке и \emph{отражения} относительно вертикали.}
	\label{Glider-2}
\end{figure}

Множество локальных состояний клеток автомата Life имеет вид
\Math{\lS=\set{0,1}=\set{\text{``\emph{мёртвая}''},\text{``\emph{живая}''}}.}
Поскольку локальное правило этого автомата не симметрично относительно перестановки локальных 
состояний \Math{0\leftrightarrow1}, группа внутренних симметрий \Math{\iG} тривиальна.
Поэтому \Math{\wG = \set{\id}^{\X}\rtimes\sG\cong\sG.} 
Действие \Math{\wG} на функции \Math{\sigma\vect{x}\in\lS^{\X}} в соответствии с формулой
\eqref{wreathaction} имеет вид 
\Math{\sigma\vect{x}g=\sigma\vect{xf^{-1}}}, где \Math{g = \vect{\id,f},~f\in\sG}.
На Рис. \ref{Glider-2} воспроизведены в предположении \Math{N>4} четыре шага эволюции глайдера.
\subsubsection{Об обратимости в дискретных детерминистических системах.} 
Типичная дискретная детерминистическая система \emph{необратима}. 
Фазовый портрет такой системы похож на представленный на Рис. \ref{PhasePortrait}.
Мы видим, что там имеются несколько изолированных и несколько предельных циклов%
\footnote{Неподвижные точки можно считать циклами единичной длины.}.
В непрерывном случае можно в принципе, увеличивая точность описания, восстановить предысторию
любой траектории, стремящейся к предельному циклу, поскольку точка континуума
может содержать потенциально бесконечное количество информации.
В дискретном случае информация о состояниях, в которых была система до перехода на 
предельный цикл, безвозвратно теряется. Это означает, что по прошествии некоторого времени 
предельные циклы становятся физически неотличимыми от изолированных и динамическая система
ведёт себя как обратимая. Возможно подобные соображения могут быть использованы для объяснения 
наблюдаемой обратимости фундаментальных законов природы.
\par 
Именно с таких позиций Г. `т Хоофт в ряде своих работ (см., например, \cite{tHooft99,tHooft06})
пытается разрешить конфликт между необратимостью гравитации (потеря информации на горизонтах чёрных
дыр) и обратимостью (унитарностью) стандартной квантовой механики. Подход `т Хоофта основан на следующих
предположениях: 
\begin{itemize}
	\item
физические системы на микроскопических (планковских) масштабах описываются 
\emph{дискретными степенями свободы};
	\item
состояния этих степеней свободы образуют \emph{первичный} (\emph{primordial}) базис гильбертова
пространства (в котором возможна неунитарная	эволюция);
	\item
первичные состояния образуют \emph{классы эквивалентности}: два состояния эквивалентны, если по 
истечении некоторого времени они переходят в одно и то же состояние;	 
	\item
эти классы эквивалентности по построению образуют базис гильбертова пространства с унитарной эволюцией,
описываемой обратимым во времени уравнением Шрёдингера.	 
\end{itemize}
В нашей терминологии это соответствует переходу к предельным циклам. За конечное число шагов по времени
система полностью ``забывает'' свою ``доцикловую'' предысторию.
\par
Если приведённые рассуждения имеют какое-либо отношение к физической реальности, то необратимости такого типа вряд ли можно наблюдать экспериментально. Система, вероятно, должна проводить вне
цикла время порядка планковского, т.е. примерно \Math{10^{-44}} секунды. 
Наиболее короткий интервал времени,
фиксируемый экспериментально в настоящее время, равен приблизительно \Math{10^{-18}} секунды
или \Math{10^{26}} планковских единиц.  
\section{Конструктивное описание квантового поведения}
Этот раздел основан на работах \cite{Kornyak11a,Kornyak11b,Kornyak12a,Kornyak12b}.
\par
\begin{figure}[!h]
\centering
\includegraphics[width=0.85\textwidth]{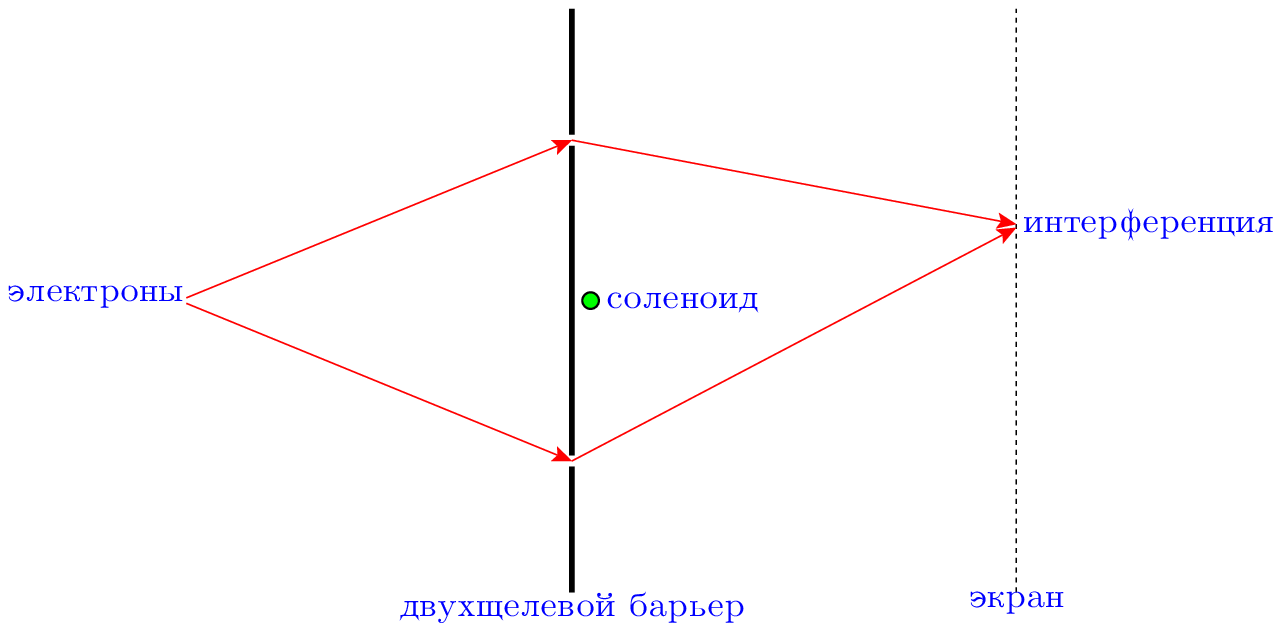}
\caption{Эффект Ааронова--Бома. Интерференционная картина сдвигается при включении
соленоида несмотря на отсутствие электромагнитных сил вне соленоида.}
	\label{Aharonov-Bohm}
\end{figure}

Наиболее известной эмпирической демонстрацией квантового поведения считается \emph{двухщелевой эксперимент}. 
Если провести двухщелевой эксперимент с заряженными
частицами в установке, снабжённой соленоидом, то можно наблюдать \emph{эффект Ааронова--Бома} 
(Рис. \ref{Aharonov-Bohm}), демонстрирующий роль калибровочных связностей в квантовой механике.
Заряженные частицы движутся в области, содержащей идеально экранированный тонкий соленоид.
Интерференционная картина меняется при включении соленоида, несмотря на отсутствие электромагнитных сил,
действующих на частицы. 
Эффект возникает из-за того, что работающий соленоид создаёт вокруг себя \Math{\UG{1}}-связность, 
изменяющую фазы волновых функций частиц.    
\par
Для квантования динамических систем с пространственной структурой
наиболее удобен фейнмановский метод интегралов по траекториям \cite{Feynman}, фактически возникший
из изучения двухщелевого эксперимента%
\footnote{Известно высказывание Р. Фейнмана о том, что ``вся квантовая механика может быть извлечена
из тщательного обдумывания следствий одного этого эксперимента''.}.
Согласно подходу Фейнмана амплитуда квантового перехода системы из одного состояния в 
другое вычисляется с помощью суммирования амплитуд вдоль всех возможных классических траекторий, 
соединяющих эти состояния.
Амплитуда вдоль отдельной траектории вычисляется как произведение амплитуд переходов между
ближайшими последовательными состояниями на траектории. 
А именно, амплитуда
вдоль траектории представляется в виде экспоненты от действия вдоль траектории
\begin{equation}
A_{\mathrm{U}(1)}~=~A_0\exp\vect{iS}=A_0\exp\vect{i\int\limits_0^T{}Ldt}.
\label{amplclass}
\end{equation}
Функция \Math{L}, зависящая от \emph{первых производных} состояний по времени, называется
\emph{лагранжианом}.
 В дискретном времени экспонента от интеграла переходит
в произведение: \Math{\exp\vect{i\int{}Ldt}\rightarrow
	\e^{{iL_{0,1}}}\ldots\e^{{iL_{t-1,t}}}
	\ldots\e^{{iL_{T-1,T}}}} и выражение для амплитуды принимает вид
\begin{equation*}
	A_{\mathrm{U}(1)}~=~A_0
	\e^{{iL_{0,1}}}\ldots\e^{{iL_{t-1,t}}}
	\ldots\e^{{iL_{T-1,T}}}.
\end{equation*}
Сомножители \Math{\Partransportx_{t-1,t}=\e^{{iL_{t-1,t}}}} этого произведения 
представляют собой элементы \emph{связности} 
со значениями в \emph{одномерном} унитарном представлении \Math{\UG{1}} окружности, 
т. е. коммутативной группы Ли
\Math{\iG=S^1\equiv\R/\Z}, являющейся в данном случае группой внутренних симметрий.
В соответствии с нашим взглядом на квантовое поведение это предполагает наличие некоторого множества 
\Math{\lS} \emph{локальных} состояний 
на котором действует группа Ли \Math{S^1}, а её представление \Math{\UG{1}} действует 
на линейном пространстве (в данном случае \Math{\C}) \emph{амплитуд} состояний из \Math{\lS}.
Далее при обсуждении квантового поведения с общей точки зрения мы уточним связь между 
состояниями и их амплитудами, т.е. числовыми весами, приписываемыми состояниям.
Здесь, как обычно, амплитуды на которых действует представление \Math{\UG{1}}
(и, соответственно, состояния на которых действует \Math{S^1}) ненаблюдаемы, а наблюдаемы
только инварианты этих действий. 
Непрерывной версией того, что элементы связности определены на парах состояний, 
соответствующих начальным и конечным моментам элементарных интервалов времени, 
т.е. на рёбрах абстрактного графа%
\footnote{Напомним, что определение связности как функции на \emph{парах} точек необходимо для 
обеспечения нетривиальности связности.}, является зависимость лагранжиана от первых производных по времени.
\par 
Естественное обобщение возникает из предположений, что
группа \Math{\iG} не обязательно окружность и что её унитарное представление 
\Math{\Rep{\iG}} не обязательно одномерно.	
В этом случае амплитуда представляет собой многокомпонентный вектор, что удобно для описания
частиц с более сложно устроенными внутренними степенями свободы. Значение такой многокомпонентной 
амплитуды на траектории принимает вид%
\footnote{В некоммутативном случае важно соблюдать порядок операторов, согласованный
в данном случае с традицией писать матрицы слева от векторов.}
\begin{equation}
	A_{\Rep{\iG}}=\Rep{\alpha_{T,T-1}}\ldots\Rep{\alpha_{t,t-1}}
	\ldots\Rep{\alpha_{1,0}}A_0,\hspace*{10pt}\alpha_{t,t-1}\in\iG.\label{amplgen}
\end{equation}
В соответствии с нашей идеологией мы будем предполагать, что \Math{\iG} --- конечная группа. 
В этом случае нет необходимости заботиться об унитарности, поскольку линейные 
представления конечных групп автоматически унитарны. 
Ясно, что стандартное квантование \eqref{amplclass} можно 
аппроксимировать с помощью одномерных представлений конечных циклических групп
достаточно большого периода.
\par
Правила Фейнмана, сформулированные в абстрактной форме, фактически совпадают 
с правилами умножения матриц. Согласно правилам Фейнмана, для того, чтобы получить
амплитуду перехода из одного состояния в другое, необходимо соединить эти состояния
всеми возможными траекториями, перемножить амплитуды вдоль этих траекторий и суммировать
полученные произведения. Совпадение этих инструкций с правилами умножения матриц очевидно
из иллюстрации на которой два шага эволюции  квантовой системы с двумя 
состояниями (``однокубитный регистр'') представлены параллельно в фейнмановской 
и матричной формах:
\begin{center}
\begin{tabular}[t]{ccc}
\includegraphics[height=0.24\textwidth]{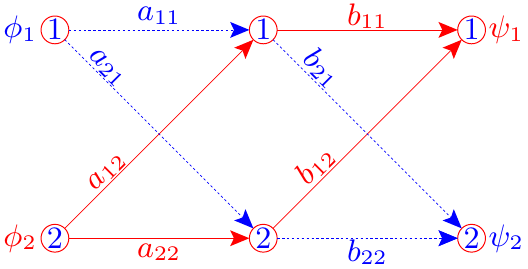}
&\raisebox{0.1\textwidth}{\Math{\sim}}
&
\includegraphics[height=0.24\textwidth]{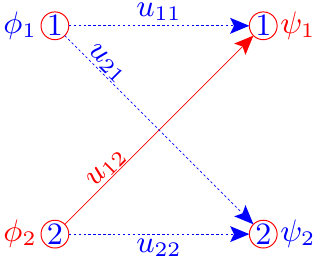}
\\
\Math{\Updownarrow}&&\Math{\Updownarrow}
\\
\Math{
BA=
\bmat
{\color{blue}b_{11}a_{11}+b_{12}a_{21}}
&
{\color{red}b_{11}a_{12}+b_{12}a_{22}}
\\[5pt]
{\color{blue}b_{21}a_{11}+b_{22}a_{21}}
&
{\color{blue}b_{21}a_{12}+b_{22}a_{22}}
\emat
}~~~~~~~~~
& \Math{\sim}~~ &
\Math{
U=
\bmat
{\color{blue}u_{11}}&{\color{red}u_{12}}
\\[5pt]
{\color{blue}u_{21}}&{\color{blue}u_{22}}
\emat
}~~~~~~~~
\end{tabular}
\end{center}
В соответствии с фейнмановскими правилами переход, скажем, из состояния 
\Math{\phi_2} в \Math{\psi_1} определяется суммой по двум путям
 \Math{b_{11}a_{12}+b_{12}a_{22}}.
Но это же выражение является элементом \Math{u_{12}} произведения матриц \Math{U=BA}.
Общий случай произвольного числа состояний и произвольного числа шагов по времени легко 
выводится из этого элементарного примера математической индукцией 
по указанным числам.
\par
В случае некоммутативной связности формула \eqref{amplgen} соответствует неабелевой калибровочной теории.
Приведённое выше рассуждение о соответствии фейнмановского квантования матричному
умножению здесь также применимо. Нам достаточно лишь трактовать эволюционные матрицы
\Math{A, B} и \Math{U} как блочные матрицы с некоммутативными элементами,
являющимися матрицами из представления \Math{\Rep{\iG}}.
\par
Представление матричного умножения правилами Фейнмана: 
``\emph{перемножить последовательные события}'' и ``\emph{суммировать альтернативные истории}'', 
пригодно 
для матриц с произвольными элементами. В обычной формулировке квантования Фейнмана
элементы матриц имеют очень специальный вид. А именно, все элементы матрицы представляют собой комплексные
числа \Math{\e^{iLdt}}, равные по модулю единице. Эти числа можно интерпретировать как континуальные аналоги
корней из единицы. То что они не могут обращаться в нуль, означает, что невозможно реализовать системы,
в которых запрещены переходы между некоторыми состояниями --- например, между состояниями 
в далеко отстоящих друг от друга точках пространства. Более того, выпадает важный случай 
тривиальной группы внутренних симметрий \Math{\iG=\set{\id}}. В этом случае квантовая эволюция описывается 
матрицами перестановок. Матрица перестановок размера \Math{n\times{}n} содержит \Math{n} единиц и
\Math{n^2-n} нулей. Нуль,  разумеется, невозможно представить  экспонентой какого-либо числа.
\par
Изучение квантового поведения дискретных динамических систем с пространственной структурой
комбинаторно трудная задача. Построение соответствующих моделей существенно зависит от ряда допущений, 
которые необходимо наложить на структуру пространства и свойства 
квантовых переходов, чтобы сделать модель поддающейся изучению.  Один из подходов к построению 
квантовых моделей на 
дискретных пространствах, описан в Приложении \ref{quantumonregulargraphs}. 
Квантовые переходы в этих моделях определены локально, а именно, на окрестностях
регулярных графов, что делает их в определённом смысле похожими на клеточные автоматы.
\par
Далее мы будем рассматривать проблему квантового поведения дискретных систем в общем виде,
т.е. игнорируя возможное наличие пространства в структуре полного множества состояний. 
Заметим, что существует много физических проблем, например,  квантовые вычисления,
для которых выделение в множестве состояний какой-либо пространственной структуры несущественно.
\subsection{Квантование дискретных систем}
Таким образом, мы будем придерживаться традиционной формулировки квантовой механики, 
в которой эволюция квантовой динамической системы из начального состояния в конечное
описывается \emph{матрицей эволюции} \Math{U}:
\Math{\barket{\psi_{\tin}}\rightarrow\barket{\psi_{\tfin}}=U\barket{\psi_{\tin}}}.
Матрица эволюции может быть представлена в виде произведения матриц, соответствующих
элементарным временн\'{ы}м шагам:
\Math{U=U_{\tfin\leftarrow\tfin-1}\cdots{}U_{t\leftarrow{}t-1}\cdots{}U_{1\leftarrow0}.}
\subsubsection{Стандартная и ``конечная'' версии квантовой механики}
\par
Для более формального обоснования законности замены в физических задачах бесконечных 
групп конечными можно воспользоваться понятием финитной аппроксимируемости.
(Бесконечная) группа \Math{G} 
называется \emph{финитно аппроксимируемой}%
\footnote{В англоязычной литературе используется термин \emph{residually finite},
ещё более подчёркивающий ``конечный'' аспект таких групп.} \cite{Magnus}, 
если для каждого её элемента \Math{g\neq\id} существует такой гомоморфизм 
\Math{\phi: G\rightarrow{}H} 
в \emph{конечную} группу \Math{H}, что \Math{\phi\vect{g}\neq\id}.
Это означает, что любые соотношения между элементами группы \Math{G} 
можно смоделировать соотношениями между элементами конечной группы.
Здесь имеется аналогия с широко используемым в физике приёмом, когда для решения 
некоторой задачи бесконечное пространство заменяется, например, тором размер 
которого достаточен для того, чтобы вместить данные задачи.
На практике все конструктивные бесконечные группы, используемые в физике,
являются финитно аппроксимируемыми. Типичным примером финитно аппроксимируемой группы
является популярная в 
теоретической физике \emph{группа кос}. Согласно стандартной конструкции \cite{PraSos} 
группа кос 
с \Math{n} нитями \Math{B_n} порождается образующими \Math{b_1,\ldots,b_{n-1}},
удовлетворяющими соотношениям:
\begin{align}
b_{i}b_{i+1}b_{i}&=b_{i+1}b_{i}b_{i+1},~~i=1,\ldots,n-1~~
\text{--- \textit{уравнение Янга-Бакстера},}\label{YB}\\
b_{i}b_{j}&=b_{j}b_{i},~~\cabs{i-j}\geq2~~
\hspace{55pt}\text{--- \textit{дальняя коммутативность}.}\label{DC}	
\end{align}
Заметим, что группа кос является ближайшим бесконечным ``родственником'' группы 
перестановок: группа перестановок \Math{\SymG{n}} получается из  \Math{B_n}, если к \eqref{YB}
и \eqref{DC} добавить соотношения \Math{b_{i}^2=\id}.
\par
Обратимся теперь к унитарным группам, играющим центральную роль в квантовой
физике. Здесь, поскольку мы начинаем с континуума, требуется два шага для перехода от 
бесконечного к конечному:
\begin{enumerate}
	\item 
Из теории квантовых вычислений \cite{Nielsen} известно, 
что существуют универсальные конечные наборы
унитарных операторов, называемых \emph{квантовыми вентилями}, с помощью которых можно порождать
множества операторов, всюду плотных в множестве всех унитарных операторов, действующих
в соответствующем гильбертовом пространстве.
Квантовые вентили действуют в гильбертовых пространствах небольших фиксированных 
размерностей (однокубитные, двухкубитные и т.д. вентили),
но из них можно построить конечные наборы унитарных матриц,
действующих в гильбертовых пространствах произвольной размерности. 
Эти матрицы можно рассматривать как образующие конечно порождённой группы, 
которая аппроксимирует с произвольной точностью (плотно вкладывается в) общую унитарную группу.
	\item
Согласно теореме А.И. Мальцева \cite{Malcev}, всякая конечно порождённая 
группа матриц над (любым) полем \emph{финитно аппроксимируема}.
\end{enumerate} 
Таким образом, унитарные группы 
можно заменить конечными безо всякого ущерба для физики как эмпирической науки.
\par
Для того, чтобы воспроизвести квантовую механику в конструктивной ``конечной''
постановке мы будем придерживаться принципа Оккама: вводить новые элементы описания
только если они действительно необходимы.
Приведём сравнение основных элементов стандартной квантовой механики с их конструктивными 
аналогами.
\begin{enumerate}
	\item
\begin{enumerate}
	\item Стандартная квантовая механика имеет дело с \emph{унитарными операторами} \Math{U},
действующими в \emph{гильбертовом пространстве} \Math{\Hspace} над полем комплексных
чисел \Math{\C}. Элементы \Math{\barket{\psi}\in\Hspace} этого пространства
называются  ``\emph{состояниями}'', ``\emph{векторами состояний}'', 
``\emph{волновыми функциями}'', ``\emph{амплитудами}'' и т. д.
Операторы \Math{U} являются элементами общей унитарной группы \Math{\Aut{\Hspace}}, 
действующей в \Math{\Hspace}.
	\item 
В ``конечной'' квантовой механике%
\footnote{Термин ``конечная квантовая механика'' используется в литературе в 
существенно другом смысле. 
Так называют направление (см., например, \cite{AthaFQM,FlorFQM,FlorNiFQM}), 
возникшее из предложения Г. Вейля \cite{WeylFQM}
заменить в квантовой механике фазовое пространство \Math{\vect{p,q}} конечным дискретным тором.
Мы используем здесь этот термин за неимением другого достаточно краткого и осмысленного.}
гильбертово пространство над полем  \Math{\C}
заменяется \Math{\adimH}-мерным гильбертовым пространством \Math{\Hspace_\adimH}
над \emph{абелевым числовым полем} \Math{\NF} --- расширением рациональных чисел
\Math{\Q} с \emph{абелевой группой Галуа}  \cite{Shafarevich}.
Операторы \Math{U} принадлежат теперь унитарному представлению \Math{\repq}
в пространстве \Math{\Hspace_\adimH} некоторой \emph{конечной группы} \Math{\wG=\set{\wg_1,\ldots,\wg_{\wGN}}}.
Поле \Math{\NF} определяется структурой группы \Math{\wG} и её представлением
\Math{\repq}. Фактически, поле \Math{\NF} представляет собой 
\emph{минимальное расширение натуральных чисел}, применимое для описания квантового поведения.
\end{enumerate}
	\item
В обеих версиях квантовой механики квантовые \emph{частицы} ассоциируются с \emph{унитарными представлениями}
 определённых групп симметрий.
Представления, в соответствии с их размерностями, называются ``\emph{синглетами}'',
 ``\emph{дублетами}'', ``\emph{триплетами}'' и т. д.
Многомерные представления, в частности, описывают  \emph{спин}.  
	\item
Квантовая \emph{эволюция}	описывается унитарным преобразованием
\emph{начального} вектора состояния \Math{\barket{\psi_{in}}} в \emph{конечный} 
\Math{\barket{\psi_{out}}=U\barket{\psi_{in}}}.
\begin{enumerate}
	\item 
В стандартной квантовой механике элементарный шаг эволюции в непрерывном времени
описывается \emph{уравнением Шрёдингера}	
\Mathh{\displaystyle{}i\frac{\mathrm{d}}{\mathrm{d}t}\barket{\psi}=H\barket{\psi},}
где \Math{H} --- \alert{эрмитов} оператор, называемый  \alert{оператором энергии} или 
 \alert{гамильтонианом}.
	\item 
В конечной квантовой механике имеется лишь конечное число возможных эволюций:	
\Mathh{U_j\in\set{\repq\vect{\wg_1},
\ldots,\repq\vect{\wg_j},\ldots,\repq\vect{\wg_\wGN}}.}
Очевидно, что здесь вообще нет необходимости в каком-либо аналоге уравнения Шрёдингера.
Впрочем, формально всегда можно ввести гамильтониан с помощью формулы
\Math{H_j=i\ln{}U_j\equiv\sum\limits_{k=0}^{p-1}\lambda_k{}U_j^k}, где \Math{p} ---
период оператора \Math{U_j} (т.e. минимальное \Math{p>0} такое, что \Math{U_j^p=\idmat}),
 \Math{\lambda_k} --- легко вычисляемые коэффициенты%
\footnote{Эти коэффициенты содержат \emph{неалгебраический} элемент, именно, число
\Math{\pi}, представляющее собой \emph{бесконечную} сумму элементов из \Math{\NF}.
Иными словами, \Math{\lambda_k} --- элементы \emph{трансцендентного расширения}
поля  \Math{\NF}. Логарифмическая функция является существенно
конструкцией из непрерывной математики, имеющей дело с актуальными бесконечностями.}.
\end{enumerate}
	\item
В обеих версиях квантовой механики квантово-механический \emph{эксперимент}
(\emph{наблюдение}, ``\emph{измерение}''
\footnote{Наименее удачный термин, поскольку квантово-механические ``\emph{измерения}'' 
фактически сводятся к подсчету числа некоторых событий.})
 сводится к сравнению
состояния \emph{системы} \Math{\barket{\psi}} с состоянием \emph{прибора} 
(\emph{аппарата})
\Math{\barket{\phi}}.
	\item
В обеих версиях квантовой механики, в соответствии с \emph{правилом Борна},
вероятность зарегистрировать частицу, описываемую состоянием  \Math{\barket{\psi}}
аппаратом, настроенным на состояние \Math{\barket{\phi}}, 
равна   	
\begin{equation}
\ProbBorn{\phi}{\psi} = \frac{\textstyle{\cabs{\inner{\phi}{\psi}}^2}}
{\textstyle{\inner{\phi}{\phi}\inner{\psi}{\psi}}}.
\label{BornEn}
\end{equation}
Однако для конечной квантовой механики требуется определённое \emph{концептуальное 
уточнение}. В ``конечном'' контексте единственной осмысленной интерпретацией
вероятности может быть лишь \emph{частотная интерпретация}: вероятность --- это
отношение числа \emph{выделенных} комбинаций элементов системы
к полному числу рассматриваемых комбинаций. Поскольку мы имеем дело с комбинациями
элементов конечных множеств, мы ожидаем, что, если всё организовано правильно,
формула \eqref{BornEn} должна выдавать \emph{рациональные числа}.
Мы будем использовать это в дальнейших построениях в качестве одного из  
ведущих принципов.
	\item
Квантовые \emph{наблюдаемые} описываются \emph{эрмитовыми операторами}
в гильбертовых пространствах.
В конечной квантовой механике эти операторы можно выразить как элементы представления
\emph{групповой алгебры}
\begin{equation*}
	A=\sum\limits_{k=1}^{\wGN}\alpha_k{}\repq\vect{\wg_k}.
\end{equation*}
Разумеется, требование эрмитовости может наложить определённые ограничения на 
коэффициенты \Math{\alpha_k}. 
\end{enumerate}
Заметим, что остальные элементы конечной квантовой теории получаются из стандартной
простым переписыванием. Например, как хорошо известно, 
\emph{принцип не\-оп\-ределённости Гейзенберга} следует из \emph{неравенства
Коши}~(-Буняковского-Шварца)
\begin{equation*}
	\inner{A\psi}{A\psi}\inner{B\psi}{B\psi}\geq\cabs{\inner{A\psi}{B\psi}}^2,
\end{equation*}
справедливого для любых гильбертовых пространств над любыми полями.
Очевидно, что неравенство Коши эквивалентно стандартному свойству любой
вероятности \Math{\ProbBorn{A\psi}{B\psi}\leq1}.
\subsubsection{Перестановки, представления и числа}
Все множества, на которых конечная группа 
\Math{\wG=\set{\wg_1,\!\ldots\!,\wg_{\wGN}}} 
действует транзитивно, можно легко описать \cite{Hall}.
Любое такое множество \Math{\wS=\set{\ws_1,\!\ldots\!,\ws_\wSN}}
находится во взаимно однозначном соответствии с множеством (\emph{правых} 
\Math{H\backslash\wG} или \emph{левых} \Math{\wG/H}) смежных классов
по некоторой подгруппе \Math{H\leq\wG}. Множество \Math{\wS} называется
\emph{однородным пространством} группы \Math{\wG} 
(или, кратко, \Math{\wG}-\emph{пространством}).
Действие  \Math{\wG} на \Math{\wS} является \emph{точным}, если подгруппа \Math{H}
не содержит нормальных подгрупп группы \Math{\wG}.
Мы можем записать действие в виде перестановок 
\begin{equation}
	\pi(g)=\dbinom{\ws_i}{\ws_ig}\sim\dbinom{Ha}{Hag},
	\hspace*{20pt}g,a\in{}\wG,~~~i=1,\ldots,\wSN,
	\label{perm}
\end{equation}
или, эквивалентно, в виде матриц, состоящих из нулей и единиц 
\begin{equation}
\pi(g)\rightarrow\regrep(g)=
\Mone{\regrep(g)_{ij}},\text{~~где~~} \regrep(g)_{ij}=\delta_{\ws_ig,\ws_j};
~~ i,j=1,\ldots,\wSN.
\label{permrepEn}
\end{equation}
Здесь \Math{\delta_{\alpha,\beta}} означает дельту Кронекера на множестве
\Math{\wS}.
Отображение \eqref{permrepEn} называется \emph{перестановочным представлением}.
\par
Максимальное транзитивное множество \Math{\wS} эквивалентно множеству элементов
самой группы \Math{\wG}, т.е. множеству смежных классов по тривиальной
подгруппе \Math{H=\set{\id}}.
Соответствующие действие и матричное представление называются \emph{регулярными}.
Одна из центральных теорем теории представлений (см. Приложение \ref{irreps}) 
утверждает, что
\emph{любое неприводимое представление конечной группы содержится в регулярном представлении}. 
\par
Представление \eqref{permrepEn} имеет смысл над любой числовой системой, содержащей
0 и 1.
Наиболее естественной системой чисел является \emph{полукольцо натуральных
чисел} \Math{\N=\set{0,1,2,\ldots}.}
С помощью этого полукольца можно снабдить элементы множества \Math{\wS}
\emph{счётчиками}, интерпретируя их как ``\emph{кратности вхождения}'' 
(или ``\emph{числа заполнения}'') элементов  \Math{\ws_i} в состояние
системы, содержащей эти элементы. 
Это состояние можно записать в виде вектора с натуральными компонентами
\begin{equation}
	\barket{n} = \Vthree{n_1}{\vdots}{n_{\wSN}}.
\label{natvect}	
\end{equation}
Таким образом, мы приходим к представлению группы \Math{\wG} в 
\Math{\wSN}-мерном \emph{модуле} \Math{\natmod_\wSN} над полукольцом \Math{\N}.
Действие  \eqref{permrepEn} на векторе \eqref{natvect} сводится попросту к 
перестановке его компонент.
Для дальнейших целей нам необходимо превратить модуль \Math{\natmod_\wSN} в  
\Math{\wSN}-мерное гильбертово пространство \Math{\Hspace_\wSN} расширив
полукольцо \Math{\N} до какого-либо поля.
\par
Основным полем в теории представлений (и, следовательно, в квантовой механике)
является поле комплексных чисел \Math{\C}. Причиной такого выбора является
алгебраическая замкнутость поля \Math{\C}, означающая, в частности,
что не может возникнуть никаких проблем при решении характеристических уравнений
и, следовательно, во всей линейной алгебре. 
Тем не менее, поле \Math{\C} чрезмерно велико --- большинство его элементов
неконструктивны и поэтому бесполезны в приложениях к эмпирической
реальности. Таким образом, следует рассмотреть ситуацию более тщательно.
\par
Прежде всего нам нет необходимости решать \emph{произвольные} характеристические 
уравнения: любое линейное представление является подпредставлением некоторого
перестановочного представления, а все собственные значения последнего являются 
\emph{корнями из единицы}.
Это видно из легко вычисляемого \emph{характеристического полинома} матрицы перестановок
\eqref{permrepEn}
\begin{equation*}
\chi_{\regrep(g)}\vect{\lambda}=\det\vect{\regrep(g)-\lambda\idmat}
=\vect{\lambda-1}^{k_1}\vect{\lambda^2-1}^{k_2}\cdots\vect{\lambda^n-1}^{k_n},
\end{equation*}
где \Math{k_i} --- число циклов длины \Math{i} в перестановке \eqref{perm}.
Для обеспечения унитарности представлений в качестве нормирующих коэффициентов 
используются \emph{квадратные корни} их размерностей. На самом деле, иррациональности обоих типов,
т.е. корни из единицы и квадратные корни из целых чисел имеют одинаковую природу
--- все они являются \emph{циклотомическими целыми}, 
т.е. целочисленными линейными комбинациями корней из единицы. Используя стандартные тождества для 
корней из единицы, отрицательные коэффициенты в циклотомических целых всегда можно
заменить положительными.
Таким образом, изначальными элементами числовой системы, которую мы собираемся построить
являются \emph{натуральные числа} и линейные комбинации с натуральными коэффициентами
\emph{корней из единицы} некоторой степени \Math{\period}, зависящей от структуры 
группы \Math{\wG}. Всё остальное строится с помощью совершенно формальных математических
процедур. Степень \Math{\period} в математической литературе обычно называется 
\emph{кондуктором}. В общем случае кондуктор является кратным некоторого делителя \emph{экспоненты} группы,
которая определяется как наименьшее общее кратное порядков элементов группы 
\Math{\wG}.
\par
С помощью любого набора циклотомических целых можно построить некоторое
\emph{абелево числовое поле} \Math{\NF}, содержащее все элементы этого набора. 
В частности, минимальное абелево числовое 
поле, содержащее заданное множество иррациональностей, можно вычислить с помощью
системы компьютерной алгебры \GAP \cite{gapEn}.
Команда \texttt{\textbf{Field(\textit{gens}\!)}} этой системы возвращает
наименьшее поле, содержащее все элементы из списка иррациональностей
 \texttt{\textbf{\textit{gens}}}.
Теорема Кронекера-Вебера утверждает, что {любое абелево
(в частности, квадратичное) расширение}
рациональных чисел является подполем некоторого \emph{циклотомического поля}. 
\par
Остановимся подробнее на конструкции циклотомических полей.
\emph{Корнем из единицы} \Math{\period}-й степени называется любое решение 
\emph{циклотомического уравнения} \Math{r^\period=1}. Любой корень из единицы 
является некоторой степенью \emph{примитивного} корня из единицы, 
т.е. корня из единицы с периодом, равным в точности \Math{\period}. 
Все примитивные корни из единицы (и только они) являются решениями
уравнения \Math{\Phi_\period\vect{r}=0}, где \Math{\Phi_\period\vect{r}} ---
неприводимый над  \Math{\Q} делитель бинома \Math{r^\period-1}, называемый
\emph{циклотомическим полиномом}.
Линейные комбинации корней из единицы с целыми коэффициентами образуют 
кольцо циклотомических целых \Math{\aNumbers_\period}.
Фактически, в качестве коэффициентов циклотомических целых достаточно ограничиться
\emph{натуральными} числами поскольку в случае \Math{\period\geq2} 
\emph{отрицательные} целые можно ввести с помощью
тождества \Math{\vect{-1}=\sum\limits_{k=1}^{p-1}\runi{}^{\frac{\period}{p}k}},
где \Math{p} --- произвольный делитель периода \Math{\period}. Введение отрицательных
чисел не является концептуально необходимым. Однако оно даёт 
технические преимущества, позволяя использовать развитые методы полиномиальной
алгебры.
\Math{\period}-е \emph{циклотомическое поле} \Math{\Q_\period} вводится как 
\emph{поле частных} кольца \Math{\aNumbers_\period}.
Поле \Math{\Q_\period} можно представить в виде
\begin{equation}
	\Q_\period=\Q\ordset{r}/\braket{\Phi_\period\vect{r}}.
	\label{powerbas}
\end{equation}
Из этого представления и свойств циклотомических полиномов следует, что поле  \Math{\Q_\period}
представляет собой векторное пространство (более того, алгебру) размерности
\Math{\varphi\vect{\period}} над полем рациональных чисел \Math{\Q}.
Здесь \Math{\varphi\vect{\period}} обозначает \emph{функцию Эйлера} (другое 
название этой функции --- \emph{тотиент}),
определяемую как число положительных целых не превосходящих  \Math{\period} и 
взаимно простых
с \Math{\period} (по определению число 1 считается взаимно простым с любым целым).
Наиболее естественным для представления \eqref{powerbas} является 
\emph{степенной базис} векторного пространства \Math{\Q_\period}, т.е. элементы
поля \Math{\Q_\period} представляются в виде линейных комбинаций с 
рациональными коэффициентами степеней примитивного корня меньших 
степени циклотомического полинома. Этот базис является \emph{целочисленным} в том смысле,
что любое циклотомическое целое может быть представлено в нём как целочисленный 
вектор.
Заметим, что в математике часто используются
другие целочисленные базисы для  циклотомических чисел.
\par
Абелево числовое поле \Math{\NF} представляет собой подполе поля \Math{\Q_\period},
фиксируемое дополнительными симметриями, называемыми \emph{автоморфизмами Галуа}.
Автоморфизм Галуа \Math{*k} это отображение, задаваемое преобразованием
\Math{\runi{\period}\rightarrow\runi{\period}^k}
примитивного корня \Math{\period}-й степени из единицы.
Здесь \Math{k} --- взаимно простое с  \Math{\period} целое  в пределах
\Math{1\leq{}k<\period}.
\par 
Циклотомические числа можно погрузить в поле комплексных 
чисел \Math{\C}, однако в этом нет никакой необходимости. 
\begin{figure}
	\centering
\begin{tabular}{cc}
\hspace*{-10pt}\Math{\period=12}&\hspace*{-20pt}\Math{\period=7}\\[-10pt]
\hspace*{-10pt}\includegraphics[width=0.507\textwidth]{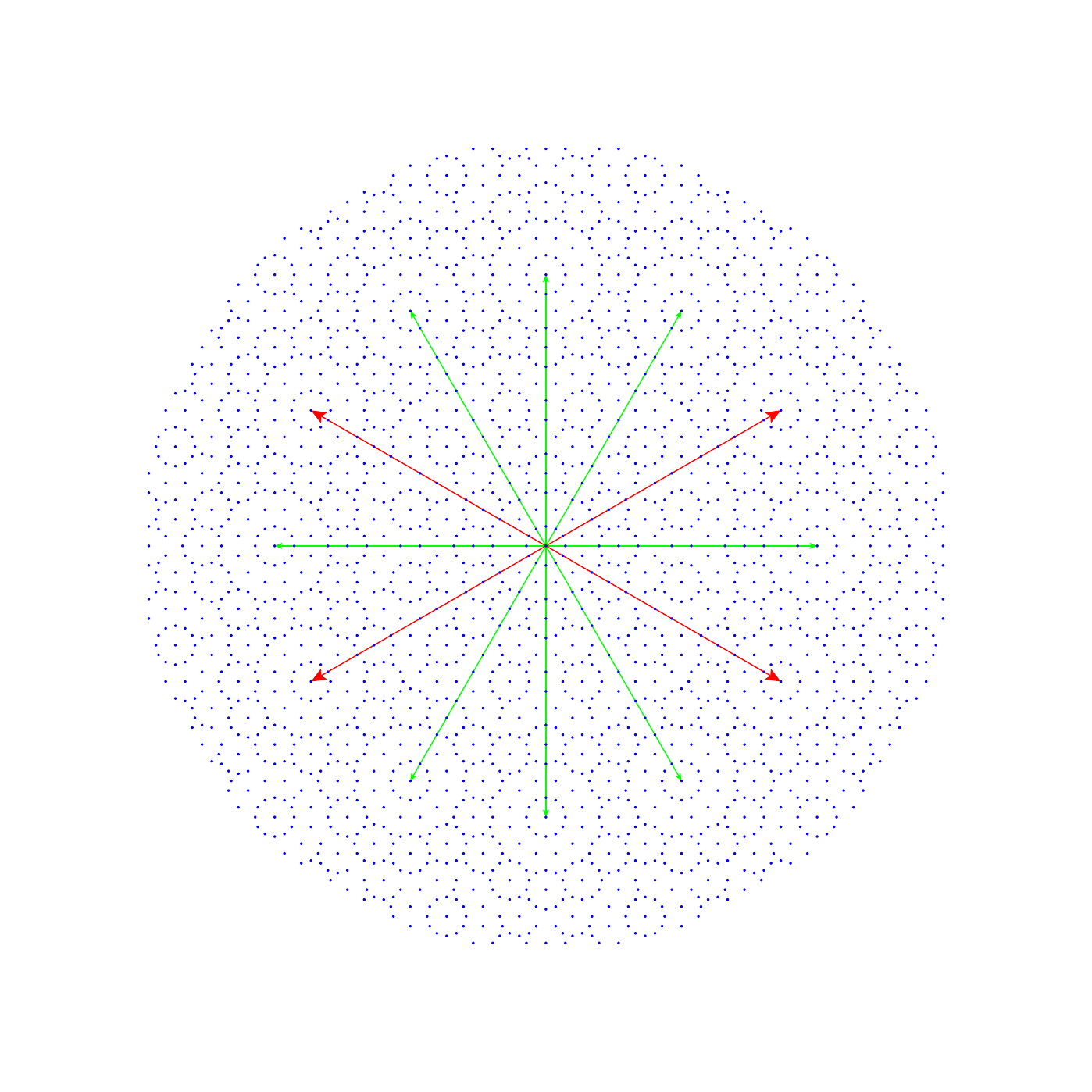}
&
\hspace*{-20pt}\includegraphics[width=0.507\textwidth]{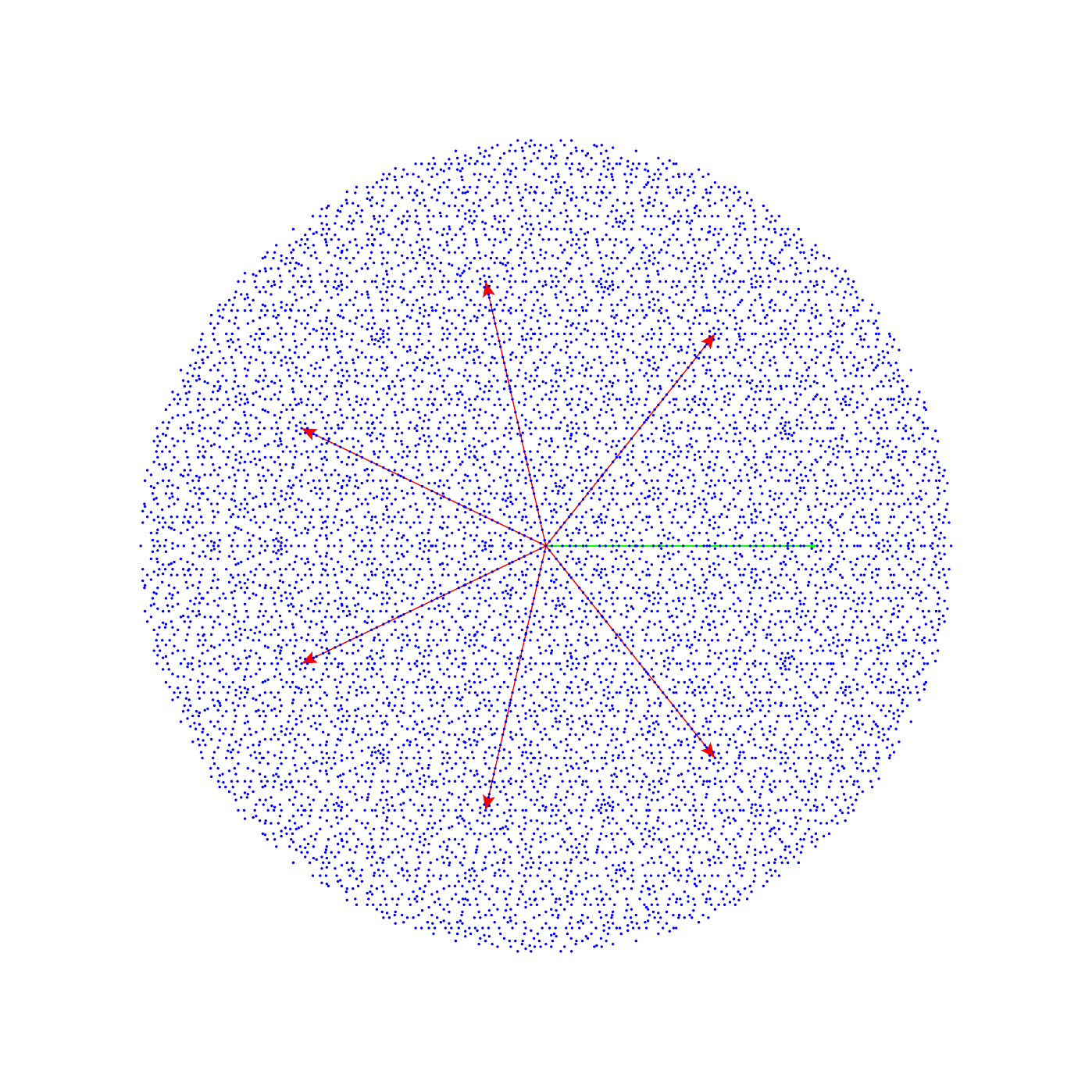}\\[-10pt]
\end{tabular}
\caption{Погружение \Math{\aNumbers_\period} в \Math{\C}.
Примитивные корни выделены более жирными стрелками.} 
\label{inC}		
\end{figure}
Чисто алгебраических свойств циклотомических чисел достаточно для всех манипуляций 
в гильбертовом пространстве \Math{\Hspace_\wSN} и его подпространствах.
Например, комплексное сопряжение циклотомического числа определяется преобразованием
\Math{\cconj{\runi{\period}^m}=\runi{\period}^{\period-m}}.
Однако, для иллюстрации, мы приводим на Рис. \ref{inC} примеры погружения 
циклотомических 
 целых в
комплексную плоскость. Ясно, что циклотомические дроби, т.е. элементы поля \Math{\Q_\period}, 
при \Math{\period\geq3} заполняют комплексную плоскость всюду плотно. 
Отсюда, в частности, следует тривиальное объяснение причины появления комплексных чисел в квантовой механике.
\par
Все иррациональности являются промежуточными элементами квантового описания,
исчезающими в окончательных выражениях для квантовых наблюдаемых.
Это утверждение представляет собой конструктивное уточнение взаимосвязи между комплексными 
и вещественными числами в стандартной квантовой механике, где промежуточные величины могут
быть комплексными, но наблюдаемые обязаны быть вещественными.
\subsubsection{Представление квантовой эволюции перестановками}
Из сказанного выше следует, что любое \Math{\adimH}-мерное представление \Math{\repq}
может быть расширено до \Math{\wSN}-мерного представления \Math{\widetilde{\repq}}
соответствующего \emph{перестановочному действию} группы \Math{\wG} на 
\Math{\wSN}-элементном множестве \Math{\wS=\set{\ws_1,\ldots,\ws_{\wSN}}}.
Это означает, что \Math{\transmatr^{-1}\regrep\transmatr=\widetilde{\repq}},
где \Math{\regrep} --- перестановочное представление \eqref{permrepEn}, а
\Math{\transmatr}  --- матрица преобразования. Ясно, что \Math{\wSN\geq\adimH}.
Если \Math{\wSN=\adimH}, то представление \Math{\repq} само по себе является
перестановочным.
Если \Math{\wSN>\adimH}, то перестановочное представление имеет следующую структуру
\Mathh{
\transmatr^{-1}\regrep\transmatr
=\Vtwo{
\left.
\begin{aligned}
\!\IrrRep{1}&\\[-2pt]
&\hspace{8pt}\mathrm{V}
\end{aligned}
\right\}\Hspace_{\wSN-\adimH}
}{
\left.
\hspace{29pt}
{\repq}
\right\}\Hspace_{\adimH}
},\hspace{10pt}
\Hspace_{\wSN} = \Hspace_{\wSN-\adimH}\oplus\Hspace_{\adimH},
}	
где \Math{\IrrRep{1}} --- тривиальное одномерное представление,
являющееся обязательной компонентой любого перестановочного представления.
Компонента \Math{\mathrm{V}} может отсутствовать.
\par
Данные в пространствах \Math{\Hspace_{\adimH}} и \Math{\Hspace_{\wSN-\adimH}} \emph{взаимно независимы}, поскольку
оба пространства являются инвариантными подпространствами в  \Math{\Hspace_{\wSN}}. 
Т.е., мы можем  рассматривать данные в \Math{\Hspace_{\wSN-\adimH}}  как ``скрытые параметры'' по отношению к данным в \Math{\Hspace_{\adimH}}. Таким образом, \emph{любая квантовая проблема}
в \Math{\adimH}-мерном гильбертовом пространстве может быть переформулирована в терминах перестановок \Math{\wSN} предметов.
\par
Приняв тривиальное предположение о том, что компоненты векторов состояния
могут быть произвольными элементами основного поля  \Math{\NF},
мы можем установить произвольные (например, нулевые) данные в подпространстве
\Math{\Hspace_{\wSN-\adimH}}, дополнительном к \Math{\Hspace_{\adimH}}.
Однако такой подход неинтересен, поскольку он не фальсифицируем в рамках стандартной квантовой механики.
Фактически он воспроизводит стандартную квантовую механику с точностью до физически несущественного различия между ``\emph{конечным}'' и ``\emph{бесконечным}''. Отличие этого подхода от стандартной форимулировки чисто техническое --- можно заменить линейную алгебру в  \Math{\adimH}-мерном пространстве \Math{\Hspace_{\adimH}} перестановками  \Math{\wSN} элементов%
\footnote{Можно было бы предположить, что замена операций линейной алгебры гораздо более простыми манипуляциями с перестановками может дать вычислительные преимущества. 
Однако разница между \Math{\adimH} и \Math{\wSN} может быть существенной.
Например, наибольшая спорадическая простая группа
--- ``\emph{монстр Фишера-Грисса}'' ---
имеет минимальное точное неприводимое представление размерности
\Math{\adimH\approx2\cdot10^5},
а степень её минимального перестановочного представления равна
\Math{\wSN\approx10^{20}}. 
}.
\par
Более перспективный подход требует некоторых изменений в концепции квантовых амплитуд.
Мы предполагаем \cite{Kornyak12b}, что квантовые амплитуды представляют собой проекции в инвариантные подпространства векторов кратностей (``чисел заполнения'') элементов множества \Math{\wS}, на котором группа  \Math{\wG} действует перестановками. 
Таким образом, мы предполагаем, что квантовые состояния
системы и аппарата в перестановочном базисе являются векторами с натуральными
компонентами:
\Mathh{
\barket{n} = \Vthree{n_1}{\vdots}{n_{\wSN}} \text{~и~}
\barket{m} = \Vthree{m_1}{\vdots}{m_{\wSN}}.}
В соответствии с правилом Борна, вероятность обнаружить состояние \Math{\barket{n}}
аппаратом, настроенным на состояние \Math{\barket{m}}, равна
\begin{equation*}
    \ProbBorn{m}{n}=\frac{\vect{\sum_i{m_i}n_i}^2}{\sum_i{m_i}^2\sum_i{n_i}^2}.
\end{equation*}
Ясно, что значение этого выражения на натуральных векторах \Math{\barket{n}} и \Math{\barket{m}} является неотрицательным рациональным числом.
Деструктивная квантовая интерференция между ненулевыми натуральными векторами перестановочного пространства возможна только если для любого \Math{i} либо \Math{m_i=0} либо \Math{n_i=0}.
В собственных \emph{инвариантных подпространствах} перестановочного представления деструктивная интерференция возможна и для векторов у которых все компоненты являются натуральными числами строго большими нуля.
Рассмотрим вначале простой пример.
\subsection{Примеры конечных квантовых систем}
\subsubsection{Подробный пример: наименьшая некоммутативная группа \Math{\SymG{3}}}
Мы выбрали группу  \Math{\SymG{3}} в качестве примера
из-за простоты анализа. Тем не менее, эта группа имеет важные приложения в физике.
В частности, она используется для описания так называемого \emph{три-бимаксимального
смешивания} в нейтринных осцилляциях \cite{HPS02,HS03}. 
(Более подробное обсуждение см. в Приложении \ref{appflavor}.)
Группа состоит из шести
элементов, которые можно представить следующими перестановками 3-х объектов
\begin{equation*}
\wg_1=\vect{}\!,~\wg_2=\vect{2,3}\!,~\wg_3=\vect{1,3}\!,~\wg_4=\vect{1,2}\!,
    ~\wg_5=\vect{1,2,3}\!,~\wg_6=\vect{1,3,2}.
\end{equation*}
Экспонента группы \Math{\SymG{3}} равна шести поскольку степени её элементов 
равны 2 и 3.
Группа может быть порождена различными парами своих элементов.
Например, в качестве порождающего множества можно выбрать пару  
\Math{\set{\wg_2, \wg_6}}.
Группа \Math{\SymG{3}} распадается на три класса сопряжённых элементов
\begin{equation*}
    \class{1}=\set{\wg_1},~~\class{2}=\set{\wg_2,~\wg_3,~\wg_4},
    ~~\class{3}=\set{\wg_5,~\wg_6}.
\end{equation*}
Таблица характеров (см. Приложение \ref{irreps})  этой группы имеет вид
\begin{center}
    \text{\begin{tabular}{c|crr}
    &\Math{\class{1}}&\Math{\class{2}}&\Math{\class{3}}\\\hline
    \Math{\chi_1}&1&1&1\\
    \Math{\chi_2}&1&-1&1\\
    \Math{\chi_3}&2&0&-1
    \end{tabular}\enspace.}
\end{center}
В соответствии с традицией, принятой в физической литературе, мы будем обозначать
\emph{неприводимые} представления их размерностями в жирном шрифте.
Таким образом, мы имеем три неприводимых представления \Math{\IrrRep{1}, \IrrRep{1'}} и 
\Math{\IrrRep{2}}, из которых только \Math{\IrrRep{2}} является точным.
Для обозначения \emph{перестановочных} представлений, играющих важную роль в нашем подходе,
мы будем использовать также их размерности в жирном шрифте, но с дополнительным подчёркиванием.
\par
Матрицы перестановочного представления порождающих элементов имеют вид
\begin{equation*}
    P_2=\Mthree{1}{~\cdot}{~\cdot}{\cdot}{~\cdot}{~1}{\cdot}{~1}{~\cdot}
    \text{~~и~}
    ~P_6=\Mthree{\cdot}{~\cdot}{~1}{1}{~\cdot}{~\cdot}{\cdot}{~1}{~\cdot}.
\end{equation*}
Эти матрицы имеют, соответственно, следующие наборы собственных значений:\\ 
\Math{\vect{1, 1, -1}} и \Math{\vect{1, \runi{3}, \runi{3}^2}}, где \Math{\runi{3}}
--- примитивный корень 3-й степени из единицы, определяемый циклотомическим полиномом
\Math{\Phi_3\vect{r} = 1+r+r^2}.
\par
Как было отмечено выше, любое перестановочное представление имеет одномерное
инвариантное подпространство с базисным вектором
 \Math{\vect{1,\ldots,1}^\mathrm{T}}.
Таким образом, единственным возможным разложением перестановочного представления
является \Math{\PermRep{3}\cong\IrrRep{1}\oplus\IrrRep{2}} или в явном матричном виде
\begin{equation}
    \widetilde{U}_j=\Mtwo{\IrrRep{1}}{0}{0}{U_j},~~j =1,\ldots,6,
    \label{S3permqEn}
\end{equation}
где матрицы  \Math{U_j} являются элементами точного неприводимого 
представления \Math{\IrrRep{2}}. 
\par
Для построения разложения \eqref{S3permqEn} необходимо определить матрицы 
\Math{U_j} и \Math{\transmatr} таким образом, чтобы выполнялось соотношение
\Math{\widetilde{U}_j=\transmatr^{-1}P_j\transmatr}. Дополнительно мы потребуем,
чтобы все эти матрицы были унитарными. Ясно, что для построения матриц 
\Math{U_j} и \Math{\transmatr} достаточно использовать только перестановочные
матрицы порождающих элементов%
\footnote{Заметим, \label{MeatAxePage}
 что проблема построения общих инвариантных подпространств для 
нескольких матриц или, в несколько более общем смысле, проблема расщепления модулей над 
ассоциативными алгебрами на неприводимиые подмодули алгоритмически весьма нетривиальна. 
Достаточно универсальный 
алгоритм, называемый \textbf{\emph{MeatAxe}} \cite{Holt}, 
разработан только для алгебр
над конечными полями. Алгоритм \textbf{\emph{MeatAxe}} включён в системы компьютерной алгебры 
\GAP и \Magma\!.
Однако в случае групповых алгебр, по крайней мере для небольших групп, 
необходимые построения можно провести ``методом проб и ошибок'', как правило, без особых затруднений 
и над полями нулевой характеристики, в частности, над циклотомическими полями или их подполями.}.
Разложение \eqref{S3permqEn} можно построить разными способами.
Ясно, что мы всегда можем выбрать диагональной матрицу \Math{U_j} одного из 
порождающих элементов.
\par 
Если начать с диагонализации  \Math{P_6}, то мы получим
следующий набор матриц \Math{U_j}:
\begin{align}
        U_1=\Mtwo{1}{0}{0}{1},~U_2=\Mtwo{0}{\runi{3}^2}{\runi{3}}{0},
        ~U_3=\Mtwo{0}{\runi{3}}{\runi{3}^2}{0},\nonumber\\
        \label{umatrices}\\[-12pt]
        ~U_4=\Mtwo{0}{1}{1}{0},
        ~U_5=\Mtwo{\runi{3}^2}{0}{0}{\runi{3}},
        ~U_6=\Mtwo{\runi{3}}{0}{0}{\runi{3}^2}.\nonumber
\end{align}
Матрица преобразования (с точностью до несущественного произвола в 
выборе её элементов) имеет вид 
\begin{equation}
\transmatr=\frac{1}{\sqrt{3}}
    \Mthree{1}{1}{\runi{3}^2}
     {1}{\runi{3}^2}{1}
     {1}{\runi{3}}{\runi{3}},~~~~
\transmatr^{-1}=\frac{1}{\sqrt{3}}
    \Mthree{1}{1}{1}
     {1}{\runi{3}}{\runi{3}^2}
     {\runi{3}}{1}{\runi{3}^2}.
\label{transS3monomial}
\end{equation}
\par
Минимальным абелевым числовым полем, содержащим все элементы матриц \eqref{umatrices}
и \eqref{transS3monomial}, является циклотомическое поле \Math{\Q_{12}}. 
Таким образом, мы можем переписать, скажем, матрицу \eqref{transS3monomial}
в терминах элементов из  
 \Math{\Q_{12}} 
\begin{equation}
\transmatr=\frac{1}{3}
    \Mthree{2\runi{12}+\runi{12}^{9}}{2\runi{12}+\runi{12}^{9}}
    {\runi{12}^7+\runi{12}^9}
     {2\runi{12}+\runi{12}^{9}}{\runi{12}^7+\runi{12}^9}
     {2\runi{12}+\runi{12}^{9}}
     {2\runi{12}+\runi{12}^{9}}{2\runi{12}^3+\runi{12}^7}{2\runi{12}^3+\runi{12}^7},
     \label{transS3monomialcyclo}
\end{equation}
где \Math{\runi{12}} --- примитивный 12-й корень из единицы, т.е. произвольное
(абстрактное) решение уравнения \Math{\Phi_{12}\vect{r} \equiv 1-r^2+r^4 =0}.
В \eqref{transS3monomialcyclo} мы использовали  ``натуральное'' представление
циклотомических целых. Приведение по модулю циклотомического полинома
\Math{\Phi_{12}\vect{r}} даёт  ``целое'' представление в степенном базисе, т.е.
представление с отрицательными коэффициентами, но с минимальными степенями
 \Math{\runi{12}}:
\begin{equation*}
\transmatr=\frac{1}{3}
    \Mthree{2\runi{12}-\runi{12}^3}{2\runi{12}-\runi{12}^3}
    {-\runi{12}-\runi{12}^3}
     {2\runi{12}-\runi{12}^3}{-\runi{12}-\runi{12}^3}
     {2\runi{12}-\runi{12}^3}
     {2\runi{12}-\runi{12}^3}{-\runi{12}+2\runi{12}^3}{-\runi{12}+2\runi{12}^3}.
\end{equation*}
\par
Диагонализация  \Math{P_2} приводит к другому виду разложения  \eqref{S3permqEn}.
В этом случае для матриц порождающих элементов мы имеем
\begin{equation}
U'_2=\Mtwo{1}{0}{0}{-1},
~~~~U'_6=\Mtwo{-\frac{1}{2}}{\frac{\sqrt{3}}{2}}{-\frac{\sqrt{3}}{2}}{-\frac{1}{2}}.
\label{US3}
\end{equation}
Матрица преобразования принимает вид
\begin{equation}
\transmatrprim
=\Mthree{\frac{1}{\sqrt{3}}}{~\,\,\sqrt{\frac{2}{3}}}{~0}
          {\frac{1}{\sqrt{3}}}{\,-\!\frac{1}{\sqrt{6}}}{-\!\frac{1}{\sqrt{2}}}
          {\frac{1}{\sqrt{3}}}{\,-\!\frac{1}{\sqrt{6}}}{~\,\,\frac{1}{\sqrt{2}}},
~~~~
\transmatrprim^{-1}
=\Mthree{\frac{1}{\sqrt{3}}}{~\,\frac{1}{\sqrt{3}}}{~\frac{1}{\sqrt{3}}}
        {\sqrt{\frac{2}{3}}}{\,-\!\frac{1}{\sqrt{6}}}{-\!\frac{1}{\sqrt{6}}}
        {~0}{\,-\!\frac{1}{\sqrt{2}}}{~\,\,\frac{1}{\sqrt{2}}}.
\label{transS3sqrt}
\end{equation}
Эта матрица используется в физике частиц для описании феноменологии нейтринных 
осцилляций, где она известна под именами \emph{матрица три-бимаксимального смешивания} 
или \emph{матрица Харрисона-Перкинса-Скотта}.
\par
Минимальное абелево числовое поле \Math{\NF}, содержащее все элементы матриц
\eqref{US3} и \eqref{transS3sqrt}, является подполем циклотомического поля
\Math{\Q_{24}}, фиксируемым с помощью автоморфизма Галуа
\Math{\runi{24}\rightarrow\runi{24}^{23}}. 
Здесь \Math{\runi{24}} --- примитивный 24-й корень из единицы. 
Соответствующий циклотомический полином имеет вид \Math{\Phi_{24}\vect{r}=1-r^4+r^8}.
В терминах циклотомических чисел матрицы \eqref{US3} и \eqref{transS3sqrt} имеют вид
\Mathh{U'_6=\frac{1}{2}\Mtwo{-1}{2\runi{24}^2-\runi{24}^6}{-2\runi{24}^2+\runi{24}^6}{-1}}
и 
\Mathh{\transmatrprim
=\frac{1}{6}\Mthree{4\runi{24}^2-2\runi{24}^6}{2\runi{24}+2\runi{24}^3
+2\runi{24}^5-4\runi{24}^7}{0}{4\runi{24}^2-2\runi{24}^6}
{-\runi{24}-\runi{24}^3-\runi{24}^5+2\runi{24}^7}
{-6\runi{24}+6\runi{24}^5}
{4\runi{24}^2-2\runi{24}^6}{-\runi{24}-\runi{24}^3-\runi{24}^5+2\runi{24}^7}
{6\runi{24}-6\runi{24}^5}.}
\par
Фактически информация о ``квантовом поведении'' закодирована в матрицах преобразования
таких как \eqref{transS3monomial} или \eqref{transS3sqrt}. 
\par
Пусть \Math{\barket{n} = \Vthree{n_1}{n_2}{n_3}} и
\Math{\barket{m} = \Vthree{m_1}{m_2}{m_3}} --- векторы состояний системы и аппарата
в ``перестановочном'' базисе. С помощью матрицы преобразования \eqref{transS3monomial}
мы можем записать эти векторы в ``квантовом'' базисе:
\begin{align*}
    \barket{\widetilde{\psi}}=\transmatr^{-1}\barket{n}
    &=\frac{1}{\sqrt{3}}\Vthree{n_1+n_2+n_3}
    {n_1+n_2\runi{3}+n_3\runi{3}^2}{n_1\runi{3}+n_2+n_3\runi{3}^2},\\
    \barket{\widetilde{\phi}}=\transmatr^{-1}\barket{m}
    &=\frac{1}{\sqrt{3}}\Vthree{m_1+m_2+m_3}
    {m_1+m_2\runi{3}+m_3\runi{3}^2}{m_1\runi{3}+m_2+m_3\runi{3}^2}.
\end{align*}
Проекции векторов состояний в двумерное инвариантное подпространство имеют вид: 
\begin{equation*}
    \barket{\psi} = \frac{1}{\sqrt{3}}\Vtwo{n_1+n_2\runi{3}+n_3\runi{3}^2}
    {n_1\runi{3}+n_2+n_3\runi{3}^2},~~~~
    \barket{\phi} = \frac{1}{\sqrt{3}}\Vtwo{m_1+m_2\runi{3}+m_3\runi{3}^2}
    {m_1\runi{3}+m_2+m_3\runi{3}^2}.
\end{equation*}
Соответственно, составляющие борновской вероятности \eqref{BornEn} для двумерной
подсистемы принимают вид: 
\begin{equation}
    \inner{\psi\!}{\!\psi}=\invarQ{3}{n}{n}-\frac{1}{3}\invarL{3}{n}^2,
\label{Born2den1En}
\end{equation}
\begin{equation}
    \inner{\phi\!}{\!\phi}=\invarQ{3}{m}{m}-\frac{1}{3}\invarL{3}{m}^2,
\label{Born2den2En}
\end{equation}
\begin{equation}
\cabs{\inner{\phi\!}{\!\psi}}^2=
\vect{\invarQ{3}{m}{n}-\frac{1}{3}\invarL{3}{m}\invarL{3}{n}}^2,
\label{Born2numEn}
\end{equation}
где \Math{\invarL{\wSN}{n}=\sum\limits_{i=1}^{\wSN}n_i} и
\Math{\invarQ{\wSN}{m}{n}=\sum\limits_{i=1}^{\wSN}m_in_i} 
--- линейный и квадратичный перестановочные инварианты. Эти инварианты
являются общими для любых групп перестановок.
\par
Обратим внимание на то, что
\begin{enumerate}
    \item 
Выражения \eqref{Born2den1En}--\eqref{Born2numEn} скомбинированы из
\emph{инвариантов перестановочного представления}.
    \item
\emph{Деструктивная квантовая интерференция}, 
т.е. обращение в нуль борновской вероятности, определяется уравнением    
\Mathh{3\vect{m_1n_1+m_2n_2+m_3n_3}-\vect{m_1+m_2+m_3}\vect{n_1+n_2+n_3}=0.}
Это уравнение имеет бесконечно много ``натуральных'' решений с ненулевыми компонентами, например,
\Mathh{{\barket{n} = \Vthree{1}{1}{2},~~\barket{m} = \Vthree{1}{3}{2}}.}
\par
Таким образом, простым переходом к инвариантному подпространству мы получили
существенные черты квантового поведения из ``перестановочной
динамики'' и ``натуральной'' интерпретации \eqref{natvect} квантовых амплитуд.
\end{enumerate}
\par
Этот пример можно слегка обобщить.
Любое перестановочное представление содержит \Math{\vect{\wSN-1}}-мерное инвариантное
подпространство. Скалярное произведение в этом подпространстве  в терминах
перестановочных инвариантов имеет вид
\Mathh{\inner{\phi\!}{\!\psi}=
\invarQ{\wSN}{m}{n}-\frac{1}{\wSN}\invarL{\wSN}{m}\invarL{\wSN}{n}.}
Тождество
\Math{\displaystyle\invarQ{\wSN}{n}{n}-\frac{1}{\wSN}\invarL{\wSN}{n}^2\equiv
\frac{1}{\wSN^2}\sum\limits_{i=1}^\wSN\vect{\invarL{\wSN}{n}-\wSN{}n_i}^2}
показывает явно, что\\
\Math{\inner{\psi\!}{\!\psi}>0} для \Math{\barket{n}} с \emph{различными}%
\footnote{Векторы с одинаковыми компонентами принадлежат одномерному
дополнению к рассматриваемому \Math{\vect{\wSN-1}}-мерному подпространству.}
компонентами \Math{n_i}. 
Это скалярное произведение не содержит иррациональностей, если векторы 
\Math{\barket{n}} и \Math{\barket{m}} натуральны. 
\subsubsection{Группа икосаэдра \Math{\AltG{5}}}
\begin{figure}[!h]
\centering
\includegraphics[width=0.62\textwidth]{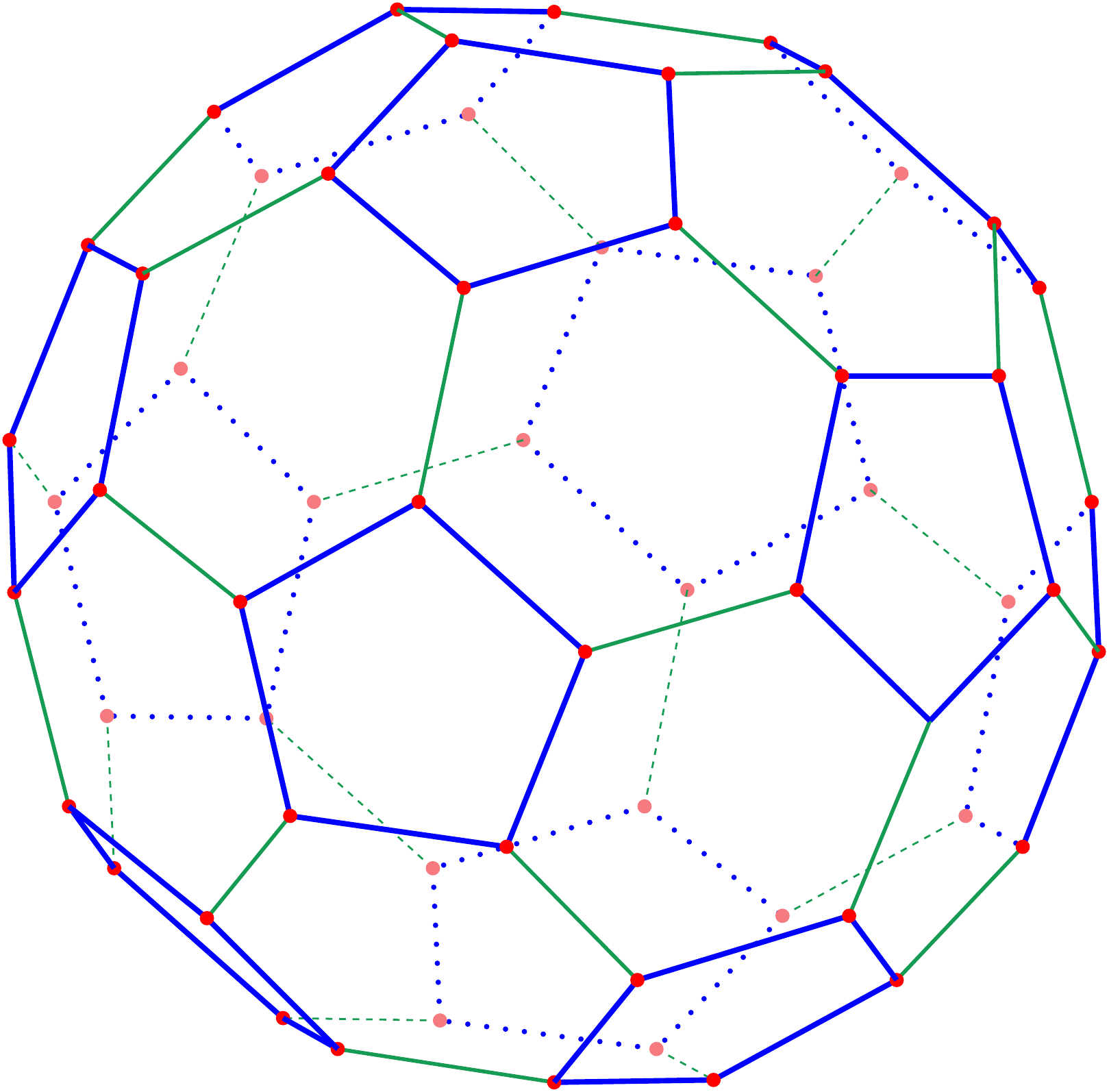}
\caption{Граф Кэли группы \Math{\AltG{5}}. 
Пятиугольники, шестиугольники и связи, соединяющие смежные пятиугольники соответствуют
реляторам \Math{a^5}, 
\Math{\vect{ab}^3} и  \Math{b^2} в представлении 
\eqref{present}.}
\label{bucky}
\end{figure}

Группа симметрий икосаэдра \Math{\AltG{5}} это наименьшая простая некоммутативная
группа. Она состоит из 60 элементов и её экспонента равна 30.
Эта группа играет настолько важную роль в математике и приложениях, что
Ф. Клейн посвятил её отдельную книгу \cite{Klein}.
В физической литературе для группы икосаэдра часто используется обозначение
\Math{\Sigma\!\vect{60}}.
Интересно отметить, что \Math{\AltG{5}} имеет ``физическое воплощение'':
молекула фуллерена \Math{C_{60}} имеет структуру графа Кэли этой группы
(см. Рис. \ref{bucky}). Это ясно из следующего \emph{представления} \Math{\AltG{5}}
с помощью двух порождающих элементов на которые наложены три \emph{соотношения}%
\footnote{Если соотношение записано в виде \Math{R=\id}, 
то его левую часть \Math{R} называют \emph{релятором}.}.
\begin{equation}
	\AltG{5}\cong\left\langle{}a, b\mid{}a^5,
	b^2, \vect{ab}^3\right\rangle.
	\label{present}
\end{equation}
\par
Группа \Math{\AltG{5}} разбивается на пять классов сопряжённых элементов 
\Math{\class{1},} \Math{\class{15},} \Math{\class{20},} \Math{\class{12},} \Math{\class{12'}}.
Для различения классов мы здесь используем их размеры в качестве нижних индексов.
Таблица характеров группы имеет вид
\begin{center}
  \begin{tabular}{c|c|c|c|c|c}
&\Math{\class{1}}&\Math{\class{15}}&\Math{\class{20}}&\Math{\class{12}}&\Math{\class{12'}}
\\\hline  
\Math{\chi_{1}}&\Math{1}&\Math{1}&\Math{1}&\Math{1}&\Math{1}
\\\hline  
\Math{\chi_{3}}&\Math{3}&\Math{-1}&\Math{0}&\Math{\phi}&\Math{1-\phi}
\\\hline  
\Math{\chi_{3'}}&\Math{3}&\Math{-1}&\Math{0}&\Math{1-\phi}&\Math{\phi}
\\\hline  
\Math{\chi_{4}}&\Math{4}&\Math{0}&\Math{1}&\Math{-1}&\Math{-1}
\\\hline  
\Math{\chi_{5}}&\Math{5}&\Math{1}&\Math{-1}&\Math{0}&\Math{0}
	\end{tabular}\enspace.
\end{center}
Здесь \Math{\phi=\frac{1+\sqrt{5}}{2}} --- ``\alert{золотое сечение}''.
Заметим что \Math{\phi} и \Math{1-\phi} являются циклотомическими 
целыми (или ``\emph{циклотомическими натуральными}''):
\Math{\phi=-\runi{5}^2-\runi{5}^3~\equiv~1+\runi{5}+\runi{5}^4} ~и~ 
\Math{1-\phi=-\runi{5}-\runi{5}^4~\equiv~1+\runi{5}^2+\runi{5}^3},
где \Math{\runi{5}} --- примитивный корень 5-й степени из единицы.
Из таблицы характеров видно, что \Math{\AltG{5}} имеет пять неприводимых
представлений: тривиальное  \Math{\IrrRep{1}} и четыре точных 
\Math{\IrrRep{3}, \IrrRep{3'}, \IrrRep{4}, \IrrRep{5}}. 
\par
Что касается перестановок, то группа имеет три \emph{примитивных} 
действия на множествах из  5, 6 и 10 элементов.
Соответствующие перестановочные представления имеют следующие разложения на неприводимые компоненты 
\Mathh{\PermRep{5}\cong\IrrRep{1}\oplus\IrrRep{4},~~
 \PermRep{6}\cong\IrrRep{1}\oplus\IrrRep{5},~~
 \PermRep{10}\cong\IrrRep{1}\oplus\IrrRep{4}\oplus\IrrRep{5}.}
Напомним, что транзитивное действие группы \Math{\wG} на множестве \Math{\wS}
называется \emph{импримитивным} \cite{Wielandt}, если имеется 
\emph{нетривиальное} разбиение
множества  \Math{\wS}, инвариантное относительно действия группы.
При этом \emph{тривиальными разбиениями} по определению являются:
разбиение на одноэлементные блоки и разбиение на блок, содержащий всё множество 
\Math{\wS}, и дополнительный к нему пустой блок.
Нетривиальное инвариантное разбиение называется \emph{системой импримитивности}
или \emph{системой блоков}.
Если инвариантными являются только тривиальные разбиения, то действие 
называется \emph{примитивным}.
Примитивные действия считаются наиболее фундаментальными среди всех 
перестановочных действий.
\par
Рассмотрим действие \Math{\AltG{5}} на множестве вершин икосаэдра \Math{\wS_{12}}. 
\begin{figure}
\centering
\includegraphics[width=0.65\textwidth]{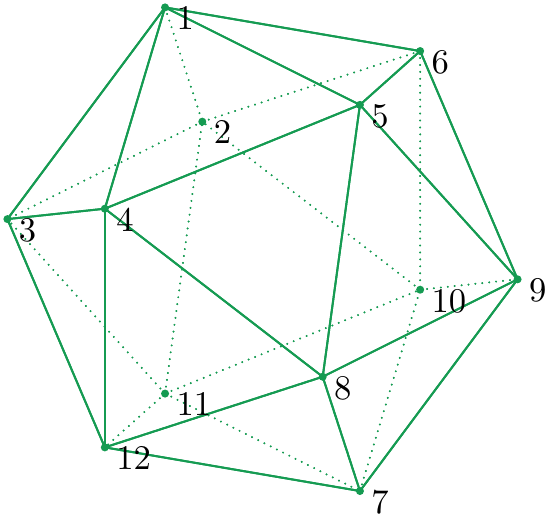}
\caption{Икосаэдр. Пары противоположных вершин образуют систему импримитивности.}
\label{ico}
\end{figure}
Это действие транзитивно, но \emph{импримитивно}.
Система импримитивности 
имеет вид
\Mathh{\set{\mid{}B_1\mid\cdots\mid{}B_i\mid\cdots\mid{}B_6\mid}\equiv
\set{\mid1,7\mid\cdots\mid{}i,i+6\mid\cdots\mid6,12\mid}}
в предположении нумерации вершин, изображённой на Рис. \ref{ico}.
Каждый блок \Math{B_i} состоит из пары противоположных вершин икосаэдра.
Группа \Math{\AltG{5}} переставляет блоки между собой как целые объекты.
Мы будем обозначать соответствие между противоположными вершинами блоков
(``комплементарность'') символом \Math{\Oppbare}, т.е. если \Math{B_i=\set{p,q}},
то  \Math{q=\Opp{p}} и  \Math{p=\Opp{q}}.
Для нумерации, используемой на Рис. \ref{ico}, комплементарность можно выразить 
формулой \Math{\Opp{p}\equiv1+\vect{p+5\mod12}}.
\par
Перестановочное представление действия \Math{\AltG{5}} на вершинах икосаэдра
имеет следующее разложение на неприводимые компоненты
\begin{equation}
	\PermRep{12}\cong\IrrRep{1}\oplus\IrrRep{3}\oplus\IrrRep{3'}\oplus\IrrRep{5}
	\text{~~или~~}\transmatr^{-1}\vect{\PermRep{12}}\transmatr
	=\IrrRep{1}\oplus\IrrRep{3}\oplus\IrrRep{3'}\oplus\IrrRep{5}.
	\label{decoico}
\end{equation}
Используя обозначения
\Mathh{\displaystyle\alpha=\frac{\phi}{4}\sqrt{10-2\sqrt{5}},~
\beta=\frac{\sqrt{5}\sqrt{10-2\sqrt{5}}}{20},~
\gamma=\frac{\sqrt{3}}{8}\vect{1-\frac{\sqrt{5}}{3}},~
\delta=-\frac{\sqrt{3}}{8}\vect{1+\frac{\sqrt{5}}{3}},}
матрицу преобразования \Math{\transmatr} из формулы \eqref{decoico}
можно записать, в частности, как
\begin{equation}
\transmatr=
\bmat
\frac{\sqrt{3}}{6}&\alpha&\beta&0&\alpha&\beta&0&\frac{1}{4}&-\frac{1}{2}&0&0&\frac{\sqrt{15}}{12}
\\
\frac{\sqrt{3}}{6}&0&\alpha&\beta&-\beta&0&\alpha&-\frac{\phi}{4}&0&-\frac{1}{2}&0&\gamma
\\
\frac{\sqrt{3}}{6}&\beta&0&\alpha&0&-\alpha&-\beta&\frac{\phi-1}{4}&0&0&-\frac{1}{2}&\delta
\\
\frac{\sqrt{3}}{6}&0&\alpha&-\beta&-\beta&0&-\alpha&-\frac{\phi}{4}&0&\frac{1}{2}&0&\gamma
\\
\frac{\sqrt{3}}{6}&-\beta&0&\alpha&0&\alpha&-\beta&\frac{\phi-1}{4}&0&0&\frac{1}{2}&\delta
\\
\frac{\sqrt{3}}{6}&\alpha&-\beta&0&-\alpha&\beta&0&\frac{1}{4}&\frac{1}{2}&0&0&\frac{\sqrt{15}}{12}
\\
\frac{\sqrt{3}}{6}&0&-\alpha&\beta&\beta&0&\alpha&-\frac{\phi}{4}&0&\frac{1}{2}&0&\gamma
\\
\frac{\sqrt{3}}{6}&\beta&0&-\alpha&0&-\alpha&\beta&\frac{\phi-1}{4}&0&0&\frac{1}{2}&\delta
\\
\frac{\sqrt{3}}{6}&-\alpha&\beta&0&\alpha&-\beta&0&\frac{1}{4}&\frac{1}{2}&0&0&\frac{\sqrt{15}}{12}
\\
\frac{\sqrt{3}}{6}&-\alpha&-\beta&0&-\alpha&-\beta&0&\frac{1}{4}&-\frac{1}{2}&0&0&\frac{\sqrt{15}}{12}
\\
\frac{\sqrt{3}}{6}&0&-\alpha&-\beta&\beta&0&-\alpha&-\frac{\phi}{4}&0&-\frac{1}{2}&0&\gamma 
\\
\frac{\sqrt{3}}{6}&-\beta&0&-\alpha&0&\alpha&\beta&\frac{\phi-1}{4}&0&0&-\frac{1}{2}&\delta
\emat.\label{TA5}
\end{equation}
Заметим, что стандартные системы компьютерной алгебры, такие как  \emph{Maple}
или \emph{Mathematica}, не в состоянии выполнять вычисления с матрицами такого типа
из-за ограниченных возможностей этих систем в упрощении сложных выражений с иррациональностями,
в особенности с вложенными корнями.
\par
Однако если выразить матричные элементы в терминах соответствующего абелевого числового поля,
то проблема упрощения сводится к простым вычислениям с полиномами от одной переменной
по модулю циклотомического полинома. 
В данном случае минимальным полем
 \Math{\NF} является подполе циклотомического поля \Math{\Q_{60}}, фиксируемое
в \Math{\Q_{60}} автоморфизмом Галуа \Math{\runi{}\rightarrow\runi{}^{59}},
где \Math{\runi{}} --- примитивный корень из единицы  60-й степени.
Соответствующий циклотомический полином имеет вид 
\Math{\Phi_{60}\vect{r}=1+r^2-r^6-r^8-r^{10}+r^{14}+r^{16}}.
Иррациональные элементы матрицы \eqref{TA5} на языке циклотомических чисел имеют вид:
\begin{align*}
\frac{\sqrt{3}}{6}&=\frac{1}{6}\vect{2\runi{}^5-\runi{}^{15}},
\\
\frac{\phi}{4}\sqrt{10-2\sqrt{5}}&=
\frac{1}{2}\vect{\runi{}+\runi{}^3-\runi{}^9-\runi{}^{11}+\runi{}^{15}},
\\
\frac{\sqrt{5}\sqrt{10-2\sqrt{5}}}{20}&=
\frac{1}{2}\vect{\runi{}+4\runi{}^3-3\runi{}^9-\runi{}^{11}+2\runi{}^{15}},
\\
-\frac{\phi}{4}&=\frac{1}{4}\vect{-\runi{}^4-\runi{}^6+\runi{}^{14}},
\\
\frac{\phi-1}{4}&=\frac{1}{4}\vect{-1+\runi{}^4+\runi{}^6-\runi{}^{14}},
\\
\frac{\sqrt{15}}{12}&=\frac{1}{12}\vect{-2\runi{}+2\runi{}^5+4\runi{}^7
+2\runi{}^9+2\runi{}^{11}-4\runi{}^{13}-3\runi{}^{15}},
\\
\frac{\sqrt{3}}{8}\vect{1-\frac{\sqrt{5}}{3}}&=
\frac{1}{12}\vect{\runi{}+2\runi{}^5-2\runi{}^7-\runi{}^9
-\runi{}^{11}+2\runi{}^{13}},
\end{align*}
\begin{align*}
-\frac{\sqrt{3}}{8}\vect{1+\frac{\sqrt{5}}{3}}&=
\frac{1}{12}\vect{\runi{}-4\runi{}^5-2\runi{}^7-\runi{}^9
-\runi{}^{11}+2\runi{}^{13}+3\runi{}^{15}}.
\end{align*}
\par
Скалярные произведения в инвариантных подпространствах представления 
\eqref{decoico} в терминах перестановочных инвариантов принимают вид:
\begin{align}
\displaystyle\inner{\Phi_{\IrrRep{1}}}{\Psi_{\IrrRep{1}}}&=
\frac{1}{12}\invarL{12}{m}\invarL{12}{n},\nonumber
\\
\displaystyle\inner{\Phi_{\IrrRep{3}}}{\Psi_{\IrrRep{3}}}&=
\frac{1}{20}\vect{5\invarQ{12}{m}{n}-5\invar{A}{m}{n}
+\sqrt{5}\vect{\invar{B}{m}{n}-\invar{C}{m}{n}}},\label{inn3}
\\
\displaystyle\inner{\Phi_{\IrrRep{3'}}}{\Psi_{\IrrRep{3'}}}&=
\frac{1}{20}\vect{5\invarQ{12}{m}{n}-5\invar{A}{m}{n}
-\sqrt{5}\vect{\invar{B}{m}{n}-\invar{C}{m}{n}}},\label{inn3'}
\\
\displaystyle\inner{\Phi_{\IrrRep{5}}}{\Psi_{\IrrRep{5}}}&=
\frac{1}{12}\vect{5\invarQ{12}{m}{n}+5\invar{A}{m}{n}
-\invar{B}{m}{n}-\invar{C}{m}{n}},\nonumber
\end{align}
где
\begin{align}
\displaystyle\invar{A}{m}{n}&
=\invar{A}{n}{m}=\sum\limits_{k=1}^{12}m_kn_{\Opp{k}},\label{Amn}\\
\displaystyle\invar{B}{m}{n}&
=\invar{B}{n}{m}=\sum\limits_{k=1}^{12}m_k\!\!\sum\limits_{i\in\Nei{k}}\!\!n_{i},\label{Bmn}\\
\displaystyle\invar{C}{m}{n}&
=\invar{C}{n}{m}=\sum\limits_{k=1}^{12}m_k\!\!\!\sum\limits_{i\in\Nei{\Opp{k}}}\!\!\!\!\!n_{i}.
\label{Cmn}
\end{align}
В формулах \eqref{Bmn} и \eqref{Cmn}  \Math{\Nei{k}} обозначает ``окрестность'' 
\Math{k}-й вершины икосаэдра, т.е. множество вершин, смежных с \Math{k}. 
Например,  \Math{\Nei{1}=\set{2,3,4,5,6}} на Рис. \ref{ico}.
Таким образом, внутренняя сумма в \eqref{Bmn} представляет собой сумму данных
на окрестности вершины \Math{k}, а аналогичная сумма в  \eqref{Cmn}
это сумма данных на окрестности вершины, противоположной к \Math{k}. 
\par
Квадратичные инварианты \eqref{Amn}--\eqref{Cmn} не независимы.
Между ними имеется следующее тождество:
\begin{equation}
	\displaystyle\invar{A}{m}{n}+\invar{B}{m}{n}+\invar{C}{m}{n}+\invarQ{12}{m}{n}
	=\invarL{12}{m}\invarL{12}{n}.
\label{A5formidentity}	
\end{equation}
\par
Скалярные произведения \eqref{inn3} и  \eqref{inn3'}, если их рассматривать по 
отдельности, приводят к упомянутой выше проблеме с вероятностями.
А именно, борновские вероятности для подпредставлений \Math{\IrrRep{3}} и 
\Math{\IrrRep{3'}} содержат иррациональности, что противоречит частотной интерпретации
вероятностей для конечных множеств. Очевидно, это является следствием импримитивности:
невозможно переместить вершину икосаэдра без одновременного перемещения 
противоположной вершины. Чтобы разрешить это противоречие необходимо рассматривать
комплементарные подпредставления \Math{\IrrRep{3}} и \Math{\IrrRep{3'}} совместно.
Скалярное произведение в подпредставлении \Math{\IrrRep{3}\oplus\IrrRep{3'}}
\begin{equation*}
\displaystyle\inner{\Phi_{\IrrRep{3\oplus3'}}}{\Psi_{\IrrRep{3\oplus3'}}}=
\frac{1}{2}\vect{\invarQ{12}{m}{n}-\invar{A}{m}{n}}	
\end{equation*}
всегда даёт рациональные борновские вероятности для векторов с натуральными
``числами заполнения''.   
\section{Перестановочно инвариантные билинейные формы}
Для вычисления квантовых интерференций 
натуральных амплитуд в соответствии с правилом Борна достаточно знать выражения для скалярных произведений 
в инвариантных подпространствах перестановочного представления. 
Ранее мы получали эти выражения прямым вычислением
с помощью матрицы перехода от ``перестановочного'' базиса к ``квантовому''.
Как было отмечено в сноске на стр. \pageref{MeatAxePage} 
проблема расщепления модуля над  ассоциативной алгеброй на неприводимые подмодули и, в частности, вычисление соответствующей матрицы перехода не имеет
удовлетворительного алгоритмического решения над важными для нас системами чисел.
Однако можно попытаться обойти эти трудности, если удастся получить выражения
для скалярных произведений непосредственно, т.е. без использования матриц перехода.
Опишем один из возможных подходов.
\subsection{Базис перестановочно инвариантных форм}
Для начала построим множество всех инвариантных билинейных форм в
перестановочном базисе. Предположим, что группа \Math{\wG\equiv\wG\vect{\wS}}, действующая 
перестановками на множестве \Math{\wS=\set{\ws_1,\ldots,\ws_{\wSN}}}
 порождается \Math{K}
элементами \Math{g_1,\ldots,g_k,\ldots,g_K}.
Ясно, что достаточно проверить инвариантность формы
относительно этих элементов.
Для удобства множество  \Math{\wS} отождествим канонически с множеством 
индексов его элементов: \Math{\wS=\onetonset{\wSN}=\set{1,\ldots,{\wSN}}}.
 Обозначим матрицы перестановочного представления
порождающих элементов символами \Math{P_k=\regrep\vect{g_k}}.
Очевидно, что \Math{\transpose{P_k\!}=\regrep\vect{g_k^{-1}}}.
Условие инвариантности билинейной формы \Math{A=\Mone{a_{ij}}} 
относительно группы \Math{\wG} 
можно записать в виде системы матричных уравнений
\begin{equation*}
A = P_kA\transpose{P_k\!},~~1\leq{}k\leq{}K.
\end{equation*}
Легко проверить, что в компонентах уравнения этой системы
имеют вид
\begin{equation}
a_{ij} = a_{ig_kjg_k}.
\label{invaracomp}
\end{equation}
Таким образом, базис всех инвариантных билинейных форм находится
во взаимно однозначном соответствии с множеством орбит 
\Math{\Delta_1,\ldots,\Delta_\baseformN}
действия группы  
\Math{\wG} на декартовом произведении \Math{\wS\times\wS}.
Орбиты группы на произведении \Math{\wS\times\wS}, называемые
\emph{орбиталами}%
\footnote{Орбиталы представляют собой частный случай \emph{когерентных конфигураций} (см., например, \cite{Cameron}, Глава 3). А именно, они являются  \emph{когерентными конфигурациями Шура}.}, 
играют важную роль в теории групп перестановок
и их представлений \cite{Cameron,Dixon}. 
Если группа транзитивна, то существует единственный орбитал, 
состоящий из всех пар вида \Math{\vect{i,i}}. 
Такой орбитал называется
\emph{тривиальным} или \emph{диагональным}.
Число орбиталов 
\Math{\baseformN} называется \emph{рангом} группы \Math{\wG}.
Каждому орбиталу \Math{\Delta_r} можно поставить в соответствие 
\begin{enumerate}
	\item 
ориентированный граф, вершинами которого являются элементы \Math{\wS}, 
а рёбрами ---
пары \Math{\vect{i,j}\in\Delta_r}, 
	\item 
матрицу \Math{A\vect{\Delta_r}} размера  \Math{\wSN\times\wSN} по правилу
\Math{
A\vect{\Delta_r}_{ij} =
\begin{cases}
1, & \text{если~} \vect{i,j}\in\Delta_r,\\
0, & \text{если~} \vect{i,j}\notin\Delta_r.
\end{cases}
}
\end{enumerate}
\par
Свойства графов орбиталов отражают важные свойства групп.
Например, транзитивная группа перестановок примитивна тогда и только тогда,
когда графы всех её нетривиальных орбиталов связны.
\par 
Множество матриц орбиталов представляет собой базис \emph{централизаторного кольца} перестановочного 
представления группы \Math{\wG}. Это кольцо играет важную роль в теории представлений групп. 
Мы будем обозначать его символом \Math{\ZR\equiv\ZR\vect{\wG\vect{\wS}}}. 
Таблица умножения базисных элементов централизаторного кольца  \Math{\ZR} имеет вид
\begin{equation}
	A\vect{\Delta_p}A\vect{\Delta_q}=\sum\limits_{r=1}^{\baseformN}\alpha^{r}_{pq}A\vect{\Delta_r},
	\label{ring}
\end{equation}
где все структурные константы \Math{\alpha^{r}_{pq}} --- натуральные числа, удовлетворяющие неравенствам \Math{0\leq{}\alpha^{r}_{pq}\leq{}\wSN}.
\par
Мы будем называть матрицы орбиталов \Math{\baseform{r}\equiv{A}\vect{\Delta_r}} 
(или, если угодно, матрицы подорбит) \emph{базисными формами},
поскольку любая перестановочно инвариантная билинейная форма 
может быть представлена их линейной комбинацией.
\par
Алгоритм вычисления базисных форм сводится к построению орбиталов в соответствии с формулой \eqref{invaracomp}. 
В нескольких словах, он сканирует элементы множества \Math{\wS\times\wS} в некотором, скажем, лексикографическом, порядке и распределяет эти элементы по классам эквивалентности.
Результом работы алгоритма является полный базис перестановочно инвариантных билинейных форм \Math{\baseform{1},\ldots,\baseform{\baseformN}}.
Алгоритм весьма прост. Наша реализация представляет собой нескольких строк на языке Си.
\par
Непосредственно из построения следует тождество
\begin{equation}
	\baseform{1}+\baseform{2}+\cdots+\baseform{\baseformN}=\transpose{L}L=
	\matones_{\wSN}\equiv
	\bmat
	1&1&\cdots&1\\
	1&1&\cdots&1\\
	\vdots&\vdots&\vdots&\vdots\\
	1&1&\cdots&1
	\emat\enspace,
	\label{aformidentity}
\end{equation}
где \Math{L} --- ковектор вида \Math{\overbrace{\vect{1,1,\ldots,1}}^{\wSN}},
а \Math{\matones_{\wSN}} --- \emph{``матрица из единиц''} размера
\Math{\wSN\times\wSN}.
Частным случаем тождества \eqref{aformidentity} является формула \eqref{A5formidentity}.
\par
Проиллюстрируем работу алгоритма на примере группы \Math{\CyclG{3}}, 
действующей на множестве \Math{\wS=\set{1,2,3}}. Группа порождается одним
элементом, например, \Math{g_1=\vect{1,2,3}}. Нам необходимо распределить
множество пар индексов
\begin{equation}
\begin{matrix}
\vect{1,1}&\vect{1,2}&\vect{1,3}\\
\vect{2,1}&\vect{2,2}&\vect{2,3}\\
\vect{3,1}&\vect{3,2}&\vect{3,3}
\end{matrix}
\label{(i,j)tab}	       
\end{equation}
по классам эквивалентности в соответствии с формулой \eqref{invaracomp}.
Если начать построение с левого верхнего угла таблицы \eqref{(i,j)tab} и искать
необработанные пары в лексикографическом порядке, то орбиталы 
будут построены в следующем порядке: 
\Mathh{\Delta_1=\set{\vect{1,1},\vect{2,2},\vect{3,3}}, 
\Delta_2=\set{\vect{1,2},\vect{2,3},\vect{3,1}}, 
\Delta_3=\set{\vect{1,3},\vect{2,1},\vect{3,2}}.}
Соответствующий полный базис инвариантных форм имеет вид
\begin{equation*}
\baseform{1} =
\Mthree{1}{\cdot}{\cdot}
	     {\cdot}{1}{\cdot}
	     {\cdot}{\cdot}{1},~~
\baseform{2} =
\Mthree{\cdot}{1}{\cdot}
	     {\cdot}{\cdot}{1}
	     {1}{\cdot}{\cdot},~~
\baseform{3} =
\Mthree{\cdot}{\cdot}{1}
	     {1}{\cdot}{\cdot}
	     {\cdot}{1}{\cdot}.
\end{equation*}
Заметим, что в общем случае, если мы условимся начинать работу алгоритма с пары
\Math{\vect{1,1}} и группа \Math{\wG} действует на \Math{\wS} транзитивно, то
первой базисной формой всегда будет единичная матрица \Math{\baseform{1} = \idmat_{\wSN}}, 
соответствующая тривиальному орбиталу. 
\par
Вводя коммутаторы \Math{\comm{\baseform{p}}{\baseform{q}}=\baseform{p}\baseform{q}-\baseform{q}\baseform{p}}, можно построить алгебру Ли, ассоциированную с централизаторным кольцом.
Таблица коммутаторов этой алгебры имеет вид
\begin{equation}
	\comm{\baseform{p}}{\baseform{q}}=\sum\limits_{r=1}^{\baseformN}\gamma^{r}_{pq}\baseform{r},
\label{Lie}
\end{equation}
где структурные константы \Math{\gamma^{r}_{pq}=\alpha^{r}_{pq}-\alpha^{r}_{qp}} представляют собой целые числа из интервала \Math{\left[-\wSN,\wSN\right]}; \Math{\alpha^{r}_{pq}} --- структурные константы из \eqref{ring}.
Алгебра Ли более удобна для наших последующих рассуждений.
\subsection{Связь перестановочно инвариантных форм с разложением на неприводимые 
компоненты}
Несмотря на простоту получения перестановочно инвариантные формы достаточно информативны.
Рассмотрим разложение перестановочного представления на неприводимые компоненты с помощью 
матрицы преобразования \Math{\transmatr}. 
Это разложение для транзитивных групп перестановок имеет вид
\begin{equation}
    \transmatr^{-1}\regrep(g)\transmatr=
    \bmat
       \IrrRep{1} &&&&&
    \\[5pt]
    &
    \hspace*{-8pt}
    \idmat_{m_2}\otimes\repq_2(g)
    &&&&
    \\
    &&
    \hspace*{-45pt}
    \ddots&&&\\
    &&&
    \hspace*{-38pt}
    \idmat_{m_{k}}\otimes\repq_{k}(g)&&\\
    &&&&
    \hspace*{-40pt}
    \ddots&\\
    &&&&&
    \hspace*{-45pt}
    \idmat_{m_{\irrepsN}}\otimes\repq_{\irrepsN}(g)\\
    \emat,~~~~g\in\wG\enspace.
		\label{repdecomp}
\end{equation}
Здесь \Math{\otimes} означает \emph{кронекерово произведение} матриц,  \Math{\irrepsN} полное число различных неприводимых представлений
\Math{\repq_k} \Math{\vect{\repq_1\equiv\IrrRep{1}}}
группы \Math{\wG}, входящих в перестановочное 
представление \Math{\regrep}, \Math{m_k} --- кратность подпредставления
\Math{\repq_k} в представлении \Math{\regrep}, \Math{\idmat_m} --- единичная матрица размера \Math{m\times{}m}.
\par
Наиболее общую перестановочно инвариантную билинейную форму можно записать в виде линейной комбинации базисных форм
\begin{equation}
A=\baseformcoegen{1}\baseform{1}+\baseformcoegen{2}\baseform{2}
+\cdots+\baseformcoegen{\baseformN}\baseform{\baseformN},
\label{genAform}	
\end{equation}
где коэффициенты \Math{\baseformcoegen{i}} являются элементами некоторого
абелева числового поля \Math{\NF}. Конкретно это поле определяется в ходе
вычислений, которые будут описаны ниже.
\par
Легко показать (см. \cite{Cameron,Wielandt}), что в системе координат, расщепляющей перестановочное представление на 
неприводимые компоненты, форма \eqref{genAform} будет иметь вид
\begin{equation}
\transmatr^{-1}A\transmatr=
\bmat
\bornform{1}&&&&&\\[3pt]
&\hspace*{-15pt}\bornform{2}\otimes\idmat_{\dimirr_2}&&&&\\
&&\hspace*{-15pt}\ddots&&&\\
&&&\hspace*{-15pt}\bornform{k}\otimes\idmat_{\dimirr_k}&&\\
&&&&\hspace*{-15pt}\ddots&\\
&&&&&\hspace*{-23pt}\bornform{\irrepsN}\otimes\idmat_{\dimirr_{\irrepsN}}
\emat~.
\label{genAformdecomp}	
\end{equation}
Здесь
\Math{\bornform{k}} --- матрица размера \Math{m_k\times{}m_k} элементы которой
представляют собой линейные комбинации коэффициентов \Math{\baseformcoegen{i}} из
формулы \eqref{genAform}, а \Math{m_k} --- кратность неприводимой компоненты \Math{\repq_k};
\Math{\idmat_{\dimirr_k}} --- единичная матрица размера \Math{\dimirr_k\times\dimirr_k}, 
где \Math{\dimirr_k} --- размерность \Math{\repq_k}.
Из структуры матрицы \eqref{genAformdecomp} видно, что ранг \Math{\rankG} группы 
(т. е. размерность централизаторного кольца) равен сумме квадратов кратностей: 
\Math{~\rankG=1+m_2^2+\cdots+m_{\irrepsN}^2}.
\par
Рассмотрим теперь определитель  \Math{\det\vect{A}}. 
В терминах переменных \Math{\baseformcoegen{1},\ldots,\baseformcoegen{\baseformN}}
этот определитель представляет собой однородный полином степени \Math{\wSN}.
Поскольку определитель формы не зависят от выбора системы координат,
а определитель блочно диагональной матрицы равен произведению определителей
её блоков, то из разложения \eqref{genAformdecomp} следует
\begin{equation*}
\det\vect{A}=\det\vect{\bornform{1}}\det\vect{\bornform{2}}^{\dimirr_2}
\cdots\det\vect{\bornform{k}}^{\dimirr_k}\cdots
\det\vect{\bornform{\irrepsN}}^{\dimirr_{\irrepsN}}.
\end{equation*}
Здесь мы воспользовались тождеством 
\Math{\det\vect{X\otimes{}Y} = \det\vect{X}^m\det\vect{Y}^n} 
для кронекерова произведения \Math{n\times{}n} матрицы \Math{X} 
на \Math{m\times{}m} матрицу \Math{Y}.
\par
Ясно, что \Math{\det\vect{\bornform{k}}} представляет собой однородный полином
степени \Math{m_k} от переменных \Math{\baseformcoegen{1},\ldots,\baseformcoegen{\baseformN}}:
\Math{~\det\vect{\bornform{k}}=\efacdetA_k\vect{\baseformcoegen{1},\ldots,\baseformcoegen{\baseformN}}.}
Таким образом, мы имеем следующее
\par 
\textbf{Утверждение}.
\emph{Определитель линейной комбинации базисных форм следующим образом разлагается
 на множители
над некоторым кольцом циклотомических целых чисел
\begin{equation}
	\det{\sum\limits_{i=1}^{\baseformN}\baseformcoegen{i}\baseform{i}}
	=\prod\limits_{k=1}^{\irrepsN}\efacdetA_k\vect{\baseformcoegen{1},\ldots,
	\baseformcoegen{\baseformN}}^{\dimirr_k},
	~~\deg{\efacdetA_k\vect{\baseformcoegen{1},\ldots,
	\baseformcoegen{\baseformN}}} = m_k.
\label{hypodet}	
\end{equation}
}
Здесь \Math{\irrepsN} --- число различных неприводимых 
представлений, входящих в состав перестановочного;
\Math{\efacdetA_k} --- неприводимый полином, соответствующий 
\Math{k}-му неприводимому представлению размерности \Math{\dimirr_k},
входящему в перестановочное представление с кратностью \Math{m_k}.
\par
Это утверждение очень похоже на \emph{теорему Фробениуса об определителе}%
\footnote{Эта теорема, открытая Дедекиндом и доказанная Фробениусом, положила начало теории теории представлений групп.
История предмета изложена, например, в  \cite{Lam}.} которую можно получить следующей модификацией приведённой выше процедуры.
Для группы \Math{\wG=\set{\wg_1,\ldots,\wg_{\wGN}}} введём линейную комбинацию
\begin{equation}
	\regrep = x_1\regrep\vect{\wg_1}+x_2\regrep\vect{\wg_2}+\cdots+x_{\wGN}\regrep\vect{\wg_{\wGN}},
	\label{gengalg}
\end{equation}
где \Math{x_1,\ldots,x_{\wGN}} --- коммутирующие переменные.
Из структуры разложения \eqref{repdecomp} следует
\begin{equation}
	\det{\sum\limits_{i=1}^{\wGN}x_i\regrep\vect{\wg_i}}
	=\prod\limits_{k=1}^{N_{\chi}}F_k\vect{x_1,\ldots,x_{\wGN}}^{m_k},
	~~\deg{F_k\vect{x_1,\ldots,x_{\wGN}}} = \dimirr_k. 
\label{galgdet}	
\end{equation}
Теорема Фробениуса получается из \eqref{galgdet}, если в качестве \Math{\regrep} взять \emph{регулярное представление} группы \Math{\wG}.
В этом случае \Math{\dimirr_k=m_k} и \Math{\irrepsN=N_{\chi}}, где \Math{N_{\chi}} --- число неприводимых характеров (равное числу классов сопряжённых элементов) группы \Math{\wG}.
Мы видим, что факторизации  \eqref{hypodet} и  \eqref{galgdet} в определённом смысле дуальны: полином \Math{\efacdetA_k\vect{\baseformcoegen{1},\ldots,\baseformcoegen{\baseformN}}} возведён в {степень} \Math{\dimirr_k} в \eqref{hypodet},
а в \eqref{galgdet} число \Math{\dimirr_k} представляет собой {степень} полинома \Math{F_k\vect{x_1,\ldots,x_{\wGN}}}, в то время, как полином \Math{F_k\vect{x_1,\ldots,x_{\wGN}}} возведён в {степень} \Math{m_k}  в \eqref{galgdet} и \Math{m_k} --- {степень} полинома  \Math{\efacdetA_k\vect{\baseformcoegen{1},\ldots,\baseformcoegen{\baseformN}}} в \eqref{hypodet}.
\par
Сравним \eqref{hypodet} и \eqref{galgdet} с вычислительной точки зрения.
Преимущество \eqref{hypodet} заключается в том, что число переменных \Math{\baseformcoegen{1},\ldots,\baseformcoegen{\rankG}} обычно существенно меньше чем число переменных \Math{x_1,\ldots,x_{\wGN}}:
\Math{\vect{\rankG=\sum\limits_{k=1}^{\irrepsN}m_k^2}\leq\vect{\wGN=\sum\limits_{k=1}^{N_{\chi}}\dimirr_k^2}},
поскольку \Math{\irrepsN\leq{}N_{\chi}} и \Math{m_k\leq{}\dimirr_k}. Например, для действия икосаэдральной группы \Math{\AltG{5}} на вершинах икосаэдра мы имеем \Math{\rankG=4} и \Math{\wGN=60}.
\par
Кроме того, полиномы \Math{F_k\vect{x_1,\ldots,x_{\wGN}}} в  \eqref{galgdet}  нелинейны для нетривиальных размерностей неприводимых компонент и не ясно существует ли простой способ справиться с этой нелинейностью.
В случае факторизации \eqref{hypodet} мы можем свести задачу с нелинейными полиномами от \Math{\rankG} переменных к задаче с линейными полиномами от \Math{\sum\limits_{k=1}^{\irrepsN}m_k} переменных с помощью подхода, основанного на 
``огрублении'' соответствующих когерентных конфигураций, как это будет описано ниже.
\par
Из  \eqref{hypodet} следует идея алгоритма вычисления инвариантных форм в неприводимых
подпространствах перестановочного представления.
\par 
Вначале необходимо вычислить полином  \Math{\det\vect{A}}, что представляет собой 
алгоритмически относительно несложную задачу. Стандартные алгоритмы вычисления определителей  
имеют кубическую (или несколько меньшую) сложность по размеру матрицы.
\par
Затем необходимо разложить полином  \Math{\det\vect{A}} на максимальное число неприводимых компонент над кольцом циклотомических целых,
кондуктор которого является некоторым делителем экспоненты группы.
Как хорошо известно, практические алгоритмы факторизации полиномов представляют собой алгоритмы типа Лас Вегас%
\footnote{Алгоритмом Лас Вегаса называется \cite{Babai} вероятностный алгоритм который либо выдаёт правильный результат либо сообщает о неудаче (в отличие от алгоритмов Монте Карло, которые могут выдавать ошибочные результаты), вероятность которой \Math{\leq1/2}.
Вероятность неудачи снижается до величины \Math{\leq2^{-t}} после \Math{t} прогонов алгоритма.} с полиномиальной временн\'{о}й сложностью. 
\par
Таким образом, построение разложения \eqref{hypodet} является алгоритмически реализуемой задачей.
Решив эту задачу мы имеем полную информацию о размерностях и кратностях всех неприводимых 
подпредставлений. 
Это даёт определённую конкретизацию структуры разложения перестановочного представления
на неприводимые компоненты.
\par
Следующий естественный шаг --- явное вычисление  
инвариантных скалярных произведений  \Math{\bornform{k}} в неприводимых подпространствах
перестановочного представления. Для этого необходимо исключить из рассмотрения множитель
\Math{\efacdetA_k\vect{\baseformcoegen{1},\ldots,\baseformcoegen{\baseformN}}}, 
относящийся к компоненте \Math{\bornform{k}} и приравнять к нулю остальные множители,
т. е. необходимо записать систему уравнений
\begin{equation}
\efacdetA_1
 =
\cdots = 
\widehat{\efacdetA_k} =
\cdots = 
\efacdetA_{\irrepsN}
 = 0.
\label{EkEquations}	
\end{equation}
\subsubsection{Однократные неприводимые компоненты.}
\label{subsectmult1}
Если все кратности \Math{m_i=1} (при соблюдении этого условия \Math{\rankG=\irrepsN} и централизаторное
кольцо коммутативно), то все полиномы \Math{\efacdetA_i} линейны.
В этом случае вычисление скалярных произведений в инвариантных подпространствах без труда может быть
доведено до конца, поскольку сводится к решению системы линейных уравнений.
\par 
Рассмотрим для примера группу \Math{\SL{2,3}}, определяемую как группа 
специальных линейных преобразований двумерного пространства над полем из
трёх элементов \Math{\F_3}.
Эта группа используется в физике элементарных частиц, где её часто
обозначают символом \Math{\mathrm{T}'} поскольку она является двухкратной 
накрывающей группы симметрий тетраэдра, сохраняющих ориентацию \Math{\mathrm{T}\cong\AltG{4}}.
Мы рассмотрим её точное перестановочное действие степени 8, которое можно породить, например, следующими
двумя перестановками
\Mathh{g_1 = \vect{1,5,3,2,6,4}\vect{7,8}\text{~~и~~}g_2 = \vect{1,3,7,2,4,8}\vect{5,6}.}
Перестановочное представление размерности 8 мы будем обозначать символом \Math{\PermRep{8}}.
\par
Следующие четыре матрицы, полученные вычислением орбиталов, составляют базис кольца 
\Math{\ZR\equiv\ZR\vect{\SL{2,3}\vect{\onetonset{8}}}} перестановочно инвариантных форм
\begin{equation*}
		\baseform{1} = 
	\bmat
	1&\cdot&\cdot&\cdot&\cdot&\cdot&\cdot&\cdot\\
	\cdot&1&\cdot&\cdot&\cdot&\cdot&\cdot&\cdot\\
	\cdot&\cdot&1&\cdot&\cdot&\cdot&\cdot&\cdot\\
	\cdot&\cdot&\cdot&1&\cdot&\cdot&\cdot&\cdot\\
	\cdot&\cdot&\cdot&\cdot&1&\cdot&\cdot&\cdot\\
	\cdot&\cdot&\cdot&\cdot&\cdot&1&\cdot&\cdot\\
	\cdot&\cdot&\cdot&\cdot&\cdot&\cdot&1&\cdot\\
	\cdot&\cdot&\cdot&\cdot&\cdot&\cdot&\cdot&1
	\emat,~~ 
		\baseform{2} =
	\bmat
	\cdot&1&\cdot&\cdot&\cdot&\cdot&\cdot&\cdot\\
	1&\cdot&\cdot&\cdot&\cdot&\cdot&\cdot&\cdot\\
	\cdot&\cdot&\cdot&1&\cdot&\cdot&\cdot&\cdot\\
	\cdot&\cdot&1&\cdot&\cdot&\cdot&\cdot&\cdot\\
	\cdot&\cdot&\cdot&\cdot&\cdot&1&\cdot&\cdot\\
	\cdot&\cdot&\cdot&\cdot&1&\cdot&\cdot&\cdot\\
	\cdot&\cdot&\cdot&\cdot&\cdot&\cdot&\cdot&1\\
	\cdot&\cdot&\cdot&\cdot&\cdot&\cdot&1&\cdot
	\emat,
\end{equation*}
\begin{equation*}
		\baseform{3} = 
	\bmat
	\cdot&\cdot&1&\cdot&1&\cdot&1&\cdot\\
	\cdot&\cdot&\cdot&1&\cdot&1&\cdot&1\\
	\cdot&1&\cdot&\cdot&\cdot&1&1&\cdot\\
	1&\cdot&\cdot&\cdot&1&\cdot&\cdot&1\\
	\cdot&1&1&\cdot&\cdot&\cdot&\cdot&1\\
	1&\cdot&\cdot&1&\cdot&\cdot&1&\cdot\\
	\cdot&1&\cdot&1&1&\cdot&\cdot&\cdot\\
	1&\cdot&1&\cdot&\cdot&1&\cdot&\cdot
	\emat,~~
		\baseform{4} =
	\bmat
	\cdot&\cdot&\cdot&1&\cdot&1&\cdot&1\\
	\cdot&\cdot&1&\cdot&1&\cdot&1&\cdot\\
	1&\cdot&\cdot&\cdot&1&\cdot&\cdot&1\\
	\cdot&1&\cdot&\cdot&\cdot&1&1&\cdot\\
	1&\cdot&\cdot&1&\cdot&\cdot&1&\cdot\\
	\cdot&1&1&\cdot&\cdot&\cdot&\cdot&1\\
	1&\cdot&1&\cdot&\cdot&1&\cdot&\cdot\\
	\cdot&1&\cdot&1&1&\cdot&\cdot&\cdot
	\emat.
\end{equation*}
Экспонента группы \Math{\SL{2,3}} равна 12. Тем не менее, кольца циклотомических целых \Math{\N_3} достаточно для ``\emph{абсолютной факторизации}''
определителя линейной комбинации
\Math
{A=\baseformcoegen{1}\baseform{1}+\baseformcoegen{2}\baseform{2}
+\baseformcoegen{3}\baseform{3}+\baseformcoegen{4}\baseform{4}}:
\begin{align}
\det{A} =& 	
  \vect{\baseformcoegen{1}+\baseformcoegen{2}
	     +3\baseformcoegen{3}+3\baseformcoegen{4}}
	    \nonumber\\
	&\set{\baseformcoegen{1}-\baseformcoegen{2}
	     +\vect{1+2\runi{}}\baseformcoegen{3}
	     -\vect{1+2\runi{}}\baseformcoegen{4}}^2
	     \nonumber\\
	&\set{\baseformcoegen{1}-\baseformcoegen{2}
	     -\vect{1+2\runi{}}\baseformcoegen{3}
	     +\vect{1+2\runi{}}\baseformcoegen{4}}^2
	     \label{decoSL23}\\
	&\vect{\baseformcoegen{1}+\baseformcoegen{2}
	     -\baseformcoegen{3}-\baseformcoegen{4}}^3,\nonumber
\end{align}
где \Math{\runi{}} --- примитивный корень из единицы 3-й степени.
Из \eqref{decoSL23}  непосредственно видна структура разложения перестановочного представления \Math{\PermRep{8}} на неприводимые компоненты:
\Mathh{\PermRep{8}=
\IrrRep{1}\oplus\IrrRep{2}\oplus\IrrRep{2'}\oplus\IrrRep{3}.}
Исключая линейные множители, соответствующие 
последовательно представлениям 
\Math{k = \IrrRep{1}, \IrrRep{2}, \IrrRep{2'}, \IrrRep{3}},
и приравнивая к нулю остальные множители мы получим четыре
системы из трёх линейных уравнений от четырёх переменных.
\par
Рассмотрим для примера подпредставление \Math{\IrrRep{2}}. 
Для этой компоненты система уравнений \eqref{EkEquations} принимает вид
\begin{align}
\baseformcoegen{1}+\baseformcoegen{2}+3\baseformcoegen{3}+3\baseformcoegen{4}&=0,
\label{eqex1}\\
-\baseformcoegen{1}+\baseformcoegen{2}+\vect{1+2\runi{}}\baseformcoegen{3}
-\vect{1+2\runi{}}\baseformcoegen{4}&=0,\\
\baseformcoegen{1}+\baseformcoegen{2}-\baseformcoegen{3}-\baseformcoegen{4}&=0.
\label{eqex3}	
\end{align}
\par
Линейные системы такого типа состоят из \Math{\rankG-1} уравнений, но содержат  
\Math{\rankG} переменных.
Поскольку билинейная форма \eqref{genAform} описывает невырожденное скалярное произведение,
коэффициент \Math{\baseformcoegen{1}} при диагональной базисной форме не может обращаться в нуль.
Поэтому \Math{\baseformcoegen{1}} можно рассматривать как параметр и решать систему уравнений 
относительно остальных переменных.
В принципе, коэффициент \Math{\baseformcoegen{1}} может быть произвольным (ненулевым) параметром.
Для удобства мы всегда будем предполагать, что \Math{\baseformcoegen{1} = 1}.
На самом деле разумно, для уменьшения числа переменных, установить \Math{\baseformcoegen{1} = 1} перед факторизацией полинома \Math{\det\vect{A}}.
\par 
Решая линейную систему (\ref{eqex1}---\ref{eqex3}), установив \Math{\baseformcoegen{1} = 1}, мы получим
\Mathh{\baseformcoegen{2}=-1,~\baseformcoegen{3}=-\frac{1+2\runi{}}{3},~\baseformcoegen{4}=\frac{1+2\runi{}}{3}.}
Для скалярного произведения в  двумерном пространстве подпредставления \Math{\IrrRep{2}} мы имеем следующее выражение
 \Mathh{\bornform{\IrrRep{2}} = C_{\IrrRep{2}}\vect{\baseform{1}-\baseform{2}-\frac{1+2\runi{}}{3}\baseform{3}+\frac{1+2\runi{}}{3}\baseform{4}}.}
Здесь \Math{C_{\IrrRep{2}}} --- произвольный нормирующий коэффициент. 
Борновская вероятность не зависит от его значения.
Тем не менее, разумно выбирать величины
\begin{equation}
	C_{\IrrRep{k}} = \dimirr_k/\wSN
	\label{coea1}
\end{equation}
для каждой неприводимой компоненты.
При такой нормировке сумма скалярных произведений в инвариантных подпространствах будет равна стандартному скалярному произведению в пространстве перестановочного представления. 
\par
Применяя аналогичную процедуру ко всем неприводимым компонентам мы приходим к следующему набору скалярных произведений в инвариантных подпространствах
\begin{align*}
	\bornform{\IrrRep{1}} =& \frac{1}{8}\vect{\baseform{1}+\baseform{2}+\baseform{3}+\baseform{4}}
\equiv\frac{1}{8}\matones_8,\\
	\bornform{\IrrRep{2}} =& \frac{1}{4}\vect{\baseform{1}-\baseform{2}-\frac{1+2\runi{}}{3}\baseform{3}+\frac{1+2\runi{}}{3}\baseform{4}},
	\\
	\bornform{\IrrRep{2'}} =& \frac{1}{4}\vect{\baseform{1}-\baseform{2}+\frac{1+2\runi{}}{3}\baseform{3}-\frac{1+2\runi{}}{3}\baseform{4}},
	\\
	\bornform{\IrrRep{3}} =& \frac{3}{8}\vect{\baseform{1}+\baseform{2}-\frac{1}{3}\baseform{3}
	-\frac{1}{3}\baseform{4}}.
\end{align*}
Легко проверить, что  нормировка \eqref{coea1} обеспечивает тождество
\Math{\bornform{\IrrRep{1}}+\bornform{\IrrRep{2}}+\bornform{\IrrRep{2'}}+\bornform{\IrrRep{3}} = \baseform{1}\equiv\idmat_8.}
Напомним, что во всех подобных задачах \Math{\bornform{\IrrRep{1}}} не требует вычисления, поскольку
скалярное произведение в подпространстве тривиального представления всегда
имеет вид \Math{\frac{1}{\wSN}\matones_\wSN.}

\subsubsection{Нетривиальные кратности подпредставлений.}
\label{subsectmultm}
В случае кратных подпредставлений ситуация существенно усложняется ввиду нелинейности уравнений \eqref{EkEquations}.
Вопрос о том, можно ли разработать общий алгоритм для случая кратных подпредставлений требует более глубокого дополнительного изучения. 
\par
Источником нелинейности является избыточное число параметров, от конкретных значений которых борновские вероятности не зависят.
Из структуры разложения \eqref{genAformdecomp} видно, что каждый блок кратных компонент \Math{\bornform{k}\otimes\idmat_{\dimirr_k}} содержит \Math{m_k\times{}m_k} таких параметров.
Наше соглашение \eqref{coea1} фиксирует \Math{m_k} диагональных элементов матрицы \Math{\bornform{k}}.
Следовательно, нам нужно как-то зафиксировать \Math{m_k^2-m_k} оставшихся параметров.
В этом может помочь изучение структуры централизаторного кольца.
Напомним, что если централизаторное кольцо коммутативно, то все кратности равны единице.
Таким образом, естественной идеей устранения лишних степеней свободы представляется поиск подходящей \emph{коммутативной} подалгебры размерности \Math{1+m_2+\cdots+m_{\irrepsN}} в алгебре Ли размерности \Math{\rankG=1+m_2^2+\cdots+m_{\irrepsN}^2}, определяемой таблицей коммутаторов \eqref{Lie}. 
\par
Опишем кратко простой комбинаторный подход к поиску таких подалгебр.
Этот подход хорошо работает на всех примерах транзитивных групп перестановок, которые нам удалось проверить.
Хотя в настоящее время у нас нет доказательства универсальности этого подхода, его можно использовать, по крайней мере, как эффективный предварительный шаг в решении задач с кратными подпредставлениями.
Базисные матрицы, получаемые из ординалов, представляют собой \Math{\vect{0,1}}-матрицы с непересекающимися множествами единичных элементов. Сложение таких матриц приводит к матрице этого же типа.
В терминах когерентных конфигураций такая процедура называется \emph{огрублением}. 
Таким образом, представляется естественной идея заменить базис \Math{\baseform{1},\ldots,\baseform{\baseformN}} меньшим множеством взаимно коммутирующих \Math{\vect{0,1}}-матриц, содержащим некоторые суммы матриц исходного базиса.
\par
В качестве иллюстрации рассмотрим группу Коксетера типа \Math{A_2}, являющуюся также группой Вейля группы Ли, например, \Math{\SU{3}}.
Мы рассмотрим натуральное действие  \Math{A_2} на её системе корней \Math{\wS = \set{1,2,\cdots,6}}.
Векторы этой корневой системы можно представить на плоскости вершинами правильного шестиугольника.
Натуральное действие порождается, например, перестановками \Mathh{g_1=(1,4)(2,3)(5,6) \text{~~и~~} g_2=(1,3)(2,5)(4,6)}.
\par
Вычисление орбиталов приводит к следующему базису централизаторного кольца
\begin{align*}
		&	\baseform{1} = 
		\bmat
		1&\cdot&\cdot&\cdot&\cdot&\cdot\\
		\cdot&1&\cdot&\cdot&\cdot&\cdot\\
		\cdot&\cdot&1&\cdot&\cdot&\cdot\\
		\cdot&\cdot&\cdot&1&\cdot&\cdot\\
		\cdot&\cdot&\cdot&\cdot&1&\cdot\\
		\cdot&\cdot&\cdot&\cdot&\cdot&1
		\emat, 
			\baseform{2} =
		\bmat
		\cdot&1&\cdot&\cdot&\cdot&\cdot\\
		\cdot&\cdot&\cdot&\cdot&\cdot&1\\
		\cdot&\cdot&\cdot&\cdot&1&\cdot\\
		\cdot&\cdot&1&\cdot&\cdot&\cdot\\
		\cdot&\cdot&\cdot&1&\cdot&\cdot\\
		1&\cdot&\cdot&\cdot&\cdot&\cdot
		\emat,
			\baseform{3} =
		\bmat
		\cdot&\cdot&1&\cdot&\cdot&\cdot\\
		\cdot&\cdot&\cdot&1&\cdot&\cdot\\
		1&\cdot&\cdot&\cdot&\cdot&\cdot\\
		\cdot&1&\cdot&\cdot&\cdot&\cdot\\
		\cdot&\cdot&\cdot&\cdot&\cdot&1\\
		\cdot&\cdot&\cdot&\cdot&1&\cdot
		\emat,
\end{align*}
\begin{align*}
	&
			\baseform{4} = 
		\bmat
		\cdot&\cdot&\cdot&1&\cdot&\cdot\\
		\cdot&\cdot&\cdot&\cdot&1&\cdot\\
		\cdot&\cdot&\cdot&\cdot&\cdot&1\\
		1&\cdot&\cdot&\cdot&\cdot&\cdot\\
		\cdot&1&\cdot&\cdot&\cdot&\cdot\\
		\cdot&\cdot&1&\cdot&\cdot&\cdot
		\emat, 
			\baseform{5} =
		\bmat
		\cdot&\cdot&\cdot&\cdot&1&\cdot\\
		\cdot&\cdot&1&\cdot&\cdot&\cdot\\
		\cdot&1&\cdot&\cdot&\cdot&\cdot\\
		\cdot&\cdot&\cdot&\cdot&\cdot&1\\
		1&\cdot&\cdot&\cdot&\cdot&\cdot\\
		\cdot&\cdot&\cdot&1&\cdot&\cdot
		\emat,
			\baseform{6} =
		\bmat
		\cdot&\cdot&\cdot&\cdot&\cdot&1\\
		1&\cdot&\cdot&\cdot&\cdot&\cdot\\
		\cdot&\cdot&\cdot&1&\cdot&\cdot\\
		\cdot&\cdot&\cdot&\cdot&1&\cdot\\
		\cdot&\cdot&1&\cdot&\cdot&\cdot\\
	\cdot&1&\cdot&\cdot&\cdot&\cdot
	\emat.\nonumber
\end{align*}
Выполняя абсолютную факторизацию%
\footnote{Экспонента группы \Math{A_2\cong\SymG{3}} равна 6. Ввиду изоморфизма колец \Math{\N_6\cong\N_3}, достаточно факторизовать над кольцом циклотомических целых \Math{\N_3}.}
определителя линейной комбинации 
\Mathh
{A=\baseformcoegen{1}\baseform{1}+\baseformcoegen{2}\baseform{2}
+\baseformcoegen{3}\baseform{3}+\baseformcoegen{4}\baseform{4}+\baseformcoegen{5}\baseform{5}+\baseformcoegen{6}\baseform{6},}
мы получаем  разложение
\begin{align*}
\det{A} =& 	
  \vect{\baseformcoegen{1}+\baseformcoegen{2}
	     -\baseformcoegen{3}-\baseformcoegen{4}
	     -\baseformcoegen{5}+\baseformcoegen{6}}
	    \nonumber\\
  &\vect{\baseformcoegen{1}+\baseformcoegen{2}
	     +\baseformcoegen{3}+\baseformcoegen{4}
	     +\baseformcoegen{5}+\baseformcoegen{6}}
	    \nonumber\\
	&\left\{\baseformcoegen{1}^2+\baseformcoegen{2}^2-\baseformcoegen{3}^2
	-\baseformcoegen{4}^2-\baseformcoegen{5}^2+\baseformcoegen{6}^2\right.
	\\
	&\left.~-\baseformcoegen{1}\baseformcoegen{2}
	-\baseformcoegen{1}\baseformcoegen{6}
	-\baseformcoegen{2}\baseformcoegen{6}
	+\baseformcoegen{3}\baseformcoegen{4}+\baseformcoegen{3}\baseformcoegen{5}
	+\baseformcoegen{4}\baseformcoegen{5}\right\}^2,\nonumber
\end{align*}
из которого следует структура перестановочного действия группы Вейля
\Math{A_2} на своей корневой системе в терминах неприводимых компонент:
\Math{\PermRep{6}=
\IrrRep{1}\oplus\IrrRep{1'}\oplus\vect{\IrrRep{2}\oplus\IrrRep{2}}.}
Поскольку мы имеем четыре неприводимые компоненты, а централизаторное кольцо шестимерно,
нам необходимо устранить две лишних степени свободы.
\par
Коммутаторы алгебры Ли имеют вид
\begin{align}
&
\ZRcomm{1}{2}=\ZRcomm{1}{3}=\ZRcomm{1}{4}=\ZRcomm{1}{5}=\ZRcomm{1}{6}=\ZRcomm{2}{6}=0,
\nonumber\\[3pt]
&
\ZRcomm{2}{3}=\ZRcomm{3}{6}=\baseform{4}-\baseform{5},
\label{A2com23}\\
&
\ZRcomm{2}{4}=\ZRcomm{4}{6}=-\baseform{3}+\baseform{5},
\\
&
\ZRcomm{2}{5}=\ZRcomm{5}{6}=\baseform{3}-\baseform{4},
\label{A2com56}\\[3pt]
&
\ZRcomm{3}{4}=-\ZRcomm{3}{5}=\ZRcomm{4}{5}=-\baseform{2}+\baseform{6}.\nonumber
\end{align}
Из \eqref{A2com23}-\eqref{A2com56}  следует, что \Math{\ZRcomm{2}{X}=\ZRcomm{6}{X}=0}, 
где \Math{\baseform{X}=\baseform{3}+\baseform{4}+\baseform{5}}. 
Легко проверить, что матрицы \Math{\baseform{1}, \baseform{2}, \baseform{6}, \baseform{X}}
образуют базис четырёхмерной коммутативной алгебры Ли. 
Теперь определитель матрицы 
\Mathh
{A'=\baseformcoegen{1}\baseform{1}+\baseformcoegen{2}\baseform{2}
+\baseformcoegen{6}\baseform{6}+\baseformcoegen{X}\baseform{X}}
разлагается над кольцом \Math{\N_3} на линейные множители:
\begin{align*}
\det{A'} =& 	
  \vect{\baseformcoegen{1}+\baseformcoegen{2}
	     +\baseformcoegen{6}-3\baseformcoegen{X}}
	    \nonumber\\
  &\vect{\baseformcoegen{1}+\baseformcoegen{2}
	     +\baseformcoegen{6}+3\baseformcoegen{X}}
	    \\
	&\set{\baseformcoegen{1}+\runi{}\baseformcoegen{2}
	     -\vect{1+\runi{}}\baseformcoegen{6}}^2\nonumber\\
&\set{\baseformcoegen{1}-\vect{1+\runi{}}\baseformcoegen{2}
	     +\runi{}\baseformcoegen{6}}^2,\nonumber
\end{align*}
где \Math{\runi{}} --- примитивный корень третьей степени из единицы.
После точно таких же манипуляций, как в разделе \ref{subsectmult1},  мы получаем следующий набор скалярных произведений в инвариантных подпространствах
\begin{align*}
	\bornform{\IrrRep{1}} =& \frac{1}{6}\vect{\baseform{1}+\baseform{2}+\baseform{6}+\baseform{X}}
\equiv\frac{1}{6}\matones_6,\\
	\bornform{\IrrRep{1'}} =& \frac{1}{6}\vect{\baseform{1}+\baseform{2}+\baseform{6}-\baseform{X}},\\
	\bornform{\IrrRep{2}} =& 
\frac{1}{3}\set{\baseform{1}-\vect{1+
	\runi{}}\baseform{2}+\runi{}\baseform{6}},
	\\
	\bornformequi{\IrrRep{2}} =& \frac{1}{3}\set{\baseform{1}+\runi{}\baseform{2}-\vect{1+
	\runi{}}\baseform{6}}.
\end{align*}
Здесь \Math{\bornform{\IrrRep{2}}} и \Math{\bornformequi{\IrrRep{2}}} --- различные координатные представления одной и той же формы, относящиеся к неприводимой компоненте  \Math{\IrrRep{2}}.
\section*{Заключение}
В основе данной статьи лежит представление о том, что любая проблема, имеющая осмысленное
эмпирическое содержание, может быть сформулирована в конструктивных, более того, конечных терминах.
Исключение актуальных бесконечностей из описания физической реальности, освобождая от множества
технических трудностей,  позволяет сконцентрироваться на содержательной стороне физических задач.
Кроме того, конструктивность --- необходимое требование для самой возможности построения компьютерных
моделей физических систем. Мы также придерживались принципа экономного введения новых элементов
в описания, т.е. математические понятия вводились только тогда, когда они были реально необходимы
(конечно, с точностью до нашего понимания проблемы)
--- бесконтрольное введение новых математических структур (или расширение старых просто 
из-за возможности обобщения) затрудняет
отделение содержательных
элементов описания от артефактов. 
\par
Придерживаясь этих позиций мы рассмотрели классические и квантовые динамические системы с нетривиальными
симметрийными свойствами. Особое внимание было уделено одному из центральных принципов в физике ---
принципу калибровочной инвариантности, который мы переформулировали в ``конечном'' виде.
Мы рассмотрели детерминистические динамические системы и обсудили групповую природу 
формирования сохраняющих форму движущихся структур, часто наблюдаемых в динамике таких систем.
Мы обсудили также каким образом в детерминистической динамике может возникнуть унитарность, присущая
фундаментальным законам природы и лежащая в основе квантово-механического поведения.
\par 
Конструктивный анализ квантово-механического поведения приводит к следующим выводам
\begin{enumerate}
\item
Квантовая механика по-существу является \emph{априорной математической схемой}
в основе которой лежит принципиальная невозможность проследить идентичность
неразличимых объектов в процессе эволюции совокупности таких объектов. 
Фактически это раздел
комбинаторики, который можно назвать \emph{``исчислением неразличимых''}.
\item
Любая квантово-механическая проблема может быть сведена к \emph{перестановкам}.
\item
\emph{Квантовые интерференции} это явления, наблюдаемые в инвариантных
 подпространствах перестановочных представлений и выражаемые в терминах 
 \emph{перестановочных инвариантов}.
\item
Натуральная интерпретация \emph{квантовых амплитуд} (\emph{``волн''}) как векторов
кра\-тностей вхождения 
элементов, 
подвергаемых перестановкам, (\emph{``частиц''}) 
в совокупность таких элементов, приводит к 
\emph{рациональным} квантовым вероятностям. 
Это согласуется с \emph{частотной интерпретацией} вероятности для конечных множеств.
\end{enumerate}
Идея натуральных квантовых амплитуд выглядит весьма привлекательной.
В частности, она позволяет интерпретировать элементы множества неразличимых объектов, 
на которых группа симметрий действует перестановками, как ``\emph{частицы}'', а
векторы кратностей вхождения этих объектов в ансамбль как ``\emph{волны}''.
Эта идея ведёт к простой и самосогласованной картине квантового поведения.
В частности, она позволяет ``вывести'' комплексные числа, которые в стандартной квантовой механике постулируются.
Идея, однако, требует проверки.
Если идея верна, то квантовые явления в различных инвариантных подпространствах
представляют собой различные проявления --- видимые в различных 
\emph{``наблюдательных (экспериментальных) установках''} --- единого процесса перестановок
одной и той же совокупности объектов.
Данные, относящиеся к различным инвариантным подпространствам  перестановочного
представления, требуют  интерпретации.
Например, тривиальное одномерное подпредставление любого перестановочного представления
можно интерпретировать как \emph{``счётчик частиц''}:
перестановочный инвариант \Math{\invarL{\wSN}{n}}, соответствующий этому подпредставлению,
представляет собой полное число частиц в ансамбле.
Интерпретация данных, относящихся к другим инвариантным подпространствам, требует
дальнейшего изучения.
\section*{Благодарности}
Автор благодарен С.А.~Абрамову, Ю.А.~Блинкову, С.И.~Виницкому, В.П.~Гердту, В.В.~Иванову, 
В.С.~Мележику и 
А.А.~Михалёву за обсуждение результатов, изложенных в статье.
Работа частично поддерживалась грантами 
Российского Фонда Фундаментальных Исследований № 13-01-00668 
и Министерства Образования и Науки РФ № 3802.2012.2.
\renewcommand{\appendixname}{Приложение}
\appendix
\section{Объединение пространственных и внутренних\\ симметрий}
\label{symmetryunification}
В общем случае, в качестве симметрий множества состояний \Math{\wS}
можно рассматривать произвольную подгруппу симметрической группы \Math{\SymG{\wS}}.
Однако, если требуется, чтобы для динамических систем с пространством и локальными состояниями, 
группа симметрий \Math{\wG} была сконструирована
из групп \Math{\sG} и \Math{\iG}, то необходимо снабдить групповой структурой
декартово произведение \Math{\iGX\otimes\sG}. Здесь \Math{\iGX} --- множество 
функций на пространстве  \Math{\X} со значениями в группе внутренних симметрий \Math{\iG}.
Таким образом, каждый элемент \Math{u\in\wG} представляет собой пару \Math{u = \welem{\alpha(x)}{a},}
где  \Math{\alpha(x)\in\iGX} и \Math{a\in\sG.} 
Естественно предположить, что результатом действия 
пространственных симметрий на функции из \Math{\iGX} являются также 
элементы  \Math{\iGX}. Это означает, что группа \Math{\iGX} должна быть 
\emph{нормальным делителем} искомой группы \Math{\wG}, 
т.е. \Math{\wG} представляет собой
\emph{расширение} группы \Math{\sG} с помощью группы \Math{\iGX}.
Если далее предположить, что симметрии пространства вкладываются в полную группу  \Math{\wG}
как подгруппа, то получается частный тип расширений называемых \emph{разложимыми}.
\par 
Наиболее важный пример разложимого расширения представляет собой \emph{сплетение},
введённое на стр.~\pageref{wreathpage}. 
Это конструкция адекватна принципу калибровочной инвариантности.
В физических теориях обычно предполагается независимость пространственных и
внутренних симметрий, что приводит к более простой конструкции
разложимого расширения, а именно, к 
\emph{прямому произведению} \Math{\wG\cong\iGX\times\sG}.
Действие и групповое умножение для прямого произведения имеют вид
\begin{align}
	\sigma(x)\welem{\alpha\vect{x}}{a}
	&=
	\sigma\vect{x}\alpha\vect{x},\nonumber\\[-7pt]
	&\label{dir}\\[-7pt]
	\welem{\alpha\vect{x}}{a}\wmult\welem{\beta\vect{x}}{b}
	&=
	\welem{\alpha\vect{x}\beta\vect{x}}{ab}.\nonumber
\end{align}
Следующее утверждение обобщает обе эти конструкции:\\
\emph{Любой антигомоморфизм
\Math{\mu: \sG \rightarrow \sG} (термин ``антигомоморфизм'' означает, 
что \Math{\mu(a)\mu(b)=\mu(ba)}) группы пространственных 
симметрий определяет класс эквивалентности разложимых групповых расширений
\begin{equation*}
	\id\rightarrow\iGX\rightarrow\wG\rightarrow\sG\rightarrow\id.
\end{equation*}
Эквивалентность описывается произвольной функцией \Math{\kappa: \sG \rightarrow \sG.}
Основные групповые операции --- действие на \Math{\lSX}, групповое умножение и 
взятие обратного элемента --- имеют следующий явный вид:}
\begin{align*}
		\sigma(x)\welem{\alpha\vect{x}}{a}
		&=\sigma\vect{x\mu(a)}\alpha\vect{x\kappa(a)}\\[3pt]
		\welem{\alpha\vect{x}}{a}\wmult\welem{\beta\vect{x}}{b}
		&=
	\welem{\alpha\vect{x\kappa(ab)^{-1}\mu(b)\kappa(a)}\beta\vect{x\kappa(ab)^{-1}
	\kappa(b)}}{ab}, \\[3pt]
		\welem{\alpha(x)}{a}^{-1}
		&=
		\welem{\alpha\vect{x\kappa\vect{a^{-1}}^{-1}\mu(a)^{-1}\kappa(a)}^{-1}}
		{a^{-1}}.
\end{align*}
\par
Это утверждение получается специализацией общей конструкции разложимых расширений
группы \Math{\sG} группой \Math{H} (см., например, \cite{Kirillov}) на случай, когда
\Math{H} представляет собой группу \Math{\iG}-значных функций на \Math{\X}, а \Math{\sG} действует
на аргументах этих функций. Эквивалентность расширений с одним и тем же антигомоморфизмом
\Math{\mu} но с различными функциями \Math{\kappa} выражается следующей коммутативной диаграммой
\begin{equation*}
\begin{diagram}
\node{\id}
\arrow[1]{e}
\node{\iGX}
\arrow[1]{e}
\arrow[1]{s,=}
\node{\wG}
\arrow[1]{e}
\arrow[1]{s,l}{K}
\node{\sG}
\arrow[1]{e}
\arrow[1]{s,=}
\node{\id}\\
\node{\id}
\arrow[1]{e}
\node{\iGX}
\arrow[1]{e}
\node{~\wG'}
\arrow[1]{e}
\node{\sG}
\arrow[1]{e}
\node{\id}
\end{diagram},
\end{equation*}
где отображение  \Math{K} имеет вид:~~ 
\Math{K: \welem{\alpha(x)}{a}\mapsto\welem{\alpha\vect{x\kappa(a)}}{a}.}
\par
Стандартные сплетение и прямое произведение получаются из этой общей конструкции
выбором антигомоморфизмов \Math{\mu(a)=a^{-1}} и \Math{\mu(a)=\id}, соответственно.
Что касается произвольной функции \Math{\kappa,} в математической литературе 
обычно используются \Math{\kappa(a)=a^{-1}} для сплетения и  \Math{\kappa(a)=\id} 
для прямого произведения. В формулах (\ref{wreathaction}--\ref{wreathinverse}) и 
\eqref{dir} мы использовали как раз эти функции.
\section{Структурный анализ дискретных отношений}
\label{discreterelations}
Методы анализа совместности, такие как вычисление базисов Грёбнера или приведение в инволюцию,
широко используются для изучения систем полиномиальных и дифференциальных уравнений. 
В этом приложении мы развиваем аналогичную технику для дискретных систем \cite{Kornyak05,Kornyak06a}. Для иллюстрации 
возможностей нашего подхода мы приводим его приложения к изучению клеточных автоматов.
\par
Рассмотрим декартово произведение \Math{\Sigma^n=\Sigma_1\times{}\Sigma_2\times\cdots\times{}\Sigma_n},
т.е. множество кортежей (векторов) вида \Math{\vect{\sigma_1,\sigma_2,\ldots,\sigma_n}}, 
где \Math{\sigma_i\in{}\Sigma_i} для каждого \Math{i}. 
\emph{\Math{n}-арным отношением} называется
любое подмножество \Math{n}-мерного гиперпараллелепипеда \Math{\Sigma^n}. Мы предполагаем, что все
\Math{\Sigma_i} --- конечные множества, состоящие из \Math{q_i = \cabs{\Sigma_i}} элементов, 
которые мы будем называть \emph{состояниями}.
\par
Мы можем интерпретировать \Math{n} измерений гиперпараллелепипеда \Math{\Sigma^n} как 
множество точек \Math{X=\vect{x_1,x_2,\ldots,x_n}}. Для того, чтобы превратить это изначально аморфное
множество в ``пространство'' нужно снабдить \Math{X} некоторой структурой, определяющей степень 
``близости'' друг к другу различных точек. Подходящей для этих целей математической конструкцией является
абстрактный симплициальный комплекс. Натуральная концепция пространства предполагает однородность его точек.
Это означает, что существует группа симметрий, действующая транзитивно на \Math{X}, т.е. имеется возможность 
``перемещать'' свободно точки друг в друга. Однородность возможна только если все \Math{\Sigma_i} эквивалентны.
 Пусть \Math{\Sigma} означает класс эквивалентности. Мы можем представить
\Math{\Sigma} канонически в виде \Math{\Sigma=\vect{0,\ldots,q-1}}, \Math{q = \cabs{\Sigma}}.
\par
Если число состояний --- степень простого числа, т.е. \Math{q=p^m}, мы можем снабдить множество
\Math{\Sigma} дополнительно структурой поля Галуа \Math{\F_q}. Воспользовавшись функциональной полнотой
полиномов над конечными полями \cite{Lidl}, мы можем представить любое \Math{k}-арное отношение 
на \Math{\Sigma} в виде множества нулей некоторого полинома из кольца 
\Math{\F_q\ordset{x_1,x_2,\ldots,x_n}}. Таким образом, множество отношений можно реализовать в виде
системы полиномиальных уравнений. Хотя такое представление не является необходимым (и не работает,
если \Math{\Sigma_i} --- различные множества или  \Math{q} не является степенью простого числа), 
оно полезно ввиду нашей привычки к работе с полиномами и возможности использовать развитые средства
полиномиальной алгебры, в частности, базисы Грёбнера.
\subsection{Основные определения и конструкции} 
Помимо симплексов, являющихся выделенными подмножествами множества \Math{X}, мы будем 
рассматривать \emph{произвольные} множества точек из  \Math{X}. Для краткости множества, содержащие
\Math{k} точек мы будем называть \Math{k}-множествами. Имея дело с системами отношений, определённых
на различных множествах точек, нам необходимо установить соответствие между точками и размерностями 
гиперкуба \Math{\Sigma^k}. Для этого мы будем использовать экспоненциальные обозначения.
Обозначение \Math{\Sigma^{\set{x_i}}} фиксирует \Math{\Sigma} как множество значений точки
\Math{x_i}. Для \Math{k}-множества \Math{\delta=\set{x_{i_1},\ldots,x_{i_k}}} мы вводим обозначение
\Math{\Sigma^\delta=\Sigma^{\set{x_{i_1}}}\times\cdots\times\Sigma^{\set{x_{i_k}}}}. Мы будем называть
множество точек \Math{\delta} \emph{областью определения} отношения \Math{R^\delta}, если 
\Math{R^\delta\subseteq\Sigma^\delta}. Весь гиперкуб \Math{\Sigma^\delta} мы будем называть
\emph{тривиальным} отношением. Заметим, что отношения весьма экономно представляются
в памяти компьютера битовыми строками, а манипуляции с отношениями эффективно реализуются
битовыми операциями из базового набора процессорных команд.
\subsubsection{Отношения} 
Таким образом, мы имеем
\par
\textbf{Определение 1} (отношение).
\emph{Отношением} \Math{R^\delta} на множестве точек \\
 \Math{\delta=\set{x_{i_1},\ldots,x_{i_k}}} называется любое подмножество гиперкуба \Math{\Sigma^\delta},
т.е. \Math{R^\delta\subseteq\Sigma^\delta}.\\
Отношение \Math{R^\delta} можно отождествить с булевозначной (характеристической)\\ функцией 
\Math{R^\delta: \Sigma^\delta\rightarrow\set{0,1}}. 
\par
Важный специальный случай отношения:
\par
\textbf{Определение 2} (функциональное отношение). 
Отношение \Math{R^\delta} на множестве точек
\Math{\delta=\set{x_{i_1},\ldots,x_{i_k}}} называется \emph{функциональным}, если существует
позиция\\ \Math{p\in\vect{1,\ldots,k}}, такая, что для любых
\Math{\sigma_{i_1},\ldots,\sigma_{i_{p-1}},\sigma_{i_{p+1}},\ldots,\sigma_{i_k},\varsigma,\tau\in\Sigma}
\\из 
\Math{\vect{\sigma_{i_1},\ldots,\sigma_{i_{p-1}},
\varsigma,\sigma_{i_{p+1}},\ldots,\sigma_{i_k}}\in{}R^\delta}
и 
\Math{\vect{\sigma_{i_1},\ldots,\sigma_{i_{p-1}},
\tau,\sigma_{i_{p+1}},\ldots,\sigma_{i_k}}\in{}R^\delta}\\ следует \Math{\varsigma=\tau}.
\\
Если интерпретировать точки пространства как переменные, то функциональное отношение можно
написать в виде
\Mathh{x_{i_p}=F\vect{x_{i_1},\ldots,x_{i_{p-1}},x_{i_{p+1}},\ldots,x_{i_k}},
\text{~где~}F: \Sigma^{\delta\setminus\set{x_{i_p}}}\rightarrow\Sigma\enspace.}
\par
Нам потребуется возможность расширять отношения с множества точек на его надмножество:
\par
\textbf{Определение 3} (расширение отношения). Для данного множества точек \Math{\delta}, его подмножества
\Math{\tau\subseteq\delta} и отношения \Math{R^\tau} на \Math{\tau} мы определяем \emph{расширение}
отношения \Math{R^\tau} следующим образом
\Mathh{R^\delta=R^\tau\times\Sigma^{\delta\setminus\tau}.}
Это определение позволяет, в частности, привести отношения \Math{R^{\delta_{1}},\ldots,R^{\delta_{m}}}, 
определённые на разных областях, к общей области определения, т.е. к объединению
\Math{\delta_{i_1}\cup\cdots\cup\delta_{i_m}}.
\par
Логические следствия отношений определяются естественным образом: 
\par
\textbf{Определение 4} (следствие отношения). Отношение \Math{Q^\delta} называется \emph{следствием}
отношения \Math{R^\delta}, если \Math{R^\delta\subseteq{}Q^\delta\subseteq{}\Sigma^\delta}, 
т.е. \Math{Q^\delta}  --- произвольное \emph{надмножество} множества \Math{R^\delta}.
\par
Отношение \Math{R^\delta} может иметь много различных следствий: их полное число равно, очевидно,
\Math{2^{\cabs{\Sigma^\delta}-\cabs{R^\delta}}}.
\par
При анализе структуры отношения важно выделить следствия, которые можно свести к отношениям на меньших
множествах точек:
\par
\textbf{Определение 5} (собственное следствие). 
\emph{Нетривиальное} отношение \Math{Q^\tau} называется \emph{собственным следствием} отношения
\Math{R^\delta}, если \Math{\tau} --- \emph{собственное подмножество} множества \Math{\delta}
(т.е. \Math{\tau\subset\delta}) и отношение \Mathh{Q^\tau\times\Sigma^{\delta\setminus\tau}}
является следствием отношения \Math{R^\delta}.
\par
Отношения, не имеющие собственных следствий, мы будем называть \emph{простыми отношениями}.
\subsubsection{Совместность систем отношений} 
Совместность нескольких отношений естественно определить как пересечение расширений этих
отношений на их общую область определения:
\par
\textbf{Определение 6} (базисное отношение).  \emph{Базисным отношением} системы отношений
 \Math{R^{\delta_1},\ldots,R^{\delta_m}} называется отношение
\begin{equation}
	R^\delta=\bigcap\limits_{i=1}^m{}R^{\delta_i}\times\Sigma^{\delta\setminus\delta_i},
	\text{~где~} \delta=\bigcup\limits_{i=1}{\delta_i}\enspace.
	\label{basisrelation}
\end{equation}
В полиномиальном случае, когда \Math{q=p^n}, стандартным средством анализа совместности является метод
базисов Грёбнера. Относительно связи базисов Грёбнера с нашим определением можно сделать
следующие замечания:
\begin{itemize}
	\item
В отличие от базиса Грёбнера, который обычно представляет собой систему полиномов, 
условие совместности, выражаемое базисным отношением,  всегда можно представить 
\emph{единственным} полиномом.	 
	\item
Любой	возможный базис Грёбнера системы отношений \Math{R^{\delta_1},\ldots,R^{\delta_m}},
представленных полиномами, является некоторой комбинацией \emph{следствий} базисного отношения.
\end{itemize}
\subsubsection{Декомпозиция отношений} 
Если отношение имеет собственные следствия, то можно попытаться представить его в виде комбинации
этих следствий, т.е. отношений на меньших множествах точек. Для этого мы вводим следующее определение.
\par
\textbf{Определение 7} (каноническая декомпозиция). \emph{Канонической декомпозицией} отношения
\Math{R^\delta}, имеющего собственные следствия  \Math{Q^{\delta_1},\ldots,Q^{\delta_m}},
называется представление \Math{R^\delta} в виде
\begin{equation}
	R^\delta=P^\delta\bigcap\vect{\bigcap\limits_{i=1}^m{}Q^{\delta_i}
	\times\Sigma^{\delta\setminus\delta_i}}\enspace,
	\label{canondeco}
\end{equation}
где фактор \Math{P^\delta} определяется следующим образом:
\par
\textbf{Определение 8} (главный фактор). \emph{Главным фактором} отношения
\Math{R^\delta}, имеющего собственные следствия  \Math{Q^{\delta_1},\ldots,Q^{\delta_m}},
называется отношение
\begin{equation*}
	P^\delta=R^\delta\bigcup\vect{\Sigma^{\delta}\setminus\bigcap\limits_{i=1}^m{}Q^{\delta_i}\times
	\Sigma^{\delta\setminus\delta_i}}\enspace.
\end{equation*}
Главный фактор представляет собой максимально ``свободное'', т.е. наиболее близкое к тривиальному,
отношение, позволяющее в комбинации с собственными следствиями восстановить исходное отношение.
\par
Если главный фактор в канонической декомпозиции тривиален, то отношение можно полностью свести
к отношениям на меньших множествах точек.
\par
\textbf{Определение 9} (приводимое отношение). Отношение \Math{R^\delta} называется \emph{приводимым},
если его можно представить в виде
\begin{equation}
	R^\delta=\bigcap\limits_{i=1}^m{}Q^{\delta_i}\times\Sigma^{\delta\setminus\delta_i}\enspace,
	\label{reducibledeco}
\end{equation}
где \Math{\delta_i} --- собственные подмножества \Math{\delta}.
\par
Это определение даёт возможность ввести ``\emph{топологию}'', т.е. 
структуру абстрактного симплициального
комплекса с соответствующими теориями гомологий, когомологий и т.д.,  на \emph{произвольном}
\Math{n}-арном отношении \Math{R\subseteq\Sigma^n}. Для этого нужно
\begin{enumerate}
	\item
назвать измерения гиперкуба \Math{\Sigma^n} ``точками'' \Math{x_1,\ldots,x_n\in{}X};	 
	\item
разложить отношение \Math{R} (которое теперь можно обозначить как \Math{R^X}) на 
\emph{неприводимые} компоненты;
	\item
объявить области  определения неприводимых компонент
отношения \Math{R^X} \emph{максимальными симплексами}	симплициального комплекса.
\end{enumerate}
\subsection{Приложение к исследованию клеточных автоматов} 
\subsubsection{Клеточный автомат Дж. Конвея Life} 
Автомат Конвея Life входит в семейство 2-мерных бинарных (т.е. \Math{\Sigma=\set{0,1}; q=2}) клеточных
автоматов с правилами, заданными на окрестности Мура \Math{3\times3}. Квадратная решётка с такой
окрестностью может быть представлена 8-валентным (8-регулярным) графом. Правило автомата Life описывается
следующим образом: клетка (вершина графа) ``рождается'', если имеет ровно 3 ``живых'' соседа, ``выживает'',
если имеет 2 или 3 ``живых'' соседа и ``умирает'' во всех остальных случаях. Символически это правило 
можно записать в виде списка ``рождение''/``выживание'' как B3/S23. Другие примеры автоматов из этого
семейства --- HighLife (правило B36/S23) Day\&Night (правило B3678/S34678).
\par
Для обобщения этого типа правил мы определяем \emph{\Math{k}-валентное правило} Life как \emph{бинарное}
правило на окрестности \Math{k}-валентного графа (мы будем обозначать центральную вершину окрестности
символом \Math{x_{k+1}}, а смежные с ней вершины, соответственно, \Math{x_{1},\ldots,x_{k}}),
задаваемое двумя \emph{произвольными} подмножествами множества \Math{\set{0,1,\ldots,k}}.
Эти подмножества \Math{B,S\subseteq\set{0,1,\ldots,k}} содержат условия переходов за один шаг по времени
\Math{x_{k+1}\rightarrow{}x'_{k+1}} вида \Math{0\rightarrow1} и \Math{1\rightarrow1}, соответственно.
Поскольку число подмножеств любого конечного множества \Math{A} равно \Math{2^{\cabs{A}}} а \emph{различные}
пары \Math{B/S} определяют  \emph{различные} правило, то полное число различных правил равно
\begin{equation}
	N_{B/S,k} = 2^{k+1}\times2^{k+1}=2^{2k+2}\enspace.\label{liferule}
\end{equation}
\par
Введём другое определение: \emph{\Math{k}-симметричным \Math{q}-арным правилом} называется правило на 
\Math{k}-валентной окрестности, симметричное относительно группы \Math{\SymG{k}} всех перестановок
\Math{k} внешних вершин окрестности. Нетрудно подсчитать полное число всех различных 
\Math{k}-симметричных \Math{q}-арных правил:
\begin{equation}
	N_{q,\SymG{k}}=q^{\binom{k+q-1}{q-1}q}\enspace.\label{symrule} 
\end{equation}
Мы видим, что при \Math{q=2} числа \eqref{liferule} и \eqref{symrule} совпадают: \Math{N_{B/S,k}=N_{2,\SymG{k}}}. 
Таким образом, любое \Math{k}-симметричное бинарное правило
можно записать в форме списка ``рождение''/``выживание''.
\par
Локальное отношение автомата Life, которое мы будем обозначать символом \Math{R^\delta_{\text{Life}}},
определено на 10-множестве \Math{\delta=\set{x_1,\ldots,x_{10}}}:
\begin{center}
\includegraphics[width=0.5\textwidth]{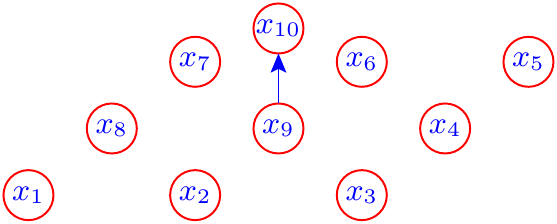}\enspace,
\end{center}
где точка \Math{x_{10}\equiv{}x'_9} отождествляется с ``точкой \Math{x_9} в следующий момент времени''.
По построению, точки 10-мерного гиперкуба \Math{\Sigma^\delta} принадлежат отношению
\Math{R^\delta_{\text{Life}}}, т.е. \Math{\vect{x_1,\ldots,x_{10}}\in{}R^\delta_{\text{Life}}}, 
в следующих случаях:
\begin{enumerate}
	\item
\Math{\vect{\sum_{i=1}^8x_i=3}\wedge\vect{x_{10}=1}\enspace,}	 
	\item 
\Math{\vect{\sum_{i=1}^8x_i=2}\wedge\vect{x_{9}=x_{10}}\enspace,}	 
	\item
\Math{x_{10}=0}, если не выполняется ни одно из предыдущих условий.	 
\end{enumerate}
Число элементов отношения равно \Math{\cabs{R^\delta_{\text{Life}}}=512.}
Отношение \Math{R^\delta_{\text{Life}}}, как и в случае любого клеточного автомата, является
\emph{функциональным}: состояние точки \Math{x_{10}} однозначно определяется состояниями других
точек. Если дополнительно снабдить множество \Math{\Sigma=\set{0,1}} структурой поля \Math{\F_2},
то отношение \Math{R^\delta_{\text{Life}}} можно записать в виде полинома из кольца 
\Math{\F_2\ordset{x_1,\ldots,x_{10}}}. Это позволяет провести исследование автомата как нашим методом
структурного анализа, так и полиномиальными методами. Приведём сравнение нашего подхода с методом базисов Грёбнера.
\par
Полиномиальное представление отношения  \Math{R^\delta_{\text{Life}}} имеет вид
\begin{equation}
P_{\text{Life}}= x_{10}
+x_9
\vect{
\esp_7
+\esp_6
+\esp_3
+\esp_2
}
+\esp_7
+\esp_3,
\label{polylife}
\end{equation}
где \Math{\esp_k\equiv\esymm{k}{x_1,\ldots,x_8}} ---
\Math{k}-й \emph{элементарный симметрический полином} определяемый 
для  \Math{n} переменных
\Math{x_1,\ldots,x_{n}} формулой:
\Mathh{\esymm{k}{x_1,\ldots,x_{n}} =
\sum\limits_{1\leq{}i_1<i_2<\cdots<i_{k}\leq{}n}x_{i_1}x_{i_2}\cdots x_{i_{k}}.}
Далее мы будем использовать следующие обозначения:
\Mathh{\esp_k\equiv\esymm{k}{x_1,\ldots,x_{8}},~~~~
\esp_k^i\equiv\esymm{k}{x_1,\ldots,\widehat{x_i},\ldots,x_{8}},}
\Mathh{\esp_k^{ij}\equiv\esymm{k}{x_1,\ldots,\widehat{x_i},\ldots,\widehat{x_j},\ldots,x_{8}}.}
\par 
Применяя написанную нами программу для структурного анализа дискретных отношений к  
\Math{R^\delta_{\text{Life}}} мы обнаруживаем, что это отношение \emph{приводимо} и его 
каноническая декомпозиция имеет вид
\begin{equation}
R^\delta_{\text{Life}} =
R_2^{\delta\setminus\set{x_9}}\bigcap
\vect{\bigcap\limits_{k=1}^7
R_1^{\delta\setminus\set{x_{i_k}}}}\enspace,
\label{lifedecomp}
\end{equation}
где \Math{\vect{i_1,\ldots,i_7}} --- произвольное 7-элементное подмножество множества \Math{\vect{1,\ldots,8}}.
Для краткости мы отбросили в \eqref{lifedecomp} тривиальные факторы \Math{\Sigma^{\set{x_{i_k}}}},
входящие в общую формулу \eqref{reducibledeco}.
\par
Восемь отношений \Math{R_1^{\delta\setminus\set{x_{i}}}} (\Math{1\leq{}i\leq8}; для построения 
декомпозиции \eqref{lifedecomp} достаточно взять любые семь из них) 
имеют следующую полиномиальную форму:
\Mathh{x_9x_{10}
\vect{\esp^i_6+\esp^i_5+\esp^i_2+\esp^i_1}
+x_{10}\vect{\esp^i_6+\esp^i_2+1}
+x_9\vect{\esp^i_7+\esp^i_6+\esp^i_3+\esp^i_2}=0.}
Полиномиальная форма отношения \Math{R_2^{\delta\setminus\set{x_9}}} имеет вид
\Mathh{x_{10}\vect{\esp_7+\esp_6+\esp_3+\esp_2+1}+\esp_7+\esp_3=0.}
Отношения \Math{R_1^{\delta\setminus\set{x_{i}}}} и \Math{R_2^{\delta\setminus\set{x_9}}}
являются \emph{неприводимыми} но \emph{не простыми} и могут быть разложены в соответствии с формулой
\eqref{canondeco}.
Продолжая итерации декомпозиций мы окончательно приходим к максимально упрощённой 
системе отношений эквивалентной исходному отношению \Math{R^\delta_{\text{Life}}}.
Для наглядности приведём их полиномиальную форму:
\begin{align}
	x_9x_{10}\vect{\esp^i_2+\esp^i_1}
	+x_{10}\vect{\esp^i_2+1}
	+x_9\vect{\esp^i_7+\esp^i_6+\esp^i_3+\esp^i_2}&=0,
	\label{poly1red}
	\\
	x_{10}\vect{\esp_3+\esp_2+1}+\esp_7+\esp_3&=0,
	\label{poly2red}
	\\
	\vect{x_9x_{10}+x_{10}}\vect{\esp^{ij}_3+\esp^{ij}_2+\esp^{ij}_1+1}&=0,
	\label{poly11red}
	\\
	x_{10}\vect{\esp^{i}_3+\esp^{i}_2+\esp^{i}_1+1}&=0,
	\label{poly12red}
	\\
	x_{10}x_{i_1}x_{i_2}x_{i_3}x_{i_4}&=0.
	\label{poly0123red}
\end{align}
Простейшие отношения \eqref{poly0123red} легко интерпретируются: 
если состояние точки \Math{x_{10}} равно 1, тогда по крайней мере одна из 
любых четырёх точек, окружающих \Math{x_{9}}, должна быть в состоянии 0.
\par
Приведённый выше анализ отношения  \Math{R^\delta_{\text{Life}}} выполняется менее чем за 1 секунду
на среднем ноутбуке.
Для вычисления базиса Грёбнера к полиному \eqref{polylife} нужно добавить десять полиномов
\Mathh{x_i^2+x_i,~~i=1,\ldots,10\enspace,}
в соответствии с тождеством \Math{x^q=x}, которому удовлетворяют все элементы конечного поля  \Math{\F_q}.
Вычисление с помощью функции встроенной в систему \textbf{Maple} выполняется примерно за 1 час.
Результат зависит от упорядочения переменных, которое необходимо задавать при использовании метода базисов Грёбнера:
\begin{itemize}
	\item 
чисто лексикографическое упорядочивание с порядком переменных
\Math{x_{10}\succ x_9\succ\cdots\succ x_1}
не даёт новой информации, оставляя исходный полином 
\eqref{polylife} неизменным;
	\item
чисто лексикографическое упорядочивание с порядком переменных
\Math{x_1\succ x_2\succ\cdots\succ x_{10}}
и упорядочивание по степеням с обратной лексикографией
воспроизводят отношения
\eqref{poly1red}---\eqref{poly0123red} с точностью до нескольких 
дополнительных полиномиальных редукций, нарушающих симметрию полиномов.
\end{itemize}
\subsubsection{Элементарные клеточные автоматы} 
Простейшие бинарные одномерные клеточные автоматы были названы 
С. Вольфрамом \emph{элементарными клеточными автоматами}.
В этом разделе мы будем придерживаться обозначений и терминологии книги Вольфрама \cite{Wolfram}.
Элементарные автоматы существенно проще чем Life и мы можем при их исследовании уделить больше 
внимания топологическим аспектам структурного анализа.
\par
Локальные правила элементарных клеточных автоматов определяются на 4-мно\-жестве
\Math{\delta=\set{p,q,r,s}}, которое можно изобразить картинкой
\begin{center}
\includegraphics[width=0.15\textwidth]{ECA-Set-pqrs-Color}.
\end{center}
Локальное правило это бинарная функция вида \Math{s=f(p,q,r)}.
Полное число таких функций равно \Math{2^{2^3}=256}.
Следовательно, каждое правило можно индексировать 8-битным бинарным числом.
\par
Наши вычисления показывают, что 256 отношений, соответствующих этим правилам, подразделяются на
118 \emph{приводимых} и 138 \emph{неприводимых}. Только два отношения из неприводимых оказались
\emph{простыми}, а именно, правила 105 и 150 в нумерации Вольфрама%
\footnote{Битовые строки значений функции \Math{s=f(p,q,r)} при лексикографически возрастающем порядке комбинаций
значений переменных \Math{p,q,r} интерпретируются как числа с бинарными
разрядами, идущими в убывающем порядке.}. 
Простые правила 105 и 150 имеют линейные полиномиальные формы: \Math{s=p+q+r+1} и \Math{s=p+q+r}, соответственно.
\par
Пространство-время элементарных автоматов представляет собой  решётку 
с целочисленными координатами \Math{(x,t)\in\Z\times\Z}. Разумеется, мы можем без всяких проблем 
рассматривать конечную решётку  \Math{\Z_m\times\ordset{\tin,1,\ldots,\tfin}}, выбрав достаточно большие
 \Math{m} и \Math{\tfin}.
Состояние решётки мы будем представлять функцией \Math{u(x,t)\in \Sigma=\set{0,1}}.
Если представить решётку в виде графа, то в общем случае мы предполагаем, что точки 
решётки связаны между собой, как показано на рисунке
\begin{center}
\includegraphics[width=0.3\textwidth]{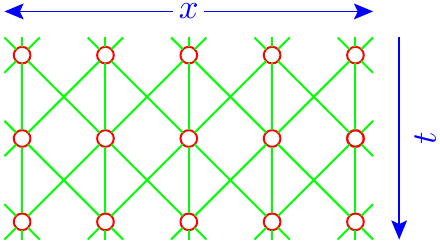}.
\end{center}
Отсутствие горизонтальных связей выражает независимость  ``пространственно-по\-доб\-ных''
точек в клеточных автоматах.
\paragraph{Приводимые автоматы.}
Анализ показывает, что некоторые автоматы с приводимыми локальными отношениями можно представить
в виде объединения автоматов, определённых на несвязных (непересекающихся) подкомплексах:
\par
\begin{itemize}
	\item  
Два автомата 0 и 255 определяются унарными отношениями \Math{s=0} и \Math{s=1}
на несвязном множестве точек:
\begin{center}
\includegraphics[width=0.2\textwidth]{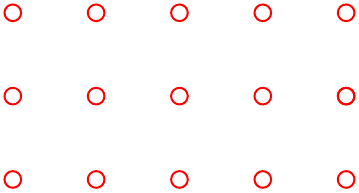}.
\end{center}
Заметим, что в математике унарные отношения обычно называют \emph{свойствами}. 
\par
	\item  
Шесть автоматов 15, 51, 85, 170, 204 и 240 фактически представляют собой наборы несвязных 
пространственно нульмерных автоматов, т.е. изолированные точки, эволюционирующие во времени.	
Для примера, рассмотрим автомат 15.
Его локальное правило, определенное на множестве
\begin{center}
\includegraphics[width=0.15\textwidth]{ECA-Set-pqrs-Color},
\end{center} 
можно представить строкой битов 0101010110101010.
Это отношение сводится к отношению на грани
\begin{center}
\includegraphics[width=0.09\textwidth]{ECA-Set-ps-Color}
\end{center}
с битовой строкой 0110. 
Пространственно-временн\'ая решётка расщепляется следующим образом:
\begin{center}
\includegraphics[width=0.24\textwidth]{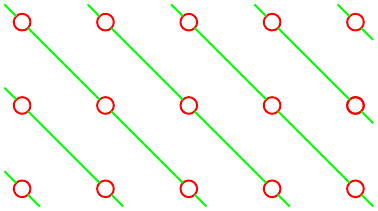}.
\end{center}
Строка битов 0110 означает, что состояния точек \Math{p} и \Math{s} всегда противоположны,
поэтому можно сразу написать общее решение для автомата 15:
\Mathh{
u(x,t) = a(x-t)+t\mod 2,
}
где \Math{u(x,0)\equiv{}a(x)} --- произвольное начальное условие.
	\item
Каждый из десяти автоматов 5, 10, 80, 90, 95, 160, 165, 175,
245, 250 распадается на два идентичных автомата.
Для примера рассмотрим правило 90.
Этот автомат известен тем, что он порождает фрактал топологической размерности 1 и хаусдорфовой
размерности \Math{\ln3/\ln2\approx1.58} известный 
как ``\emph{треугольник Серпинского}'' (а также как ``\emph{решётка}'' или 
``\emph{салфетка}'' Серпинского).
Локальное отношение правила 90 на множестве
\begin{center}
\includegraphics[width=0.15\textwidth]{ECA-Set-pqrs-Color} 
\end{center}
представляется битовой строкой 1010010101011010.
Это отношение, в соответствии с разложением \eqref{reducibledeco},
сводится к отношению с битовой строкой
\begin{equation}
10010110 \text{~~~~на грани~~
\raisebox{-0.02\textwidth}{\includegraphics[width=0.15\textwidth]{ECA-Set-prs-Color}}}.
\label{bt90}
\end{equation}
Из структуры этой грани  видно, что пространственно-временн\'ая решётка распадается на два идентичных
независимых комплекса:
\begin{center}
\includegraphics[width=0.8\textwidth]{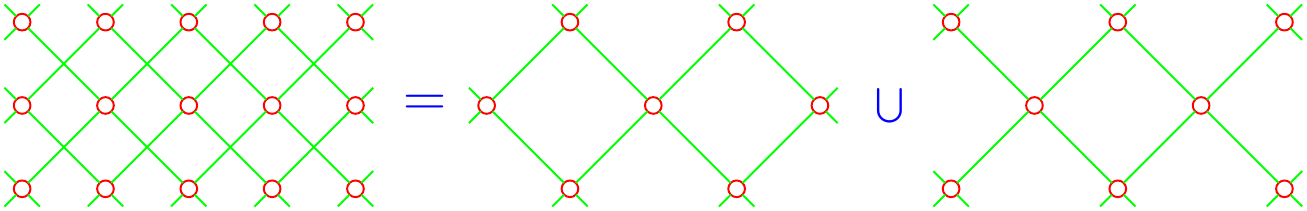}\enspace.
\end{center}
Для построения общего решения автомата 90 удобно использовать полиномиальную форму \Math{s+p+r=0} 
отношения \eqref{bt90}.
Из этого линейного выражения легко выводится общее решение:
\Mathh{
u(x,t)=\sum\limits_{k=0}^t \binom{t}{k}a(x-t+2k)\mod 2,\qquad{} u(x,0)\equiv{}a(x).
}
\end{itemize}
\paragraph{Использование собственных следствий.}
Собственные следствия, даже если они не являются функциональными, содержат полезную
информацию о глобальном поведении клеточного автомата.
\par
Например, 64 автомата%
\footnote{Приведём полный список этих автоматов в нумерации Вольфрама:
  2,   4,   8,  10,  16,  32,  34,  40,
 42,  48,  64,  72,  76,  80,  96, 112,
128, 130, 132, 136, 138, 140, 144, 160,
162, 168, 171, 174--176, 186, 187,
190--192, 196, 200, 205, 206, 208,
220, 222--224, 234--239,
241--254. Автоматы из этого списка имеют как приводимые, так и неприводимые локальные отношения.}
имеют собственные следствия с битовой строкой
\begin{equation}
1101
\label{finiterod}
\end{equation}
на, по крайней мере, одной из граней
\begin{equation}
\text{\includegraphics[width=0.09\textwidth]{ECA-Set-ps-Color}
~~~~
\includegraphics[width=0.035\textwidth]{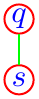}
~~~~
\includegraphics[width=0.09\textwidth]{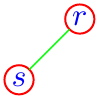}}.
\label{finiterodsets}
\end{equation}
Отношения \eqref{finiterod} имеют на гранях \eqref{finiterodsets}
полиномиальные формы
\Mathh{ps+s=0,~~qs+s=0,~~rs+s=0,}
соответственно.
\par
\begin{figure}[!h]
\begin{center}
\includegraphics[width=0.6\textwidth]{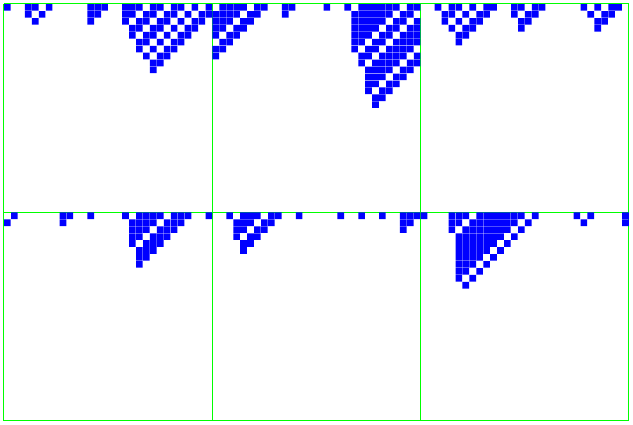}
\caption{Правило 168. Несколько эволюций со случайными начальными условиями.\label{figu}}
\end{center}
\end{figure}

Отношение \eqref{finiterod} \emph{не функционально}, 
и, следовательно, не может описывать какую бы то ни было детерминистическую эволюцию.
Тем не менее, оно накладывает жёсткие ограничения на поведение соответствующих автоматов.
Характерные черты такого поведения отчётливо
видны на многих представленных в атласе \cite{site} иллюстрациях к компьютерным вычислениям
эволюции автоматов.
Типичная картина из этого атласа, на которой представлены несколько эволюций автомата 168, воспроизведена на Рис. \ref{figu}.
На рисунке нули и единицы изображены пустыми и заполненными квадратными клетками, соответственно.
Заметим, что авторы вычислений использовали периодичность по пространственному измерению: \Math{x\in\Z_{30}}.
\par
Локальное отношение автомата 168 в полиномиальной форме имеет вид  \Math{pqr+qr+pr+s=0}.
Это отношение имеет собственное следствие \Math{rs+s=0}, означающее, что если \Math{r} находится в состоянии 1,
то  \Math{s} может быть в любом из двух состояний 0 или 1, однако, если состояние \Math{r} равно 0, то 
\Math{s} \emph{обязано} быть в состоянии 0. Символически:
\begin{align*}
	r=1\Rightarrow{}&s=0\vee{}s=1,\\
	r=0\Rightarrow{}&s=0.
\end{align*}
На Рис. \ref{figu} видно, что любая эволюция состоит из обрывающихся диагоналей, направленных влево вниз.
Каждая диагональ начинается с нескольких идущих подряд единиц, но после первого появления нуля 
все последующие клетки диагонали могут быть только нулями.
\paragraph{Сравнение канонической декомпозиции с базисами Грёбнера.}
Ясно, что наша каноническая декомпозиция \eqref{canondeco} 
--- более общий метод исследования дискретных отношений, чем базисы Грёбнера.
Однако в полиномиальном случае их можно сравнивать.
В качестве примера приведём базисы Грёбнера и канонические декомпозиции  двух элементарных
клеточных автоматов.
Мы вычисляли базисы Грёбнера в упорядочении по полной степени и обратной лексикографии.
Тривиальные полиномы  \Math{p^2+p,} \Math{q^2+q,} \Math{r^2+r} и \Math{s^2+s} исключены из описаний базисов Грёбнера.
\begin{itemize}
\item 
\textbf{Автомат 30} известен своим хаотическим поведением. Он даже используется
как генератор случайных чисел в системе \textbf{\emph{Mathematica}}.
\par
Отношение: \Math{1001010101101010} ~или~  \Math{qr+s+r+q+p=0}.
\par
\textbf{Каноническая декомпозиция:}
\par
Собственные следствия:
\par
\begin{tabular}{lll}
грань&
\includegraphics[width=0.09\textwidth]{ECA-Set-pqs-Color}
&
\includegraphics[width=0.15\textwidth]{ECA-Set-prs-Color}
\\[10pt]
строка битов&11011110 & 11011110\\[10pt]
полином \hspace*{20pt}&\Math{qs+pq+q}\hspace*{20pt} & \Math{rs+pr+r}.\\[10pt]
\end{tabular}
\par
Главный фактор: \Math{1011111101111111} ~или~ \Math{qrs+pqr+rs+qs+pr+pq+s+p=0.}
\par
\textbf{Базис Грёбнера:}
\Math{\set{qr+s+r+q+p,~qs+pq+q,~rs+pr+r}.}\\[10pt]
Таким образом, для автомата 30 множество полиномов канонической декомпозиции совпадает 
(с точностью до очевидных полиномиальных подстановок в главный фактор) с базисом Грёбнера.
\item 
 \textbf{Автомат 110} является, подобно машине Тьюринга, \emph{универсальным}, т.е. с его помощью
можно реализовать любой вычислительный процесс, в частности, воспроизвести поведение любого другого
клеточного автомата.
\par
Отношение: \Math{1100000100111110} ~или~ \Math{pqr+qr+s+r+q=0.}
\par
\textbf{Каноническая декомпозиция:}
\par
Собственные следствия:
\par
\begin{tabular}{llll}
грань&
\includegraphics[width=0.09\textwidth]{ECA-Set-pqs-Color}
&
\includegraphics[width=0.15\textwidth]{ECA-Set-prs-Color}
&
\includegraphics[width=0.09\textwidth]{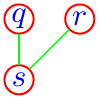}
\\[10pt]
строка битов&11011111 & 11011111 &10010111\\[10pt]
полином&
\Math{pqs+qs+pq+q} &
\Math{prs+rs+pr+r} & \Math{qrs+s+r+q}.\\[10pt]
\end{tabular}
\par
Главный фактор:
\Math{1111111111111110} ~или~ \Math{pqrs=0.}
\par
\textbf{Базис Грёбнера:}
\Mathh{\set{prs+rs+pr+r,~qs+rs+r+q,~qr+rs+s+q,~pr+pq+ps}.}
В этом случае полиномы базиса Грёбнера не совпадают с полиномами
канонической декомпозиции.
Система отношений, определяемых базисом Грёбнера имеет вид:
\begin{eqnarray*}
R_1^{\set{p,r,s}}&=&11011111=\vect{prs+rs+pr+r=0},
\\
R_2^{\set{q,r,s}}&=&10011111=\vect{qs+rs+r+q=0},
\\
R_3^{\set{q,r,s}}&=&10110111=\vect{qr+rs+s+q=0},
\\
R_4^{\set{p,q,r,s}}&=&1110101110111110=\vect{pr+pq+ps=0}.
\end{eqnarray*}
\end{itemize}
В целом, можно упомянуть следующие различия между нашим подходом и методом базиса Грёбнера.
\begin{itemize}
	\item 
В отличие от базиса Грёбнера, базисное отношение	\eqref{basisrelation},
определяемое как пересечение условий, согласуется с натуральным для логики и теории множеств
понятием совместности.
	\item
В отличие от канонической декомпозиции, базис Грёбнера может выглядеть	вне полиномиального
контекста как набор случайных надмножеств.
	\item
Тем не менее, некоторая аналогия между базисами Грёбнера и каноническими декомпозициями имеется.
Фактически, в наших вычислениях соответствующие наборы полиномов совпадали примерно в половине случаев.
	\item
Каноническая декомпозиция более эффективна в проблемах,	в которых полиномы могут иметь 
произвольные степени --- иллюстрацией этого являются приведённые выше вычисления с автоматом Конвея.
  \item
Для задач с полиномами малых степеней и большим числом  \Math{n} переменных базис Грёбнера
превосходит каноническую декомпозицию --- число полиномов ограниченной степени полиномиально
зависит от \Math{n}, а алгоритм канонической декомпозиции сканирует экспоненциально большое число 
\Math{q^{\textstyle{n}}}  точек гиперкуба \Math{\Sigma^n}.
\end{itemize}
  
\section{Локальные квантовые модели на регулярных графах}
\label{quantumonregulargraphs}
В стандартной квантовой механике амплитуда вдоль пути пропорциональна экспоненте от действия:
\Math{A\propto\exp\vect{\frac{i}{\hbar}S}}. Ясно, что это выражение никогда не обращается в нуль.
Поэтому в стандартной формулировке в принципе возможны квантовые переходы между любыми состояниями 
системы. Здесь мы рассмотрим более удобные для исследования модели с ограниченными множествами
возможных переходов. 
А именно, в наших моделях \cite{Kornyak09c,Kornyak10}, определённых на регулярных графах,
переходы будут возможны только в пределах окрестностей вершин графов.
\subsection{Общее определение квантовой модели на графе} 
Определение \emph{локальной квантовой модели на \Math{k}-валентном графе} включает следующее:
\begin{enumerate}
\item
\emph{Пространство} \Math{\X=\set{\x_1,\ldots, \x_\XN}}
 представляет собой \Math{k}-валентный граф.
 Вершины графа, смежные с данной вершиной \Math{\x_i}, мы будем обозначать символами
 \Math{\x_{1,i},\ldots,\x_{k,i}}.
\item
\emph{Схема локальных переходов} \Math{E_i=\set{e_{0,i}, e_{1,i},\cdots,e_{k,i}}} это множество
рёбер, смежных с вершиной \Math{\x_i}, дополненное ребром	
\Math{e_{0,i}=\vect{\x_i\rightarrow{}\x_i}\equiv\vect{\x_i\rightarrow{}\x_{0,i}}}.  
Каждое ребро \Math{e_{m,i}=\vect{\x_i\rightarrow{}\x_{m,i}}} символизирует переход за один шаг времени
из вершины  \Math{\x_i} в некоторую вершину окрестности или в саму \Math{\x_i}.
\item
Мы предполагаем, что группа симметрий пространства \Math{\sG=\mathrm{Aut}\vect{\X}}
действует транзитивно на множестве схем локальных переходов \Math{\set{E_1,\cdots,E_{\XN}}}.
\item
Определим группу \emph{локальных симметрий пространства} как подгруппу \Math{\sG} фиксирующую вершину \Math{\x_i}:
\Math{\sGloc=\mathrm{Stab}_{\sG}\vect{\x_i}\leq\sG}. Ввиду транзитивности действия \Math{\sG} на
множестве окрестностей, группа \Math{\sGloc} не зависит от выбора вершины.
\item
 \Math{\Omega_i=\set{\omega_{0,i},\omega_{1,i},\cdots,\omega_{h,i}}}	
	означает \emph{множество орбит} \Math{\sGloc} на \Math{E_i}. Ясно, что эти множества
	изоморфны для всех вершин \Math{\x_i}.
\item
Мы предполагаем, что с каждой вершиной ассоциировано множество внутренних степеней свободы
(локальных состояний) \Math{\lS} на котором действует группа внутренних симметрий \Math{\iG}.
\item
Мы будем называть \emph{эволюционным правилом} \Math{R} функцию на \Math{E_i} со значением в некотором
представлении \Math{\Rep{\iG}} группы внутренних симметрий. Это правило предписывает 
\Math{\Rep{\iG}}-значные веса переходам за один шаг по времени между вершинами окрестности \Math{\x_i}.
Из соображений симметрии правило \Math{R} должно быть функцией на орбитах из множества
\Math{\Omega_i}, т.е. \Math{R\vect{e_{m,i}g}=R\vect{e_{m,i}}},  если \Math{g\in\sGloc}.
\end{enumerate}
Для иллюстрации этих конструкций рассмотрим локальную квантовую модель на графе фуллерена \Math{C_{60}}.
Этот граф изображён на Рис. \ref{NanoCarbons} и \ref{bucky}.
В данном случае группой симметрий пространства \Math{X=\set{\x_1,\cdots,\x_{60}}} является  
\Math{\sG=\mathrm{Aut}\vect{\X}=\AltG{5}\times\CyclG{2}}.
Окрестность вершины графа \Math{\x_i} устроена следующим образом 
\begin{center}
\includegraphics[width=0.2\textwidth]{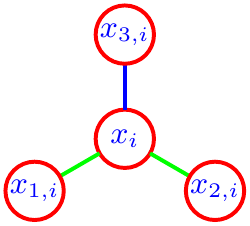}.
\end{center}
Схема локальных переходов имеет вид \Math{E_i=\set{e_{0,i}, e_{1,i}, e_{2,i}, e_{3,i}}}, 
где
\begin{align*}
		 e_{0,i}&=\vect{\x_i\rightarrow{}\x_i},\\
		 e_{1,i}&=\vect{\x_i\rightarrow{}\x_{1,i}},\\
		 e_{2,i}&=\vect{\x_i\rightarrow{}\x_{2,i}},\\
		 e_{3,i}&=\vect{\x_i\rightarrow{}\x_{3,i}}.
\end{align*}
Заметим, что рёбра окрестности не равнозначны: рёбра \Math{e_{1,i}} и \Math{e_{2,i}}
принадлежат пятиугольнику, примыкающему к вершине \Math{\x_i}, а ребро \Math{e_{3,i}} разделяет
два шестиугольника; или, с точки зрения химии, ребро \Math{e_{3,i}} соответствует двойной связи
в молекуле углерода \Math{C_{60}}, а остальные рёбра представляют простые связи.  
Поэтому группа локальных симметрий пространства, т.е. стабилизатор вершины \Math{\x_i}, не является 
максимальной для подобных окрестностей группой \Math{\DihG{6}}, а имеет вид \Math{\sGloc=\mathrm{Stab}_{\sG}\vect{\x_i}=\CyclG{2}}. 
Множество орбит \Math{\sGloc} на \Math{E_i} состоит из трёх элементов (подмножеств):
 \Mathh{\Omega_i=\set{\omega_{0,i}=\set{e_{0,i}}, \omega_{1,i}
=\set{e_{1,i}, e_{2,i}}, \omega_{2,i}=\set{e_{3,i}}},}
т.e. стабилизатор переставляет рёбра \Math{\vect{\x_i\rightarrow{}\x_{1,i}}} и
\Math{\vect{\x_i\rightarrow{}\x_{2,i}}} но не двигает  \Math{\vect{\x_i\rightarrow{}\x_i}}
и
 \Math{\vect{\x_i\rightarrow{}\x_{3,i}}}. 
\par
Учитывая эти симметрии правило эволюции принимает вид:
\begin{align*}
&R\vect{\x_i\rightarrow{}\x_i}~~=\Rep{\alpha_0},\\
R\vect{\x_i\rightarrow{}\x_{1,i}}=
&R\vect{\x_i\rightarrow{}\x_{2,i}}=\Rep{\alpha_1},\\
&R\vect{\x_i\rightarrow{}\x_{3,i}}=\Rep{\alpha_2},	
\end{align*}
где \Math{\alpha_0,\alpha_1,\alpha_2\in\iG}. 
\subsection{Иллюстрация: дискретная модель квантовой свободной\\ частицы} 
\par
Рассмотрим квантование свободной частицы, движущейся в одном измерении.
Такая частица описывается лагранжианом
 \Math{L = \frac{\textstyle{m\dot{x}^2}}{\textstyle{2}}.}
Предполагая, что имеются квантовые переходы только в ближайшие точки дискретизированного
пространства мы приходим к следующему правилу переходов за один шаг времени   
\begin{center}
\includegraphics[width=0.20\textwidth]{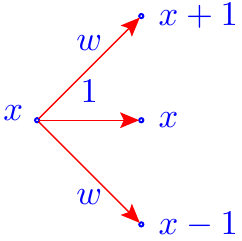}\hspace*{20pt}
\raisebox{34pt}{\begin{tabular}{l}
\Math{e^{\textstyle{\frac{i}{\hbar}\frac{m\left\{(x+1)-x\right\}^2}{2}}}
=e^{\textstyle{i\frac{m}{2\hbar}}}}\\[10pt]
\Math{e^{\textstyle{\frac{i}{\hbar}\frac{m\vect{x-x}^2}{2}}}\hspace*{17pt}=~\id}\\[10pt]
\Math{e^{\textstyle{\frac{i}{\hbar}\frac{m\left\{(x-1)-x\right\}^2}{2}}}=e^{\textstyle{i\frac{m}{2\hbar}}},}
\end{tabular}}
\end{center}
т.е. правило эволюции \Math{R} представляет собой  функцию со значениями в 
 унитарном представлении окружности \Math{\UG{1}}:
\begin{align*}
		R\vect{x\rightarrow{}x}&=1~~\in~~\UG{1},\\
		R\vect{x\rightarrow{}x-1}=R\vect{x\rightarrow{}x+1}
		&=w=e^{\textstyle{i\frac{m}{2\hbar}}}~~\in~~\UG{1}.
\end{align*}
Далее предположим, что \Math{w} --- элемент некоторого одномерного  представления конечной группы \Math{\iG}.
Одномерность представления предполагает, что без потери общности можно считать группу циклической  
\Math{\iG=\Z_M}, а \Math{w} --- примитивным корнем из единицы \Math{M}-й степени.
Реорганизуя \emph{мультиномиальные коэффициенты} --- \emph{триномиальные} в данном конкретном случае ---
нетрудно написать сумму амплитуд по всем возможным путям между пространственно-временн\'ыми
точками \Math{\vect{0,0}} и \Math{\vect{x,t}}
\begin{equation*}
A_x^t\vect{w}=\sum\limits_{\tau=0}^t\frac{\tau!}{\vect{\frac{\tau-x}{2}}!\vect{\frac{\tau+x}{2}}!}
	\times
	\frac{t!}{\tau!\vect{t-\tau}!}~w^{\tau}.
\end{equation*}
Заметим, что значение \Math{x} должно лежать в пределах \Math{\pm{}t}: \Math{x\in\left[-t,t\right].}
\par
Можно усложнить модель добавив, например, калибровочные связности, действующие вдоль путей, вводя 
ограничения на возможные пути, как, например, экран в ``двухщелевом эксперименте'' и т.п.
В любом случае амплитуда \Math{A\vect{w}} будет полиномиальной функцией от  \Math{w}. 
Для того, чтобы воспроизвести, скажем, деструктивную интерференцию  --- одно из ярких проявлений 
квантового поведения --- необходимо решить систему полиномиальных уравнений: 
\Math{A\vect{w}=\Phi_M\vect{w}=0}, где \Math{\Phi_M\vect{w}} --- \Math{M}-й циклотомический полином.
В данной модели наименьшей группой, при которой появляется деструктивная интерференция,
является   \Math{\CyclG{4}}.
\begin{figure}[!h]
\centering
\includegraphics[width=0.95\textwidth]{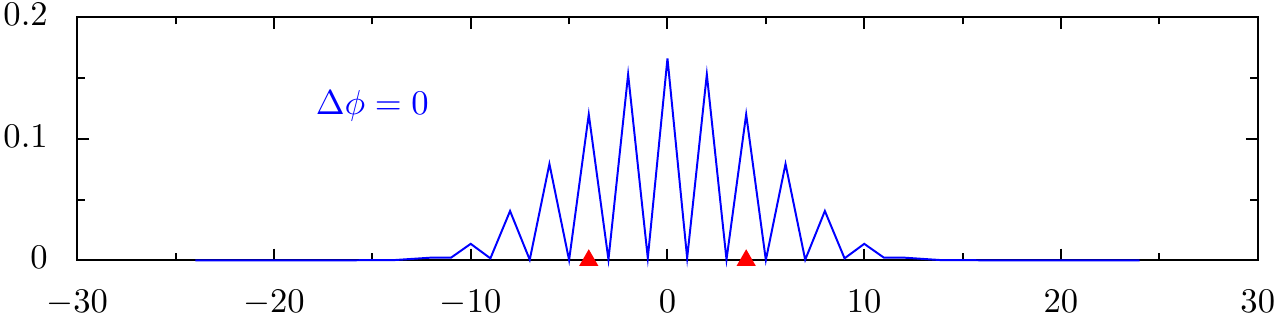}\\
\includegraphics[width=0.95\textwidth]{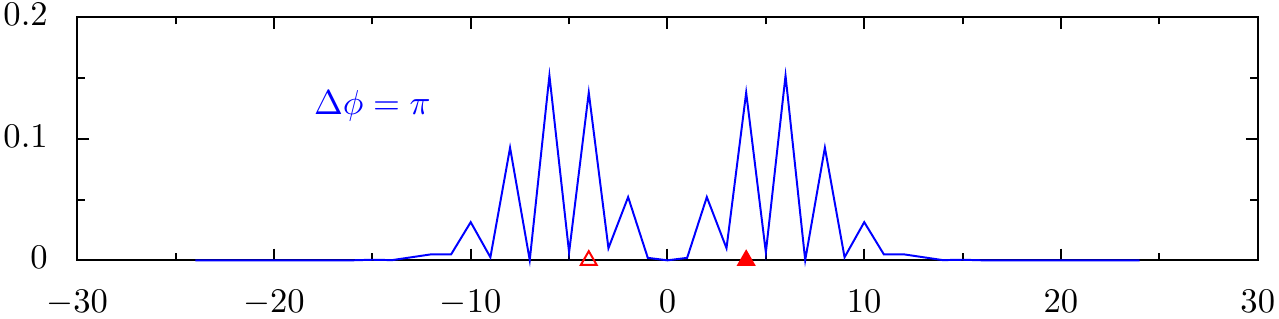}
\caption{Группа \Math{\CyclG{4}}. Интерференция от двух источников в позициях -4 и 4. 
Число шагов по времени \Math{T=20}. \Math{\Delta\phi=\phi_4-\phi_{-4}} --- разности фаз между
источниками.}
	\label{Interf2}
\end{figure}
\par
На Рис. \ref{Interf2} показаны интерференционные картины --- нормализованные квадраты модулей амплитуд
(``вероятности'') ---  от двух источников, помещённых в точках  \Math{x=-4} и \Math{x=4} 
для 20 шагов по времени.
Верхний и нижний графики показывают интерференции в случаях, когда фазы источников одинаковы
(\Math{\Delta\phi=0}) и противоположны
(\Math{\Delta\phi=\pi}), соответственно.
\section{Линейные представления конечных групп}
\label{irreps}
Любое линейное представление конечной группы эквивалентно унитарному,
поскольку из произвольного скалярного произведения с помощью ``усреднения по группе''
можно построить инвариантное. Например, в качестве исходного скалярного произведения 
в \Math{\adimH}-мерном гильбертовом пространстве \Math{\Hspace}  можно взять
стандартное  
\begin{equation}
    \innerstandard{\phi}{\psi}\equiv\sum\limits_{i=1}^{\adimH}\cconj{\phi^i}\psi^i.
\label{innerstdEn}
\end{equation}
Усреднив \eqref{innerstdEn} по группе мы приходим к \emph{инвариантному скалярному произведению} 
\begin{equation*}
\inner{\phi}{\psi}\equiv\frac{\textstyle{1}}{\textstyle{\cabs{G}}}\sum\limits_{g\in{}G}
\!\innerstandard{U\!\vect{g}\phi}{U\vect{g}\psi},
\end{equation*}
обеспечивающему унитарность представления \Math{U} группы \Math{G} в пространстве \Math{\Hspace}.
\par
Важным преобразованием элементов группы --- аналогом замены системы координат
в физике --- является сопряжение:
\Math{a^{-1}ga\rightarrow{}g',} \Math{g, g'\in\wG,} \Math{a\in\SymG{\wG}}.
Сопряжение с помощью элемента самой группы, т.е., если
 \Math{a\in\wG}, называется \emph{внутренним автоморфизмом}.
Классы эквивалентности относительно внутренних автоморфизмов называются \emph{классами
сопряжённых элементов}. Отправной точкой изучения представлений группы
является её разложение на классы сопряжённых элементов
\Mathh{\wG=\class{1}\sqcup\class{2}\sqcup\cdots\sqcup\class{\classN}.}
\par
Групповое умножение индуцирует \emph{умножение на классах}.
Произведение классов \Math{\class{i}} и \Math{\class{j}} представляет собой 
разложенное на классы \emph{мультимножество} всех возможных произведений
\Math{ab,~a\in\class{i},~b\in\class{j}}. Очевидно, что это произведение
коммутативно, поскольку \Math{ab} и \Math{ba} принадлежат одному и тому же
классу: \Math{ab\sim{}a^{-1}\vect{ab}a=ba}.
Таким образом, мы имеем следующую таблицу умножения классов
\begin{equation*}
    \class{i}\class{j}= \class{j}\class{i} = \sum\limits_{k=1}^{\classN}c_{ijk}\class{k}.
\end{equation*}
\emph{Натуральные целые} \Math{c_{ijk}}
--- кратности вхождения классов в мультимножества  --- называются \emph{коэффициентами классов}. 
Полученная таким образом \emph{алгебра классов} содержит всю информацию,
необходимую для построения представлений группы.
\par
Приведём краткий список основных свойств линейных представлений конечных групп:
\begin{enumerate}
    \item
Любое неприводимое представление содержится в регулярном.
Более конкретно, существует матрица  преобразования \Math{\transmatr} одновременно приводящая
все матрицы регулярного представления к виду  
\begin{equation}
    \transmatr^{-1}\regrep(g)\transmatr=
    \bmat
            \repirr_1(g) &&&
    \\[5pt]
    &
    \hspace*{-27pt}d_2\left\{
    \begin{matrix}
    \repirr_2(g)&&\\
    &\hspace*{-10pt}\ddots&\\
    &&\hspace*{-7pt}\repirr_2(g)
    \end{matrix}
    \right.
    &&
    \\
    &&\hspace*{-10pt}\ddots&\\
    &&&
    \hspace*{-25pt}d_{\classN}\left\{
    \begin{matrix}
    \repirr_{\classN}(g)&&\\
    &\hspace*{-10pt}\ddots&\\
    &&\hspace*{-7pt}\repirr_{\classN}(g)
    \end{matrix}
    \right.
    \emat,
\label{regrepdecompEn}
\end{equation}
и любое неприводимое представление является одним из \Math{\repirr_j}.
Число неэквивалентных неприводимых представлений равно числу классов 
сопряжённых элементов.
Число \Math{d_j} это одновременно размерность неприводимой компоненты 
\Math{\repirr_j} и кратность её вхождения в регулярное представление.
Из \eqref{regrepdecompEn} видно, что для размерностей неприводимых  представлений
справедливо равенство
\Math{d^2_1+d^2_2+\cdots+d^2_{\classN}=\cabs{\wG}=\wGN.} Кроме того,
размерности неприводимых представлений делят порядок группы:
\Math{d_j\mid\wGN.}
    \item
Любое неприводимое представление \Math{\repirr_j} определяется однозначно
своим \emph{характером} \Math{\chi_j}, т.е. набором следов матриц представления:
\Math{\chi_j\vect{g}=\mathrm{Tr}\repirr_j\vect{g}}. Этот набор является фактически
функцией на классах сопряжённых элементов, поскольку в силу стандартных свойств следов
матриц
\Math{\chi_j\vect{g}=\chi_j\vect{a^{-1}ga}}.
Очевидно, что \Math{\chi_j\vect{\id}=d_j}.
    \item
Компактным средством перечисления всех неприводимых представлений является    
\emph{таблица характеров}.
Столбцы этой таблицы пронумерованы классами сопряжённых элементов,
а строки содержат значения характеров неэквивалентных представлений.
\begin{table}[h]
\begin{center}
\begin{tabular}{c|cccc}
&\Math{\class{1}}&\Math{\class{2}}&\Math{\cdots}&\Math{\class{\classN}}\\\hline
\Math{\chi_1}&1&1&\Math{\cdots}&1\\
\Math{\chi_2}&\Math{\chi_2\vect{\class{1}}=d_2}&\Math{\chi_2\vect{\class{2}}}&
\Math{\cdots}&\Math{\chi_2\vect{\class{\classN}}}\\
\Math{\vdots}&\Math{\vdots}&\Math{\vdots}&&\Math{\vdots}\\
\Math{\chi_{\classN}}&\Math{\chi_{\classN}\vect{\class{1}}=d_{\classN}}&
\Math{\chi_{\classN}\vect{\class{2}}}&\Math{\cdots}&\Math{\chi_{\classN}\vect{\class{\classN}}}
\end{tabular}.
\end{center}
\caption{Таблица характеров.}
\label{chartab}
\end{table}
Согласно традиции, 1-й столбец соответствует классу групповой единицы,
а 1-я строка содержит \emph{тривиальное} представление. Как строки так и столбцы
попарно\emph{ ортогональны}.
Важная информация о группе и её представлениях может быть считана непосредственно
с таблицы характеров. Например, неприводимое представление является \emph{точным}
тогда и только тогда, когда значение характера в 1-м столбце (т.е. размерность)
не повторяется более нигде в соответствующей строке.
Таблицы характеров почти полностью определяет группы.
Они  не различают лишь изоклинные группы  \cite{atlas}.
Группы называются \emph{изоклинными}, если их фактор-группы по их центрам
изоморфны. 
Изоклинные группы имеют идентичные таблицы характеров поскольку характеры
``игнорируют'' центры. Примером изоклинных групп являются 8-элементные
диэдральная \Math{\DihG{8}=\set{\text{\textit{симметрии квадрата}}}}
и кватернионная \Math{\QuatG{8}=\set{\pm1,\pm{}i,\pm{}j,\pm{}k}} группы.
Здесь \Math{i,j,k} --- кватернионные
мнимые единицы.
\end{enumerate}
\section{Конечные группы симметрий и феноменология\\ элементарных частиц}
\label{appflavor}
В настоящее время все экспериментальные данные \cite{RevPartPhys}, 
касающиеся фундаментальных частиц, согласуются со Стандартной Моделью, которая
представляет собой калибровочную теорию с группой внутренних (калибровочных) 
симметрий \Math{\iG}, являющейся прямым произведением 
групп \Math{\UG{1}}, \Math{\SU{2}} и \Math{\SU{3}}. 
В контексте Теории Большого
Объединения предполагается, что  \Math{\iG} --- подгруппа некоторой большей
(предположительно простой) группы. Относительно пространственно-временн\'ых симметрий
элементарные частицы подразделяются на два класса: \emph{бозоны}, отвечающие за 
физические силы (грубо говоря, они соответствуют элементам калибровочной группы)
и \emph{фермионы}, интерпретируемые как частицы материи. Фермионы
Стандартной Модели составляют три \emph{поколения} \emph{кварков} и \emph{лептонов}
представленных в таблице \ref{smfermions}.
\begin{table}[h]
\begin{center}
\begin{tabular}{l|c|c|c}
Поколения&1&2&3
\\\hline
Верхние кварки&\Math{u}&\Math{c}&\Math{t}
\\
Нижние кварки&\Math{d}&\Math{s}&\Math{b}
\\\hline
Заряженные лептоны&\Math{e^-}&\Math{\mu^-}&\Math{\tau^-}
\\
Нейтрино&\Math{\nu_e}&\Math{\nu_\mu}&\Math{\nu_\tau}
\end{tabular}
\end{center}
\caption{Фермионы стандартной модели. Античастицы для краткости опущены.}
\label{smfermions}
\end{table}
Частицы различных поколений отличаются друг от друга только массами и квантовым свойством,
называемым \emph{ароматом}.
Физические процессы, изменяющие ароматы, такие как, слабые распады кварков 
или осцилляции нейтрино, описываются унитарными \emph{матрицами смешивания} 
размера  \Math{3\times3}.
Экспериментальные данные позволяют вычислить абсолютные величины элементов этих матриц.
\par
В случае кварков (``\emph{в кварковом секторе}''),
матрицей перехода между ``верхними'' и ``нижними'' кварками
является так называемая \emph{матрица Кабиббо--Кобаяши--Маскава} (CKM)
\Mathh {
V_{\text{CKM}}=\Mthree{V_{ud}}{V_{us}}{V_{ub}}
       {V_{cd}}{V_{cs}}{V_{cb}}
       {V_{td}}{V_{ts}}{V_{tb}},
}
где \Math{\cabs{V_{\alpha\beta}}^2} представляет собой вероятность того, 
кварк (имеющий аромат) \Math{\beta} при слабом процессе переходит в кварк \Math{\alpha}.
Полученные к настоящему времени экспериментальные данные дают следующие 
величины для модулей матричных элементов: 
\Mathh{
\Mthree{\cabs{V_{ud}}}{\cabs{V_{us}}}{\cabs{V_{ub}}}
       {\cabs{V_{cd}}}{\cabs{V_{cs}}}{\cabs{V_{cb}}}
       {\cabs{V_{td}}}{\cabs{V_{ts}}}{\cabs{V_{tb}}}
=\Mthree{0.974}{~0.225}{~0.004}
        {0.225}{~0.974}{~0.041}
        {0.009}{~0.040}{~0.999}.
}
Мы использовали округление до трёх десятичных цифр ---
более точные величины можно найти в \cite{RevPartPhys}.
\par
В ``\emph{лептонном секторе}'' слабые процессы описываются матрицей
смешивания \emph{Пон\-текорво--Маки--Накагава--Саката} (PMNS)
\Mathh
{
U_{\text{PMNS}}=\Mthree{U_{e1}}{U_{e2}}{U_{e3}}
       {U_{\mu1}}{U_{\mu2}}{U_{\mu3}}
       {U_{\tau1}}{U_{\tau2}}{U_{\tau3}}.
}
Здесь индексы \Math{e, \mu, \tau} соответствуют ароматам нейтрино. Это означает,
что в слабых процессах нейтрино \Math{\nu_e, \nu_\mu, \nu_\tau} возникают совместно с 
\Math{e^+, \mu^+, \tau^+} (или порождают \Math{e^-, \mu^-, \tau^-}), соответственно. 
Индексы \Math{1, 2, 3} соответствуют собственным значениям оператора массы.
Т.е.  символы \Math{\nu_1, \nu_2, \nu_3} означают нейтрино, имеющие определённые
массы \Math{m_1, m_2, m_3}.
Многочисленные наблюдения солнечных и атмосферных нейтрино и эксперименты с
нейтрино, получаемых с помощью реакторов и ускорителей, выявляют дискретные 
(более того, конечные) симметрии, которые невозможно вывести из Стандартной Модели.
Феноменологическая картина с разумной точностью выглядит следующим образом \cite{Smirnov}:
\begin{enumerate}
    \item  Ароматы \Math{\nu_\mu} и
\Math{\nu_\tau} распределены с равными весами по всем
трём массам
\Math{\nu_1, \nu_2, \nu_3} (это называется ``\emph{би-максимальным смешиванием}''):
\Math{\cabs{U_{\mu{i}}}^2=\cabs{U_{\tau{i}}}^2,} \Math{i=1,2,3};
    \item
Все три аромата представлены равномерно в собственном значении \Math{\nu_2} 
оператора массы (``\emph{три-максимальное смешивание}''):
\Math{\cabs{U_{e{2}}}^2=\cabs{U_{\mu{2}}}^2=\cabs{U_{\tau{2}}}^2};
    \item  \Math{\nu_e} отсутствует в \Math{\nu_3}: \Math{\cabs{U_{\mu3}}^2=0}.
\end{enumerate}
Эти соотношения совместно с условиями нормализации вероятностей приводят к следующим
значениям квадратов модулей матричных элементов:
\begin{equation}
    \Mone{\cabs{U_{l{i}}}^2}
    =\Mthree{~\fra{2}{3}}{~~\fra{1}{3}}{~~0~}
            {\\[-8pt]~\fra{1}{6}}{~~\fra{1}{3}}{~~\fra{1}{2}~}
            {\\[-8pt]~\fra{1}{6}}{~~\fra{1}{3}}{~~\fra{1}{2}~}.
\label{tribi2}
\end{equation}
Частный вид унитарной матрицы удовлетворяющей данным \eqref{tribi2} 
был предложен Харрисоном, Перкинсом и Скоттом в \cite{HPS02}:
\begin{equation*}
    U_{\text{TB}}
    =\Mthree{~\,\sqrt{\fra{2}{3}}}{~~\fra{1}{\sqrt{3}}}{~~0}
            {-\fra{1}{\sqrt{6}}}{~~\fra{1}{\sqrt{3}}}{~-\!\fra{1}{\sqrt{2}}}
            {-\fra{1}{\sqrt{6}}}{~~\fra{1}{\sqrt{3}}}{~~~~\,\fra{1}{\sqrt{2}}}.
\end{equation*}
Эта, так называемая \emph{трибимаксимальная} (TB) матрица смешивания 
совпадает --- с точностью до тривиальной перестановки двух столбцов,
соответствующей переименованию \Math{\nu_1\rightleftarrows\nu_2} двух массовых состояний
--- с матрицей преобразования \eqref{transS3sqrt}, разлагающей 3-мерное 
перестановочное представление группы \Math{\SymG{3}} на неприводимые компоненты.
Это означает, что мы можем отождествить базис ароматов с базисом перестановок
трёх элементов, а массовый базис с базисом в котором перестановочное 
представление разложено на неприводимые компоненты.
В \cite{HS03} Харрисон и Скотт подробно исследуют связь оператора массы нейтрино
с таблицей характеров и алгеброй классов группы \Math{\SymG{3}}.
\par
В настоящее время активно разрабатываются и изучаются модели, в основе которых
лежат конечные группы симметрий, связанных с ароматами (обзоры этой деятельности представлены,
например, в работах  \cite{Ishimori,Ludlgen}). При конструировании 
этих моделей наиболее популярны следующие группы:
\begin{itemize}
    \item \Math{\mathsf{T}=\AltG{4}} --- группа (симметрий) тетраэдра,
    \item   \Math{\TprimG} --- двойная накрывающая группы \Math{\AltG{4}},
    \item \Math{\mathsf{O}=\SymG{4}} --- группа октаэдра,
    \item \Math{\mathsf{I}=\AltG{5}} --- группа икосаэдра,
    \item \Math{\DihG{N}} --- диэдральная группа (\Math{N} --- чётное),
    \item \Math{\QuatG{N}} --- кватернионная группа (4 делит \Math{N}),
    \item   \Math{\Sigma\vect{2N^2}}
--- группы этой серии имеют структуру 
\Math{\vect{\CyclG{N}\times\CyclG{N}}\rtimes\CyclG{2}},
    \item   \Math{\Delta\vect{3N^2}} --- структура
\Math{\vect{\CyclG{N}\times\CyclG{N}}\rtimes\CyclG{3}},
    \item   \Math{\Sigma\vect{3N^3}}
--- структура
\Math{\vect{\CyclG{N}\times\CyclG{N}\times\CyclG{N}}\rtimes\CyclG{3}},
    \item   \Math{\Delta\vect{6N^2}} --- структура
\Math{\vect{\CyclG{N}\times\CyclG{N}}\rtimes\SymG{3}}.
\end{itemize}
\par
В кварковом секторе наблюдения не дают такой отчётливой картины как в случае лептонов.
Имеется ряд работ посвящённых поиску конечных симметрий в кварковом секторе и симметрий,
общих для кварков и лептонов.
Например, в работе \cite{BlumHagedorn} группа \Math{\DihG{14}} использовалась для
объяснения величины угла Кабиббо (один из параметров матрицы смешивания CKM), однако,
без какой-либо связи с лептонными симметриями. В целом, естественные попытки найти
дискретные симметрии, объединяющие лептоны и кварки, до сих пор остаются не слишком
успешными. Тем не менее, имеются некоторые обнадёживающие наблюдения: например,
\emph{кварк-лептонная комплементарность} --- экспериментальный факт
приближённого равенства  \Math{\pi/4} суммы углов смешивания для кварков и лептонов.
\par
Происхождение конечных симметрий между фундаментальными частицами в настоящее время
не ясно.
Имеются различные попытки объяснения, иногда
нес\-колько переусложнённые и искусственные. Например, эти симметрии
интерпретируются как симметрии многообразий, возникающих в результате
компактификации дополнительных измерений в теориях  
с пространствами высоких размерностей \cite{Altarelli}.
С нашей точки зрения, идея того, что симметрии наиболее фундаментального уровня 
являются конечными сами по себе, выглядит более привлекательной.
При таком подходе, унитарные группы, используемые в физических теориях
представляют собой своего рода ``резервуары'' в которых собраны ``на всякий случай'' 
все возможные конечные группы, имеющие точные представления соответствующей размерности.
Например, группа \Math{\SU{n}} в качестве подгрупп содержит все конечные группы, 
имеющие точные \Math{n}-мерные представления с определителем равным единице.
Более того, ввиду избыточности поля \Math{\C}, группа \Math{\SU{n}} не является
минимальной группой, содержащей все конечные группы с указанными свойствами.
\par
Скорее всего, такие небольшие группы, как \Math{\SymG{3},~\AltG{4}} и т.п.,
представляют собой лишь следы больших комбинаций (расширений, прямых и 
полупрямых произведений) более фундаментальных конечных симметрий.
Можно ожидать, что естественный масштаб, на котором такие фундаментальные симметрии
проявляются, это масштаб Большого Объединения.
Можно попытаться для направленного поиска фундаментальных симметрий воспользоваться
симметриями Стандартной Модели. Например, конечные подгруппы группы  
\Math{\SU{3}}, имеющие точные 3-мерные представления, полностью описаны:
их список содержит несколько бесконечных серий и несколько отдельных групп.
Однако такая попытка вряд ли будет достаточно успешной, поскольку Стандартная 
Модель, представляет собой компактную форму описание феноменологии, полученной
при масштабах энергии около \Math{10^{4}} ГэВ, тогда как масштаб Большого 
Объединения равен примерно \Math{10^{16}} ГэВ, что довольно близко к 
планковскому масштабу (\Math{10^{19}} ГэВ) и далеко выходит за пределы
 экспериментальных возможностей.
По-видимому, единственным практическим подходом является конструирование моделей 
и сравнение следствий из них 
с доступными экспериментальными данными.
В такого рода исследованиях могут оказаться
полезными методы вычислительной теории групп \cite{Holt}.
\renewcommand{\refname}{References}

\end{document}